\documentclass[iop,revtex4]{emulateapj}
\usepackage{graphics,graphicx}
\usepackage{helvet}
\usepackage[utf8]{inputenc}
\usepackage[T1]{fontenc}
\usepackage{longtable}
\usepackage{soul}
\usepackage{hyperref}
\usepackage{amsmath, amssymb, gensymb}
\usepackage{ulem}
\usepackage{natbib}
\usepackage{microtype}
\usepackage{multirow}
\usepackage{morefloats}
\def\eg {e.g.,} %e.g.,
\def\ie {i.e.,} %i.e.,
\def\Lsol {$\hbox{L}_\odot$}
\def\Msol {$\hbox{M}_\odot$}
\def\Msolperyr {$\hbox{M}_\odot$\,yr$^{-1}$}
\def\HII {\ion{H}{2}} % H II
\def\kms {km\,s$^{-1}$}
\def\Kkms {K\,km\,s$^{-1}$}

\def\mjybm {mJy\,bm$^{-1}$}
\def\mjybmvert {$\left(\frac{\textrm{mJy}}{\textrm{bm}}\right)$}

\def\etal {\textit{et al.}}
\newcommand{\mir}[1]{\textbf{\fontfamily{lmvtt}\selectfont #1}} % font for code, e.g., MIRIAD
\newcommand{\numSFR}{8}

\newcommand{\numcore}{30}

\begin{document}

% Plot directory
\graphicspath{ {python_plots/} }

\slugcomment{\textit{Submitted to ApJS on 24 October 2013; accepted 11 March 2014.}}

\title{\vspace{-.2in} TADPOL: A 1.3\,\lowercase{mm} Survey of Dust Polarization in Star-forming Cores and Regions}
\shorttitle{TADPOL Data Release}

\author{Charles~L.~H.~Hull,\altaffilmark{1} Richard~L.~Plambeck,\altaffilmark{1} Woojin~Kwon,\altaffilmark{13} 
Geoffrey~C.~Bower,\altaffilmark{1,18} John~M.~Carpenter,\altaffilmark{2} Richard~M.~Crutcher,\altaffilmark{4} Jason~D.~Fiege,\altaffilmark{9}  
Erica~Franzmann,\altaffilmark{9} Nicholas~S.~Hakobian,\altaffilmark{4} Carl~Heiles,\altaffilmark{1} Martin~Houde,\altaffilmark{10,3}
A.~Meredith~Hughes,\altaffilmark{14} James~W.~Lamb,\altaffilmark{2} 
Leslie~W.~Looney,\altaffilmark{4} Daniel~P.~Marrone,\altaffilmark{15} Brenda~C.~Matthews,\altaffilmark{11,12}
Thushara~Pillai,\altaffilmark{2} Marc~W.~Pound,\altaffilmark{5} Nurur~Rahman,\altaffilmark{16,20} 
G\"oran~Sandell,\altaffilmark{8} Ian~W.~Stephens,\altaffilmark{4,17} John~J.~Tobin,\altaffilmark{7,19}
John~E.~Vaillancourt,\altaffilmark{8} N.~H.~Volgenau,\altaffilmark{6} and Melvyn~C.~H.~Wright\altaffilmark{1}}

\shortauthors{Hull \etal}
\email{chat@astro.berkeley.edu}

\altaffiltext{1}{\scriptsize Astronomy Department \& Radio Astronomy Laboratory, University of California, Berkeley, CA 94720-3411}
\altaffiltext{2}{Department of Astronomy, California Institute of Technology, 1200 E. California Blvd., MC 249-17, Pasadena, CA 91125, USA}
\altaffiltext{3}{Division of Physics, Mathematics, \& Astronomy, California Institute of Technology, Pasadena, CA 91125, USA}
\altaffiltext{4}{Department of Astronomy, University of Illinois at Urbana-Champaign, 1002 W Green Street, Urbana, IL 61801, USA}
\altaffiltext{5}{Astronomy Department \& Laboratory for Millimeter-wave Astronomy, University of Maryland, College Park, MD 20742}
\altaffiltext{6}{Combined Array for Research in Millimeter-wave Astronomy, Owens Valley Radio Observatory, P.O. Box 968, Big Pine, CA 93513, USA}
\altaffiltext{7}{National Radio Astronomy Observatory, 520 Edgemont Rd., Charlottesville, VA 22903, USA}
\altaffiltext{8}{SOFIA Science Center, Universities Space Research Association, NASA Ames Research Center, Moffett Field, CA 94035, USA}
\altaffiltext{9}{Department of Physics \& Astronomy, University of Manitoba, Winnipeg, MB, R3T 2N2, Canada}
\altaffiltext{10}{Department of Physics \& Astronomy, University of Western Ontario, London, ON, N6A 3K7, Canada}
\altaffiltext{11}{Department of Physics \& Astronomy, University of Victoria, 3800 Finnerty Rd., Victoria, BC, V8P 5C2, Canada}
\altaffiltext{12}{National Research Council of Canada, 5071 West Saanich Rd., Victoria, BC, V9E 2E7, Canada}
\altaffiltext{13}{SRON Netherlands Institute for Space Research, Landleven 12, 9747 AD Groningen, The Netherlands}
\altaffiltext{14}{Van Vleck Observatory, Astronomy Department, Wesleyan University, 96 Foss Hill Drive, Middletown, CT 06459, USA}
\altaffiltext{15}{Steward Observatory, University of Arizona, 933 North Cherry Avenue, Tucson, AZ 85721, USA}
\altaffiltext{16}{Physics Department, University of Johannesburg, C1-Lab 140, PO Box 524, Auckland Park 2006, South Africa}
\altaffiltext{17}{Institute for Astrophysical Research, Boston University, Boston, MA 02215, USA}
\altaffiltext{18}{ASIAA, 645 N. A'ohoku Place, Hilo, HI, 96720}
\altaffiltext{19}{Hubble Fellow}
\altaffiltext{20}{South Africa SKA Fellow}

\begin{abstract}
\noindent
We present $\lambda\,$1.3\,mm CARMA observations of dust polarization toward
$\numcore$ star-forming cores and $\numSFR$ star-forming regions from the TADPOL survey.  
%All data are publicly available.
We show maps of all sources, and compare the $\sim$\,2.5$\arcsec$ resolution TADPOL maps with  
$\sim$\,20$\arcsec$ resolution polarization maps from single-dish submillimeter telescopes.
Here we do not attempt to interpret the detailed B-field morphology of each object. 
Rather, we use average B-field orientations 
to derive conclusions in a statistical sense from the ensemble of sources, 
bearing in mind that these average orientations can be quite uncertain.
We discuss three main findings: (1) A subset of the 
sources have consistent magnetic field (B-field) orientations between large ($\sim$\,20$\arcsec$) 
and small ($\sim$\,2.5$\arcsec$) scales.  Those
same sources also tend to have higher fractional polarizations
than the sources with inconsistent large-to-small-scale fields.
%possibly because there is less twisting of the B-fields to reduce the polarization fraction. 
We interpret this to mean that in at least some cases B-fields play a role in regulating the infall of material
% have higher ratios of magnetic to turbulent energy 
%from $\sim$\,100\,pc molecular cloud scales 
all the way down to the $\sim$\,1000\,AU scales of protostellar envelopes.  
(2) Outflows appear to be randomly aligned with B-fields; although, in 
sources with low polarization fractions there is a hint that outflows are \textit{preferentially perpendicular} to small-scale B-fields, 
which suggests that in these sources the fields have been 
wrapped up by envelope rotation.         
(3) Finally, even at $\sim$\,2.5$\arcsec$ resolution we see 
the so-called ``polarization hole'' effect,
where the fractional polarization drops significantly near the total intensity peak.
All data are publicly available in the electronic edition of this article.
\\
%When this effect was seen in low-resolution single-dish maps, it was 
%attributed to the averaging of unresolved structure in the plane of the sky.
%However, the higher resolution maps we present here resolve these
%twisted polarization morphologies, and yet the drop in fractional
%polarization persists, suggesting that fields are twisted along the line of sight, or
%that grain alignment is poor in dense regions with high extinction and high collision rates.

%However, in nearly all of the more massive star-forming regions, the small-scale ($\sim$\,10,000\,AU) B-fields in the tend to be twisted
%with respect to the larger-scale fields, which implies that the B-fields
%in these regions are not a dominant factor regulating the motion
%of material, but are instead dragged along and disturbed by gravitational collapse and
%turbulence.

\end{abstract}


\section{INTRODUCTION}
\label{sec:intro} 

Magnetic fields have long been considered one of
the key components that regulate star formation \citep[\eg{}][]{Shu1987,McKee1993}.
And indeed, observations of polarization in star-forming regions have shown that magnetic
fields (B-fields) often are well ordered on scales from $\sim$\,100\,pc 
%scales of molecular cloud complexes 
\citep{Heiles2000} down to $\sim$\,1\,pc, 
%scales of single clouds \citep[\eg{}][]{Franco2010}, 
which suggests that on large scales B-fields are dynamically important.
%the ratio of magnetic to turbulent energies may be higher on larger scales where the density is lower.
%are magnetically supported (``subcritical'').  
At smaller scales
ambipolar diffusion \citep[\eg][]{Mestel1956, Fiedler1993, Tassis2009} 
or turbulent magnetic reconnection diffusion \citep{Lazarian2005, Leao2012} 
are thought to allow dense cores to become ``supercritical'' \citep[see][]{Crutcher2012}, at which point gravity
overwhelms magnetic support and allows the formation of a central
protostar.  Alternatively, the cores could form as supercritical objects in a turbulent environment
\citep[\eg][]{MacLow2004}.
%Studying field morphology on smaller scales will help to discern
%which of these mechanisms ultimately regulates the formation of the central
%protostar.

Under most circumstances, spinning
dust grains align themselves with their long axes
perpendicular to the B-field \citep[\eg{}][]{Hildebrand1988, Lazarian2003, Lazarian2007, Hoang2009,
Andersson2012}, so the thermal
radiation from these grains is polarized perpendicular to the B-field. 
Ambient B-fields can be probed on scales of $\gtrsim$\,1\,pc using optical
observations of background stars \citep[\eg{}][]{Heiles2000}, whose light becomes polarized after passing
through regions of aligned dust grains.  
However, this type of observation is not possible 
%polarization observations of background stars are unable to
%probe the magnetic field morphologies 
inside the dense cores where the central
protostars and their circumstellar disks form; even at infrared wavelengths
the extinction through these dense regions is too great.

Mapping the polarized thermal emission from dust grains at millimeter and
submillimeter wavelengths is the usual means of studying the B-fields
in these regions.  The 1.3\,mm dual-polarization receiver system at CARMA (the
Combined Array for Research in Millimeter-wave Astronomy; \citealt{Bock2006}), described in \citet{Hull2011}, has allowed us to
map the dust polarization toward a sample of several dozen nearby star-forming cores and a few
 star-forming regions (SFRs) as part of the
TADPOL\footnote{ TADPOL: {\bf T}elescope {\bf A}rray {\bf D}oing {\bf POL}arization} survey---a CARMA key project.
  
%The results of our survey show that polarization remains detectable
%on small scales, as has been shown in previous polarization studies
%from both SMA (Submillimeter Array) and CARMA \citep[\eg{}][]{Girart2006,
%  Hull2012}.
%Observations of polarization on smaller scales, from $\sim$\,10,000\,AU dense
%cores \citep[\eg{}][]{Matthews2009} to the $\sim$\,1000\,AU envelopes
%surrounding the central protostars \citep[\eg{}][]{Girart2006, Hull2012},
%have shown that polarization is still detectable and is frequently well
%ordered.

Previous results from the TADPOL survey have touched on several topics
including the consistency of B-fields from large to small scales
\citep{Stephens2013}, the low levels of dust polarization in the circumstellar disks around more evolved 
Class~II sources like DG~Tau (\citealt{Hughes2013}; see Figure \ref{fig:DGTau}), and the
misalignment of bipolar outflows and small-scale B-fields in low-mass protostars \citep{Hull2012}.
The latter result has been used to place limits on the 
fraction of protostars that should harbor circumstellar disks \citep{Krumholz2013}.

Here we present the data from the full survey.
We compare these $\sim$\,2.5$\arcsec$ resolution data with  
$\sim$\,20$\arcsec$ resolution polarization maps from single-dish submillimeter telescopes
to analyze the consistency of B-field orientations down to the $\sim$\,1000\,AU scale of protostellar envelopes.
We also revisit the correlation of B-fields with bipolar outflows
and see hints that sources with low polarization fractions have outflows 
and small-scale B-field orientations that are preferentially perpendicular.  
Finally, even at $\sim$\,2.5$\arcsec$ resolution we see 
the so-called ``polarization hole'' effect,
where the fractional polarization drops significantly near the total intensity peak.

\section{SOURCE SELECTION \& OBSERVATIONS}
\label{sec:obs}

%The results from the TADPOL survey are summarized in Table \ref{table:obs},
%which includes fitted coordinates of the dust emission peaks, maximum total
%intensity $I_\textrm{max}$, maximum bias-corrected polarized intensity $P_\textrm{c,pk}$,
%average polarization fraction $\overline{P}_\textrm{frac}$, inferred large- and
%small-scale B-field orientations $\chi_{\textrm{lg}}$ and $\chi_{\textrm{sm}}$,
%angle difference $|\chi_{\textrm{lg}} - \chi_{\textrm{sm}}|$ between the large- and
%small-scale B-field orientations, source type, distance to the source,
%and average beam size (resolution element) of the maps.  See Section \ref{sec:results}
%for further discussion.

We selected sources from catalogs of young stellar objects 
\citep[e.g.,][]{Jorgensen2007, Matthews2009, Tobin2010, Enoch2011}.  
While several well known, high-mass SFRs are included
in the survey, we focus mainly on nearby ($d \lesssim 400~\textrm{pc}$) Class
0 and Class I objects that are known to have bipolar outflows, and that had been 
observed previously with the polarimeters on the JCMT (James Clerk Maxwell Telescope) 
and the CSO (Caltech Submillimeter Observatory), two submillimeter single-dish telescopes
with $\sim$\,20$\arcsec$ resolution.
See Appendix \ref{appendix:blurbs} for source descriptions.
Since the survey spanned five observing semesters, sources were selected to
cover a wide range of hour angles to allow most observations to be
scheduled during the more stable nighttime weather.  

Observations were made with CARMA between May 2011 and April 2013.
Three different array configurations were used: C (26--370\,m baselines, or
telescope spacings), D (11--148\,m), and E (8.5--66\,m), which correspond to
angular resolutions at 1.3\,mm of approximately $1\arcsec$,
$2\arcsec$, and $4\arcsec$, respectively.  

%The results from the TADPOL survey are summarized in Table \ref{table:obs},
%and are discussed in Section \ref{sec:results}.

\section{CALIBRATION \& DATA REDUCTION}
\label{sec:redux}

%The receivers comprise a single feed horn, a waveguide circular polarizer
%\citep{Plambeck_2010}, an orthomode transducer \citep{TMTT.2005.860505}, and
%two mixers.  The receivers are double-sideband; a phase-switching pattern
%applied to the local oscillator (LO) allows signals in the lower (LSB) and
%upper sidebands (USB) to be separated in the correlator.

The CARMA polarization system consists of dual-polarization receivers that are
sensitive to right- (R) and left-circular (L) polarization, and a
spectral-line correlator that measures all four cross polarizations (RR, LL,
LR, RL) on each of the 105 baselines connecting the 15 telescopes (six 
with 10\,m diameters and nine with 6\,m diameters).  Each
receiver comprises a single feed horn, a waveguide circular polarizer, an
orthomode transducer (OMT), two heterodyne mixers, and two low-noise
amplifiers, all mounted in a cryogenically cooled dewar.  The local oscillator (LO)
and sky signals are combined using a mylar beamsplitter in front of the dewar
window.

The waveguide polarizer is a two-section design with half-wave and
quarter-wave retarder sections rotated axially with respect to one another to
achieve broadband (210--270\,GHz) performance; the retarders are sections of
reduced-height, faceted circular waveguide \citep{Plambeck_2010}.  The
polarizer converts the R and L circularly polarized radiation from the sky
into orthogonal X and Y linear polarizations, which then are separated by the
OMT \citep{TMTT.2005.860505}.  The mixers use ALMA Band 6 SIS
(superconductor-insulator-superconductor) tunnel junctions fabricated at the
University of Virginia by Arthur Lichtenberger.  
Although at ALMA these devices are used in sideband-separating mixers 
\citep{Kerr2013}, at CARMA they are used in double-sideband mixers that are
sensitive to signals in two bands, one 1--9\,GHz above (upper sideband, or USB), 
and the other 1--9\,GHz\,below (lower sideband, or LSB) the LO
frequency.  A phase-switching pattern applied to the LO allows
the LSB and USB signals to be separated in the correlator.
The 1--9\,GHz intermediate frequency (IF) from each mixer is 
amplified with WBA13 low-noise amplifiers \citep{Weinreb1998, Pandian2006}.  

%SiO: 217.105\,GHz
%CO: 230.538\,GHz

For the TADPOL observations the LO frequency was 223.821\,GHz.  The correlator
was set up with three 500\,MHz-wide bands centered at IF values of 6.0,
7.5, and 8.0\,GHz, and one 31\,MHz wide band centered at 6.717\,GHz.\footnote{Some or all
of the data for the following six sources are from another CARMA project led by Kwon et al.:
L1448~IRS~2, HH~211~mm, L1527, Ser-emb~1, HH~108~IRAS, and L1165.
These observations had a different correlator setup, with an LO frequency of
228.5988\,GHz; three 500\,MHz-wide bands centered at IF values of 1.9392, 2.4392, and 2.9392\,GHz;    
and one 31\,MHz wide band centered at 1.9392\,GHz.  Dust continuum and CO($J = 2\rightarrow1$) data from
these datasets are reported in this paper.}$^,$\footnote{The following sources have narrow-band windows
with widths of 62\,MHz and corresponding channel spacings of $\sim$\,0.4\,\kms: 
W3~Main, W3(OH), OMC3-MMS5/6, OMC2-FIR3/4, G034.43+00.24~MM1, and DR21(OH).}  
The corresponding sky frequencies are equal to the difference (LSB) or the sum
(USB) of the LO and the IF.  The narrowband section allowed simultaneous
spectral line observations of the SiO($J = 5\rightarrow4$) line (217.105\,GHz) in the LSB
 and the CO($J = 2\rightarrow1$) line (230.538\,GHz) in the USB, with a
channel spacing of $\sim$\,0.2\,\kms.  These lines were used to map bipolar outflows.

%In addition to the usual gain, passband, and flux calibrations, polarization
%observations require two additional calibrations: ``XYphase'' and leakage.
%The XYphase calibration corrects for phase differences between the LCP and
%RCP channels, which are caused by delay differences in the receiver,
%underground cables, and the correlator.  XYphase errors rotate the position
%angles of linear polarization vectors.  At CARMA the XYphase calibration is
%done by moving wire grids in front of each 10\,m receiver to create an
%artificially polarized noise source. The XYphase corrections then are derived
%on a channel-by-channel basis from RL ``cross-autocorrelation'' spectra.

In addition to the usual gain, passband, and flux calibrations, two additional
calibrations are required for polarization observations: ``XYphase'' and
leakage.  The XYphase calibration corrects for the phase difference between
the L and R channels on each telescope, caused by delay differences in the
receiver, underground cables, and correlator cabling.  To calibrate the
XYphase one must observe a linearly polarized source with known position
angle.  Since most astronomical sources at millimeter wavelengths are weakly polarized
and time-variable, CARMA uses artificial linearly polarized noise sources for
this purpose.  The noise sources are created by inserting wire grid polarizers
into the beams of the 10\,m telescopes.  With the grid in place, one linear polarization
reaching the receiver originates from the sky, while the other
originates from a room temperature load.  Since the
room temperature load is much hotter than the sky, the receiver sees thermal
noise that is strongly polarized.  The L--R phase difference
is then derived, channel by channel, from the L vs. R
autocorrelation spectrum obtained with the grid in place.
One of the 10\,m telescopes is always
used as the reference for the regular passband observations, thus
transferring the L--R phase calibration to all other telescopes.

The leakage corrections compensate for cross-coupling between the L and R
channels, caused by imperfections in the polarizers and OMTs
and by crosstalk in the analog electronics that precede the correlator.
Leakages are calibrated by observing a strong source (usually the gain
calibrator) over a range of parallactic angles.  There are no moving parts in
the CARMA dual polarization receivers, so the measured leakages are stable
with time.  A typical telescope has a band-averaged leakage amplitude (\ie{} a
voltage coupling from L into R, or vice versa) of 6\%.

Observations of 3C286, a quasar known to have a very stable polarization
position angle $\chi$, yield $\chi = 41 \pm 3\degree$ (measured
counterclockwise from north), consistent with recent measurements by
\cite{2012A&A...541A.111A}: $\chi = 37.3 \pm 0.8\degree$ at $\lambda\,$3\,mm and
$\chi = 33.1 \pm 5.7\degree$ at $\lambda\,$1.3\,mm.  Our results also are
consistent with ALMA (Atacama Large Millimeter-submillimeter Array) 
commissioning results at $\lambda\,$1.3\,mm ($\chi = 39\degree$; Stuartt Corder, priv. comm., 2013), as well as with centimeter
observations compiled by \citet{Perley2013}, who showed that the polarization
position angle of 3C286 increases slowly from $\chi = 33\degree$ at
$\lambda \gtrsim 3.7$\,cm to $\chi = 36\degree$ at $\lambda = 0.7$\,cm.
The uncertainty of $\pm$\,3$\degree$ in the CARMA value is the result of systematic errors in the R--L phase correction,
and is estimated from the scatter in the $\chi$ values derived using different 10\,m 
reference antennas.

% Previous numbers: \lesssim 8 GHz, and nu = 43.5 GHz
% Because we take 4--6 hr tracks, errors in the leakages tend to average out

% RELEVANT STUFF FOR BLLAC:
% redux_110928_beamPol.py has the redux procedures
% plot_beamPol.py has the numbers
% central numbers: 14.2º and 9.1%
% min/max numbers: 11.2º/18.0º, and 8.1/9.6%

To check for variations in the instrumental polarization across the primary
beams of the telescopes, we observed BL Lac (a bright, highly polarized quasar)
at 8 offset positions, each $12\arcsec$ from the field center.  The 
deviations in position angle and polarization fraction from the field-center values were $\pm$\,4$\degree$ and
$\pm$\,8\%, respectively.  Primary beam polarization
will therefore have a relatively small effect on the results presented here,
since most of the sources in the TADPOL survey are less than $10\arcsec$ across and are
centered in the primary beam.

We perform calibration and imaging with the MIRIAD data reduction package
\citep{1995adass...4..433S}.  We calibrate the complex gains by observing a
nearby quasar every 15 minutes; the passband by observing a bright quasar for
10 minutes; and the absolute flux using observations of Uranus, Mars, or 
MWC~349.\footnote{CARMA absolute flux measurements at 1.3\,mm are estimated to be uncertain 
by $\pm$\,15\%, due in part to uncertainties in planet models, pointing, and antenna focus. 
%in all bands (1\,mm, 3\,mm, and 1\,cm)
%due to uncertainties in planet models \citep{Bauermeister2012}.  
However, these uncertainties do not affect the conclusions drawn in this paper.}
Using multi-frequency synthesis and natural weighting, we create
dust-continuum maps of all four Stokes parameters $(I, Q, U, V)$ by inverting
the calibrated visibilities, deconvolving the source image from the synthesized beam
pattern with \mir{CLEAN} \citep{Hogbom1974}, and restoring them with a
Gaussian fit to the synthesized beam.  The typical beam size is 2.5\arcsec. 

We produce polarization position-angle and intensity maps from the Stokes $I$,
$Q$, and $U$ data.  (Note that since we are searching for linear dust polarization,
we do not use the Stokes $V$ maps, which are measures
of circular polarization.) The rms noise values in the $Q$ and $U$ maps are generally
comparable, such that we define the rms noise $\sigma_P$ in the polarization maps as
$\sigma_P \approx \sigma_Q \approx \sigma_U$.  The
polarized intensity $P$ is

\begin{equation}
P = \sqrt{Q^2 + U^2}\, .
\end{equation}

\noindent
However, polarization measurements have a positive bias because the
polarization $P$ is always positive, even though the Stokes parameters $Q$ and $U$ 
from which $P$ is derived can be either positive or negative.  This bias has a significant effect 
in low signal-to-noise (SNR) measurements ($P \lesssim 3\,\sigma_P$) and 
can be taken into account by
calculating the bias-corrected polarized intensity $P_c$ (\eg{} \citealt{2006PASP..118.1340V}; 
see also \citealt{Naghizadeh1993} for a discussion of the statistics of position angles in low SNR measurements).

All of the maps we present here have been bias-corrected.  For polarization detections with $P \lesssim 5\,\sigma_P$, we
calculated $P_c$ by finding the maximum of the probability distribution function (i.e., the most probable value) of
the true polarization $P_c$ given the observed polarization $P$ \citep[see][]{2006PASP..118.1340V}.
%\begin{equation}
%\textrm{PDF}(P_c | P, \sigma_P) = \frac{P}{\sigma_P^2} \, I_0\left(\frac{PP_c}{\sigma_P^2}\right) \exp{\left[-(P^2 + P_c^2) / 2\sigma_P^2\right]}\,.
%\end{equation} 
%
%If one assumes a uniform prior for the true polarization $P_c$, then by Bayes's theorem the equation
%above is the same as equation as $\textrm{PDF}(P | P_c, \sigma_P)$ (see Equation 6 in \citealt{2006PASP..118.1340V}).
For very significant polarization detections ($P \gtrsim 5\,\sigma_P$), we used the high-SNR limit:
%(see \citealt{2006PASP..118.1340V}, Equation 12):

\begin{equation}
P_c \approx \sqrt{Q^2 + U^2 - \sigma_P^2}\,.
\end{equation}

The fractional polarization is

\begin{equation}
P_\textrm{frac} = \frac{P_c}{I}\,.
\end{equation}

The position angle $\chi$ and uncertainty $\delta\chi$ (calculated using standard error propagation) of the incoming
radiation are

\begin{equation}
\chi = \frac{1}{2} \arctan{\left(\frac{U}{Q}\right)}\,,
\end{equation}

\begin{equation}
\delta \chi = \frac{1}{2} \frac{\sigma_P}{P_c}\,.
\end{equation}

Note that polarization angles (and the B-field orientations inferred from them)
are not vectors, but are \textit{polars.}  A polar is a ``headless'' vector that has an 
orientation (not a direction) with a $180\degree$ ambiguity.

In good weather $\sigma_P \approx 0.4\,$\mjybm{} for a single 6-hour
observation, and can be as low as $\sim$\,0.2\,\mjybm{} when multiple
observations are combined.  We consider it a detection if $P_c \geq
2\,\sigma_P$ (corresponding to $\delta\chi \approx \pm 14\degree$) 
and if the location of the polarized emission coincides with a
detection of $I \geq 2\,\sigma_I$, where
$\sigma_I$ is the rms noise in the Stokes $I$ map.  
%We choose the latter criterion
%to reduce the number of spurious detections caused by noise in the Stokes $I$ maps.
We also generate
maps of the red- and blueshifted CO($J = 2\rightarrow1$) and SiO($J = 5\rightarrow4$) line wings, but we do not
attempt to measure polarization in the spectral line data because of
fine-scale frequency structure in the polarization leakages.

\section{DATA PRODUCTS \& RESULTS}
\label{sec:data}

%\textbf{Additionally, the refereed version of this article will include the data products 
%via ``Data behind the Figure.''\footnote{ Data behind the Figure: \url{authortools.aas.org/dbf/dbf.html}}}
%At that point, data will also be accessible TADPOL website\footnote{TADPOL website: \url{tadpol.astro.illinois.edu}}, 

Maps of all sources are shown in Appendix \ref{appendix:maps}. 
Note that all of the polarization orientations have been rotated by 
90$\degree$ to show the inferred B-field directions in the plane of the sky.

There are typically three plots per source:

\begin{enumerate}

\item[(a)] \textbf{Small-scale (CARMA) B-fields, with outflows overlaid.}  These plots include 
 B-field orientations as well as
%  (where $P_c > 2\,\sigma_P$)
  red- and blueshifted outflow lobes, all overlaid on the 
  total intensity (Stokes $I$) dust
  emission in gray.  The outflow data are
  CO($J = 2\rightarrow1$) for all sources except for Ser-emb~8 and 8(N) (Figure \ref{fig:SerEmb8}),
  which have more clearly defined outflows in SiO($J = 5\rightarrow4$). 
%  These plots emphasize the bipolar outflows in the lower mass sources.

\item[(b)] \textbf{Small-scale B-fields overlaid on Stokes $I$ dust contours.} 
   In these plots the B-field orientations are black for significant detections ($P_c >
  3.5\,\sigma_P$) or gray for marginal detections ($2\,\sigma_P < P_c < 3.5\,\sigma_P$),
  and are overlaid on total intensity dust emission contours.
  The B-field orientations are the same as those plotted in (a).  
  These plots zoom in on the source to provide a clearer view of the small-scale B-field morphology. 
  %with respect to the dust emission.

\item[(c)] \textbf{Comparison of large- and small-scale B-fields.} These plots include the same dust contours and
  small-scale B-field orientations as in (b), but zoomed out so that the
  large-scale B-fields from the SCUBA (orange), Hertz (light blue),
  and SHARP (purple) polarimeters (see below) can be plotted.  
%  All published SCUBA, Hertz, and SHARP detections
%  are plotted, and have $\gtrsim 2\,\sigma_P$ significance.  
  These plots show how the B-field
  morphology changes from the $\sim$\,0.1\,pc scales probed by single-dish
  submillimeter telescopes to the $\sim$\,0.01\,pc scales probed by CARMA.

\end{enumerate}

To show the B-field morphologies as clearly as possible, we have chosen to plot the lengths of the line  
segments on a square-root scale.  

 at the CSO,
the segment lengths are proportional to the square root of the polarized intensity.
%Only the SHARP data from \citet{Attard2009} are plotted with the same lengths,
%since polarized intensity was not reported.}

%and for SCUBA, Hertz, and SHARP the lengths
%are proportional to the square root of \textit{fractional polarization.}  This is because
%it is difficult to match the absolute flux scales among the three different polarimeters, two of which
%observe at 350\,$\mu$m (Hertz, SHARP), and one of which observes at 850\,$\mu$m (SCUBA).

%The longest line segment in each CARMA map is plotted in the location with the maximum
%polarized intensity $P_{\textrm{max}}$, which is reported in Table
%\ref{table:obs}.  

All maps from the TADPOL survey are publicly available as FITS images
and machine readable tables for each figure in Appendix \ref{appendix:maps}. 
For each figure we include maps of Stokes
$I$, $Q$, and $U$; bias-corrected polarization intensity $P_c$;
polarization fraction $P_{\textrm{frac}} = P_c/I$; and inferred B-field orientation $\chi_{\textrm{sm}}$.  
Additionally, we include FITS cubes of total intensity (Stokes $I$) spectral-line data, as well as
machine readable tables listing the RA, DEC, $I$, $P_c$, $P_\textrm{frac}$, $\chi_{\textrm{sm}}$, and associated uncertainties 
of each line segment plotted in the figures.  
These files are available in a .tar.gz package available via the link in the figure caption.

%text files that list RA, DEC, $I$, $P_c$, $P_\textrm{frac}$, $\chi_{\textrm{sm}}$, and their associated uncertainties 
%for each line segment plotted in the figures.  
%FITS cubes of total intensity (Stokes $I$) spectral-line
%data [either CO($J = 2\rightarrow1$) or SiO($J = 5\rightarrow4$)] are also available.

The results for each source are summarized in Table \ref{table:obs}.
We give fitted coordinates of the dust emission peaks, maximum total
intensity $I_\textrm{pk}$, maximum bias-corrected polarized intensity $P_\textrm{c,pk}$,
average polarization fraction $\overline{P}_\textrm{frac}$, 
average small-scale B-field orientation $\chi_{\textrm{sm}}$,
outflow orientation $\chi_\textrm{o}$,
source type, distance $d$ to the source,
and synthesized-beam size $\theta_\textrm{bm}$ (resolution element) of the maps.

We also tabulate the average large-scale B-field orientation $\chi_{\textrm{lg}}$ 
from the SCUBA, Hertz, and SHARP data.
%These single-dish submillimeter telescopes include the SCUBA polarimeter at the JCMT
%\citep{Matthews2009}; the Hertz polarimeter at the CSO \citep{Dotson2010}; 
%and the SHARC-II polarimeter (SHARP) at the CSO
%\citep{Attard2009, Davidson2011, Chapman2013}.  
We averaged $\chi_{\textrm{lg}}$ values
within a radius of $\sim$\,40$\arcsec$ of the CARMA field center; all of these detections
are shown in the figures in Appendix \ref{appendix:maps}.
%\textbf{Finally, we calculate the angle difference $|\chi_{\textrm{lg}} - \chi_{\textrm{sm}}|$ between the large- and
%small-scale B-field orientations.}

The values $\overline{P}_\textrm{frac}$, $\chi_{\textrm{lg}}$, and $\chi_{\textrm{sm}}$ are averages of quantities that vary across each source, and 
hence are sensitive to the weighting schemes used to derive them.  
Since the locations of the intensity and polarization peaks for each source are
not necessarily spatially coincident, we chose to calculate a measure
of fractional polarization $\overline{P}_{\textrm{frac}}$ using the mean polarized and total intensities across the entire source. 
%$\overline{P}_{\textrm{frac}} = \overline{P_c}\,/\,\overline{I}$, where
%$\overline{P_c}$ and $\overline{I}$ are the unweighted averages of the polarized and total
%intensities in locations where $P_c > 3.5\,\sigma_P$.  
%The average polarization fractions are $\overline{P}_\textrm{frac} =
%\overline{P}_{\textrm{frac}} \times 100\%$.
%The uncertainty in the fractional polarization is
%the median uncertainty in the fractional polarization used in the average.
%Each measured value of $\overline{P}_{\textrm{frac}}$ 
%%(and the associated $\overline{P}_{\textrm{\%}}$ reported in Table \ref{table:obs})
%corresponds roughly to the minimum of the ``polarization hole'' in each source
%(see Section \ref{sec:pol_hole}).  
To do this, we average only pixels where $P_c > 3.5\,\sigma_P$.
We average $I$ and $P_c$ separately over this set of pixels, and define $\overline{P}_{\textrm{frac}}$ = $\overline{P_c}\,/\,\overline{I}$.
For the typical source $P_c$ has a much flatter 
distribution than $I$ over these pixels, so that our average is biased toward the 
minimum of the ``polarization hole'' in each source (see Section \ref{sec:pol_hole}).
The uncertainty in the fractional polarization is calculated rather differently: it is the 
median of the uncertainties in the fractional polarization in each pixel.

%Fractional polarization values are only reported in Table \ref{table:obs} for sources
%with significant polarization detections ($P_c > 3.5\,\sigma_P$), as marginal data can lead to
%spuriously high percentage values.  
Note that when calculating $\overline{P}_{\textrm{frac}}$ we average only the 
magnitude of $P_c$ (and not the orientation $\chi$ of the B-field) across the source,
which makes our measurements
%rather than averaging the Stokes $Q$ and $U$ values in the original maps.  This is because
%averaging $Q$ and $U$ across the entire map is akin to averaging the source over
%one large resolution element, which would result in depolarization of sources that have tangled fields 
%in our maps.  
sensitive only to depolarization along the line of sight (LOS) 
or in the plane of the sky at scales smaller than the resolution of our CARMA maps. 

We should note that interferometric measurements of fractional polarization
can be problematic because an interferometer acts as a spatial filter, and is
insensitive to large scale structure.  This makes direct comparisons of  
fractional polarization results from single dish telescopes and interferometers extremely difficult.
For example, in cases where polarized emission
(Stokes $Q$ or $U$) is localized, but total intensity (Stokes $I$) is
extended, it is possible to overestimate the polarization fraction with an interferometer.  
The comparison of polarization angles should be less problematic, however, as it is unlikely
that Stokes $Q$ would be very localized and $U$ would be very extended, or vice versa.

To calculate $\chi_{\textrm{sm}}$ we performed a total-intensity-weighted
average of each small-scale B-field orientation $\chi$ where $P_c > 2\,\sigma_P$:

\begin{equation}
\chi_{\textrm{sm}} = \frac{\sum \chi I}{\sum I} \,.
\end{equation}

\noindent
This method gives more weight to the B-field orientations in the highest density regions of the source, 
and is the same method used in \citet{Hull2012}.

To calculate $\chi_{\textrm{lg}}$ we performed total-intensity-weighted averages of
the large-scale B-field orientations from SCUBA, Hertz, and/or SHARP.  
For sources that had detections from more than one telescope, we weighted each of the
averages by the number of detections present in the map (i.e., for a source with 40 SCUBA
and 5 Hertz detections, more weight is given to the average of the SCUBA detections).  
%Note that
%for SHARP data from \citet{Attard2009} all detections were given equal weight, since Stokes $I$ values
%were not reported.

The dispersions in $\chi_{\textrm{sm}}$ and $\chi_{\textrm{lg}}$ 
are calculated using the circular standard deviation of the  B-field orientations across
each source.  
Note that these dispersions  
reflect the spread in B-field orientations in each source, not
the uncertainty in the measurements.  For example, a source
with complicated B-field morphology such as NGC~7538~IRS~1 (see Figure \ref{fig:NGC7538}) has
a large scatter in $\chi_{\textrm{sm}}$ because of the widely varying B-field orientations across the source.
Nevertheless, any given B-field orientation in the map has an uncertainty of $\lesssim 14\degree$, since we only plot
detections where $P_c > 2\,\sigma_P$.

The value $|\chi_{\textrm{lg}} - \chi_{\textrm{sm}}|$ was used to characterize 
the consistency between large- and small-scale B-field orientations.
The dispersion in $|\chi_{\textrm{lg}} - \chi_{\textrm{sm}}|$ is equal to the dispersions in 
$\chi_{\textrm{sm}}$ and $\chi_{\textrm{lg}}$ added in quadrature.

Generally the outflow angle $\chi_\textrm{o}$ is determined by 
connecting the center of the continuum source and the intensity peaks of the
red and blue outflow lobes, and taking the average of the two position angles.  
Of course, this is somewhat arbitrary because it depends on the selected velocity ranges
for the red and blue lobes, and because outflows can have complex morphology.  
We do not report outflow orientations in sources where the morphology is extremely complex.
The outflow orientation is indicated in the first panel of most plots in Appendix \ref{appendix:maps}.

Note that as a test, we performed polarized-intensity-weighted (as opposed to total-intensity-weighted)
averages of $\chi_{\textrm{lg}}$ and $\chi_{\textrm{sm}}$ and found that our main conclusions were unchanged.
For the low-mass cores plotted in Figures \ref{fig:frac} and \ref{fig:ks}, the two weighting schemes resulted
in $\lesssim 20\degree$ differences in the consistency angle $|\chi_{\textrm{lg}} - \chi_{\textrm{sm}}|$.

% ============================== TABLE ===================================
%\begin{landscape}
\begin{deluxetable*}{lrrrrrrrrrccc} %[hbt!]
%\begin{deluxetable}{lrrrrrrrrrccc} %[hbt!]
%\tabletypesize{\scriptsize}
%\setlength{\tabcolsep}{10in}
\tablecaption{\large Observations }
\tablewidth{0pt}
\tablehead{
\colhead{Source} & \colhead{$\alpha$} & \colhead{$\delta$} & 
\colhead{$I^{\tablenotemark{a}}_\textrm{pk}$} & \colhead{$P^{\tablenotemark{a,b}}_\textrm{c,pk}$} & \colhead{$\overline{P}_\textrm{frac}$} &
\colhead{$\chi_{\textrm{lg}}$} & \colhead{$\chi^{\tablenotemark{b}}_{\textrm{sm}}$} & \colhead{$|\chi_{\textrm{lg}} - \chi_{\textrm{sm}}|$} & \colhead{$\chi_{\textrm{o}}$} &
\colhead{Type} & \colhead{$d$} & \colhead{$\theta_\textrm{bm}$}  \vspace{0.05in} \\
			% & \colhead{Dist.} \\
                                & \colhead{\scriptsize (J2000)} & \colhead{\scriptsize (J2000)} & \colhead{\scriptsize \mjybmvert{}} & 
                                \colhead{\scriptsize \mjybmvert{}} & \colhead{(\%)} & \colhead{\scriptsize ($\degree$)} & \colhead{\scriptsize ($\degree$)} &
			    \colhead{\scriptsize ($\degree$)} & \colhead{\scriptsize ($\degree$)} &  &
                                \colhead{\scriptsize (pc)} & \colhead{\scriptsize ($\arcsec$)} % & \colhead{ref.\tablenotemark{d}}
\vspace{0.05in}
}

\startdata
W3~Main & 02:25:40.6 & 62:05:51.6 & 374 & 3.3 & 2.0 (0.5) & 135 (49) & 100 (36) & 35 (60) & -----\phantom{1} & SFR & 1950 & 2.9 \\
W3(OH) & 02:27:03.9 & 61:52:24.6 & 2760 & 13.8 & 1.0 (0.4) & 22 (25) & 82 (53) & 60 (58) & -----\phantom{1} & SFR & 2040 & 2.7 \\
L1448~IRS~2 & 03:25:22.4 & 30:45:13.2 & 136 & 3.4 & 3.7 (0.9) & 148 (12) & 135 (43) & 13 (44) & 134$^*$ & 0 & 232 & 3.8 \\
L1448N(B) & 03:25:36.3 & 30:45:14.7 & 596 & 5.4 & 1.3 (0.2) & 14 (33) & 26 (37) & 12 (49) & 97$^*$ & 0 & 232 & 2.5 \\
L1448C & 03:25:38.9 & 30:44:05.3 & 186 & <\,2.4 & -----\phantom{11} & 110 (39) & 112 (32) & 2 (50) & 161\phantom{1} & 0 & 232 & 2.5 \\
L1455~IRS~1 & 03:27:39.1 & 30:13:03.0 & 43 & <\,2.0 & -----\phantom{11} & 72 (19) & 150 (24) & 78 (30) & 66\phantom{1} & I & 320 & 2.7 \\
NGC~1333-IRAS~2A$^\tablenotemark{c}$ & 03:28:55.6 & 31:14:37.0 & 322 & 3.1 & 1.8 (0.4) & 135 (56) & 70 (23) & 65 (60) & 21$^*$ & 0 & 320 & 3.5 \\
 &  &  &  &  &  &  &  &  & 98$^*$ &  &  &  \\
SVS~13 & 03:29:03.7 & 31:16:03.5 & 276 & 3.8 & 2.0 (0.5) & 171 (24) & 6 (24) & 15 (33) & -----\phantom{1} & 0/I & 235 & 3.3 \\
NGC~1333-IRAS~4A & 03:29:10.5 & 31:13:31.3 & 1680 & 46.1 & 4.5 (0.5) & 53 (25) & 56 (20) & 3 (32) & 18$^*$ & 0 & 320 & 2.4 \\
NGC~1333-IRAS~4B & 03:29:12.0 & 31:13:08.1 & 866 & 9.7 & 1.7 (0.3) & 55 (27) & 84 (34) & 29 (43) & 0$^*$ & 0 & 320 & 2.5 \\
NGC~1333-IRAS~4B2 & 03:29:12.8 & 31:13:07.1 & 244 & <\,2.0 & -----\phantom{11} & 55 (27) & 55 (20) & 0 (33) & 76\phantom{1} & 0 & 320 & 2.5 \\
HH~211~mm & 03:43:56.8 & 32:00:50.0 & 196 & 4.8 & 4.1 (1.2) & 168 (17) & 164 (32) & 4 (36) & 116$^*$ & 0 & 320 & 4.1 \\
DG~Tau & 04:27:04.5 & 26:06:15.9 & 296 & <\,2.8 & -----\phantom{11} & -----\phantom{11} & 84 (14) & -----\phantom{11} & -----\phantom{1} & II & 140 & 2.4 \\
L1551~NE & 04:31:44.5 & 18:08:31.5 & 418 & 8.3 & 2.0 (0.3) & 46 (32) & 164 (15) & 62 (35) & 67$^*$ & I & 140 & 2.6 \\
L1527 & 04:39:53.9 & 26:03:09.6 & 161 & 3.4 & 2.2 (0.3) & 38 (42) & 3 (8)\phantom{1} & 35 (42) & 92$^*$ & 0/I & 140 & 3.0 \\
CB~26 & 04:59:50.8 & 52:04:43.5 & 77 & <\,1.8 & -----\phantom{11} & 81 (21) & 87 (66) & 6 (69) & 147\phantom{1} & I & 140 & 2.5 \\
Orion-KL & 05:35:14.5 & --05:22:31.6 & 3270 & 91.7 & 5.3 (1.2) & 119 (13) & 140 (34) & 21 (36) & -----\phantom{1} & SFR & 415 & 2.7 \\
OMC3-MMS5 & 05:35:22.6 & --05:01:16.5 & 123 & 5.2 & 4.4 (0.7) & 49 (10) & 59 (12) & 10 (15) & 80$^*$ & 0 & 415 & 3.0 \\
OMC3-MMS6 & 05:35:23.4 & --05:01:30.6 & 984 & 20.2 & 3.0 (0.3) & 51 (12) & 44 (8)\phantom{1} & 7 (14) & 171$^*$ & 0 & 415 & 3.0 \\
OMC2-FIR4 & 05:35:26.9 & --05:09:55.8 & 57 & 2.2 & 7.9 (2.2) & 43 (27) & 146 (64) & 77 (69) & -----\phantom{1} & SFR & 415 & 3.0 \\
OMC2-FIR3 & 05:35:27.6 & --05:09:34.2 & 76 & 2.8 & 5.8 (1.4) & 50 (30) & 166 (7)\phantom{1} & 64 (30) & -----\phantom{1} & 0 & 415 & 3.0 \\
CB~54 & 07:04:20.8 & --16:23:22.2 & 93 & <\,2.8 & -----\phantom{11} & 173 (38) & 32 (42) & 39 (56) & 108\phantom{1} & I & 1100 & 3.0 \\
VLA~1623 & 16:26:26.4 & --24:24:30.5 & 283 & 3.8 & 1.7 (0.4) & 60 (32) & 23 (48) & 37 (57) & 120$^*$ & 0 & 125 & 3.3 \\
Ser-emb~17 & 18:29:06.2 & 00:30:43.3 & 156 & <\,2.2 & -----\phantom{11} & -----\phantom{11} & 73 (39) & -----\phantom{11} & -----\phantom{1} & I & 415 & 3.0 \\
Ser-emb~1 & 18:29:09.1 & 00:31:31.1 & 220 & <\,1.6 & -----\phantom{11} & -----\phantom{11} & 127 (52) & -----\phantom{11} & 12\phantom{1} & 0 & 415 & 3.3 \\
Ser-emb~8 & 18:29:48.1 & 01:16:43.6 & 165 & 3.5 & 3.0 (0.6) & 94 (35) & 7 (44) & 87 (56) & 129$^*$ & 0 & 415 & 2.6 \\
Ser-emb~8~(N) & 18:29:48.7 & 01:16:55.8 & 72 & 2.5 & 5.2 (1.2) & 92 (31) & 83 (15) & 9 (34) & 107$^*$ & 0 & 415 & 2.6 \\
Ser-emb~6 & 18:29:49.8 & 01:15:20.3 & 1230 & 17.1 & 1.4 (0.2) & 86 (29) & 172 (33) & 86 (43) & 135$^*$ & 0 & 415 & 2.7 \\
HH~108~IRAS & 18:35:42.1 & --00:33:18.4 & 198 & <\,2.3 & -----\phantom{11} & -----\phantom{11} & 4 (34) & -----\phantom{11} & 34\phantom{1} & 0/I & 310 & 4.1 \\
G034.43+00.24~MM1 & 18:53:18.0 & 01:25:25.4 & 1160 & 12.6 & 1.9 (0.4) & -----\phantom{11} & 41 (22) & -----\phantom{11} & 47\phantom{1} & SFR & 1560 & 2.6 \\
G034.43+00.24~MM3 & 18:53:20.6 & 01:28:26.4 & 66 & <\,2.4 & -----\phantom{11} & -----\phantom{11} & 57 (41) & -----\phantom{11} & -----\phantom{1} & SFR & 1560 & 2.6 \\
B335~IRS & 19:37:00.9 & 07:34:09.3 & 71 & <\,3.0 & -----\phantom{11} & 18 (35) & 123 (40) & 75 (53) & 99\phantom{1} & 0 & 150 & 3.5 \\
DR21(OH) & 20:39:01.1 & 42:22:49.0 & 615 & 8.5 & 2.2 (0.4) & 89 (22) & 42 (37) & 47 (43) & -----\phantom{1} & SFR & 1500 & 2.6 \\
L1157 & 20:39:06.2 & 68:02:15.8 & 197 & 7.7 & 5.8 (1.2) & 143 (23) & 147 (29) & 4 (37) & 146$^*$ & 0 & 250 & 2.2 \\
CB~230 & 21:17:38.7 & 68:17:32.4 & 104 & 2.1 & 5.4 (3.2) & 113 (34) & 96 (35) & 17 (48) & 172$^*$ & 0/I & 325 & 3.0 \\
L1165 & 22:06:50.5 & 59:02:45.9 & 128 & <\,2.9 & -----\phantom{11} & -----\phantom{11} & 113 (4)\phantom{1} & -----\phantom{11} & 52\phantom{1} & I & 300 & 3.9 \\
NGC~7538~IRS~1 & 23:13:45.4 & 61:28:10.3 & 3230 & 11.6 & 1.7 (0.8) & 145 (26) & 52 (62) & 87 (67) & -----\phantom{1} & SFR & 2650 & 2.4 \\
CB~244 & 23:25:46.6 & 74:17:38.3 & 43 & <\,1.5 & -----\phantom{11} & 168 (79) & 170 (49) & 2 (92) & 42\phantom{1} & 0 & 200 & 2.7 \\

\enddata

\tablecomments{Coordinates are fitted positions of dust emission peaks measured in the CARMA maps.
$I_\textrm{pk}$ and $P_\textrm{c,pk}$ are the maximum total intensity and bias-corrected polarized intensity, respectively.
The polarization fraction $\overline{P}_{\textrm{frac}} = \overline{P}\,/\,\overline{I}$, where
  $\overline{P}$ and $\overline{I}$ are the unweighted averages of the polarization and total intensities in locations where $P_c > 3.5\,\sigma_P$.
  %The uncertainty in $\overline{P}_\textrm{frac}$ is in parentheses, and is the median of the uncertainties in the values used in the average. 
%  A fractional polarization of $0.05 \pm 0.01$ would be represented as $5 \pm 1\%$.
The bipolar outflow orientations $\chi_{\textrm{o}}$ and the  large- and small-scale B-field orientations $\chi_{\textrm{lg}}$ and $\chi_{\textrm{sm}}$ are measured counterclockwise from north.
Sources included in Figure \ref{fig:ks} are marked with an asterisk (*) next to their outflow orientations.
$|\chi_{\textrm{lg}} - \chi_{\textrm{sm}}|$ is the angle difference between the large- and small-scale B-field orientations.
The uncertainties in $\chi_{\textrm{lg}}$ and $\chi_{\textrm{sm}}$ are in parentheses; these numbers are the circular standard deviations of the B-field orientations used in the averages, and thus reflect the dispersion of the B-field orientations in each source.  
The uncertainty in $|\chi_{\textrm{lg}} - \chi_{\textrm{sm}}|$ is equal to the uncertainties in $\chi_{\textrm{sm}}$ and $\chi_{\textrm{lg}}$ added in quadrature.
The B-field is assumed to be perpendicular to the position angle of the dust polarization.
Source types are: 0 (Class 0 young stellar object [YSO]), I (Class I YSO), II (Class II YSO), and SFR (star-forming region).
$d$ is the distance to the source.  
$\theta_\textrm{bm}$ is the geometric mean of the major and minor axes of the synthesized beam. \\ }

\tablenotetext{a}{Polarized and total intensity maxima do not necessarily coincide spatially.}
\tablenotetext{b}{Upper limits on the polarized intensity $P_{\textrm{c,pk}}$ are given for sources with $P_{\textrm{c,pk}} < 3.5\,\sigma_P$.  Because of low-level calibration artifacts, the small-scale B-field angles $\chi_{\textrm{sm}}$ for such sources are not always reliable.}
\tablenotetext{c}{NGC~1333-IRAS~2A has two well defined outflows.  Both outflow orientations are listed here, and both are included in Figure \ref{fig:ks}.}
%\tablenotetext{d}{\textbf{FIX:} Distance references.  1: \citet{Hirota2011}.  2: \citet{Loinard2007}.   
%  3: \citet{Menten2007}.  4: \citet{Loinard2008}.  5: \citet{Imai2007}.  6: \citet{Dzib2010}.  7: \citet{Looney2007}.  8: \citet{Launhardt2010}.}

\label{table:obs}
%\end{deluxetable}
\end{deluxetable*}

\section{ANALYSIS \& DISCUSSION}
\label{sec:discussion}

In this paper, we do not attempt to interpret the detailed B-field morphology of each object. 
Rather, our goal is to use average B-field orientations to derive conclusions in a 
statistical sense from the ensemble of sources.
The large uncertainties in $\chi_\textrm{lg}$ and $\chi_\textrm{sm}$ in Table \ref{table:obs}
reflect the large dispersions in the B-field orientations across each of these objects.
The mean B-field orientation is necessarily determined by detections of polarization 
in locations where the observations have sufficient signal-to-noise, and may not reflect
the B-field orientation across the entirety of the source.
Furthermore, the B-fields may have been distorted by collapse, pinching, or outflows,
and thus caution must be used when interpreting the source-averaged values that we
report in Table \ref{table:obs}.

%While our weighting schemes (described in Section \ref{sec:XXX}) influence 
%how 

%So, certain objects show big-to-small field consistency AND higher pol
%fraction.  Big-to-small consistency suggests magnetic regulation (no
%twisting), which suggests strong fields.  Higher pol fraction suggests well
%ordered fields on small scales, which also suggests strong fields.  

%One subtlety that we may have overlooked: the high mass regions have twisted
%fields at our resolution.  However, they're 10x further away, so we're
%actually probing the more diffuse field surrounding the individual cores.  We
%normally expect diffuse envelopes to be well ordered, yet they're clearly not
%in our massive sources.  Maybe this has something to do with high-mass
%regions in particular?

\subsection{Consistency of B-fields from large to small scales}
\label{sec:consistency}

%this suggests that
%the ratio of magnetic to turbulent energy is higher in these sources, and thus the B-fields are
%less subject to B-field twisting, which would tend to decrease the polarization
%signal.  

\begin{figure*} 
\begin{center}
\epsscale{0.84}
\plotone{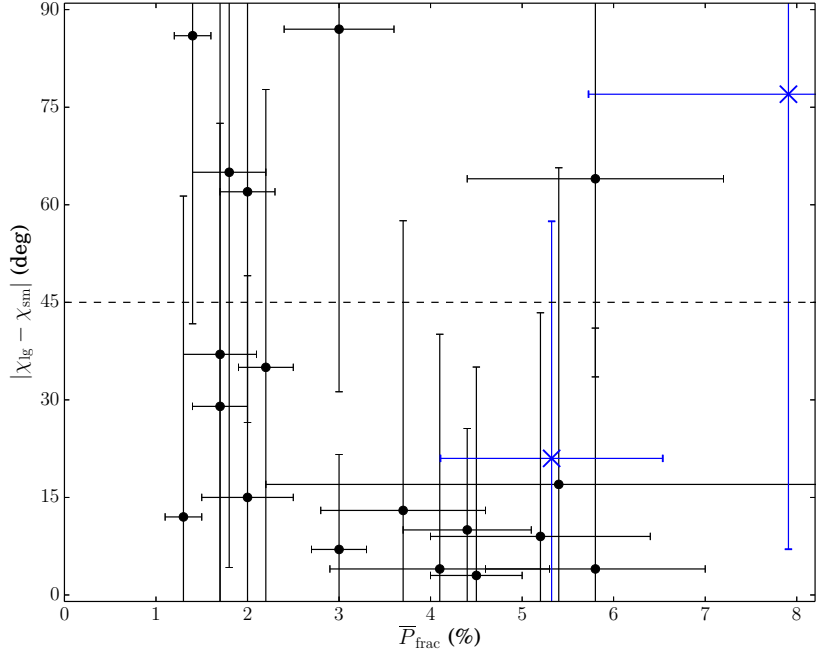}
\caption[]{ \footnotesize{ Large- vs. small-scale B-field orientation $|\chi_{\textrm{lg}} - \chi_{\textrm{sm}}|$ as a function of
    polarization fraction $\overline{P}_\textrm{frac}$.  Sources are included if they have 
  (1) B-field detections at both scales, (2) CARMA polarization detections $P_c > 3.5\,\sigma_P$,
  and (3) distances $d \lesssim 400$\,pc.  
    The plotted uncertainty in $|\chi_{\textrm{lg}} - \chi_{\textrm{sm}}|$ is equal to the uncertainties in 
    $\chi_{\textrm{sm}}$ and $\chi_{\textrm{lg}}$ added in quadrature, where those uncertainties
    reflect the dispersion of the B-field orientations in each source. 
%    A blue $\times$ denotes a star-formation region, and a black $\bullet$ denotes a star-forming core (see Sections \ref{sec:cores} and \ref{sec:SFRs}).  
%    Only sources with large-scale B-field detections and with
%    polarization detections of $P_c > 3.5\,\sigma_P$ are plotted.
%    The angle difference $|\chi_{\textrm{lg}} - \chi_{\textrm{sm}}|$ between the large- and small-scale  B-field orientations
%    $\chi_{\textrm{lg}}$ and $\chi_{\textrm{sm}}$
% is always between 0$\degree$ and 90$\degree$.
      The fractional polarization
    $\overline{P}_{\textrm{frac}} = \overline{P}\,/\,\overline{I}$, where
$\overline{P}$ and $\overline{I}$ are the unweighted averages of the polarized and total
intensities in locations where $P_c > 3.5\,\sigma_P$.  Points below the
    $45\degree$ line exhibit overall alignment between large- and small-scale
    fields.
    }}
\label{fig:frac}
\end{center}
\end{figure*}

%\subsubsection{Nearby star-forming cores}
%\label{sec:cores}

%The majority of the sources in the TADPOL sample are nearby star-forming cores.  
While $\sim$\,kpc-scale galactic B-fields do not seem to be correlated with smaller-scale
B-fields in clouds and cores \citep[e.g.,][]{Stephens2011}, 
%While \citet{Stephens2011} found no correlation between $\sim$\,kpc
%galactic-scale and $\sim$\,0.1\,pc dense-core-scale B-field orientations,
\citet{Li2009} did find evidence that B-field orientations are consistent from the $\sim$\,100\,pc scales of molecular clouds  
to the $\sim$\,0.1\,pc scales of dense cores.  We take the next step by examining 
the consistency of B-field orientations from the $\sim$\,0.1\,pc core to $\sim$\,0.01\,pc envelope scales.

In Figure \ref{fig:frac} we plot $|\chi_{\textrm{lg}} - \chi_{\textrm{sm}}|$ as a function of the
polarization fraction.  This plot is limited to sources with 
(1) B-field detections at both scales,
(2) CARMA polarization detections $P_c > 3.5\,\sigma_P$,
and (3) distances $d \lesssim 400$\,pc.

The most notable feature of the plot is the relative absence of star-forming cores
in the upper-right quadrant, i.e., sources that are strongly polarized but have 
inconsistent large-to-small-scale B-field orientations.
With the exception of OMC2-FIR3 and Ser-emb~8,
we see that the cores with high CARMA polarization fractions 
($\overline{P}_\textrm{frac} \geq \,3\%$)
have B-field orientations that are consistent from large to small scales.  
These ``high-polarization'' sources are 
%\textul{\textbf{L1455~IRS~1}} (Figure \ref{fig:L1455}),
L1448~IRS~2 (Figure \ref{fig:L1448I}),
NGC~1333-IRAS~4A (Figure \ref{fig:IRAS4A}), 
HH~211~mm (Figure \ref{fig:HH211}), 
Orion-KL (Figure \ref{fig:OrionKL}), 
OMC3-MMS5 
and MMS6 (Figure \ref{fig:MMS6}), 
OMC2-FIR3 
and 4 (Figure \ref{fig:FIR4}), 
Ser-emb~8
and 8(N) (Figure \ref{fig:SerEmb8}),
L1157 (Figure \ref{fig:L1157}), 
and CB~230 (Figure \ref{fig:CB230}).

% Of this subset, all but OMC2-FIR4 show consistency between large- and small-scale B-field orientations,
In these sources the consistency of the B-fields from large to small scales suggests
that the fields have not been twisted by turbulent motions as the material collapses to form the protostellar cores. 
This is in turn consistent with the sources' higher fractional polarization, 
because more ordered B-fields would lead to less averaging of disordered polarization along the LOS.  
In this subset of sources the B-fields appear to be dynamically important, and may play 
a role in regulating the infall of material down to $\sim$\,0.01\,pc scales.

%Such averaging, both in the synthesized beam and 
%along the LOS, tends to reduce polarization intensity.  \textbf{This repeats what we said at the end of section 5.}

%It is also worth noting that
%both the large- and the small-scale B-field orientations in these sources tend to have lower
%standard deviations, which suggests that the B-fields in these sources are indeed dynamically important.

%\textit{ All of this points to the importance of the mass-to-flux ratio.  In
%  objects like L1157 (see Figure \ref{fig:L1157}), the mass-to-flux ratio is
%  presumably smaller.  In contrast, objects that are misaligned, twisted, or
%  unpolarized would have higher mass-to-flux ratios, making the field less
%  dynamically relevant.  Should I mention that field strengths for L1157
%  ($\sim$\,3\,mG, \citealt{Stephens2013}) and IRAS~4A ($\sim$\,5\,mG,
%  \citealt{Girart2006}) are high relative to the average field of
%  $\sim$\,1\,mG reported from high-density Zeeman observations by
%  \citet{Crutcher2012}?  }

The remaining ``low-polarization'' sources ($\overline{P}_\textrm{frac} <\,3\%$) are
%B-fields that are generally inconsistent from large to small scales.  
%This subset of ``low-polarization sources'' includes
%the remainder of the sources in Table \ref{table:obs} that are not SFRs, 
%that have polarization detections on both large and small scales, and that have
%well defined outflows and significant polarization ($P_c > 3.5\,\sigma_P$).
L1448N(B) (Figure \ref{fig:L1448N}), 
NGC~1333-IRAS~2A (Figure \ref{fig:IRAS2A}), 
SVS~13 (Figure \ref{fig:SVS13}),
NGC~1333-IRAS~4B (Figure \ref{fig:IRAS4B}),
L1551~NE (Figure \ref{fig:L1551}),
L1527 (Figure \ref{fig:L1527}),
%CB~54 (Figure \ref{fig:CB54}),
%OMC2-FIR4 (Figure \ref{fig:FIR4}), 
VLA~1623 (Figure \ref{fig:VLA1623}), 
and Ser-emb~6 (Figure \ref{fig:SerEmb6}).

Unlike the high-polarization sources, 
these low-polarization sources may have low ratios of magnetic to turbulent energy, which would result
in more twisted small-scale B-fields and thus low CARMA polarization fractions.
Note that straight B-fields with a high inclination angle relative to the LOS would also result in low 
fractional polarization; however, the likelihood of observing B-fields nearly pole-on is low.

Note that we are not asserting that higher polarization is caused directly by stronger B-fields, or
that weak polarization occurs because of weak B-fields or poor grain alignment.
We simply assume that high and low polarization fractions are caused by B-fields that are less or more twisted, respectively.

We have not yet discussed the more distant sources in our sample, which are all massive star-forming regions (SFRs).
%(SFRs; denoted by $\times$ symbols in Figure \ref{fig:frac}).  
%Unlike the cores described in Section \ref{sec:cores}, these sources do not always have 
%well defined bipolar outflows; they harbor many low-, intermediate- and high-mass protostars; 
%and they tend to have complicated small-scale B-field morphologies. 
%All of the SFRs listed in Table \ref{table:obs} are at distances $d > 1$\,kpc, with the exception
%of the two sources in Orion ($d \approx 400$\,pc) that have complicated small-scale B-field structure: 
%Orion~KL (see Figure \ref{fig:OrionKL}) and OMC2-FIR3/4 (see Figure \ref{fig:FIR4}).
Four of these have been observed
previously by SCUBA, Hertz, and/or SHARP:  
W3~Main (Figure \ref{fig:W3Main}), W3(OH) (Figure \ref{fig:W3OH}), 
%Orion-KL (Figure \ref{fig:OrionKL}), OMC2-FIR3/4 (Figure \ref{fig:FIR4}), 
DR21(OH) (Figure \ref{fig:DR21OH}), and NGC~7538~IRS~1 (Figure \ref{fig:NGC7538}).
It is important to note that  
we are probing different structures in these objects
 than we are in the nearby star-forming cores:
at the distances to the more distant SFRs, the angular resolution of our CARMA maps
corresponds to a spatial resolution of $\sim$\,$0.1$\,pc.
% or the size of a typical dense core.
It is evident from our maps that at these scales the B-fields 
in the SFRs have been twisted, most likely by dynamic processes,
as high-mass SFRs are known to be highly turbulent \citep{Elmegreen2004}.  
This suggests that for massive SFRs the ratio of magnetic to turbulent energy is low
at $\sim$\,0.1\,pc scales.

%Consequently, the large- and small-scale fields orientations in these sources
%are not consistent.
%The obviously twisted small-scale B-fields and the inconsistent B-field orientations from large to small scales
%and the reduced polarization in the SFRs in our sample 

%The more distant objects ($d \gtrsim 1$\,kpc) tend to have low average polarization fractions (see Figure \ref{fig:frac}, where data for the SFRs
%are plotted as $\times$ symbols), presumably because of this twisting of the fields at smaller scales: as described in Section \ref{sec:cores}, disordered
%fields can cause low fractional polarization when the polarized radiation is averaged in
%the synthesized beam and along the LOS.

%This is especially
%evident in the higher mass star-forming regions, including
%see W3~Main (Figure \ref{fig:W3Main}), where field lines appear to be
%stretched out along the walls of an \HII{} region being blown by a massive
%forming star; and in NGC~7538 (Figure \ref{fig:NGC7538}), where field lines
%are stretched along what may be a dense filament feeding the massive O-star
%forming in the center.  Other high-mass

%The polarization in these more
%complicated objects is still very significant, but is not consistent with the
%larger-scale polarization orientations when averaged over the whole image, which
%suggests that the magnetic fields in these objects are not dynamically
%important at small scales.

\subsection{Misalignment of B-fields and bipolar outflows}
\label{sec:misalignment}

\begin{figure*} 
\begin{center}
\epsscale{1.17}
\plottwo{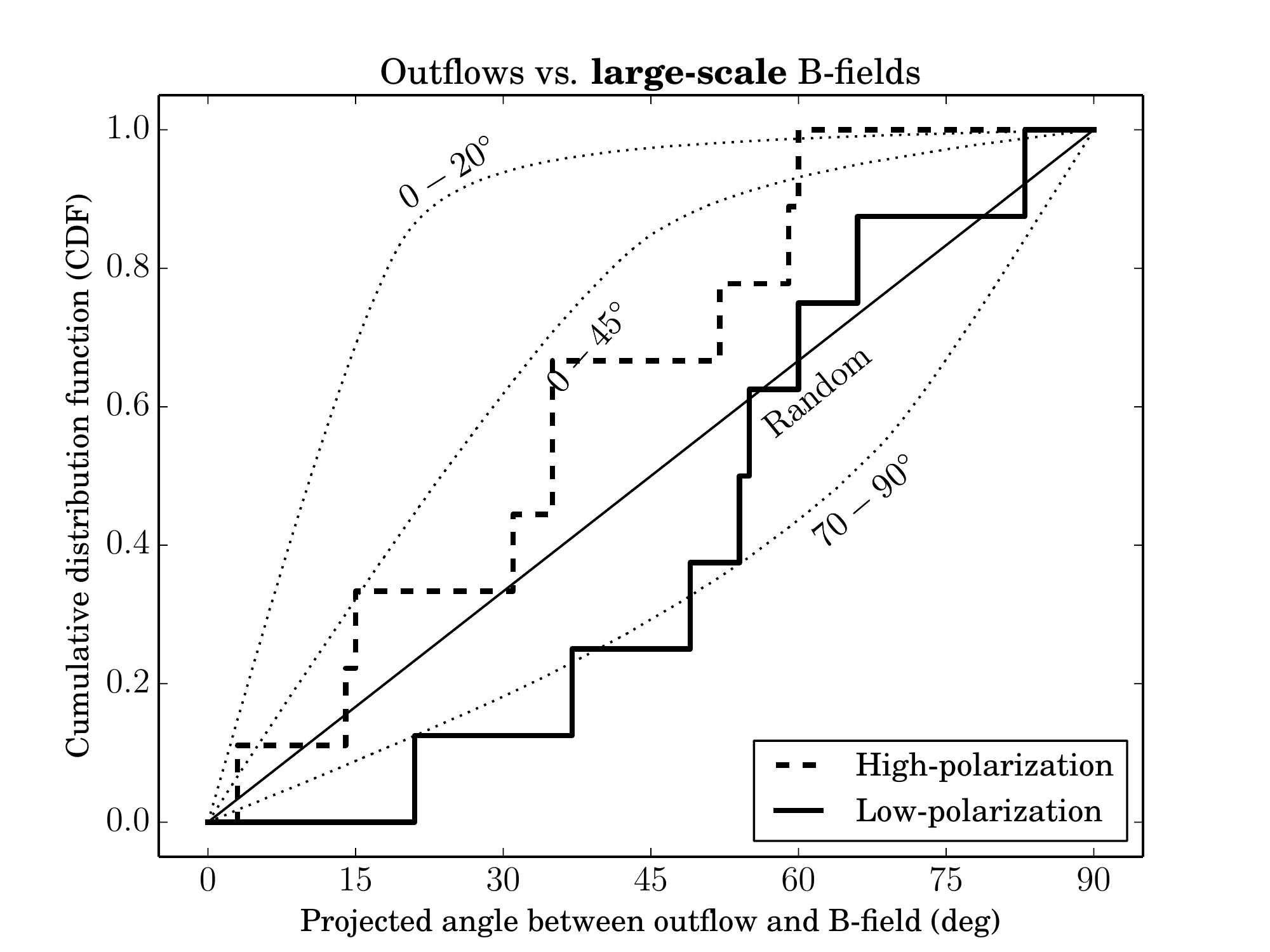}{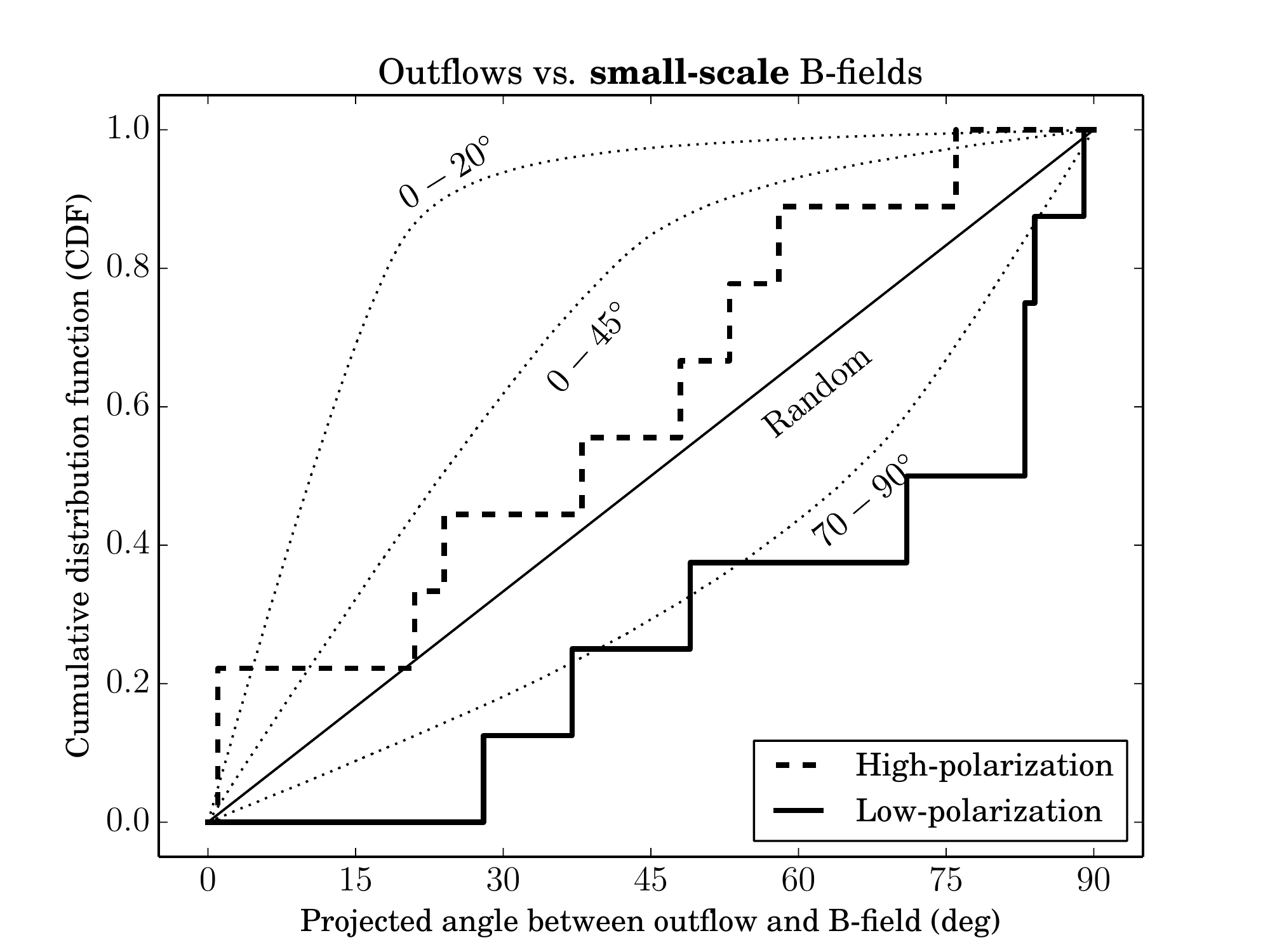}
\caption[]{ \footnotesize{ The thick, stepped curves show the cumulative
    distribution functions (CDF) of the (projected) angles between the 
    bipolar outflows and the mean
    large-scale (left) and small-scale (right) B-field orientations in the low-mass protostellar
    cores listed in Table \ref{table:obs}.  Sources included in the plot have an asterisk (*) next to their outflow orientation in the table.  
    Large-scale B-fields are from archival CSO and JCMT data, and have $\sim$\,$20\arcsec$ resolution;
    small-scale B-fields are from the CARMA data, and have $\sim$\,2.5$\arcsec$ resolution.
    The dashed curves include the ``high-polarization'' sources, and the 
    solid curves include the ``low-polarization'' sources (see Section \ref{sec:consistency} for a discussion
    of high- vs. low-polarization sources).  
    Sources are included if they have
  (1) B-field detections at both large and small scales, (2) CARMA polarization detections $P_c > 3.5\,\sigma_P$,
  (3) distances $d \lesssim 400$\,pc, and (4) well defined bipolar outflows.  
    The dotted curves
    are the CDFs from Monte Carlo simulations where the B-fields and outflows
    are oriented within $20\degree$,
    $45\degree$, and 70--90$\degree$ of one another, respectively.  The straight line is the CDF
    for random orientation.  
    The two plots show that outflows appear to be randomly aligned with B-fields; although, in 
sources with low polarization fractions there is a hint that outflows are preferentially perpendicular to small-scale B-fields, 
which suggests that in these sources the fields have been 
wrapped up by envelope rotation (see Section \ref{sec:misalignment}).
}}
\label{fig:ks}
\end{center}
\end{figure*}

We first addressed the question of B-field and outflow misalignment in \citet{Hull2012},
where we found that bipolar outflows were randomly aligned with---or perhaps preferentially
perpendicular to---the small-scale B-fields in their associated protostellar envelopes.
In this paper we use the same sample of nearby ($d \lesssim 400$\,pc) low-mass cores with well defined
outflows used by \citet{Hull2012}, 
%plus L1455~IRS~1, and 
minus IRAS~16293~A, which was not a TADPOL source.

The outflow angles are the same as those used in \citet{Hull2012}; the
values for $\chi_{\textrm{sm}}$ typically differ by a few degrees because of the inclusion of additional data.
Note that we do not include SFRs in this analysis, nor do we include sources with
complicated outflow structure such as SVS13 (Figure \ref{fig:SVS13}) and OMC2-FIR3/4 (Figure \ref{fig:FIR4}).
All sources included in Figure \ref{fig:ks} have an asterisk (*) next to their outflow orientation
in Table \ref{table:obs}.

%Here we compare not only outflow orientation vs. small-scale B-field, but also vs. large-scale B-field.

In this paper we extend this analysis to include a comparison of outflow orientations vs. 
large-scale B-fields. Additionally, for each of these comparisons we split the sources into 
high- and low-polarization subsamples and plot
a separate CDF for each.  The heavy dashed and solid curves in Figure \ref{fig:ks} correspond
to the high- and low-polarization subsamples, respectively.

%\textbf{See Figure \ref{fig:ks} for two sets of distributions:
%(1) the distribution of outflow vs. \textit{large-scale} B-field orientations for the
%high- and low-polarization subsets of sources, and (2) the distribution of outflow vs. \textit{small-scale} B-field orientations for the
%both subsets.}

%In Figure \ref{fig:ks} we plot the alignment of outflows with both large- and small-scale B-fields. 
%In each plot we divide the data into the high- and low-polarization samples.  

%In high-polarization sources, the outflows appear to be randomly aligned with the B-fields on
%both large and small scales.  However, for low-polarization sources, 
%despite having consistent large-to-small scale B-field orientations, 
%the high-polarization sources have outflows that are \textit{randomly aligned} with B-fields
%on both large and small scales.  However, the low-polarization sources have  
%outflows and small-scale B-fields that are \textit{preferentially perpendicular.}

As discussed in \cite{Hull2012}, the B-field and outflow position angles we observe are projected onto the plane of the sky.
To determine if the large scatter in position angle differences could be due to projection effects, 
we compare the results with Monte Carlo simulations where the outflows and B-fields are 
tightly aligned, somewhat aligned, preferentially perpendicular, or randomly aligned.

For the tightly aligned case, the simulation randomly selects pairs of vectors in three dimensions that are within
20$\degree$ of one another, and then projects the vectors onto the plane of the sky
and measures their angular differences.  The resulting CDF is shown in 
Figure \ref{fig:ks}.  In this case projection effects are not as problematic as one might think: to have a
projected separation larger than 20$\degree$ the two vectors must point almost along the line of sight.

For the somewhat-aligned and preferentially-perpendicular cases the simulation randomly selects pairs of vectors that are separated
by 0--45$\degree$ or 70--90$\degree$, respectively.   
  In these cases projection effects are more important and result in CDFs that are
closer to that expected for random alignment, shown by the thin straight line (see Figure \ref{fig:ks}).

%The dashed curves in Figure \ref{fig:ks} correspond to the sources with high CARMA polarization 
%fractions.  Despite having consistent large-to-small scale B-field orientations, these sources have outflows and B-fields
%that are \textit{randomly aligned} on both large and small scales. This suggests that even in cases
%where the ratio of magnetic to turbulent energy is high, B-fields still do not play
%an obvious role in setting the final angular momentum directions of protostellar outflow
%and circumstellar disks (which we presume to be perpendicular to the outflows they launch).  
%
%The solid curves correspond to the sources with low CARMA polarization 
%fractions.  There is no clear consistency between large- and small-scale B-field orientations
%in these sources (see Figure \ref{fig:frac}).  And while the outflows and large-scale B-fields are randomly
%aligned in the low-polarization sources, the outflows and small-scale B-fields in these sources are \textit{preferentially
%perpendicular.  This suggests that in sources where the ratio of magnetic to turbulent energy is low
%and the B-fields are more twisted, the B-fields may be wrapped up by envelope rotation at small, $\sim$\,1000\,AU scales.}

In all four cases in Figure \ref{fig:ks} a Kolmogorov--Smirnov (K-S) test rules out the
 scenario where outflows and B-fields are tightly aligned
(the K-S probabilities for all distributions are $< 0.002$).  This is consistent with the results from
 \citet{Hull2012}, who found that outflows and small-scale B-fields are not tightly aligned.

The K-S test also shows that all of the distributions 
are consistent with random alignment.  
However, in low-polarization sources the K-S test gives a probability of only 0.12 that 
 the outflows and small-scale B-fields are randomly aligned, hinting\footnote{We use the word ``hint'' because typically a K-S
 test is considered to be definitive only when the statistic is $< 0.1$.}
 that they may be preferentially perpendicular.
(Note that the K-S test does not take into account the dispersions in the B-field orientations
 reported in Table \ref{table:obs}.)

%However, there is a hint (K-S probability of 0.12) that 
%in low-polarization sources the outflows are \textit{not} randomly aligned with small-scale B-fields; 
%rather, they may be preferentially perpendicular.}

We speculate that the polarization fractions are low in these sources because 
B-fields have be wrapped up toroidally by envelope rotation.
Rotation at $\sim$\,1000\,AU scales has been detected in at least two of the sources: see 
N$_2$H$^+$ observations of CB~230 and CB~244 by \citet{Chen2007} using OVRO (the Owens Valley
Radio Observatory).  The envelope rotation axes are roughly aligned
with the outflow axes in both of these sources.  
%Observations of envelope rotation in our entire low-polarization subsample will be
%needed to confirm that rotation is a plausible explanation for the perpendicular outflows and small-scale B-fields.

%This implies that the angular momentum vectors of the
%circumstellar disks that ultimately form deep within the protostellar
%envelopes, and which launch the jets and outflows that are ubiquitous in
%young stars, are not aligned with the fields in the cores from which
%they formed.  This in turn suggests that different---possibly
%turbulent---star-formation mechanisms are at work at the scales where the protostar and
%its disk finally form.

This result could have important consequences for the formation of circumstellar disks within rotating envelopes,
% Azimuthal wrapping is not the same as misaligned...  While it is not
% entirely clear whether core rotation can affects fields on $\sim$\,1000\,AU
% scales,
%The B-field vs. outflow angle distribution from Figure 2 in \citet{Hull2012}
%showed hints of an increase in the
%number of sources with nearly perpendicular outflows and B-fields.
%This suggests that fields in some
%low-mass cores may be wrapped azimuthally by core rotation; 
since preferential misalignment of the B-field and the rotation axis should
allow disks to form more easily \citep{2009A&A...506L..29H,
  Krasnopolsky2012, Joos2012, Li2013}.  Objects with misaligned B-fields and rotation axes are less susceptible to
the ``magnetic braking catastrophe,'' where
magnetic braking prevents the formation of a rotationally supported Keplerian
disk \citep{Allen2003, Li2011}.  Indeed, these models suggest that misalignment may be a necessary
condition for the formation of disks (see also \citealt{Krumholz2013}).

%The preferentially perpendicular outflows and small-scale B-fields 
%in the low-polarization sources can be interpreted two ways: (1) envelope rotation wraps up
%the B-fields \textit{before} the disk and outflow form, thus making disk formation possible.  
%Or (2) the B-fields are wrapped up by envelope rotation \textit{after} disk and outflow formation because the B-fields 
%were unable to prevent disk formation.  In case (1), our results may
%have solved the magnetic braking catastrophe.  In case (2), the magnetic braking
%catastrophe may not be as large a problem as originally predicted.

What about the high-polarization population?  These could be sources where we do not have the
angular resolution to see B-field twisting and instead are seeing a bright sheath of polarized
material that has retained the ``memory'' of the global B-field.
Perhaps these are younger sources, or perhaps cores can form with a 
wide range of B-field strengths \citep[e.g.,][]{VazquezSemadeni2011} and some are strong
enough to resist twisting.

%the reason for the existence of low- and high-polarization cores
%populations is unclear.
%It could be an age effect, where the low-polarization sources are older and have had more time to 
%wrap up the small-scale B-fields and dissipate the bright sheath of polarized material that appears to have retained
%the ``memory'' of the global B-field in the high-polarization sources.  Or perhaps the two populations
%simply reflect the fact that cores can form with a wide range of B-field strengths \citep[e.g.,][]{VazquezSemadeni2011}, 
%and that the high-polarization sources happen to have formed with B-fields that are strong enough to resist twisting.

It is important to emphasize that even if we are seeing wrapped small-scale B-fields in the low-polarization sample,
the scales we are probing are $\sim$\,500--1000\,AU envelope scales, \textit{not} $\sim$\,100\,AU disk scales.
Consequently, the B-fields would have been wrapped up by the envelopes and not by the disks.
However, many simulations \citep[e.g.,][]{Machida2006, Myers2013} expect the B-fields 
in a protostar to be wrapped up at disk scales, regardless of the larger-scale
B-field morphology in the envelope and the core.  
%Higher resolution observations of dust polarization with ALMA will reveal whether B-fields are indeed 
%wrapped up by circumstellar disks.  
If this is the case, then with sufficient angular resolution ALMA should see perpendicular B-fields and outflows even
in our high-polarization sample.

%If this is the case, and if the envelope- and disk-rotation axes are aligned,
%then the B-field orientations in the low-polarization cases should be consistent from $\sim$\,1000\,AU envelope
%scales to $\sim$\,100\,AU disk scales, whereas the B-field orientations should be inconsistent in the high-polarization cores,
%whose $\sim$\,1000\,AU-scale B-fields are still parallel with the global B-field.

\smallskip

%Mention that (1) doesn't seem very reasonable, because
%you wouldn't expect field wrapping to be possible if B-fields were strong enough?  
%
%Mention that large
%disks tend to be formed in the sources with perpendicular B-fields (L1527), and that disks
%in aligned sources are tiny (L1157)?  

One possible concern with this analysis is that
outflows could disrupt the small-scale B-fields in the protostellar envelopes.  And indeed, in a few
sources we see hints that the fields are stretched along the direction of the
outflow [\eg{} NGC-1333~IRAS~2A (Figure \ref{fig:IRAS2A}), HH~211~mm (Figure
  \ref{fig:HH211}), Ser-emb~6 (Figure \ref{fig:SerEmb6}), and L1157 (Figure
  \ref{fig:L1157})].  However, these detections tend to be quite far from the
central intensity peak, where the B-field orientation is usually different.  This suggests
that while outflows may drag B-fields along with them, the outflows do not
disrupt the B-fields in the densest parts of the protostellar
envelope.

Another concern is that over time outflows could have changed direction,
%as a result of the disk being torqued by asymmetric infall of
%envelope material, 
and that deep in the core the outflows and B-fields could
actually be aligned.  
However, many sources show 
bipolar ejections with consistent position angles over parsec scales. 
Some examples of such sources from the TADPOL survey include HH~211~mm \citep{Lee2009}, 
L1448~IRS~2 \citep{Tobin2007, OLinger1999},
L1157 \citep{Gueth1996, Bachiller1997}, 
L1527 \citep{Hogerhiejde1998}, and VLA~1623 \citep{Andre1990}.

A source that helps dispel the above concerns is OMC3-MMS6,
which has a very small bipolar outflow with
a dynamical age of only 100\,yr \citep{Takahashi2012a}, too young to have either perturbed the B-field
or changed direction appreciably.  As is clear in the maps in Figure \ref{fig:MMS6}, the outflow is
not aligned with either the large- or the small-scale fields around MMS6,
suggesting that the orientation of the disk launching the
outflow truly is misaligned with the B-field in the envelope.

%For L1157 there is Gueth 1996, Bachiller 1997.
%HH211 - Lee, Chin-Fei 2009
%L1448 IRS2 Tobin et al. 2007, O'Linger et al. 1999
%L1527 Hogerhiejde et al. 1998 Figure 6
%VLA 1623 Andre et al. 1990

\subsection{Fractional polarization ``hole''}
\label{sec:pol_hole}

\begin{figure*} [hbt!]
\begin{center}
\epsscale{1.1}
\plottwo{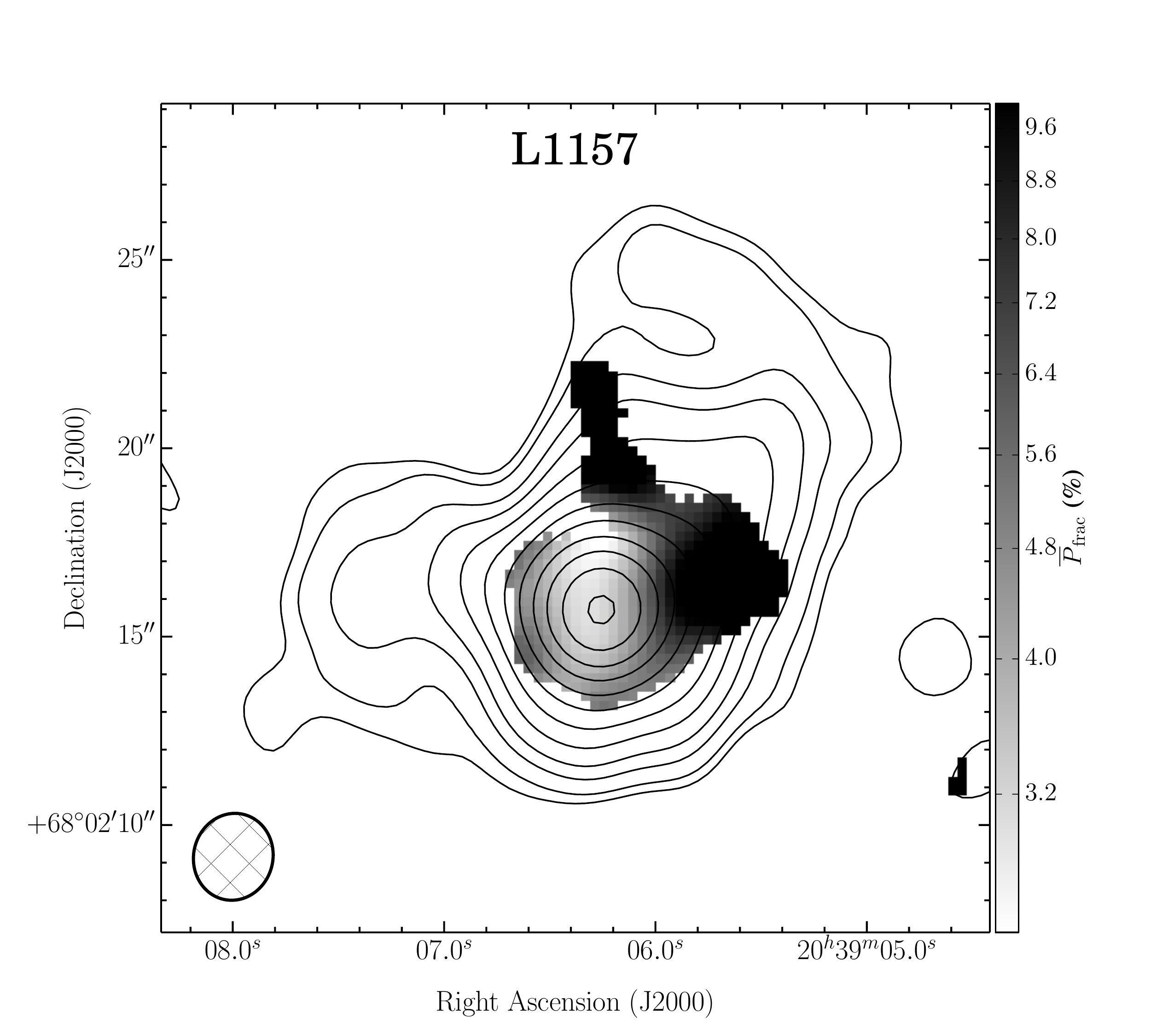}{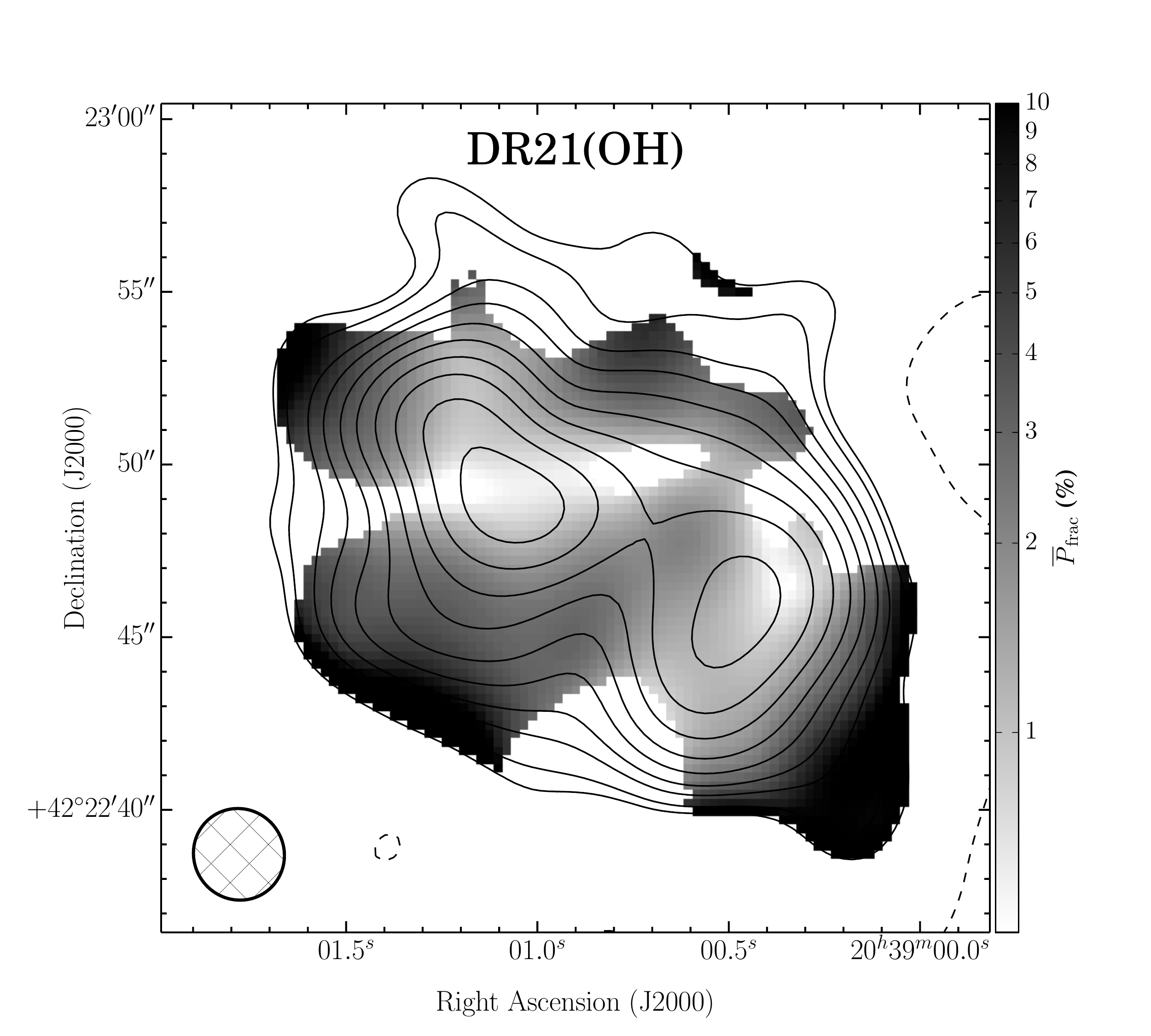}
\caption[]{ \footnotesize{ Sample maps of polarization fraction (grayscale), with dust continuum contours overlaid.   
The grayscale saturates at 10\% in order to emphasize the low polarization fraction near the center of each object; however,
the polarization fraction near the edge can be significantly higher.
The dust continuum contours in all dust maps are
--3,\,2,\,3,\,5,\,7,\,10,\,14,\,20,\,28,\,40,\,56,\,79,\,111,\,155,\,217\,$\times\,\sigma_I$ (see Appendix \ref{appendix:maps}).  Polarization fraction has only been 
plotted in locations with significant polarization detections (i.e., $P_c > 3.5\,\sigma_P$).}}
\label{fig:pct}
\end{center}
\end{figure*}

%In the CARMA maps we have chosen to scale the lengths of the line segments by
%the square root of the total polarized intensity, rather than by the fractional polarization, because it makes the B-field
%morphology easier to see.  
The ``polarization hole'' effect, where the fractional polarization of protostellar cores drops near their dust emission
peaks, is a well known phenomenon that has been seen in many previous observations 
\citep[\eg{}][]{Dotson1996,Matthews2002,Girart2006,Liu2013}.  We see the same effect in all of our maps,
for both nearby low-mass sources and distant high-mass sources; this shows that the polarization
hole effect is present across many size scales, although the reasons for the effect may be different at different scales. 
See Figure \ref{fig:pct} for sample maps of polarization fraction in L1157 and DR21(OH); these
maps show that in both cores and SFRs, the polarization fraction 
is higher at the edges and lower near the total intensity peaks.

For low resolution maps (\eg{} those with $\sim$\,20$\arcsec$ resolution from SCUBA, Hertz, and SHARP), a
plausible explanation of the polarization holes was unresolved structure that
was averaged across the beam.  However, in some of the higher resolution ($\sim$\,2.5$\arcsec$ resolution) maps presented
here and in previous interferometric observations, these
twisted plane-of-sky B-field morphologies have been resolved, and yet the drop in fractional
polarization persists.

There are multiple possible explanations. 
First, except for a very few lines of sight through the densest parts of protostellar disks,
millimeter-wavelength thermal dust emission is optically thin,
%\textbf{(should we make an estimate here?  Hua-Bai+2009 says
%cores have Av of 1000.)} 
and thus we are integrating along the LOS.
If the B-field orientation is not consistent along the LOS (due to turbulence or rotation, for example),
averaging will result in reduced fractional polarization.  
Second, there could still be unresolved B-field structure in the plane of the sky at scales
smaller than the $\sim$\,2.5$\arcsec$ resolution of the CARMA data \citep[e.g.,][]{Rao1998}.
And third, grains at the centers of cores
could be poorly aligned because grain
alignment is less efficient in regions with high extinction, or because collisions
knock grains out of alignment at higher densities.
Simulations of polarized emission from turbulent cores that include the above effects 
show the polarization hole \citep[e.g.,][]{Padoan2001, Lazarian2005, Bethell2007, Pelkonen2009}.

%It is important to note that the drop in polarization in our CARMA maps appears in both 
%the nearby objects in our sample as well as in 
%the SFRs, which tend to be $\sim$\,10 times further away.
%This suggests that the process causing de-polarization may be independent of physical scale.  

%It is important to note that the drop in polarization in our CARMA maps appears to be independent of absolute
%physical scale, as it is evident both in the nearby objects in our sample (see Appendix \ref{appendix:cores})
%as well as in the more distant SFRs (see Appendix \ref{appendix:SFRs}), which tend to be $\sim$\,10 times further away.
%This implies that the process causing de-polarization is independent of physical scale.  SFRs are known to
%be highly turbulent \citep{Elmegreen2004}, which suggests that the first explanation is more likely: that the drop in fractional
%polarization is caused by averaging disordered polarization along the LOS, with field twisting cause by turbulence.

\section{SUMMARY}
\label{sec:summary}

We have presented polarization maps of low-mass star-forming cores and 
high-mass star-forming regions from the TADPOL survey. Using
source-averaged B-field orientations and polarization fractions, we have studied the statistical 
properties of the ensemble of sources and have come to the following key conclusions:
%in a statistical sense from the ensemble of sources.}

\begin{enumerate}

\item[(1)] Sources with high CARMA polarization fractions also have consistent B-field orientations
on large ($\sim$\,20$\arcsec$) 
and small ($\sim$\,2.5$\arcsec$) scales.
We interpret this to mean that in at least some cases B-fields play a role in regulating the infall of material
all the way down to the $\sim$\,1000\,AU scales of protostellar envelopes.  

\item[(2)]  Outflows appear to be randomly aligned with B-fields; although, in 
sources with low polarization fractions there is a hint that outflows are preferentially perpendicular to small-scale B-fields, 
which suggests that in these sources the fields have been 
wrapped up by envelope rotation.

\item[(3)] Finally, even at $\sim$\,2.5$\arcsec$ resolution we see 
the so-called ``polarization hole'' effect,
where the fractional polarization drops significantly near the total intensity peak.

\end{enumerate}

As the largest survey of low-mass protostellar cores to date, the TADPOL project
sets the stage for observations with ALMA.  ALMA's unprecedented sensitivity
will allow us to answer the question of what happens to magnetic fields in very young
Class 0 protostars between the $\sim$\,1000\,AU
scales we probe in this work and the $\sim$\,100\,AU scales of the
circumstellar disks.
%where more evolved Class I and II stars like DG~Tau
%(Figure \ref{fig:DGTau}) show no significant polarization
%\citep{Hughes2009b, Hughes2013}.
The addition of ALMA data to the TADPOL sample will 
%In addition to allowing us to connect core- and disk-scale magnetic-field morphologies,
%the addition of ALMA data to the TADPOL sample 
%more CARMA data, as well as both
%new and archival SMA data, would be invaluable.  With CARMA and the SMA
%combined, the sample of interferometric polarization maps would approach
%$\sim$\,60, which 
also enable more robust statistical analyses of the
types done in both this work and in \citet{Hull2012}, and will allow us to
see trends in B-field morphology with source mass, age, environment, multiplicity, envelope
rotation, outflow velocity, and B-field strength.

% but which may have coherent fields on smaller scales if MHD turbulence in
% the disk tangles up fields in sub-scale-height turbulent cells.

% At very small scales the field morphology in the protostellar envelope
% should differ from what we see with CARMA, and should be either wrapped up
% in or perpendicular to the disk; what that scale is, however, or if the
% field is even detectable in the disk environment, remains to be seen.

% Future work could lead in other directions as well.  Several of the more
% highly polarized star-forming regions [\eg{} DR21(OH) (Figure
% \ref{fig:DR21OH}), Orion~KL (Figure \ref{fig:OrionKL}), NGC~7538 (Figure
% \ref{fig:NGC7538}), W3~Main (Figure \ref{fig:W3Main}), and W3~OH (Figure
% \ref{fig:W3OH})] are good candidates for the turbulent dispersion analysis
% method developed by \citet{Hildebrand2009} and \citet{Houde2009, Houde2011},
% which leads to a more robust estimate of magnetic field strength, and
% estimates the importance of MHD turbulence.

% Qizhou estimates:
% SMA low-mass: 30+
% SMA high-mass: <20  regions (>20 pointings)
% CARMA: 37

% Some sort of vector-by-vector analysis to determine alignment, rather than
% integrating over whole image Binarity (other stuff from Hull+12)

\acknowledgments

We would like thank the referee for the thorough and insightful comments,
which improved the paper significantly.

C.L.H.H. would like to acknowledge the advice and guidance of the members of
the Berkeley Radio Astronomy Laboratory and the Berkeley Astronomy Department.
In particular he would like to thank James Gao and James McBride, as well as
the authors of the APLpy plotting package, for helping make the Python plots of
TADPOL sources a reality.  He would also like to thank Nicholas Chapman for
helping to compile the SHARP data.

C.L.H.H. acknowledges support from an NSF Graduate Fellowship and from a Ford
Foundation Dissertation Fellowship. J.D.F. acknowledges support from an NSERC
Discovery grant.  J.J.T. acknowledges support provided by NASA through Hubble
Fellowship grant \#HST-HF-51300.01-A awarded by the Space Telescope Science
Institute, which is operated by the Association of Universities for Research
in Astronomy, Inc., for NASA, under contract NAS 5-26555.  N.R. acknowledges
support from South Africa Square Kilometer Array (SKA) Postdoctoral Fellowship
program.

Support for CARMA construction was derived from the states of California,
Illinois, and Maryland, the James S. McDonnell Foundation, the Gordon and
Betty Moore Foundation, the Kenneth T. and Eileen L. Norris Foundation, the
University of Chicago, the Associates of the California Institute of
Technology, and the National Science Foundation. Ongoing CARMA development and
operations are supported by the National Science Foundation under a
cooperative agreement, and by the CARMA partner universities.

\bibliographystyle{apj}

% \rmxaa not in emulateapj
%\let\jnl@style=\rmfamily
%\def\ref@jnl#1{{\jnl@style#1}}%
%\newcommand\rmxaa{\ref@jnl{Rev. Mexicana Astron. Astrofis.}} % Revista Mexicana de Astronomia y Astrofisica

\bibliography{ms}

%\vspace*{9in}
%\vspace*{1.4in}

\clearpage
\appendix

\section{\bf \large APPENDIX A: SOURCE MAPS}
\label{appendix:maps}

All maps from the TADPOL survey are publicly available as FITS images
and machine readable tables. 
For each figure below we include maps of Stokes
$I$, $Q$, and $U$; bias-corrected polarization intensity $P_c$;
polarization fraction $P_{\textrm{frac}} = P_c/I$; and inferred B-field orientation $\chi_{\textrm{sm}}$.  
Additionally, we include FITS cubes of total intensity (Stokes $I$) spectral-line data, as well as
machine readable tables listing the RA, DEC, $I$, $P_c$, $P_\textrm{frac}$, $\chi_{\textrm{sm}}$, and associated uncertainties 
of each line segment plotted in the figures.  
These files are available in a .tar.gz package available via the link in the figure caption.

%%% Maps of W3~Main
\begin{figure*} [tbph!]
\begin{center}
\epsscale{.86}
\plottwo{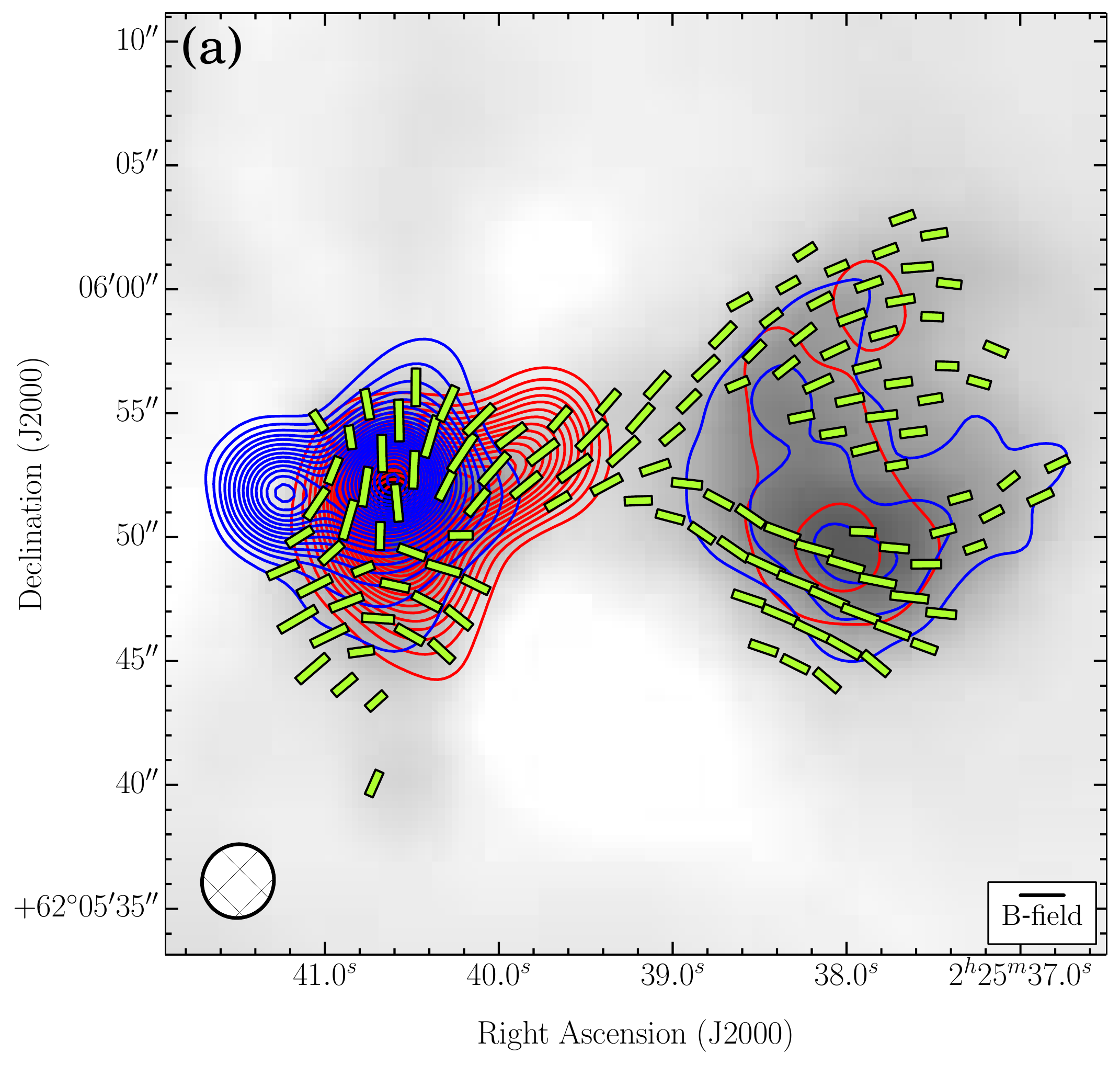}{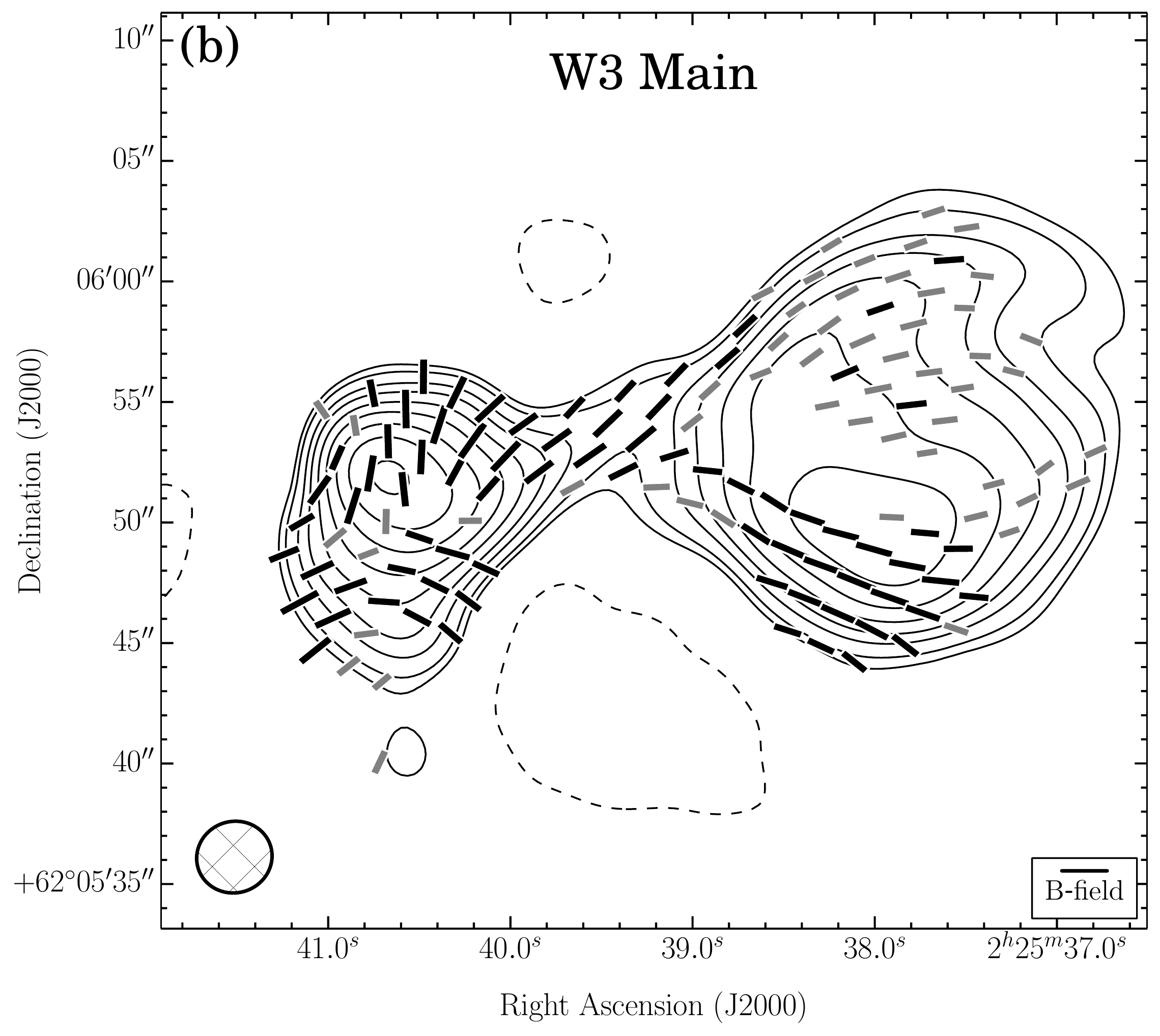}
\epsscale{0.58}
\plotone{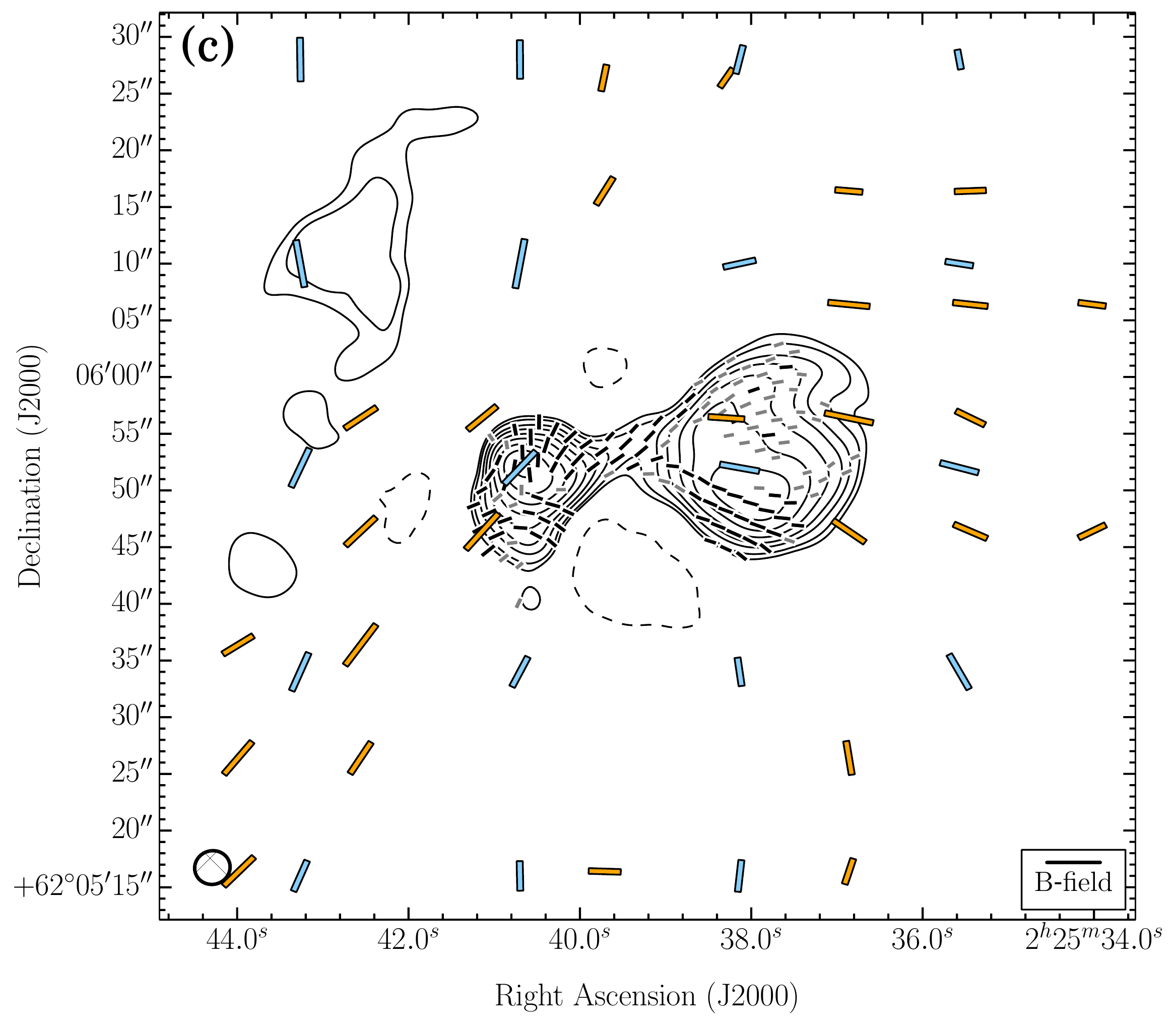}
\caption[]{ \footnotesize{ Maps of W3~Main.
    The line segments show the inferred magnetic field orientations; they
    have been rotated by 90$\degree$ relative to the polarization orientations.    
%    The longest segment is plotted in the location of the
%    maximum polarized intensity $P_{\textrm{c,max}}$ (see Table
%    \ref{table:obs}).
    Segments are
plotted twice per synthesized beam (resolution element) in locations
where $I > 2\,\sigma_I$ and $P_c > 2\,\sigma_P$, where $I$ is the total
intensity of the dust emission, $P_c$ is the bias-corrected polarized intensity, and
$\sigma_I$ and $\sigma_P$ are the rms noise values in the total and polarized
intensity maps, respectively.
%The scale bars show the peak intensity of polarized emission in \mjybm.  
\textbf{(a)}~The segment lengths are proportional to
    the square root of polarized intensity, \textit{not} fractional
    polarization.  The grayscale is proportional to the total intensity
(Stokes $I$) dust emission.   The blue and red contours are the blue- and redshifted spectral line wings.
The outflow orientation is indicated by a gray line for sources with outflows listed in Table \ref{table:obs}.
The velocity ranges of the CO($J = 2\rightarrow1$) emission in this map are
--2.3~to~--10.8~\kms{} (redshifted) and --67.9~to~--80.6~\kms{} (blueshifted).
The contour levels in all spectral line maps are 
4,\,8,\,12,\,16 and \,20,\,25,\,30\,...\,190,\,195,\,200\,$\times\,\sigma_{\textrm{SL}}$,
% 4,\,6,\,10,\,16,\,25,\,40,\,64,\,102\,$\times\,\sigma_{\textrm{SL}}$,
where $\sigma_{\textrm{SL}}$ is the rms noise measured in the spectral line
moment maps.  
In this map, $\sigma_{\textrm{SL}} = 0.50$~\Kkms.
\textbf{(b)}~Line segments
are black where $P_c > 3.5\,\sigma_P$ and gray where $2\,\sigma_P < P_c <
3.5\,\sigma_P$.  The grid on which the line segments are plotted is centered on
the polarization intensity peak $P_{\textrm{c,pk}}$, which is not necessarily
spatially coincident with the total intensity peak $I_{\textrm{pk}}$.  The
ellipses show the synthesized beams.  The dust continuum contours in all dust maps are
--3,\,2,\,3,\,5,\,7,\,10,\,14,\,20,\,28,\,40,\,56,\,79,\,111,\,155,\,217\,$\times\,\sigma_I$.
In this map, $\sigma_I = 8.8$~\mjybm{}.  
\textbf{(c)}~Same dust contours and  B-field orientations as in (b), with data
from three submillimeter polarimeters overlaid: SCUBA (in orange,
from \citealt{Matthews2009}), Hertz (in light blue, from \citealt{Dotson2010}),
and SHARP (in purple, from \citealt{Attard2009, Davidson2011, Chapman2013}). 
For SCUBA, Hertz, and SHARP data the segment lengths are proportional to the square root of the polarized intensity.
%with the exception of SHARP data from \citet{Attard2009}, where polarized intensity was not reported.
}}
\label{fig:W3Main}
\end{center}
\end{figure*}

%%% Maps of W3(OH)
\begin{figure*} [hbt!]
\begin{center}
\epsscale{1.1}
\plottwo{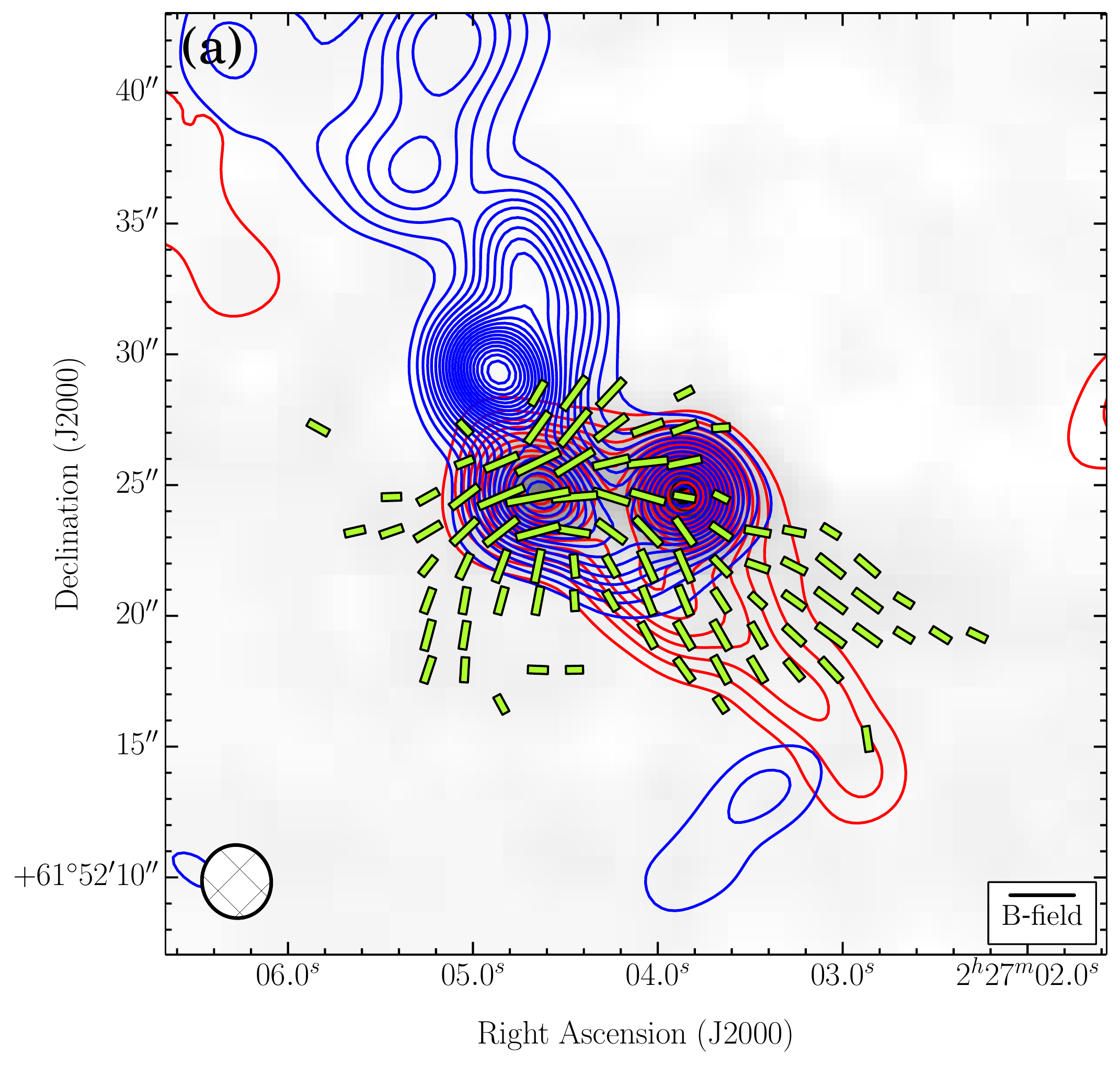}{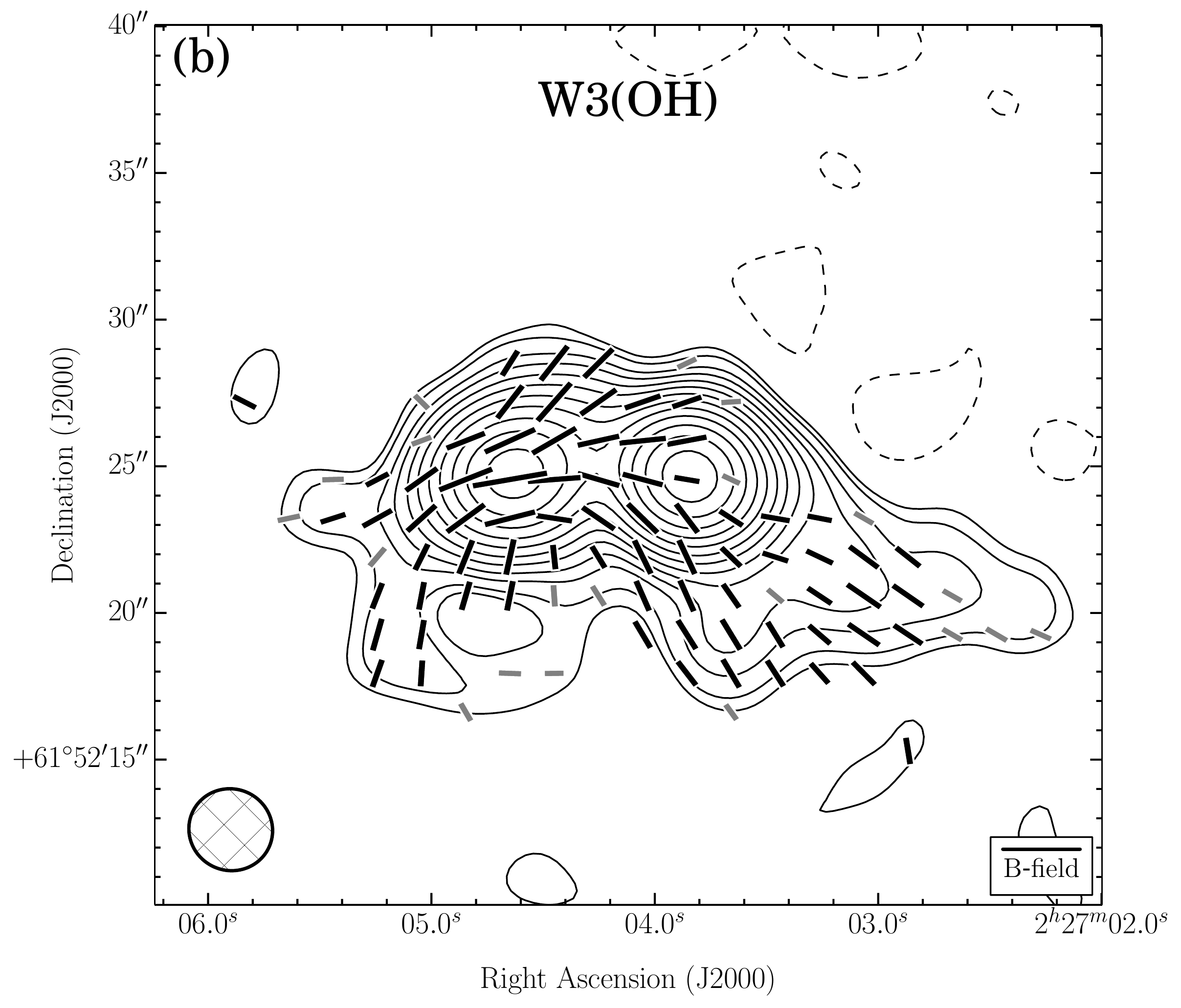}
\epsscale{0.8}
\plotone{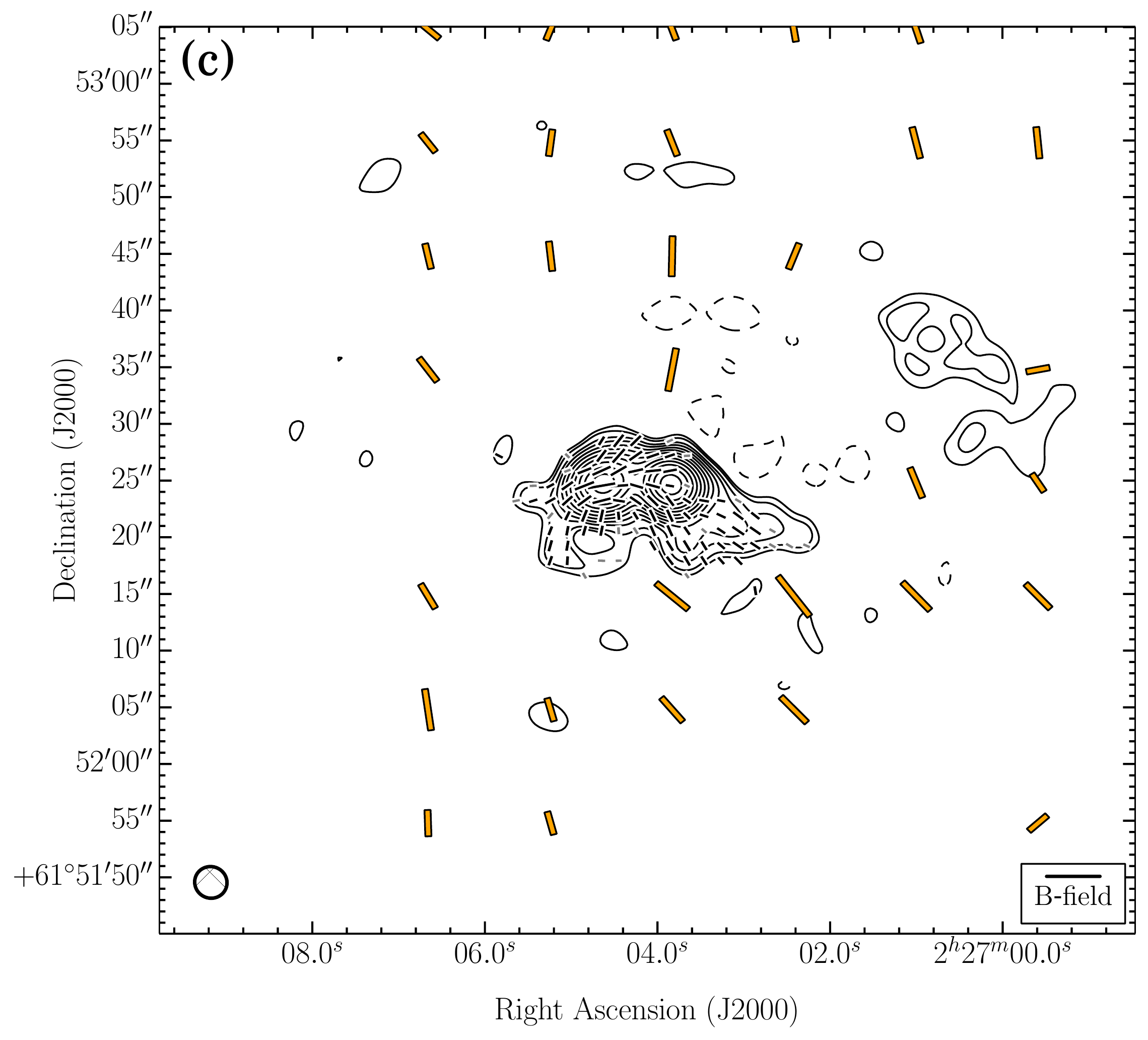}
\caption[]{ \footnotesize{
\input{W3OH_caption.txt}
}}
\label{fig:W3OH}
\end{center}
\end{figure*}

%%% Maps of L1448~IRS~2
\begin{figure*} [hbt!]
\begin{center}
\epsscale{1.10}
\plottwo{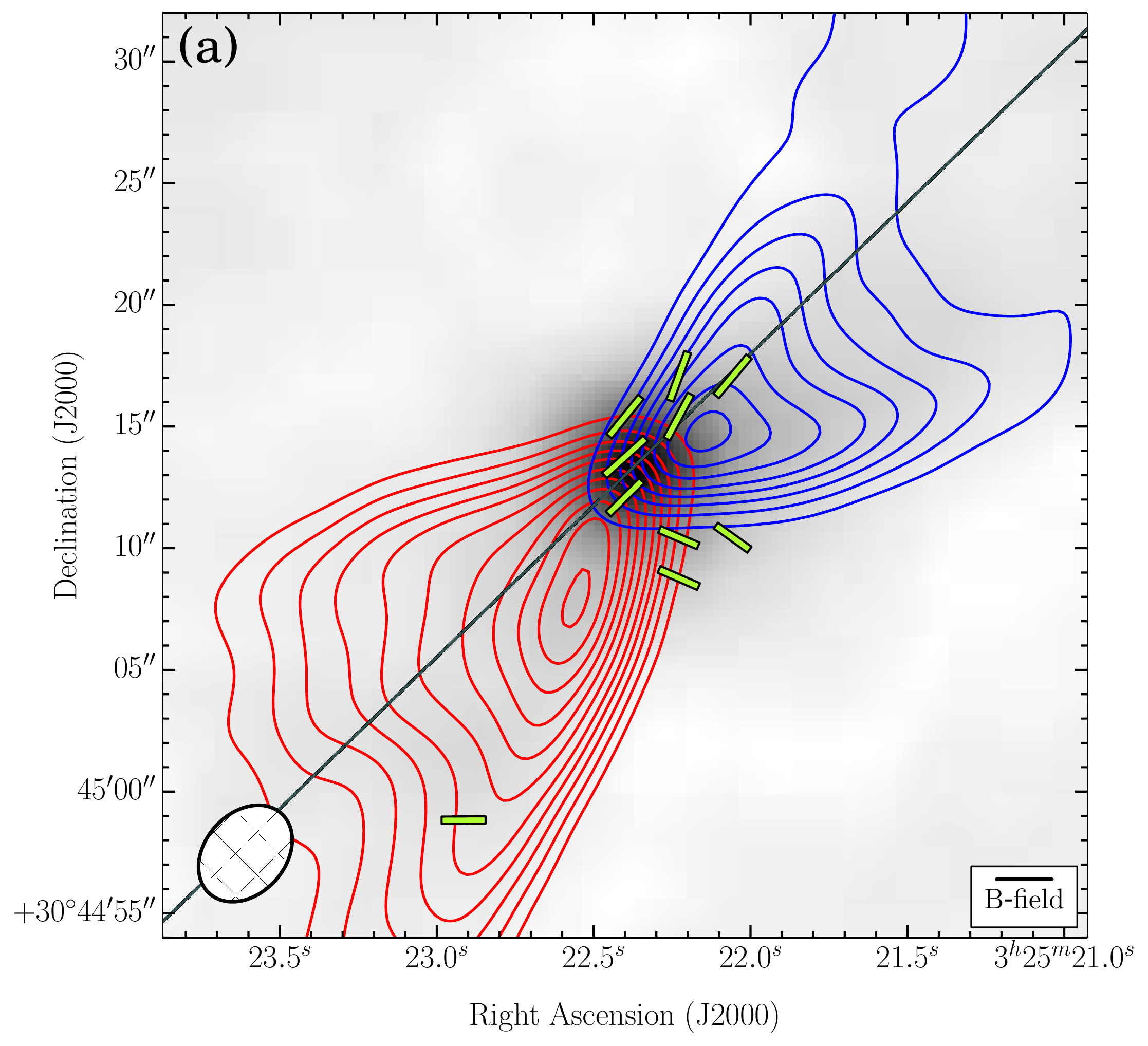}{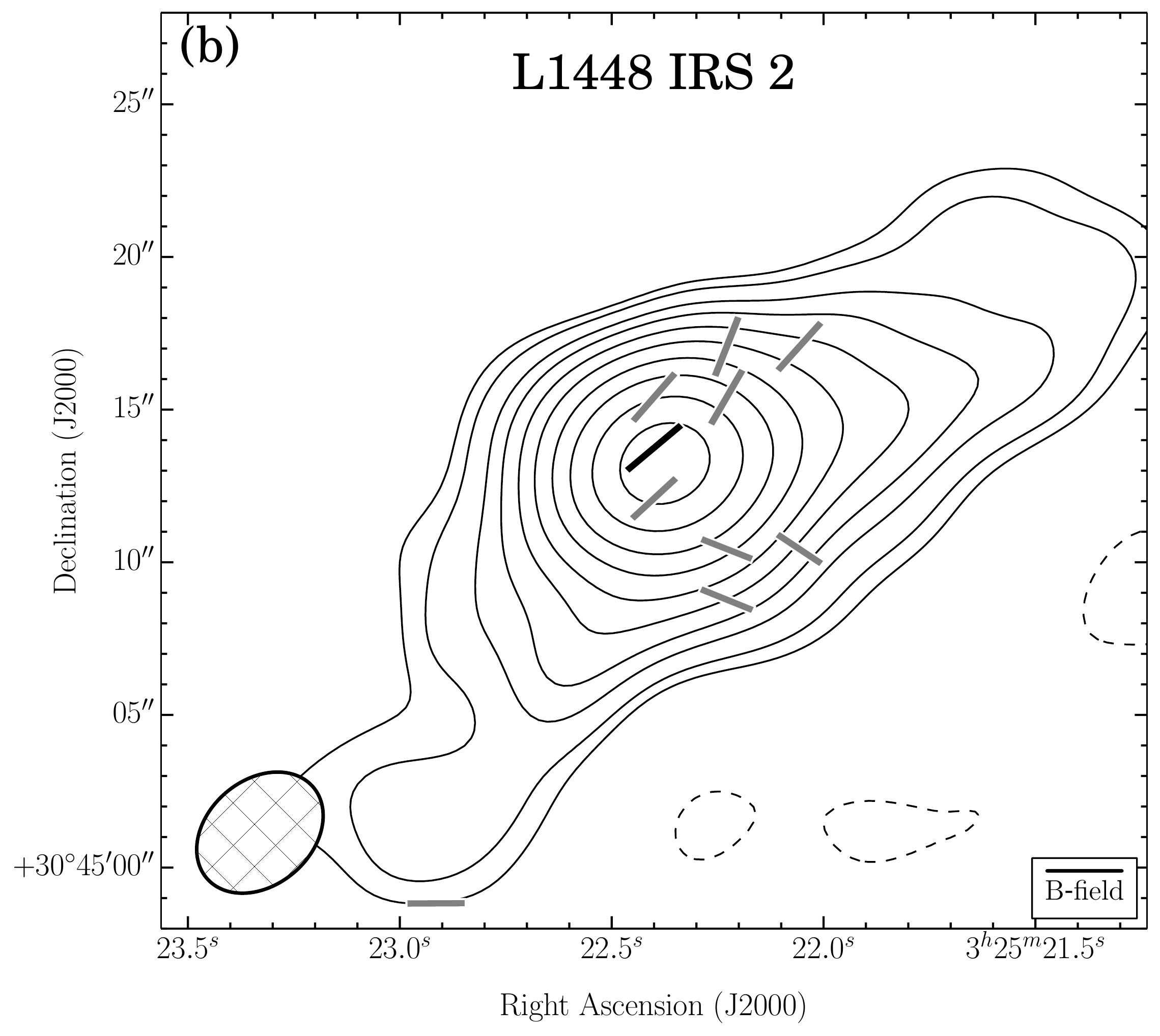}
\epsscale{0.8}
\plotone{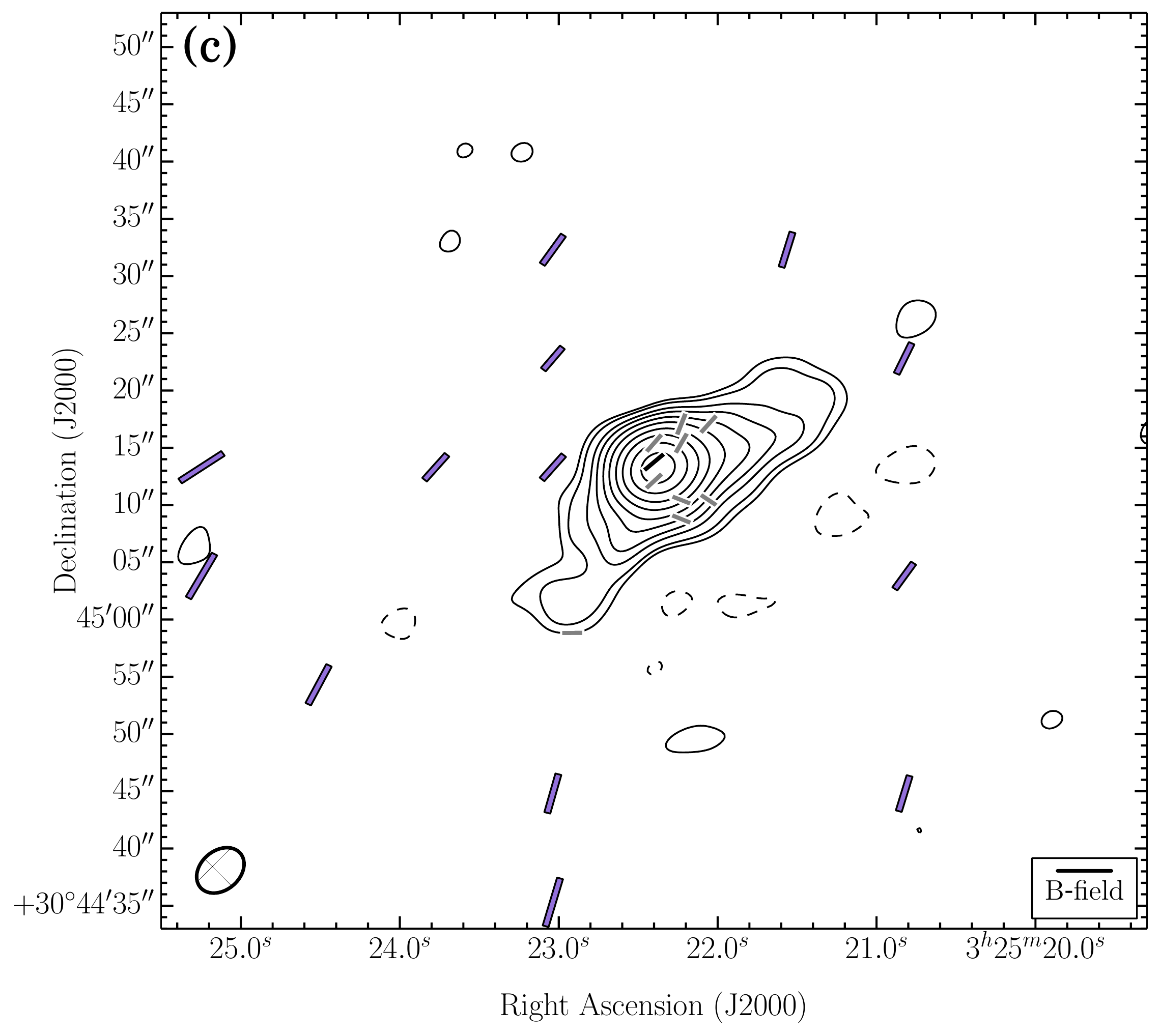}
\caption[]{ \footnotesize{
\input{L1448I_caption.txt}
}}
\label{fig:L1448I}
\end{center}
\end{figure*}

%%% Maps of L1448N(B)
\begin{figure*} [hbt!]
\begin{center}
\epsscale{1.1}
\plottwo{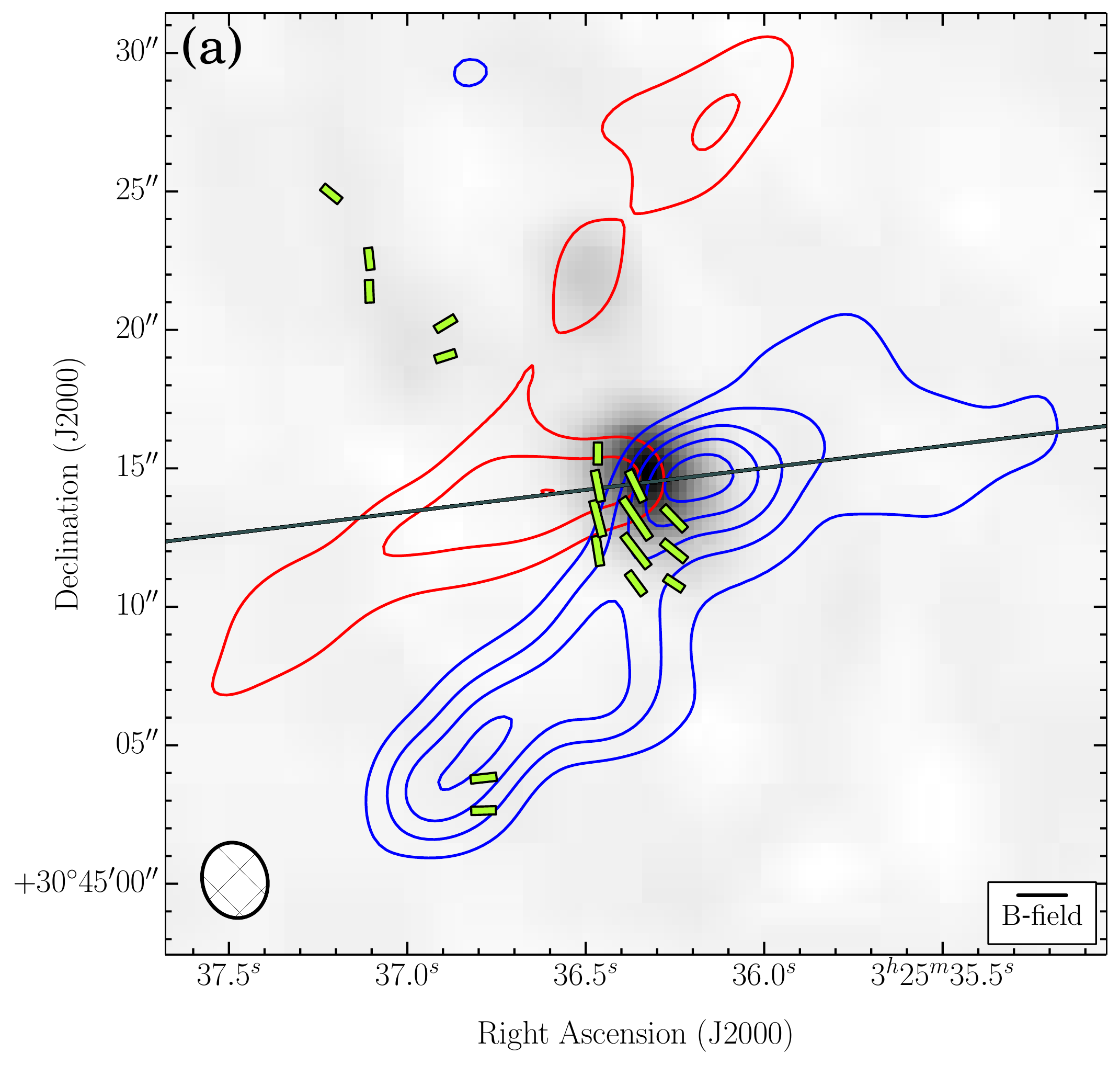}{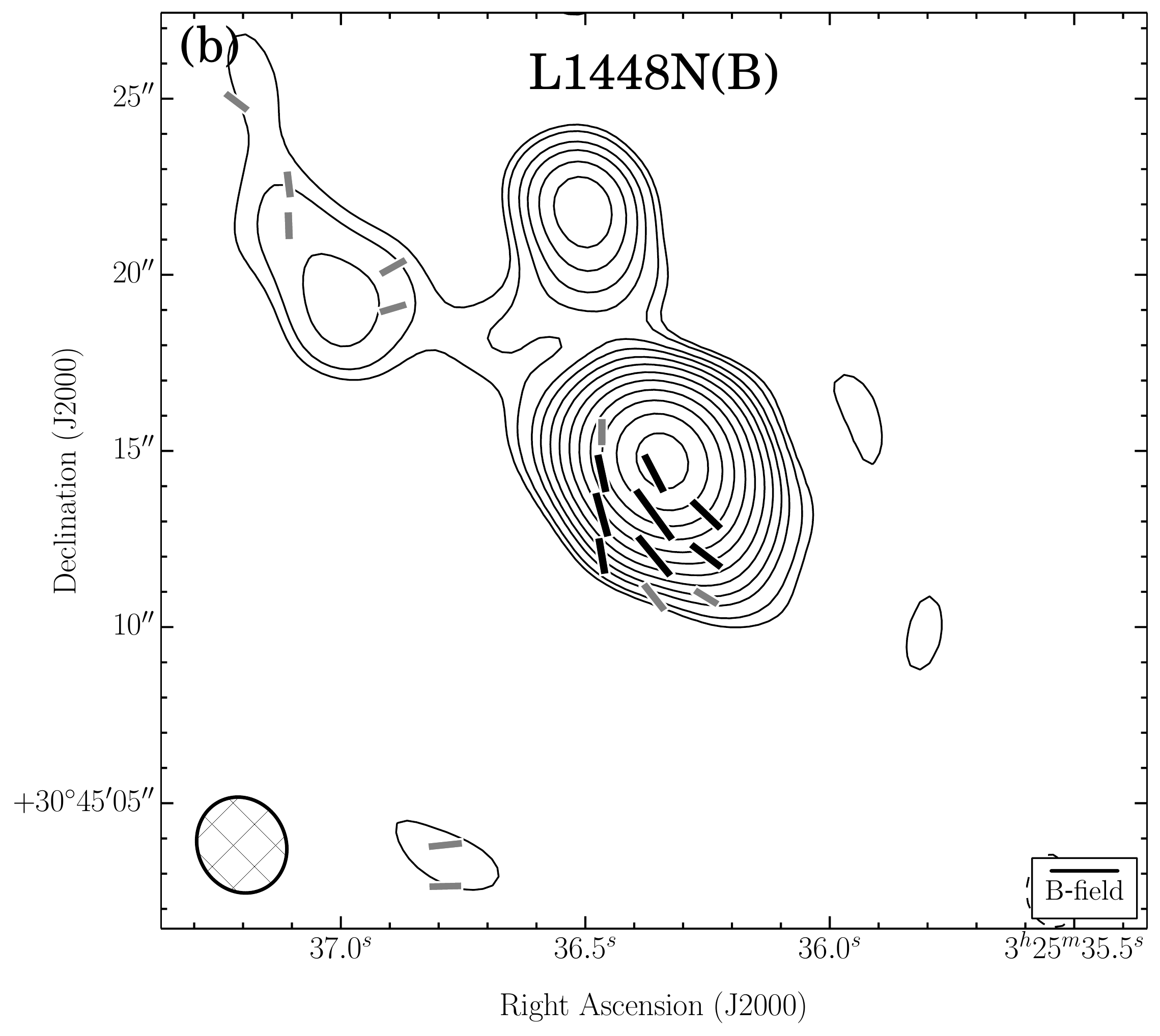}
\epsscale{0.8}
\plotone{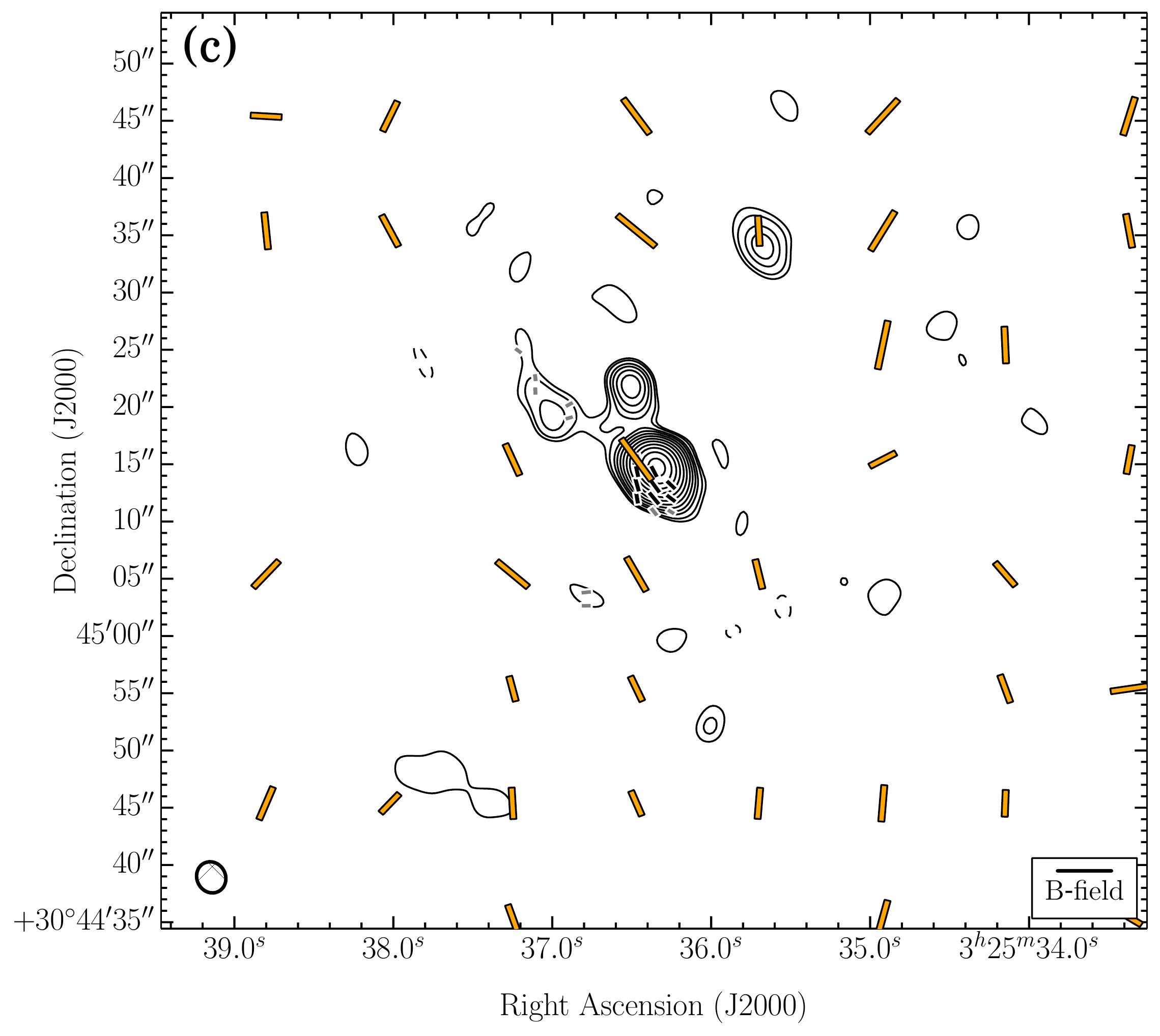}
\caption[]{ \footnotesize{
\input{L1448N_caption.txt}
}}
\label{fig:L1448N}
\end{center}
\end{figure*}

%%% Maps of L1448C
\begin{figure*} [hbt!]
\begin{center}
\epsscale{1.1}
\plottwo{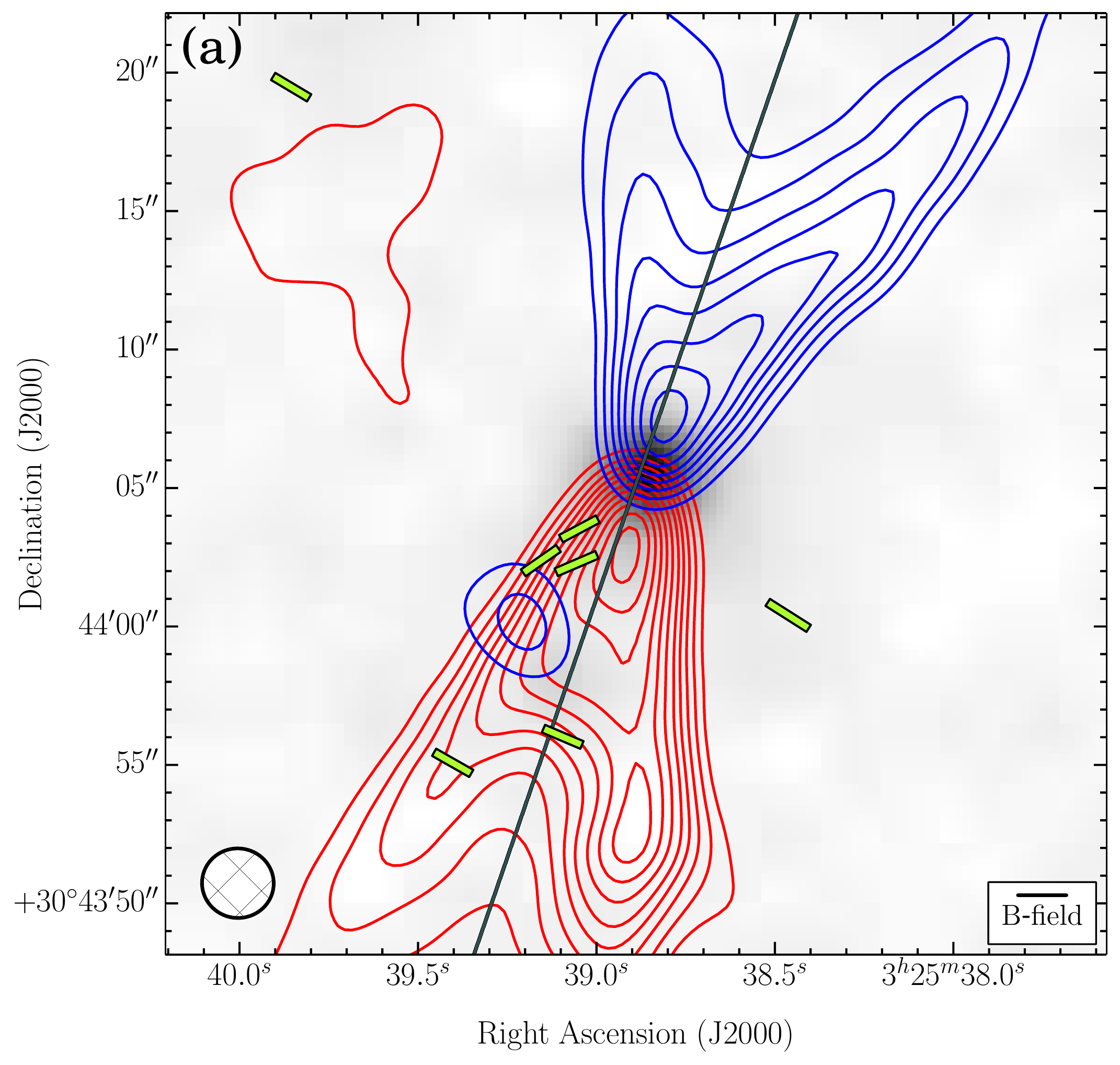}{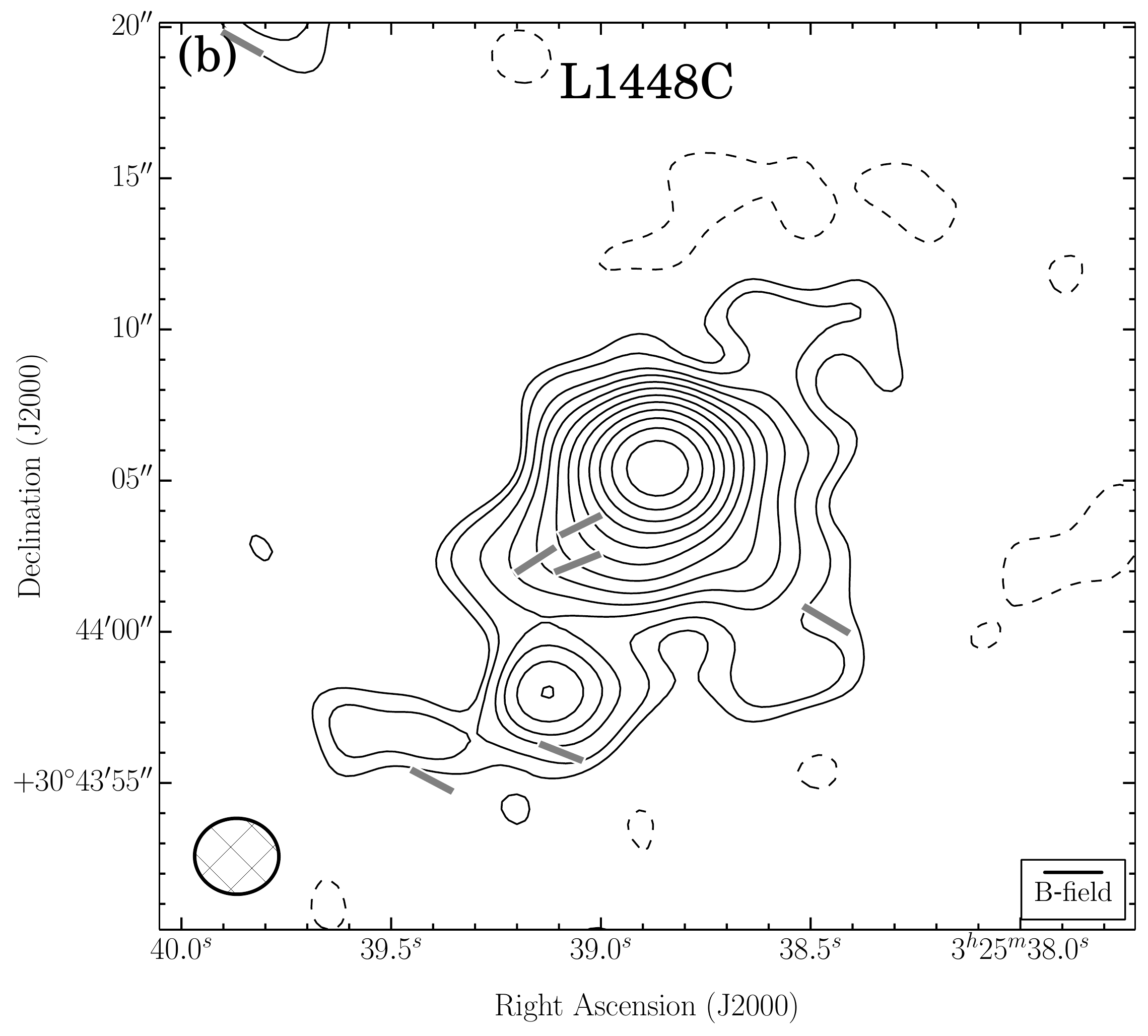}
\epsscale{0.8}
\plotone{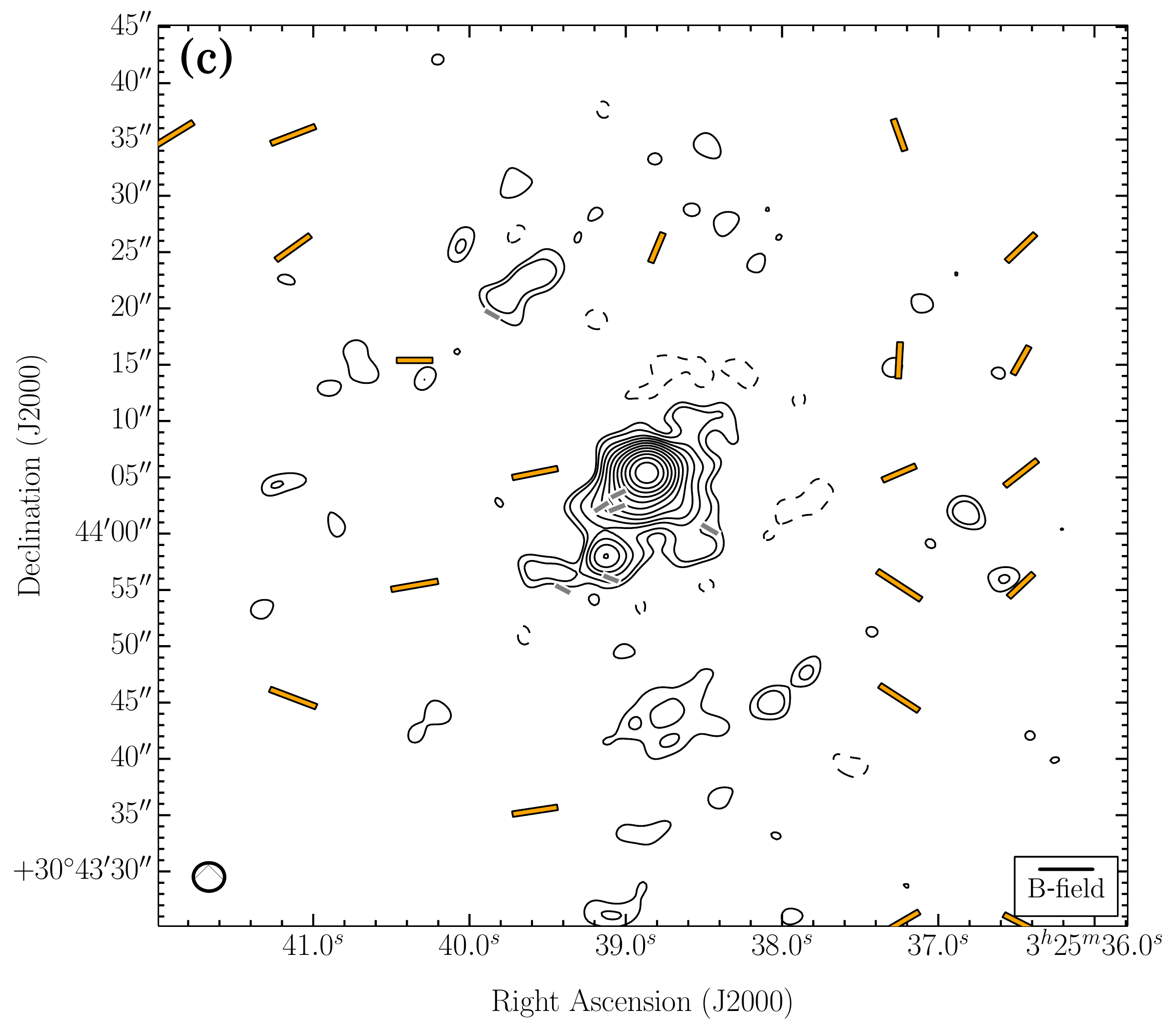}
\caption[]{ \footnotesize{
\input{L1448C_caption.txt}
}}
\label{fig:L1448C}
\end{center}
\end{figure*}

%%% Maps of L1455~IRS~1
\begin{figure*} [hbt!]
\begin{center}
\epsscale{1.1}
\plottwo{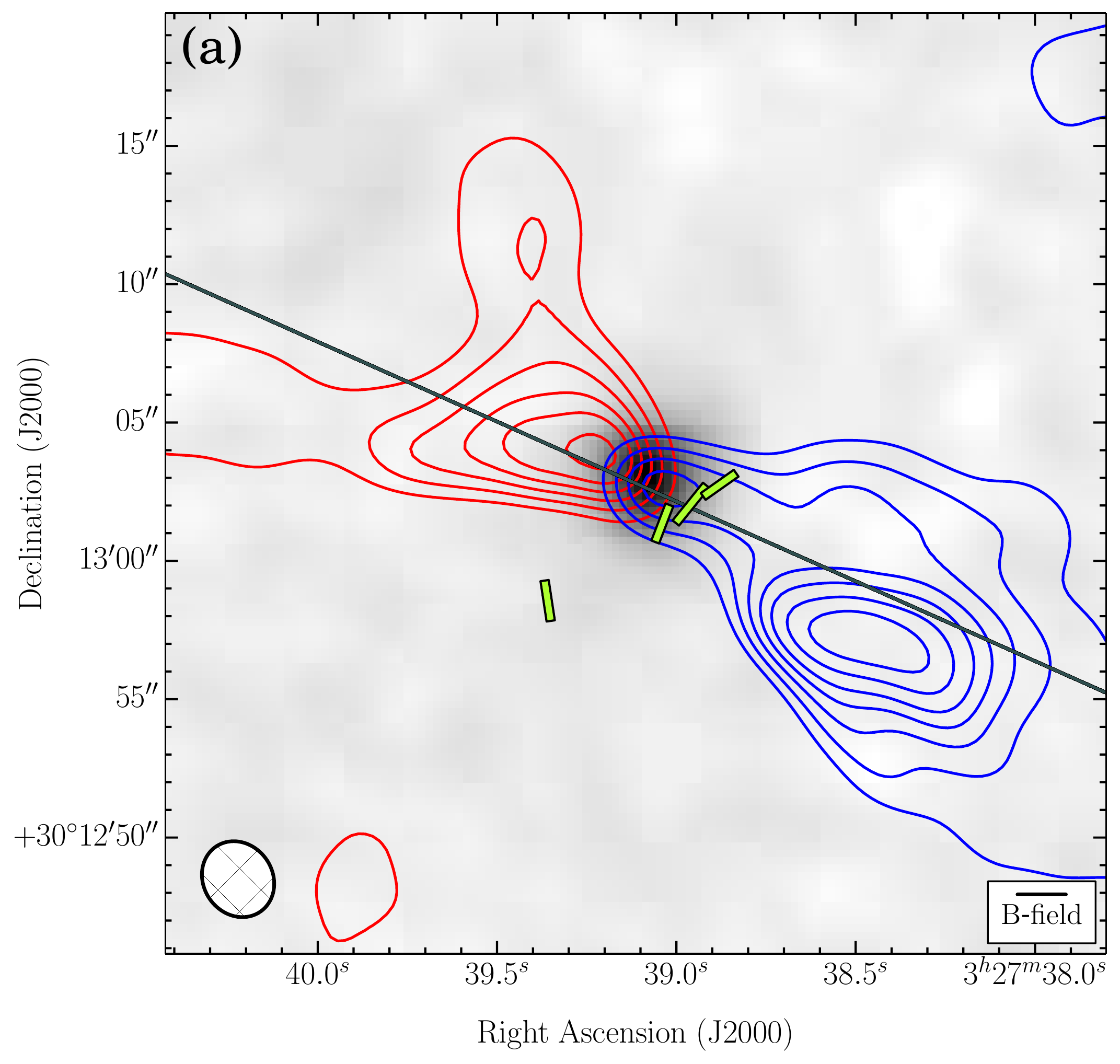}{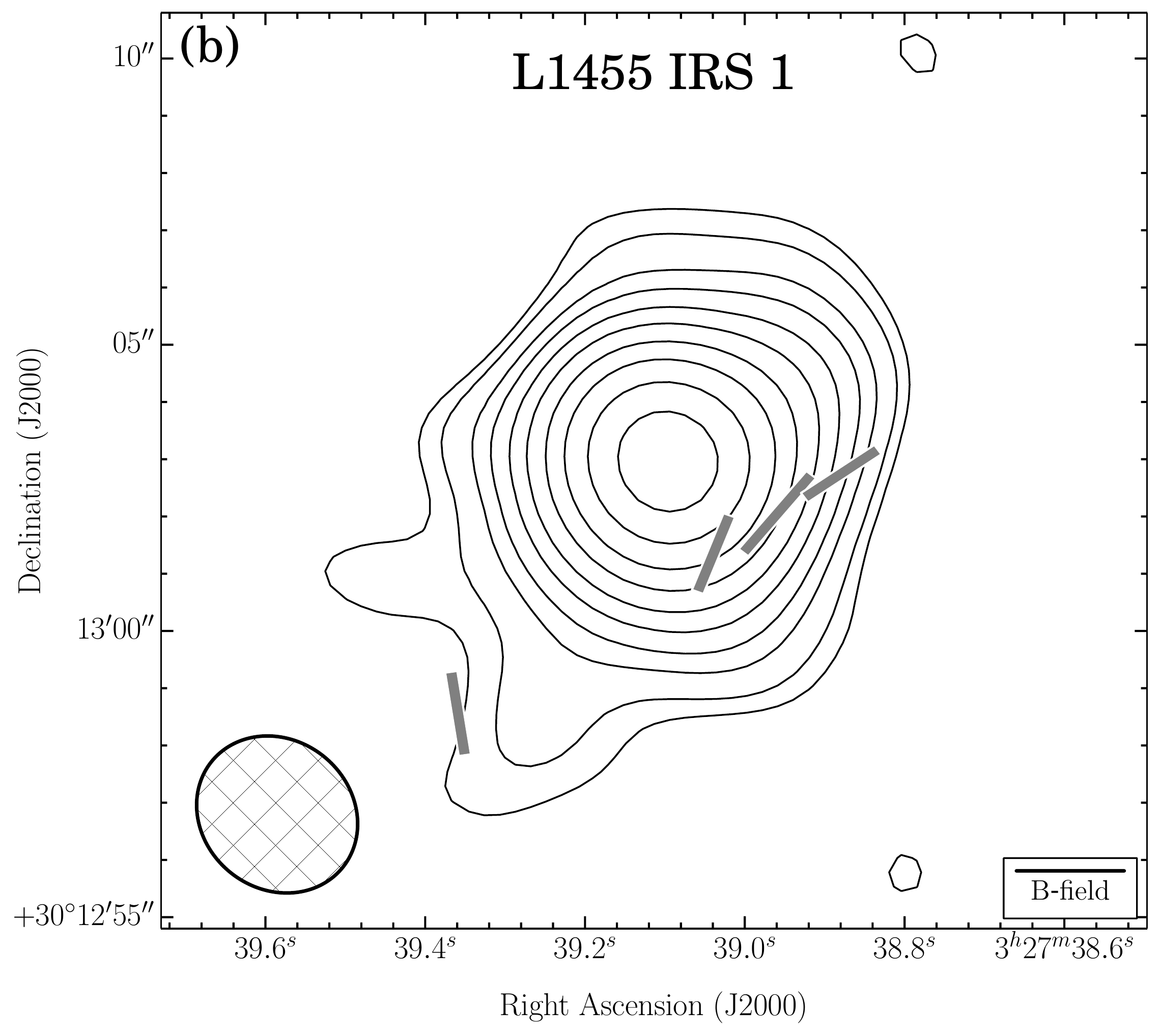}
\epsscale{0.8}
\plotone{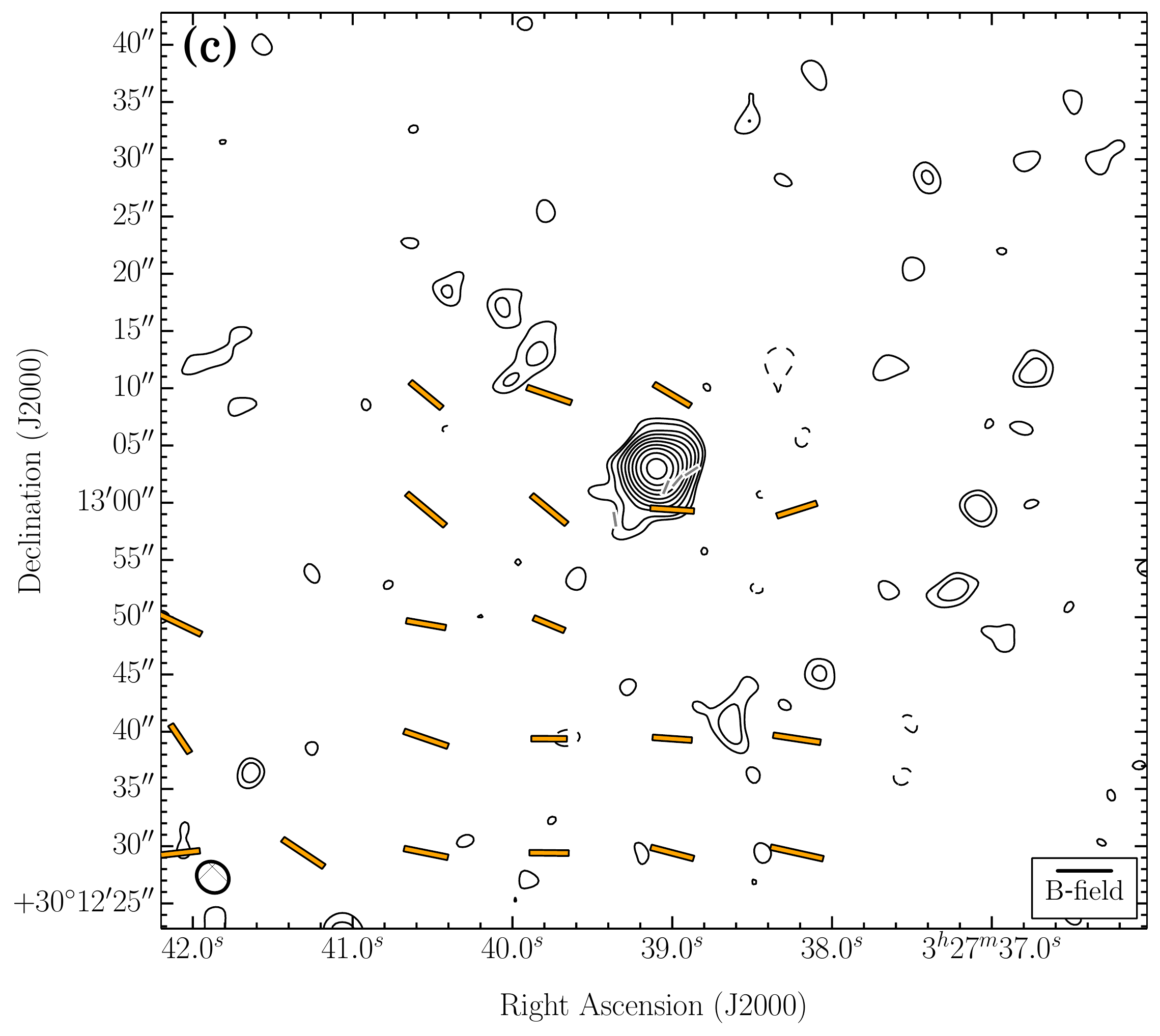}
\caption[]{ \footnotesize{
\input{L1455_caption.txt}
}}
\label{fig:L1455}
\end{center}
\end{figure*}

%%% Maps of NGC~1333-IRAS~2A
\begin{figure*} [hbt!]
\begin{center}
\epsscale{1.1}
\plottwo{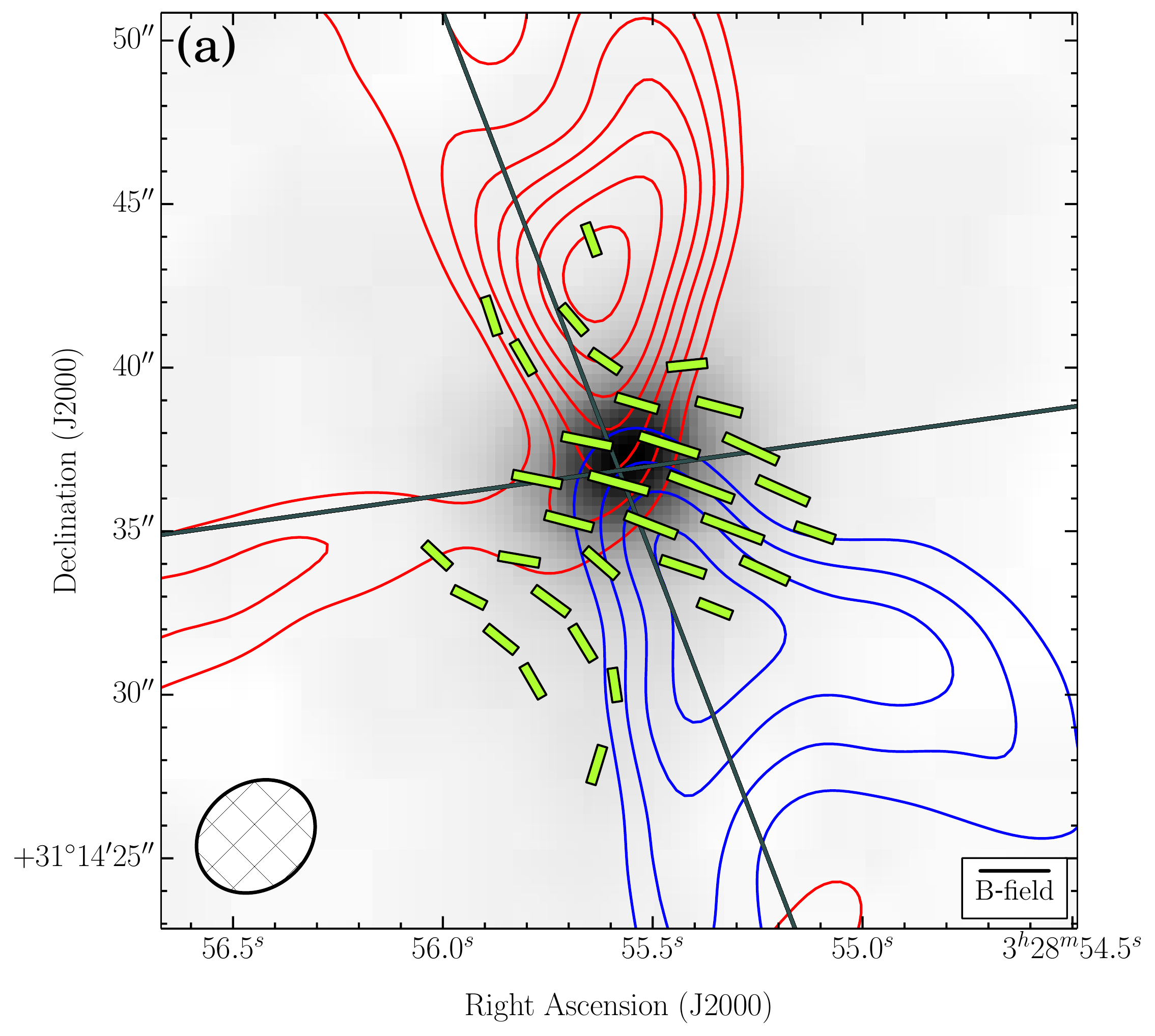}{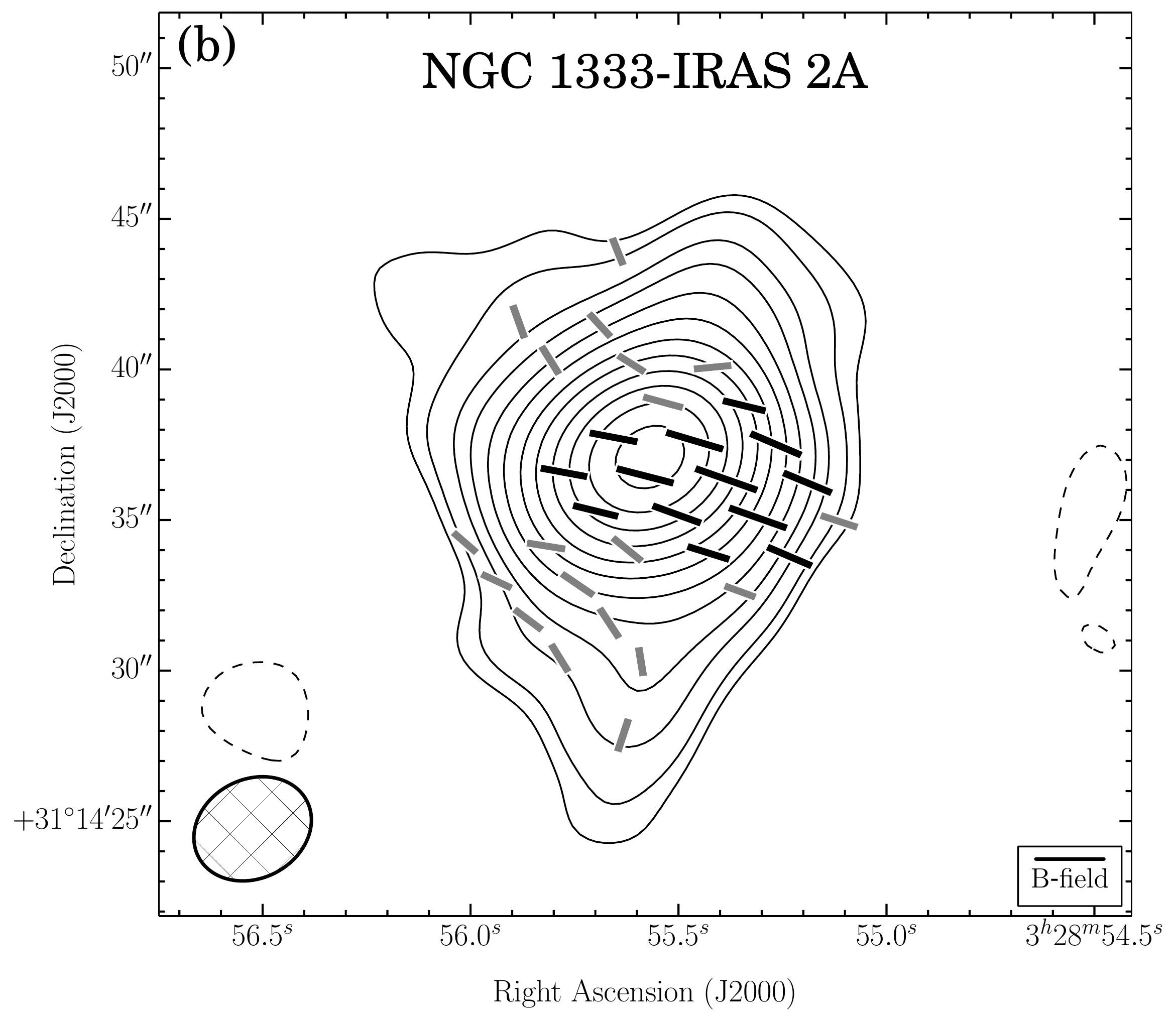}
\epsscale{0.8}
\plotone{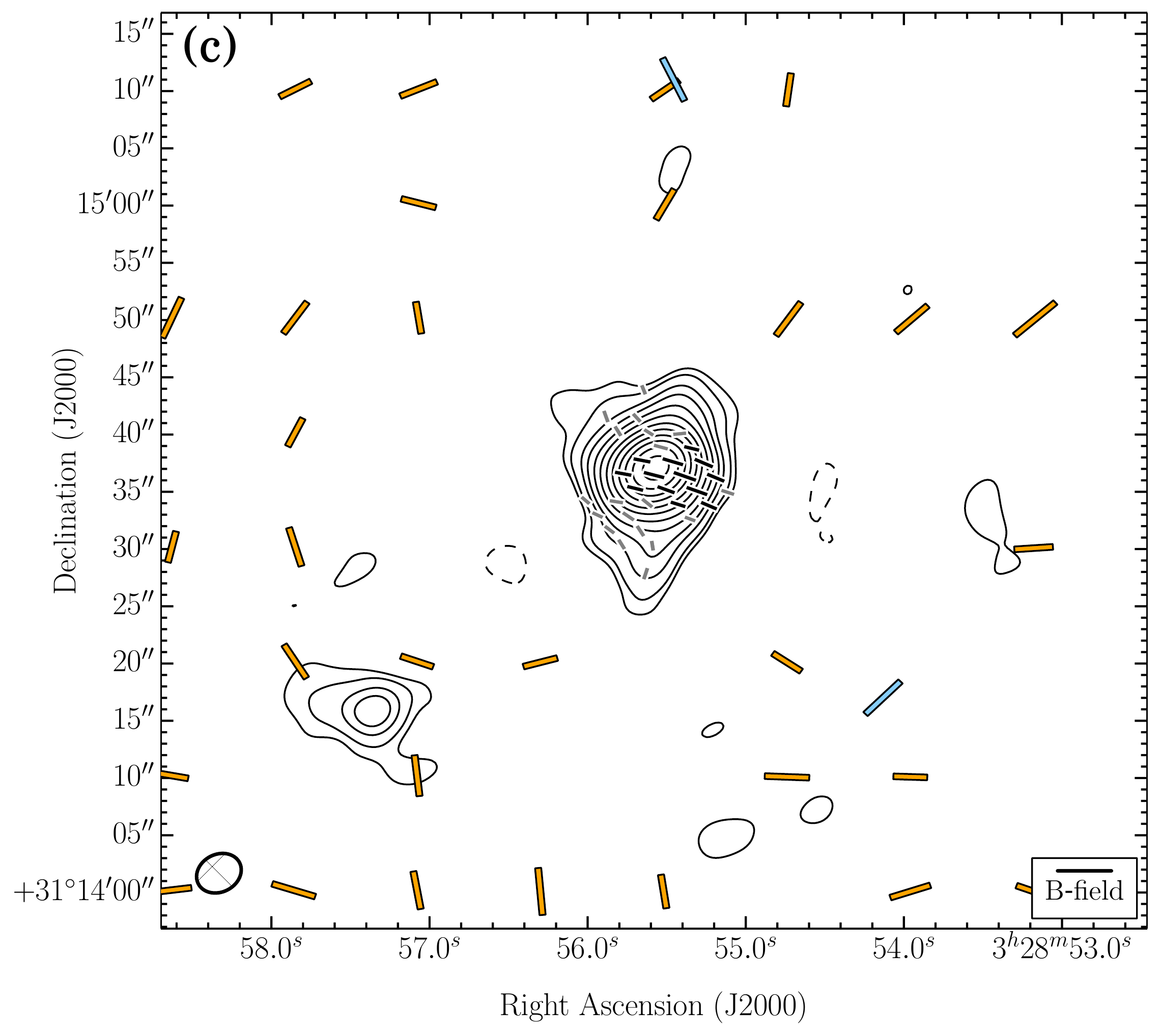}
\caption[]{ \footnotesize{
\input{IRAS2A_caption.txt}
}}
\label{fig:IRAS2A}
\end{center}
\end{figure*}

%%% Maps of SVS~13
\begin{figure*} [hbt!]
\begin{center}
\epsscale{1.1}
\plottwo{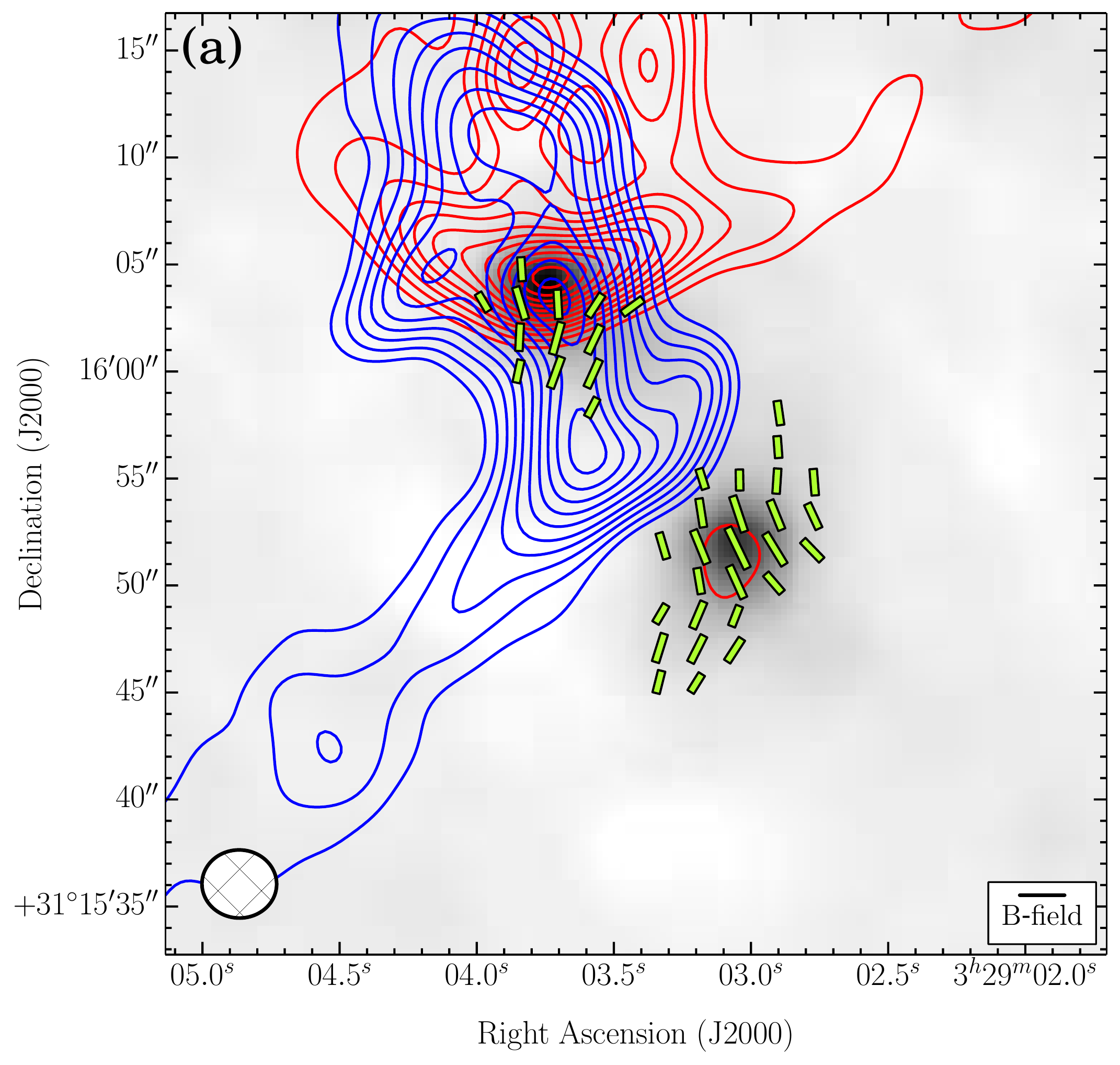}{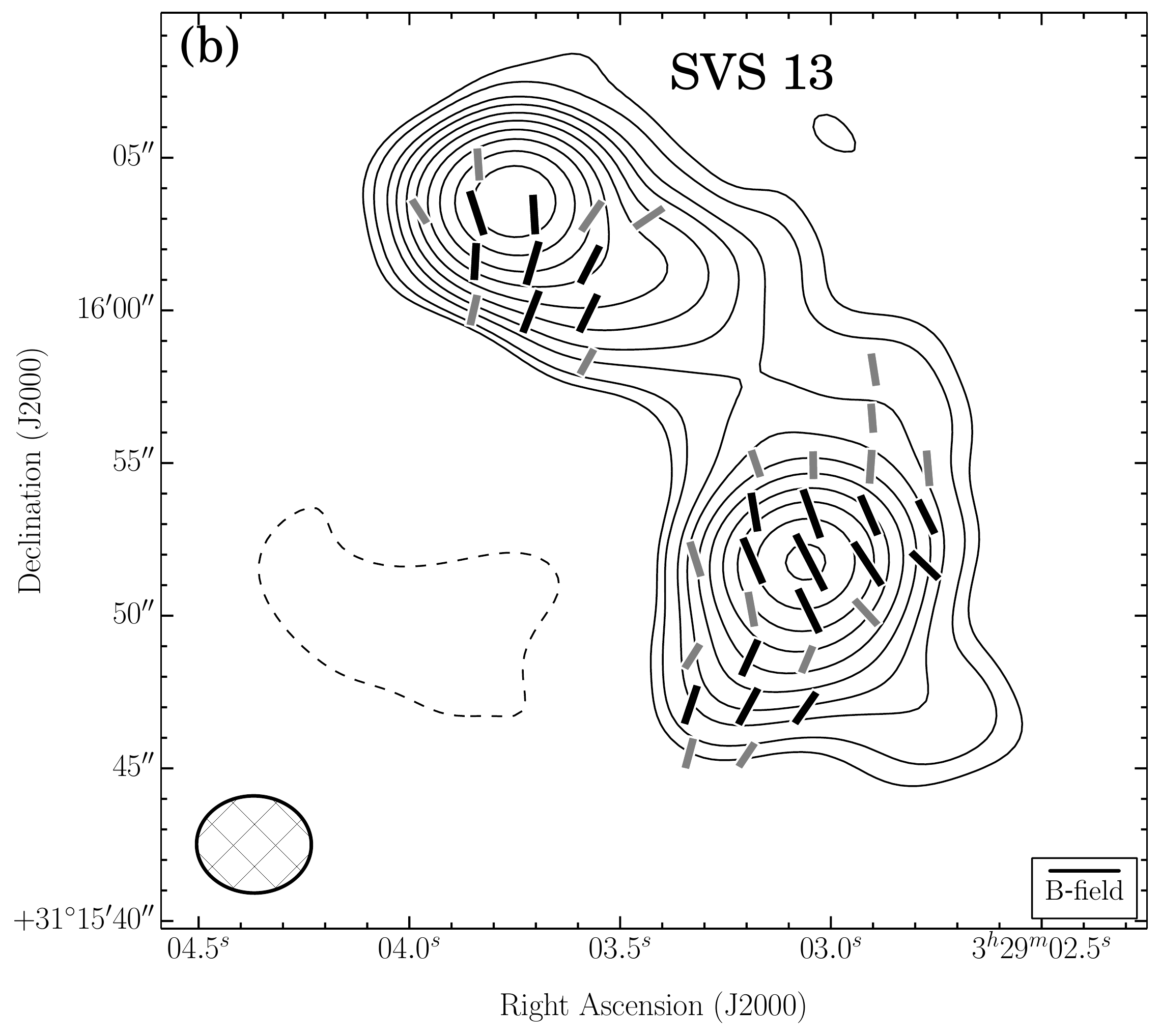}
\epsscale{0.8}
\plotone{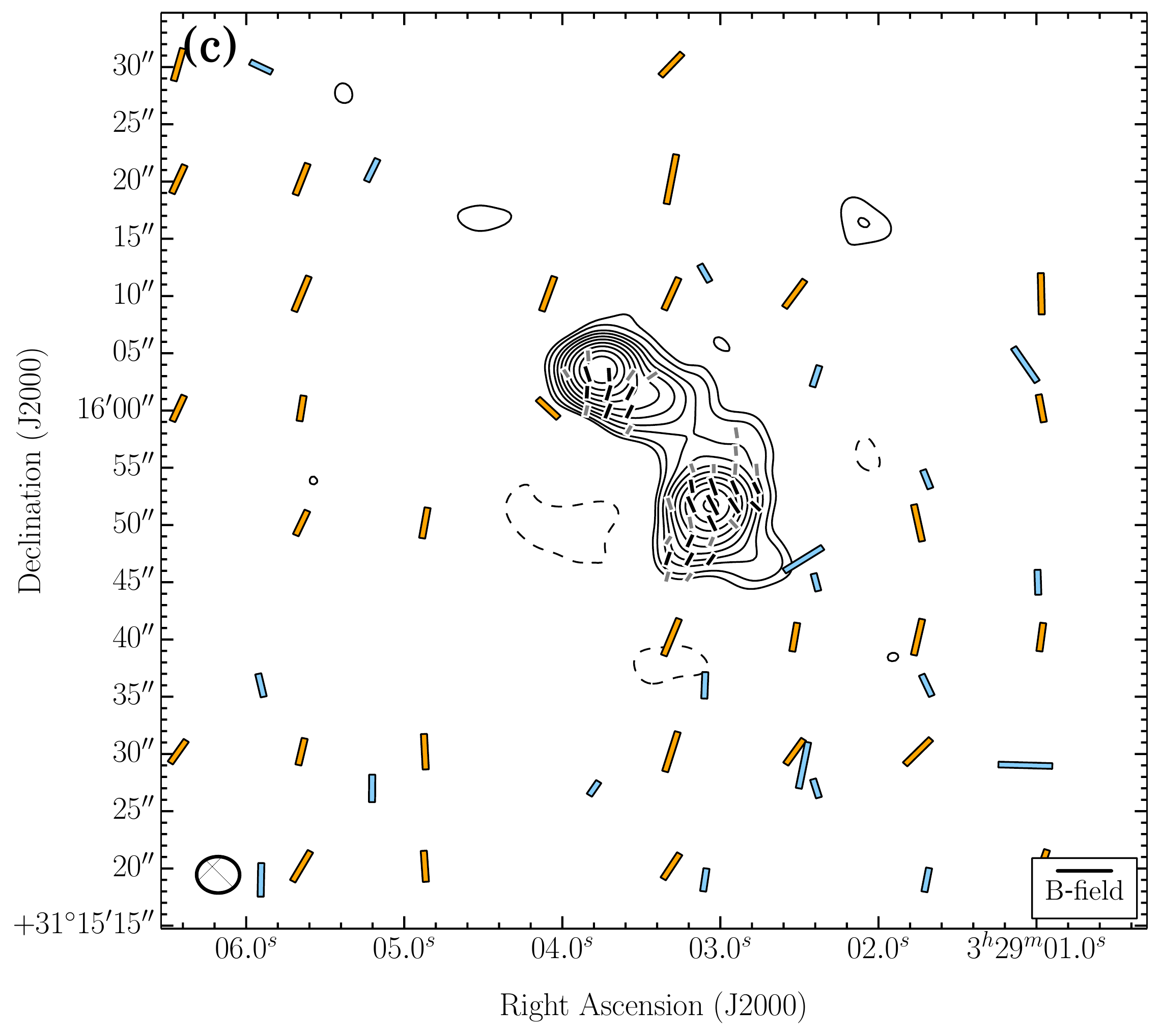}
\caption[]{ \footnotesize{
\input{SVS13_caption.txt}
}}
\label{fig:SVS13}
\end{center}
\end{figure*}

%%% Maps of NGC~1333-IRAS~4A
\begin{figure*} [hbt!]
\begin{center}
\epsscale{1.1}
\plottwo{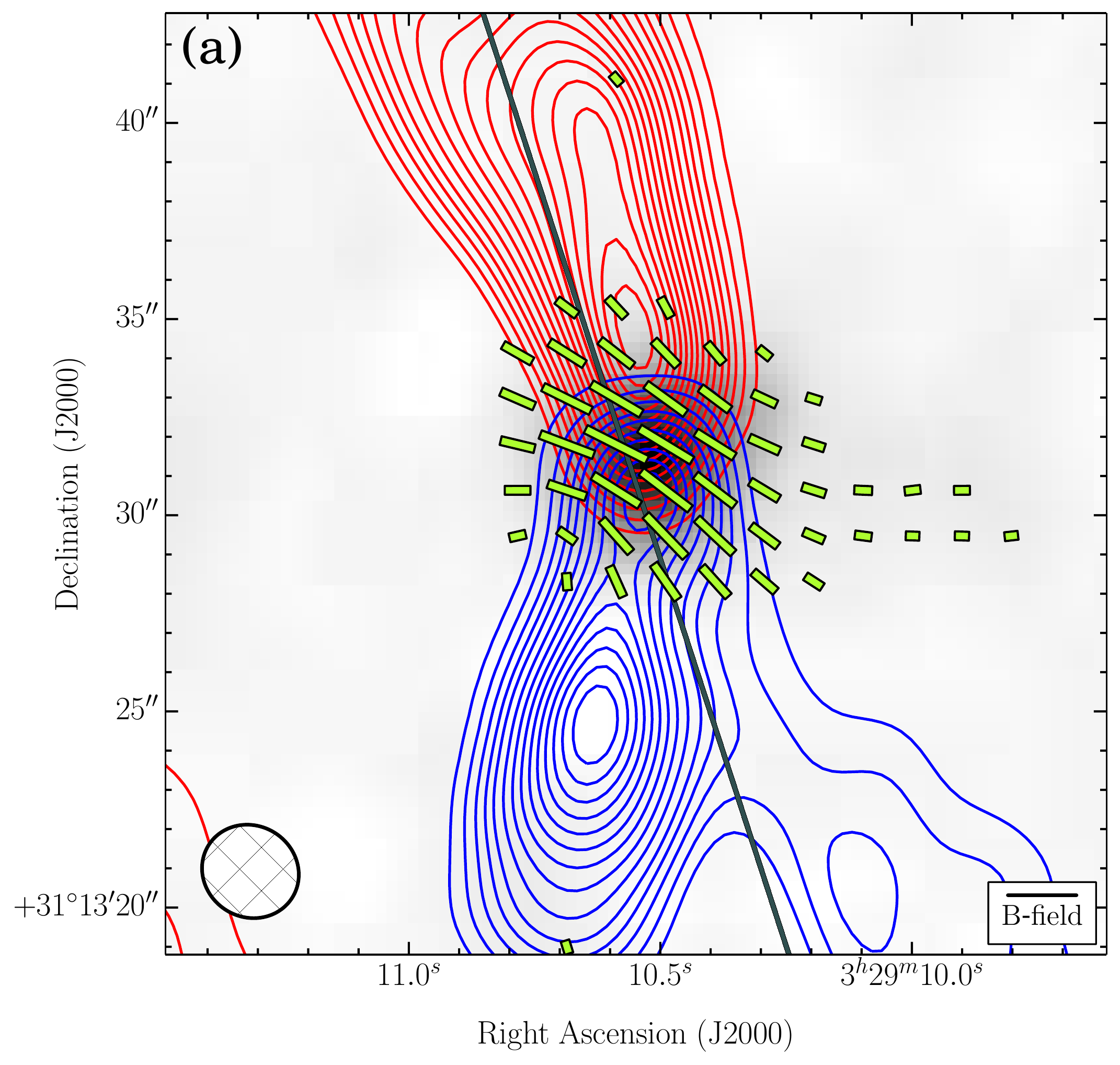}{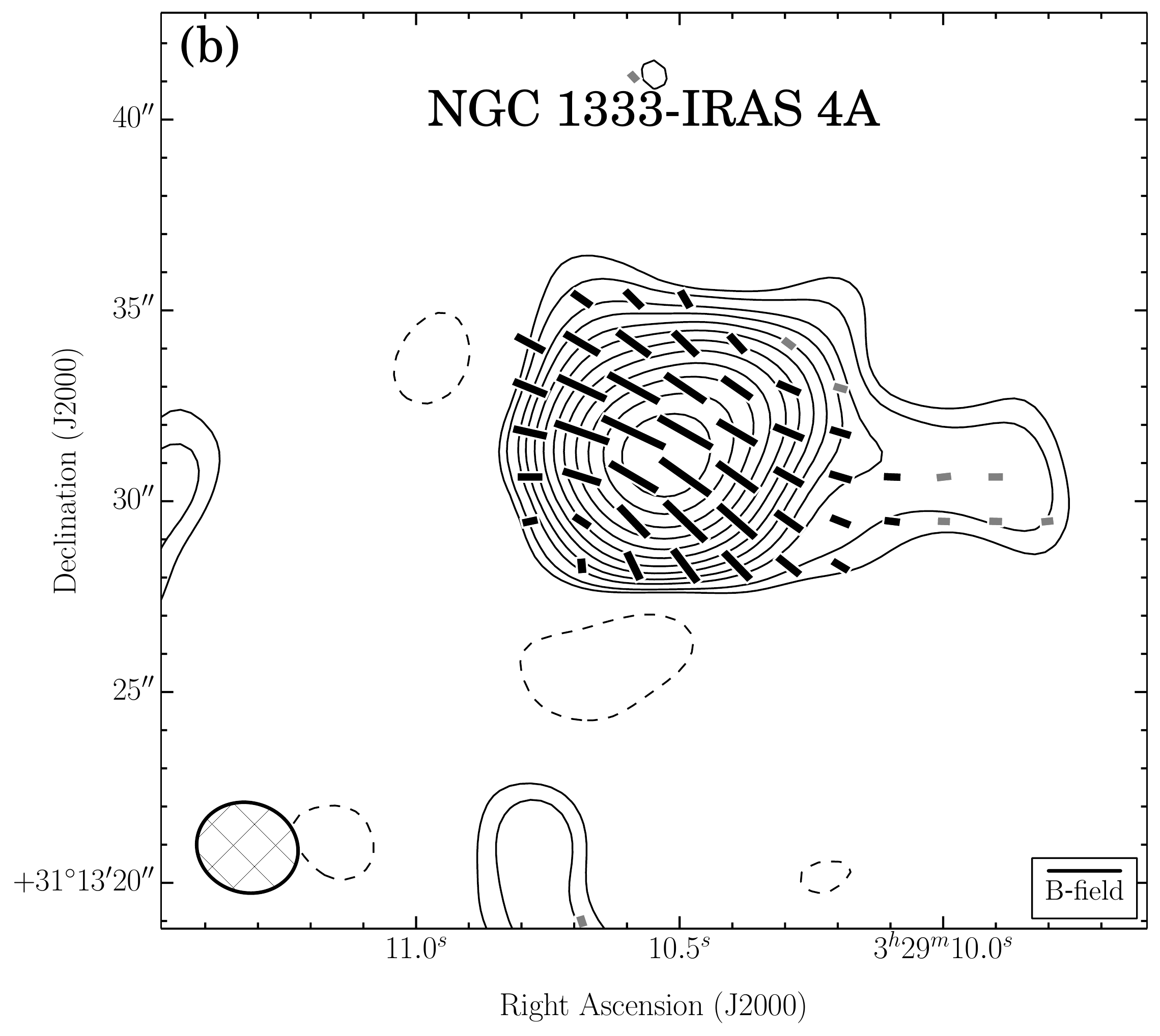}
\epsscale{0.8}
\plotone{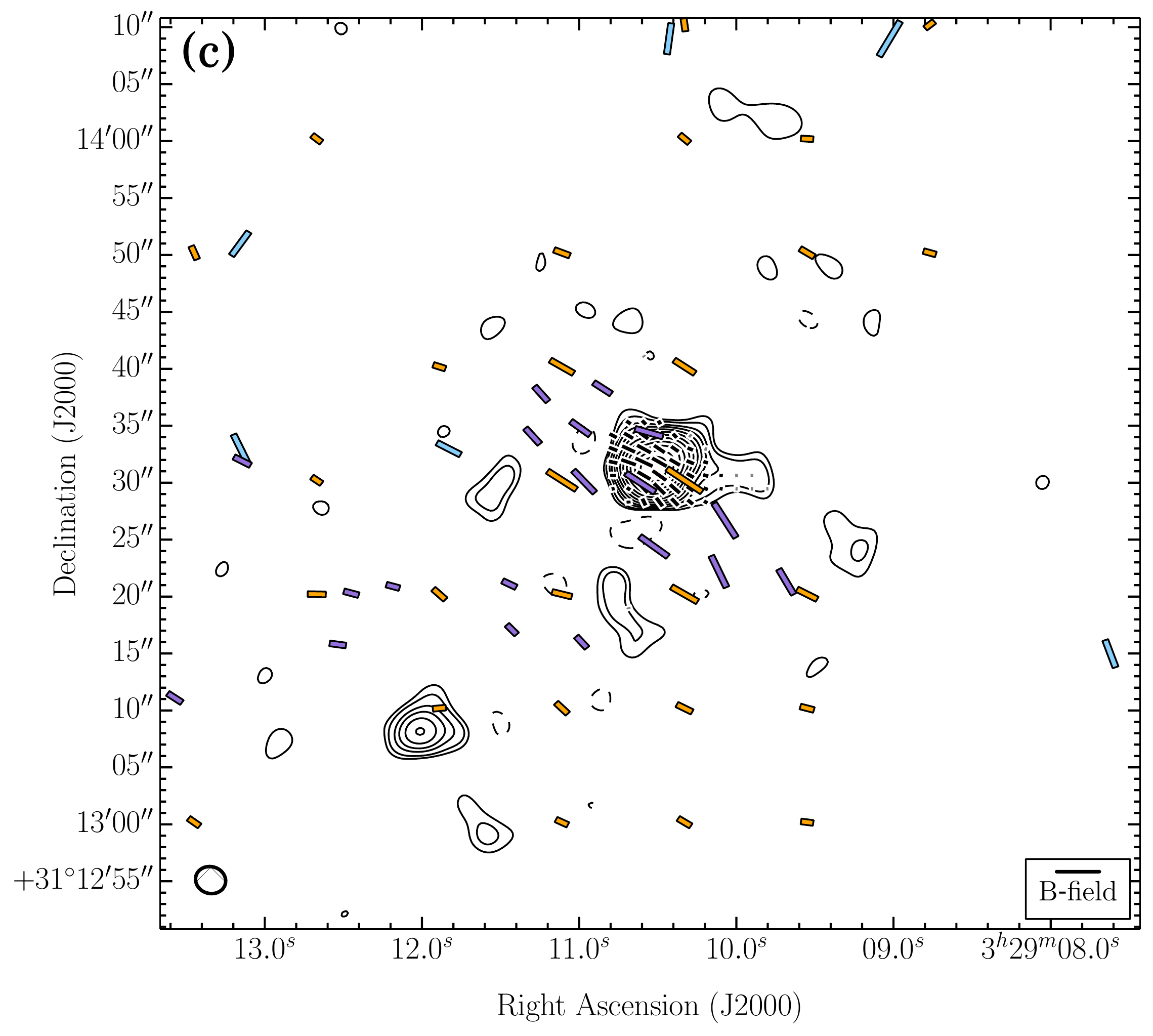}
\caption[]{ \footnotesize{
\input{IRAS4A_caption.txt}
}}
\label{fig:IRAS4A}
\end{center}
\end{figure*}

%%% Maps of NGC~1333-IRAS~4B and B2
\begin{figure*} [hbt!]
\begin{center}
\epsscale{1.1}
\plottwo{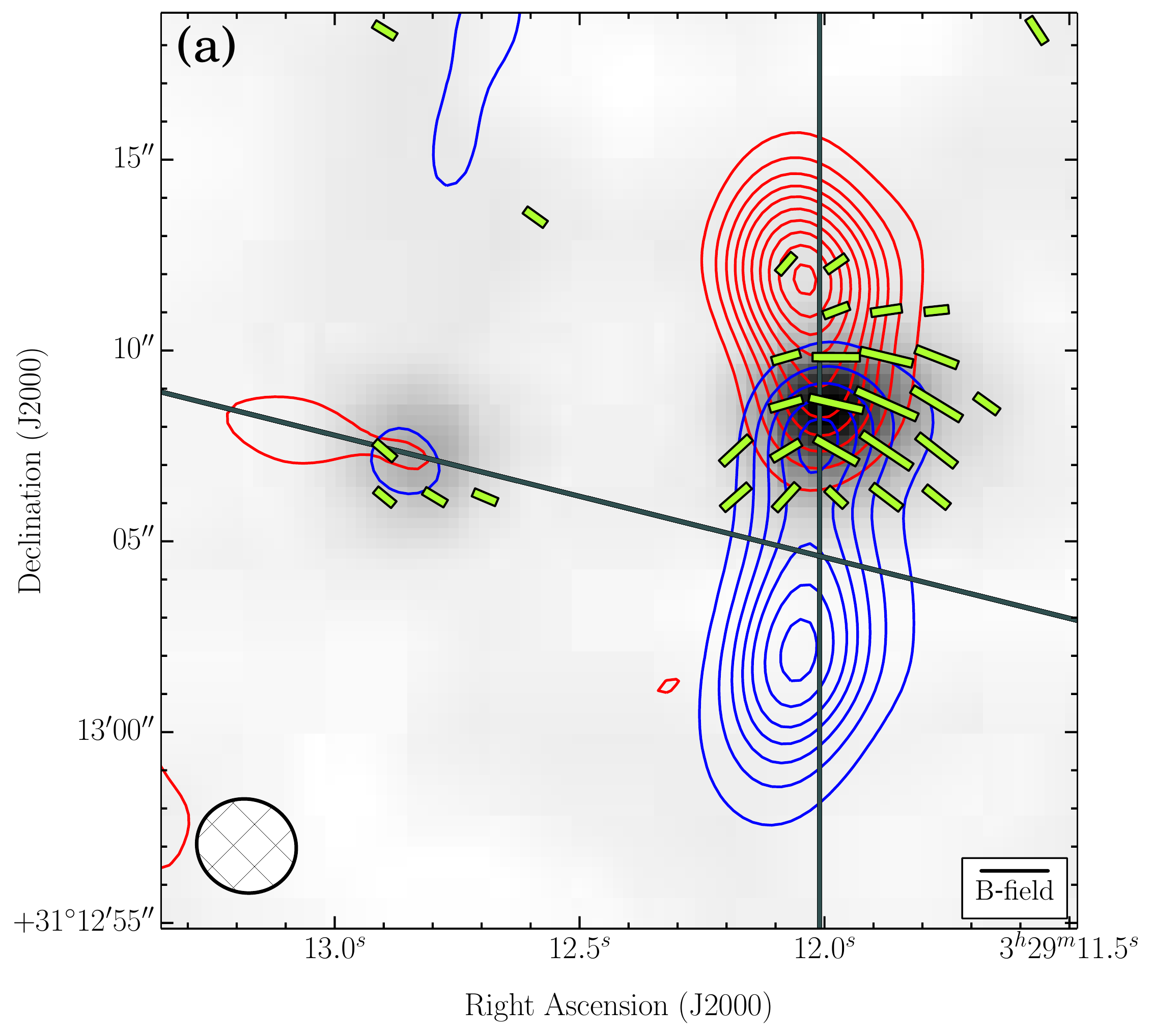}{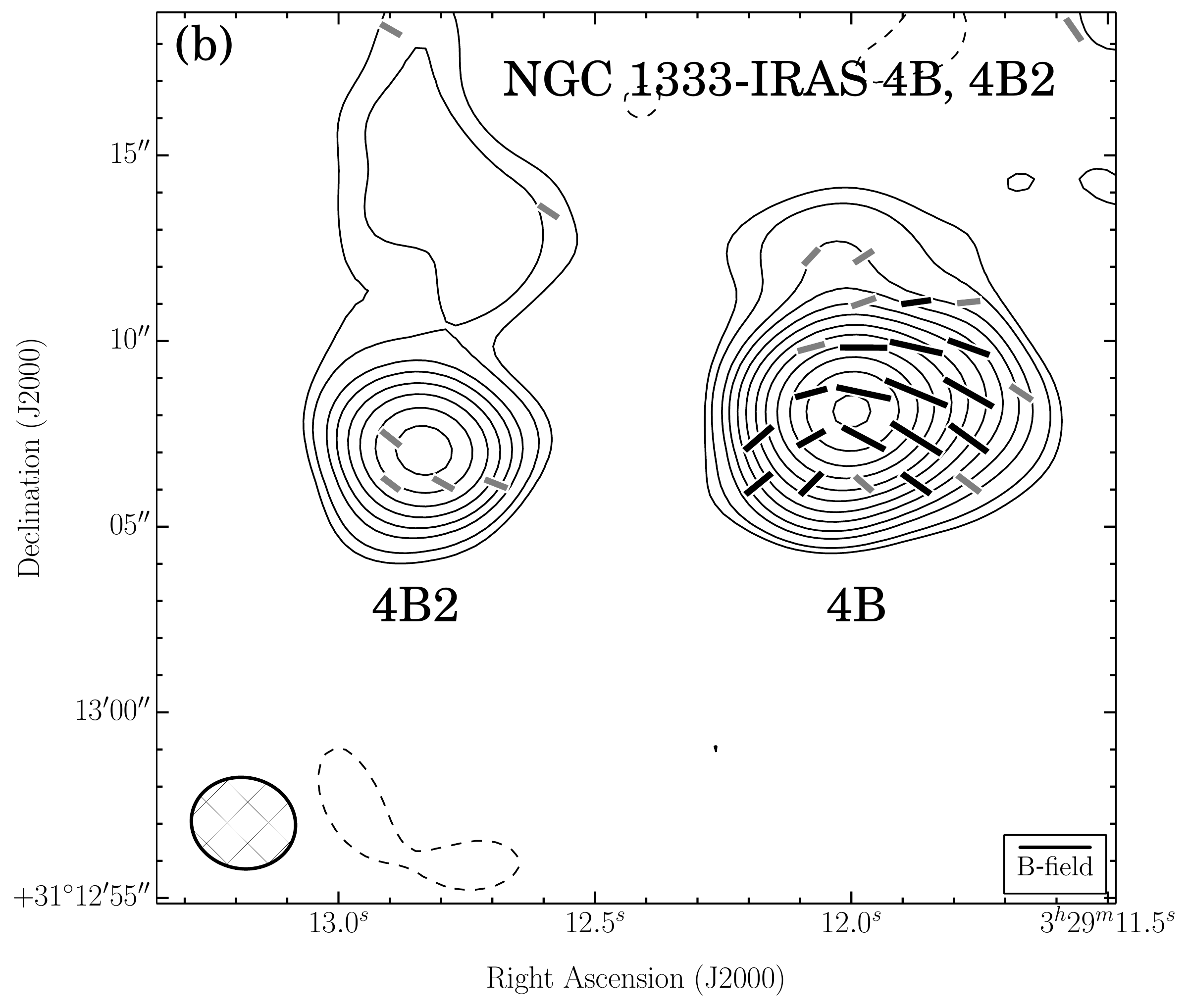}
\epsscale{0.8}
\plotone{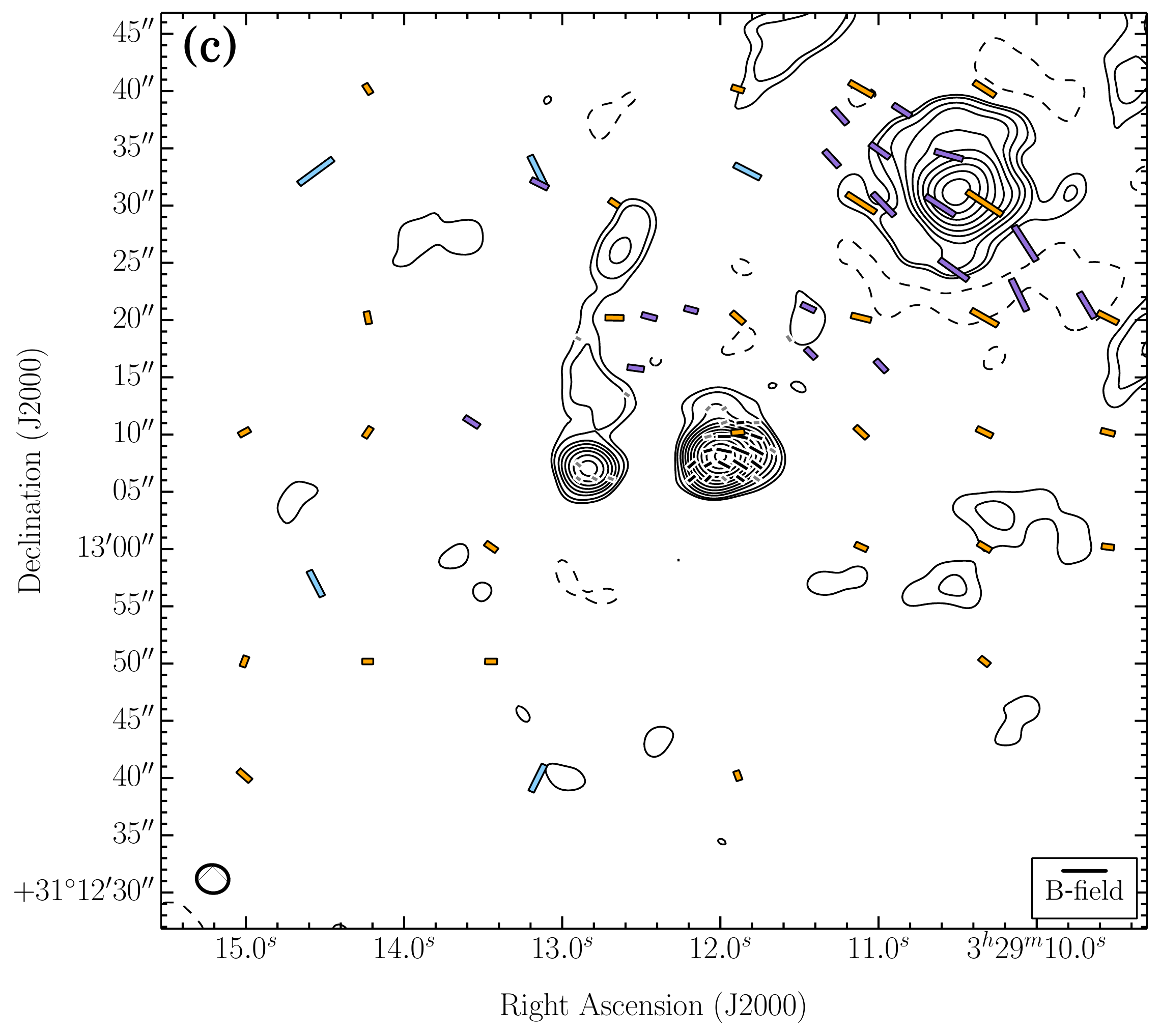}
\caption[]{ \footnotesize{
\input{IRAS4B_caption.txt}
}}
\label{fig:IRAS4B}
\end{center}
\end{figure*}

%%% Maps of HH~211~mm
\begin{figure*} [hbt!]
\begin{center}
\epsscale{1.1}
\plottwo{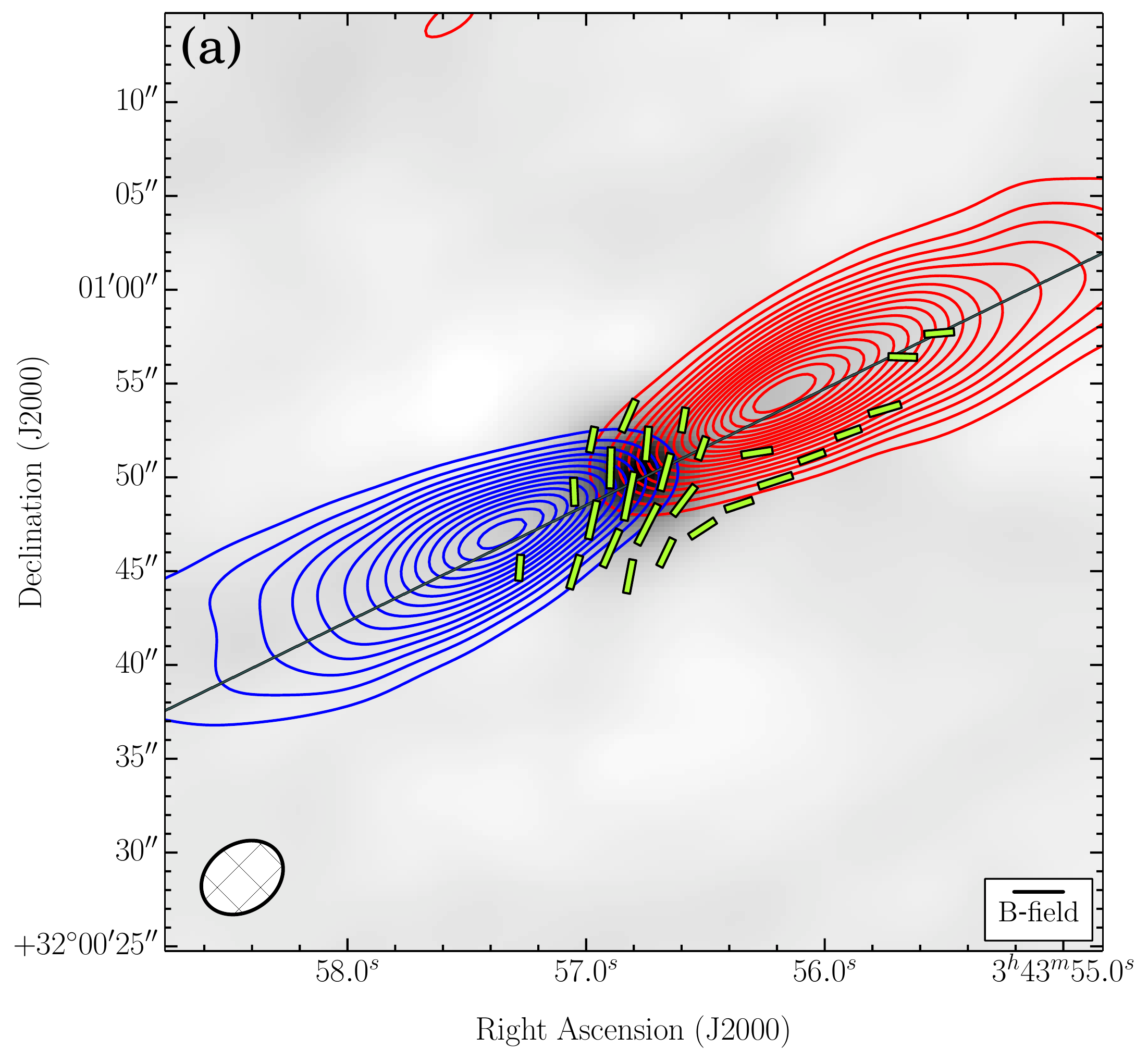}{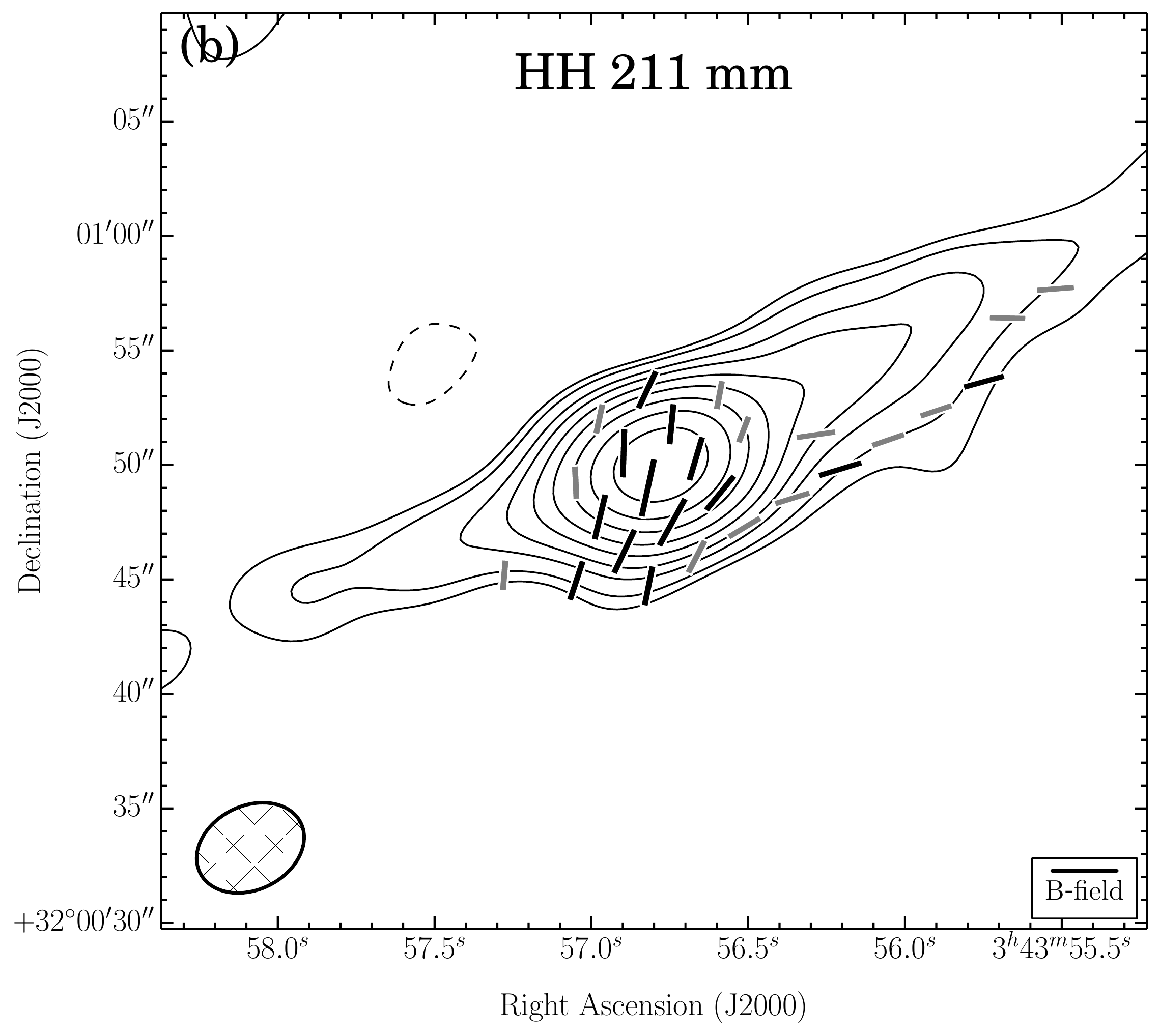}
\epsscale{0.8}
\plotone{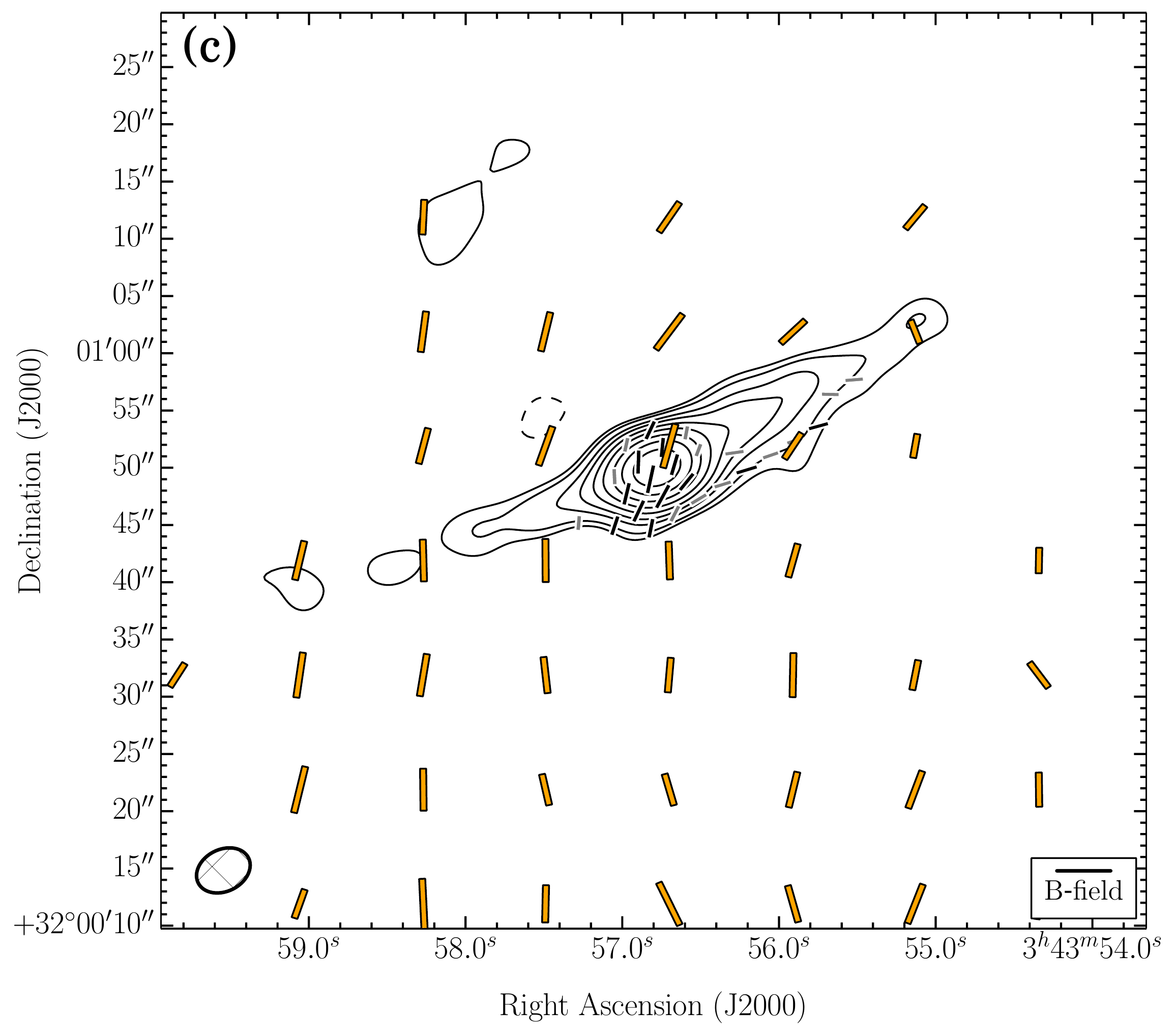}
\caption[]{ \footnotesize{
\input{HH211_caption.txt}
}}
\label{fig:HH211}
\end{center}
\end{figure*}

%%% Maps of DG~Tau
\begin{figure*} [hbt!]
\begin{center}
\epsscale{1.1}
\plottwo{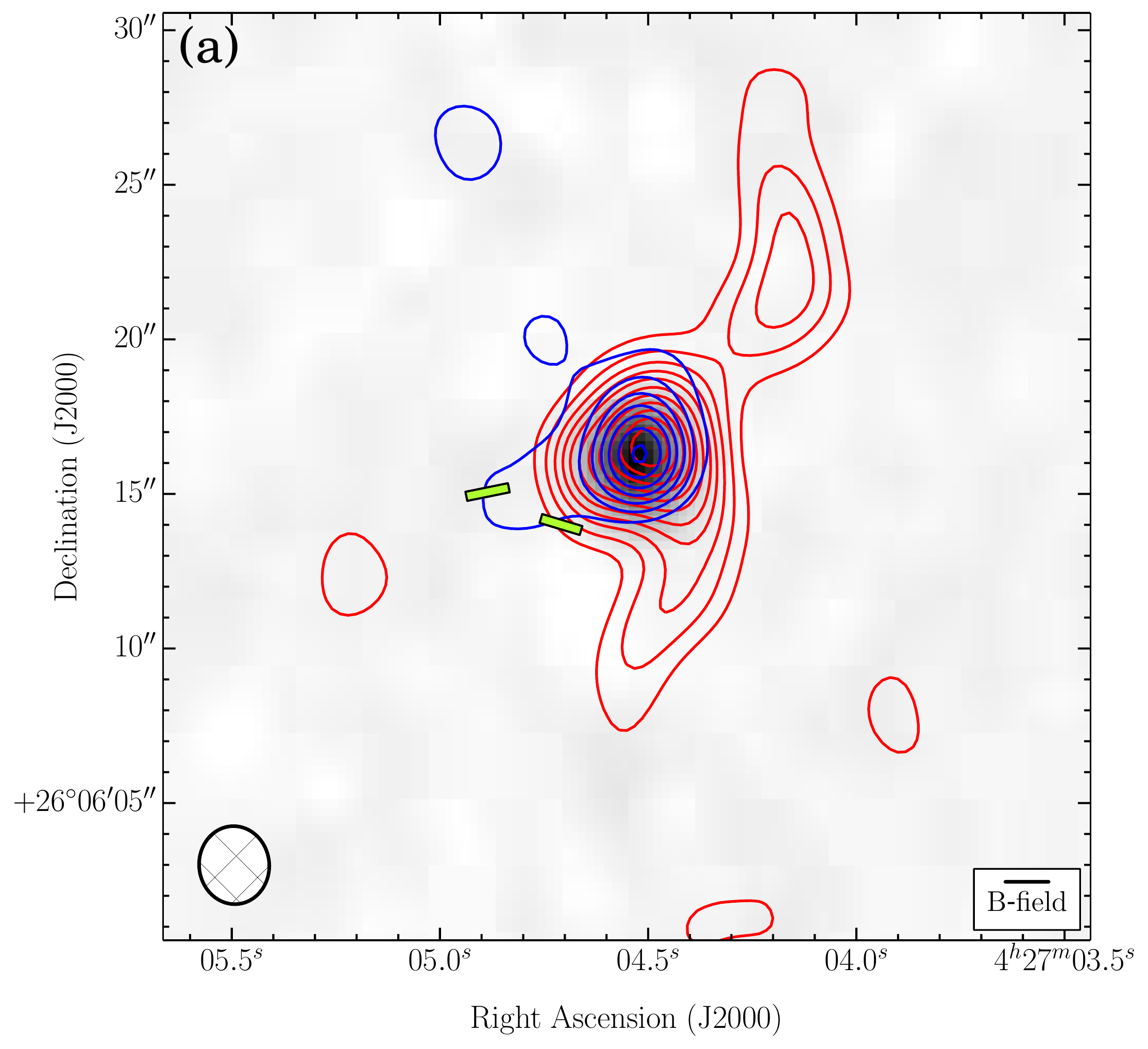}{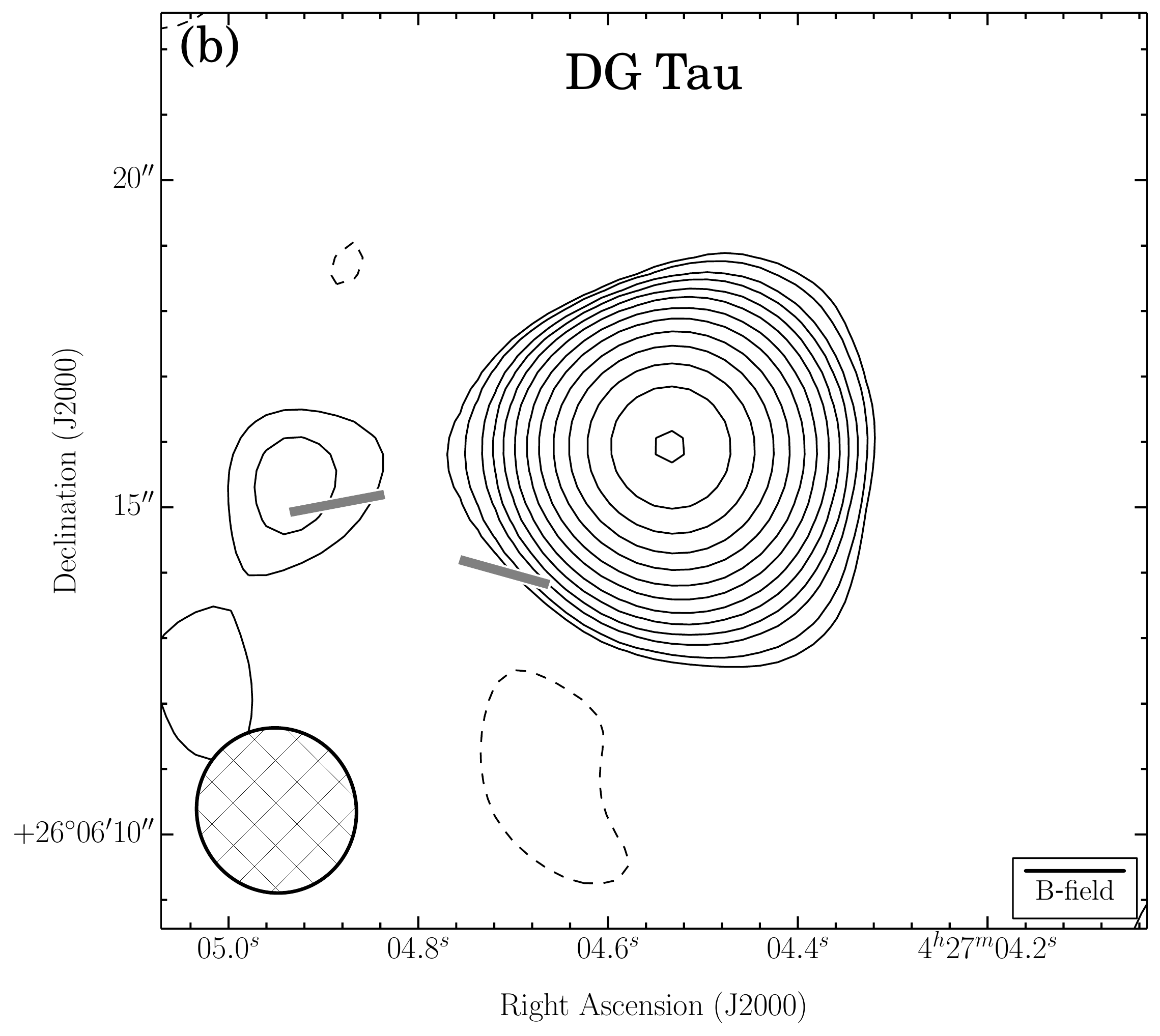}
\caption[]{ \footnotesize{
\input{DGTau_caption.txt}
There is no \textbf{(c)} plot because there were no SCUBA, SHARP, or Hertz data to overlay.
}}
\label{fig:DGTau}
\end{center}
\end{figure*}

%%% Maps of L1551~NE
\begin{figure*} [hbt!]
\begin{center}
\epsscale{1.1}
\plottwo{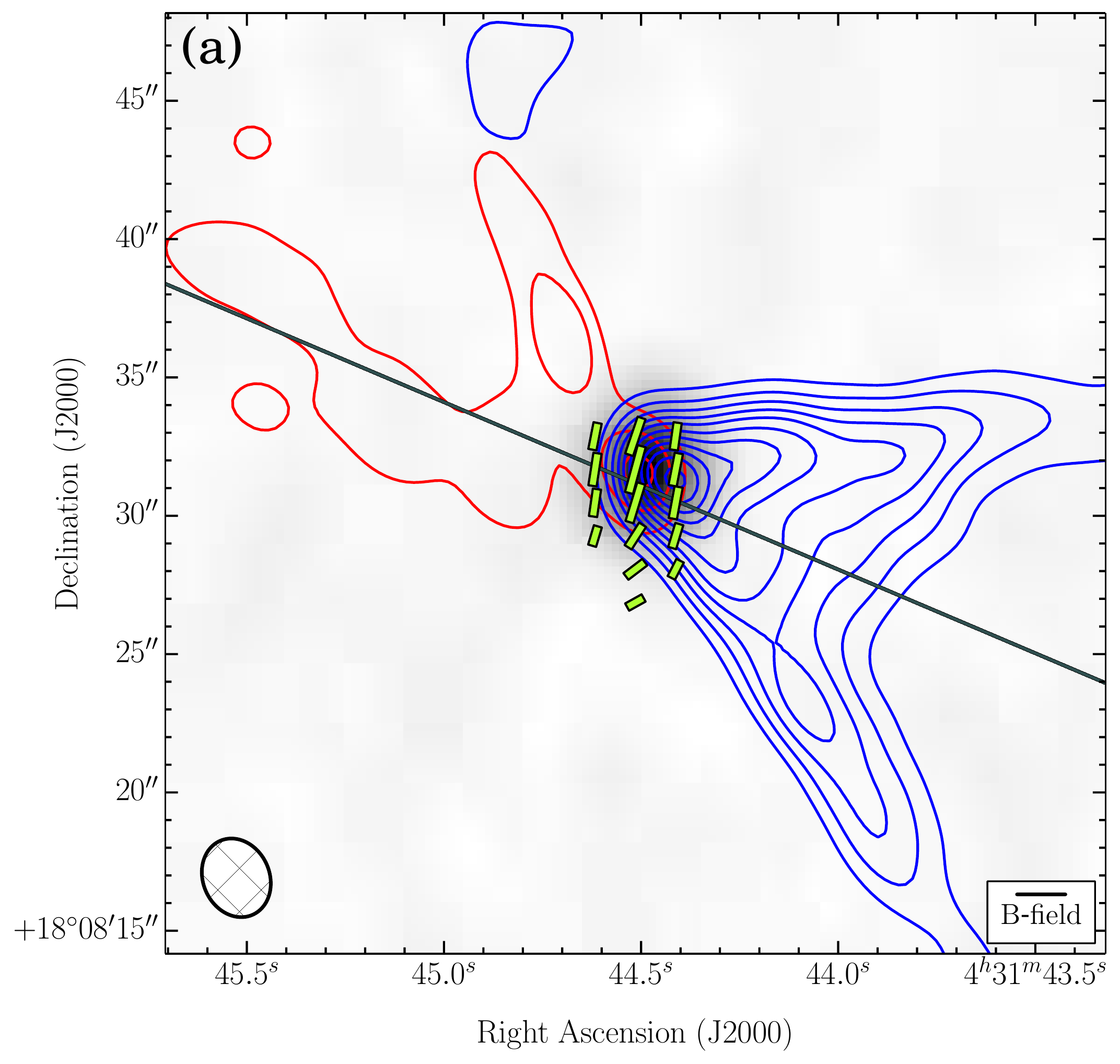}{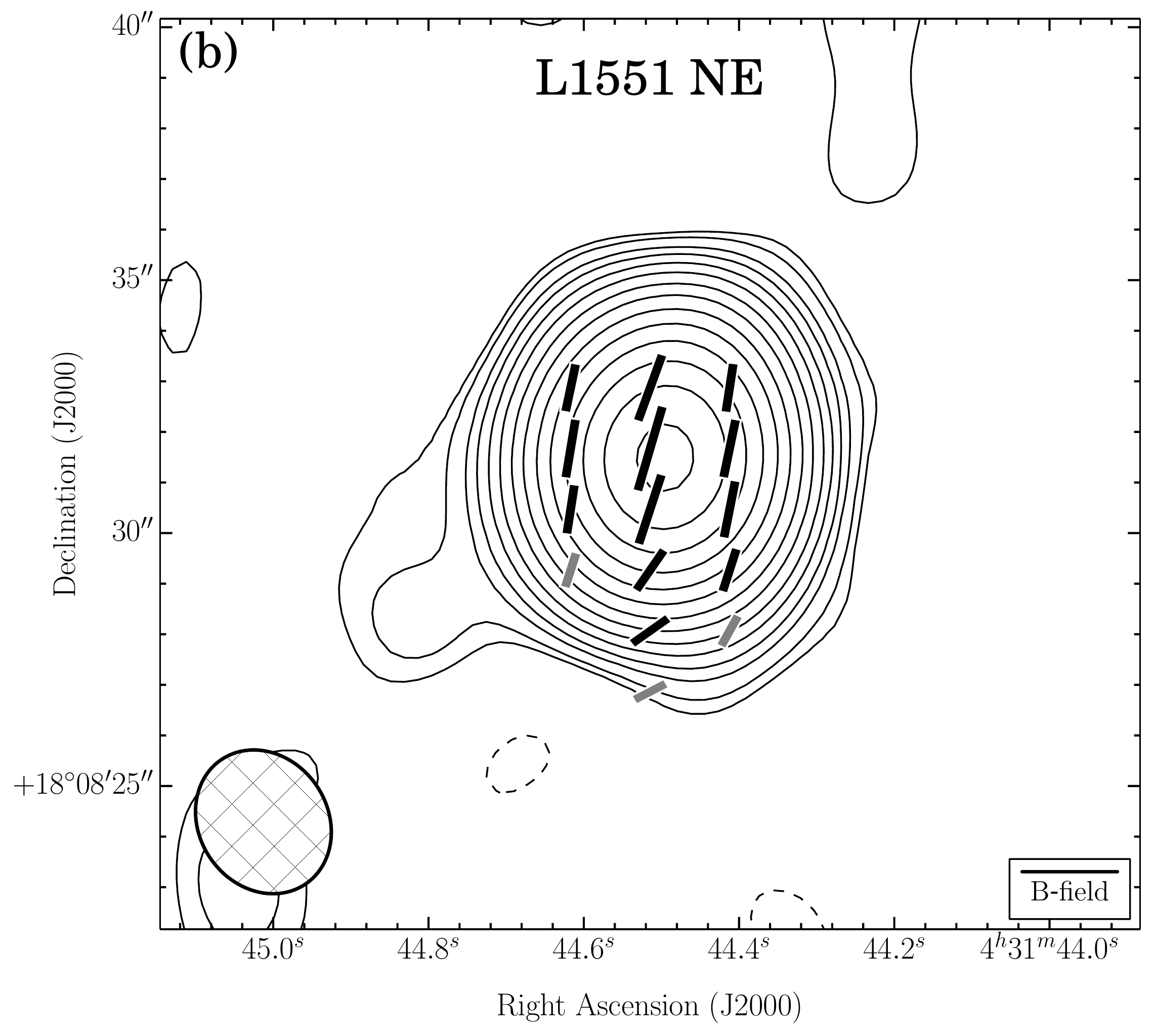}
\epsscale{0.8}
\plotone{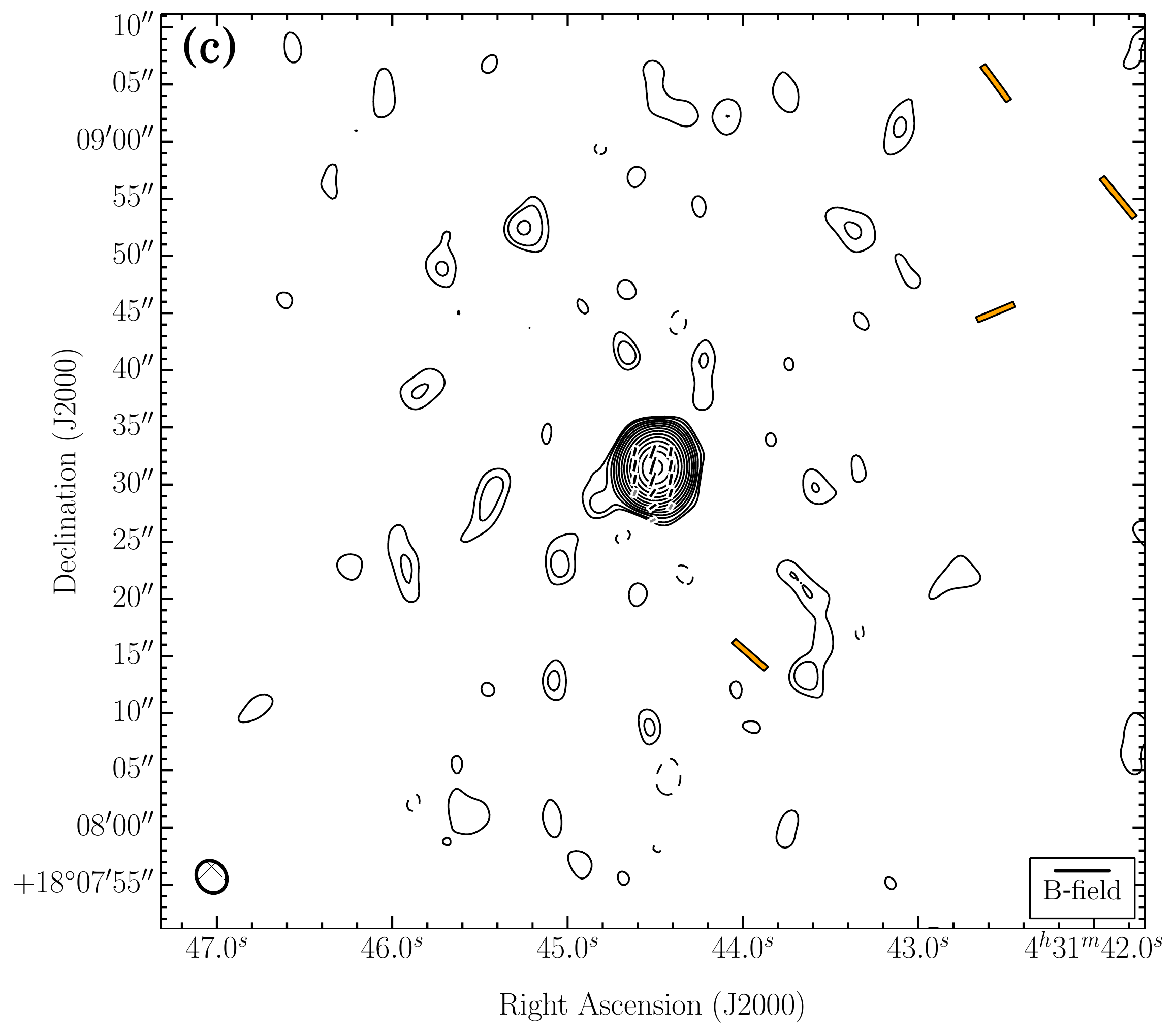}
\caption[]{ \footnotesize{
\input{L1551_caption.txt}
}}
\label{fig:L1551}
\end{center}
\end{figure*}

%%% Maps of L1527
\begin{figure*} [hbt!]
\begin{center}
\epsscale{1.1}
\plottwo{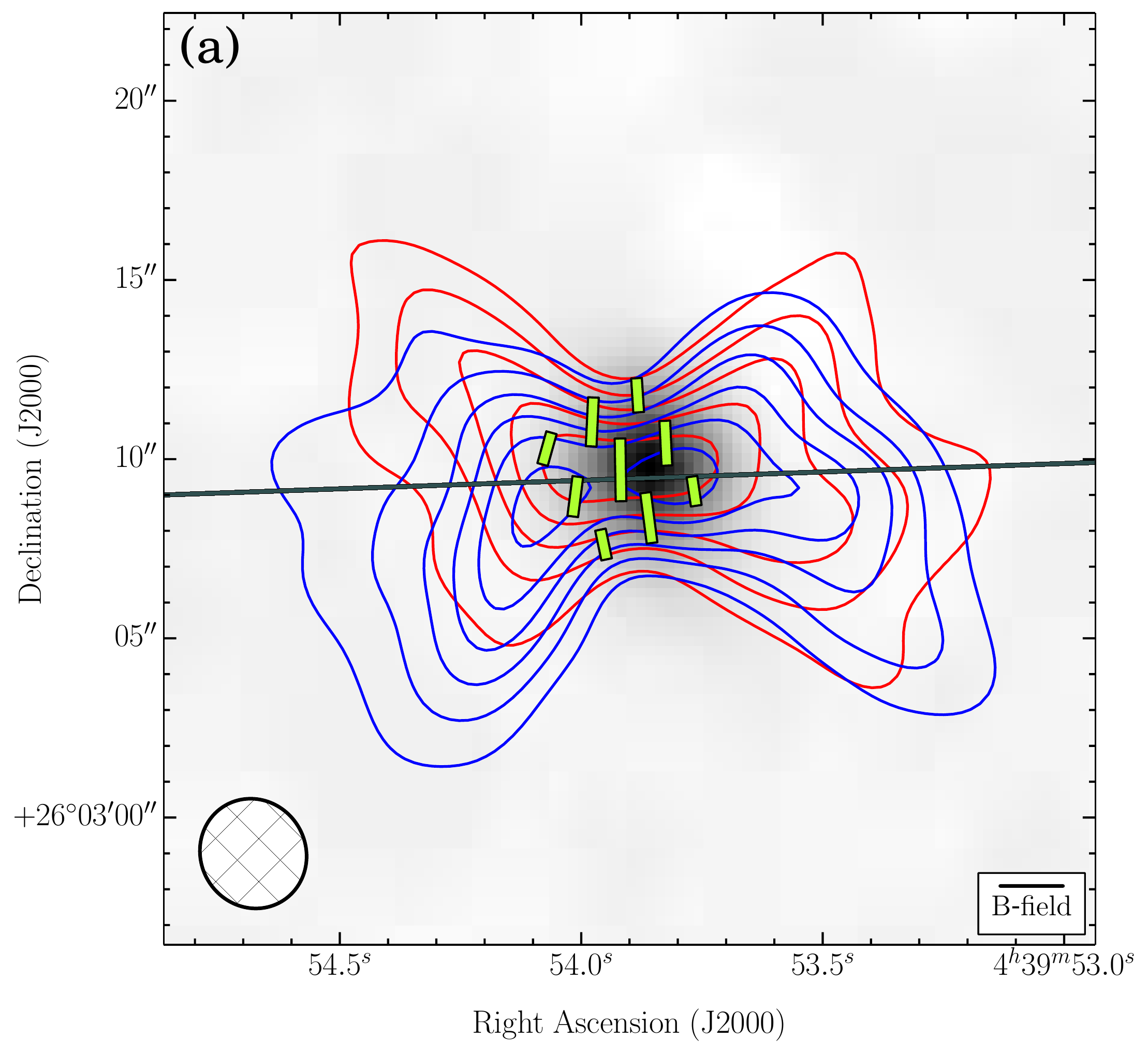}{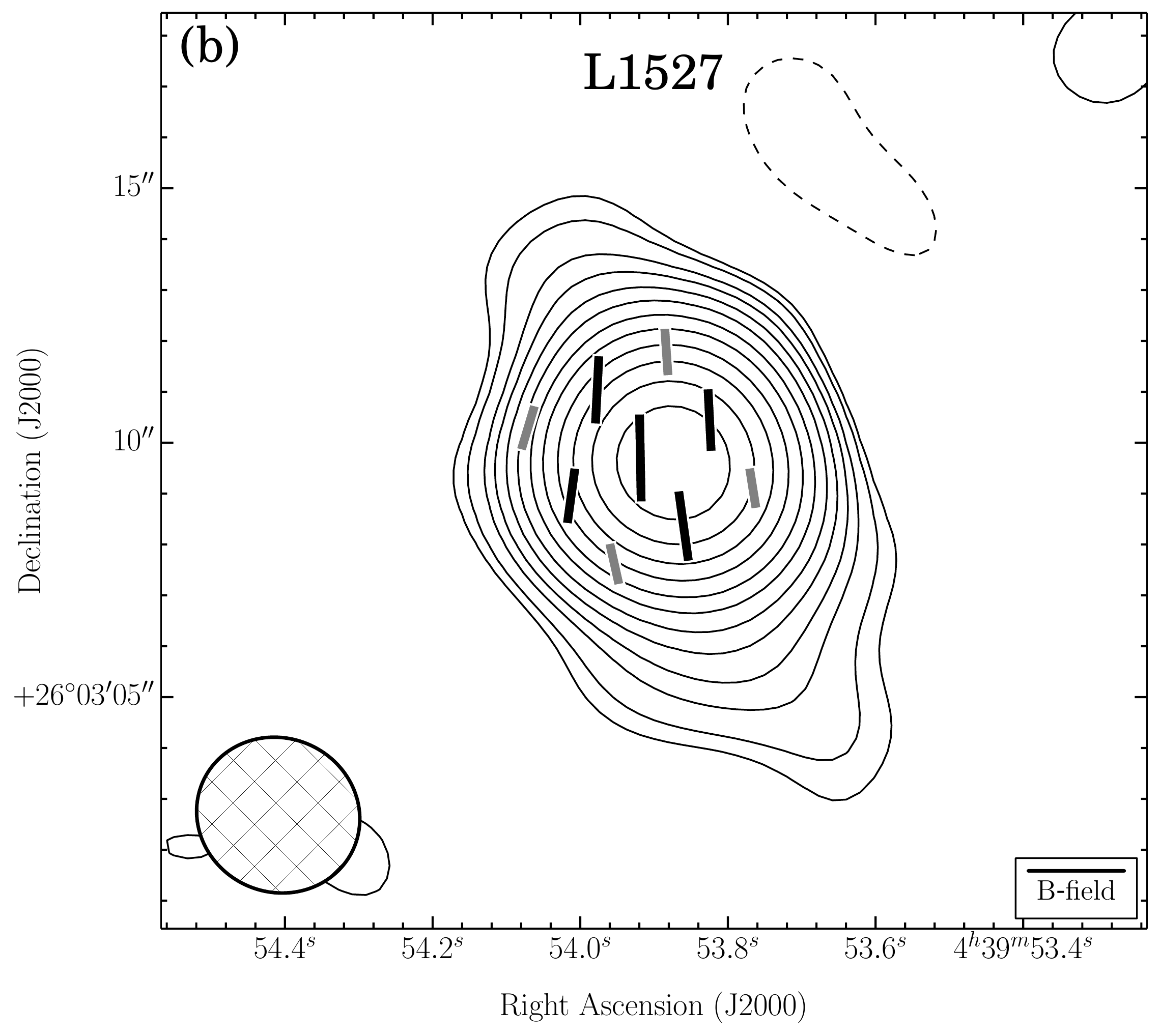}
\epsscale{0.8}
\plotone{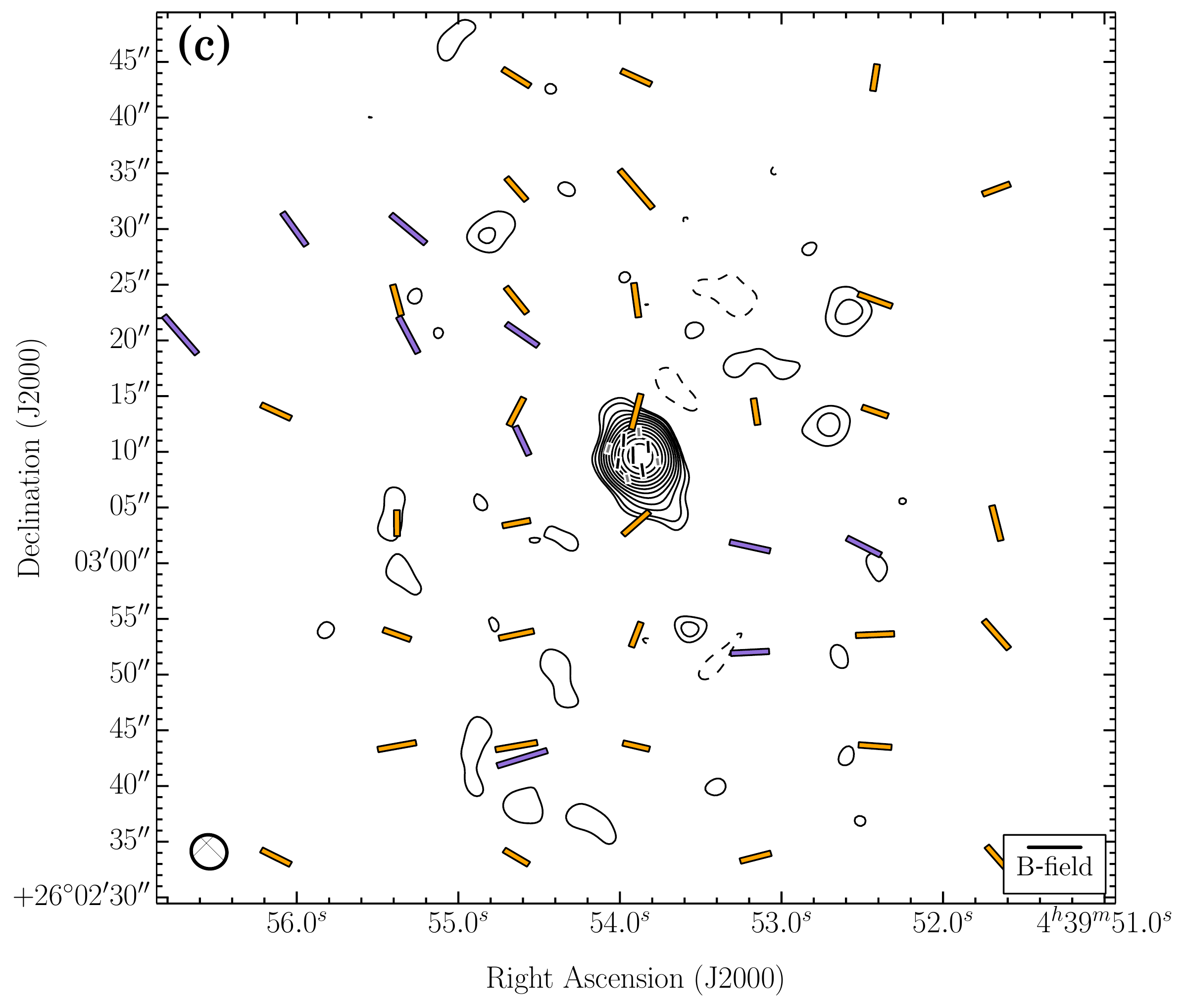}
\caption[]{ \footnotesize{
\input{L1527_caption.txt}
}}
\label{fig:L1527}
\end{center}
\end{figure*}

%%% Maps of CB~26
\begin{figure*} [hbt!]
\begin{center}
\epsscale{1.1}
\plottwo{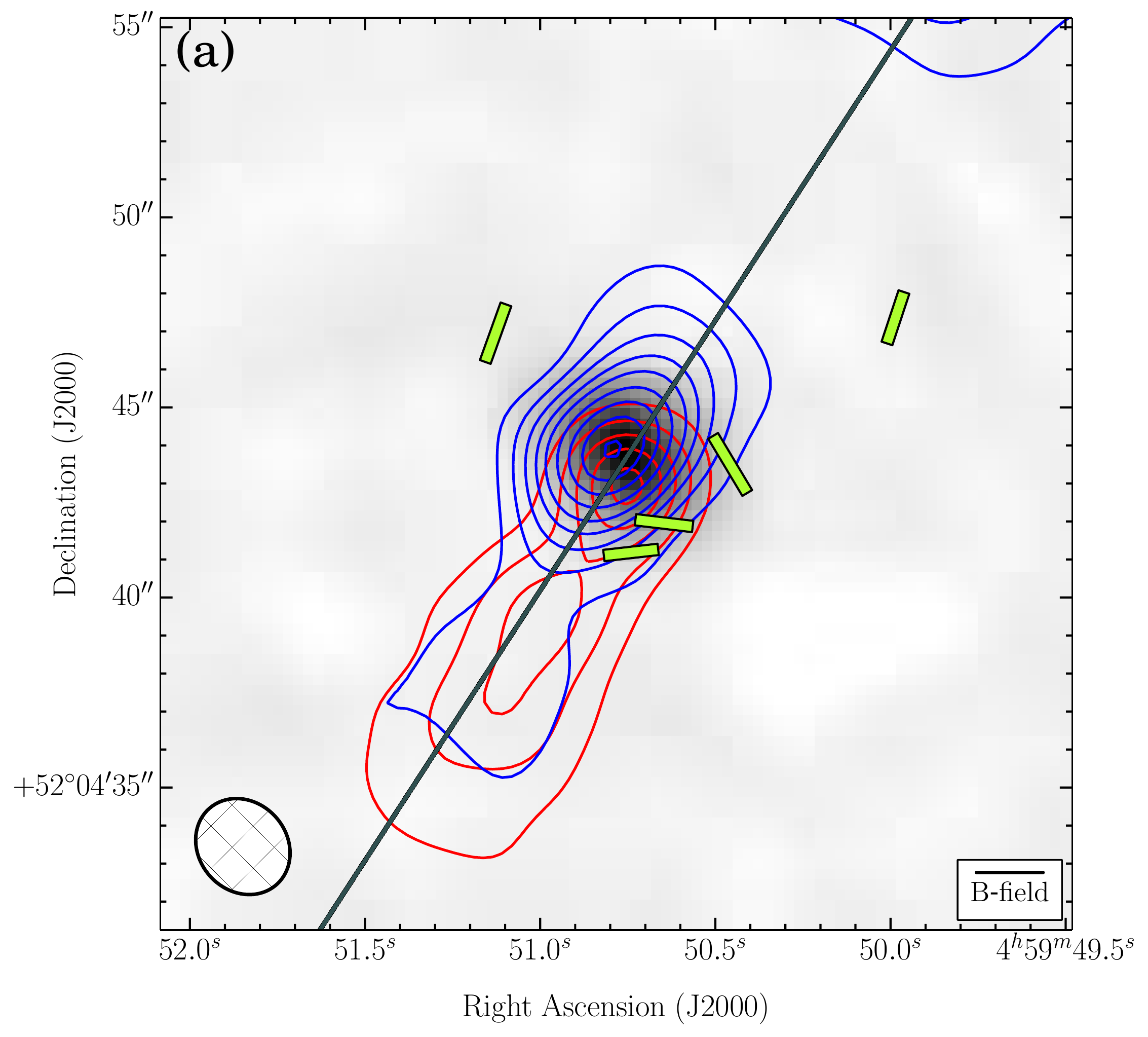}{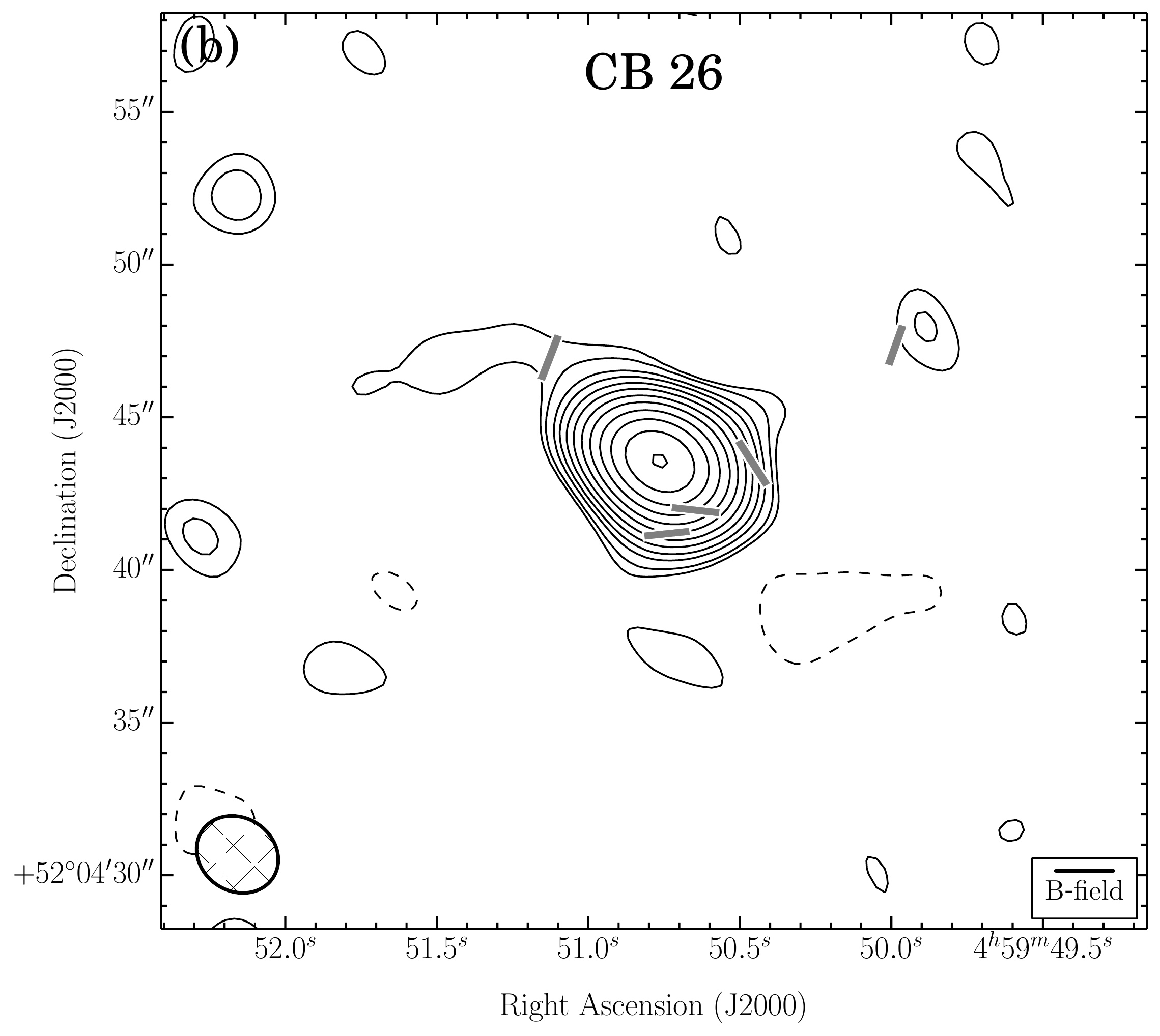}
\epsscale{0.8}
\plotone{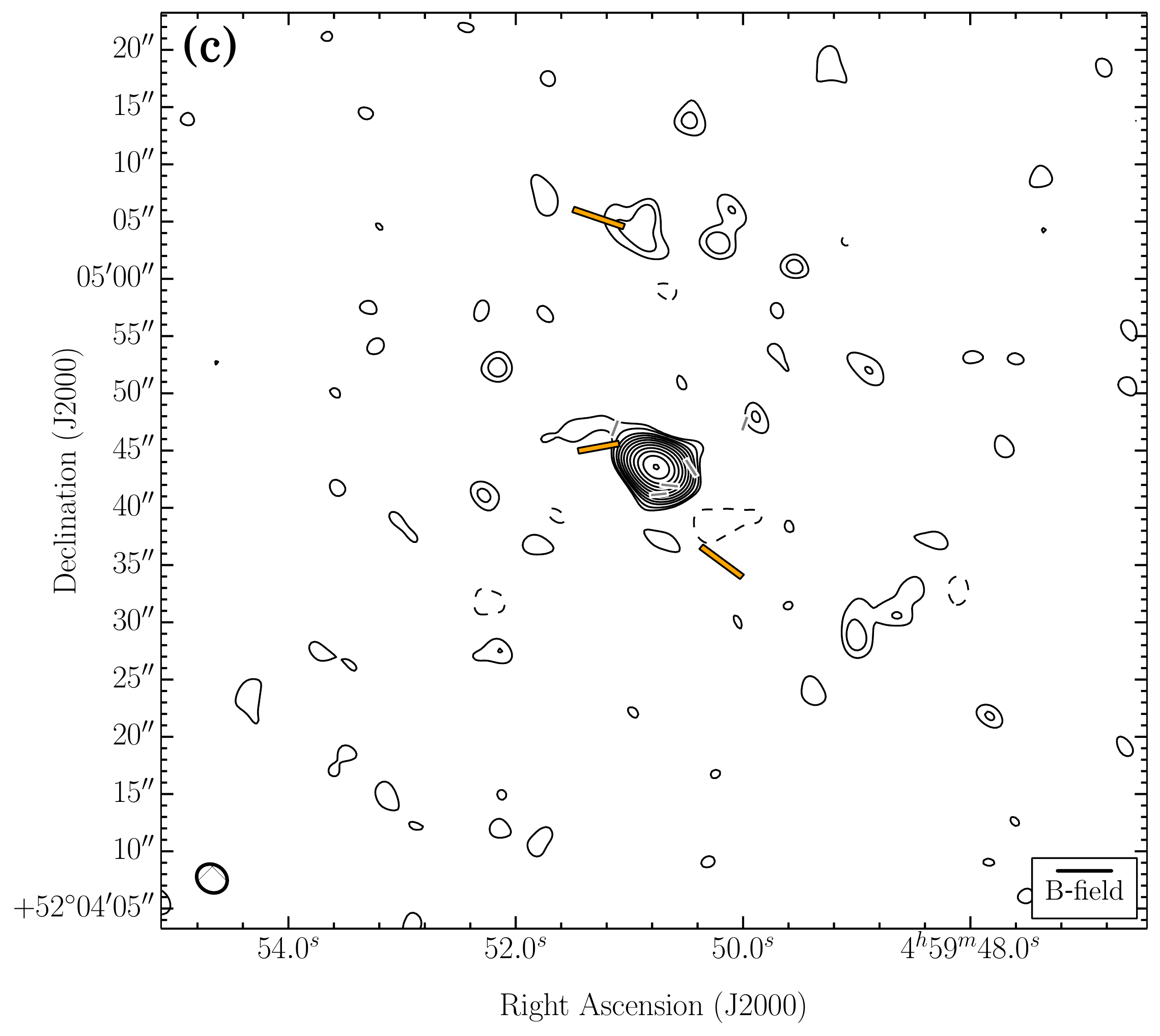}
\caption[]{ \footnotesize{
\input{CB26_caption.txt}
}}
\label{fig:CB26}
\end{center}
\end{figure*}

%%% Maps of Orion-KL
\begin{figure*} [hbt!]
\begin{center}
\epsscale{0.7}
%\plottwo{OrionKL_pol_moments}{OrionKL_vectors}
\plotone{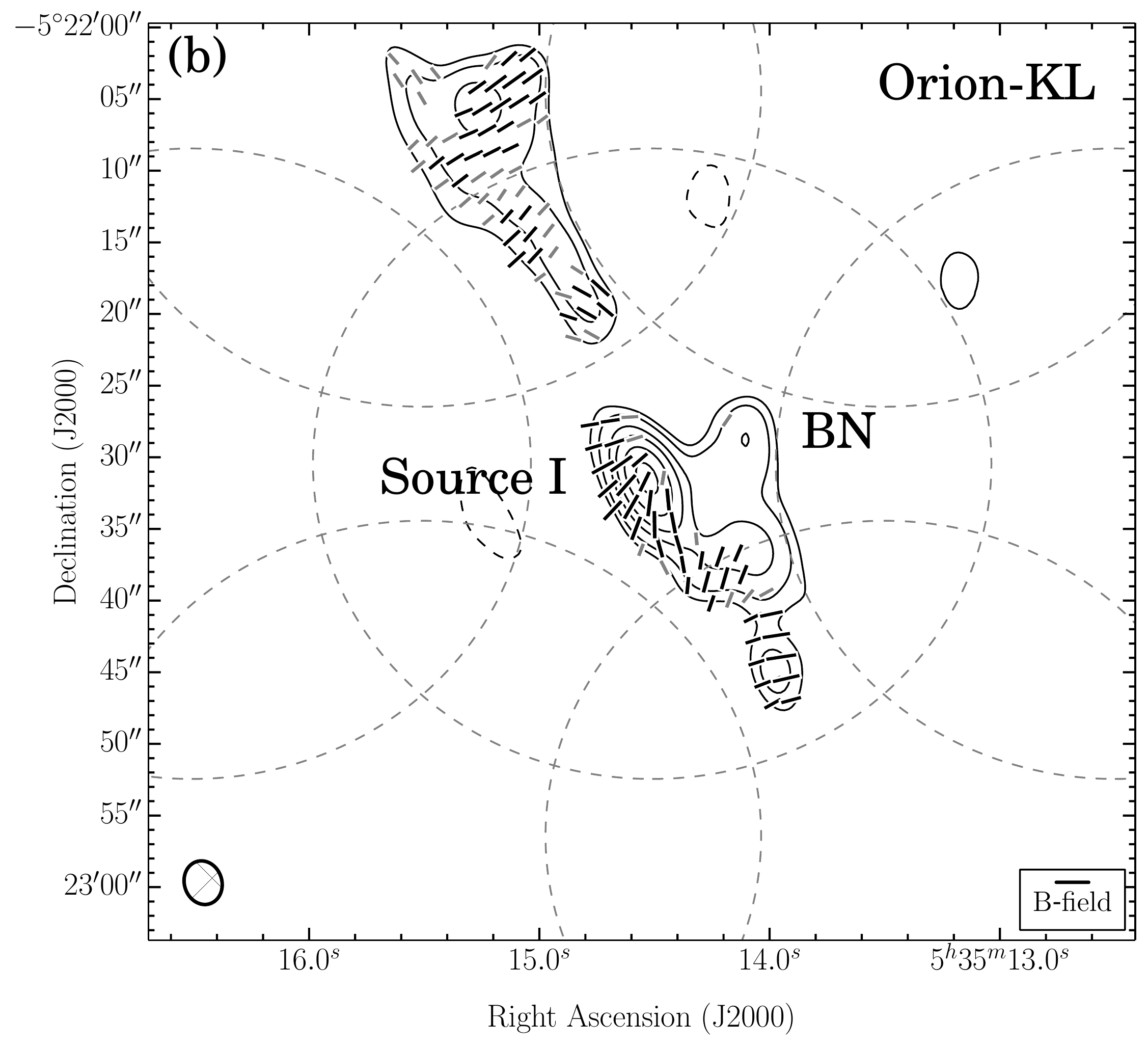}
\epsscale{0.7}
\plotone{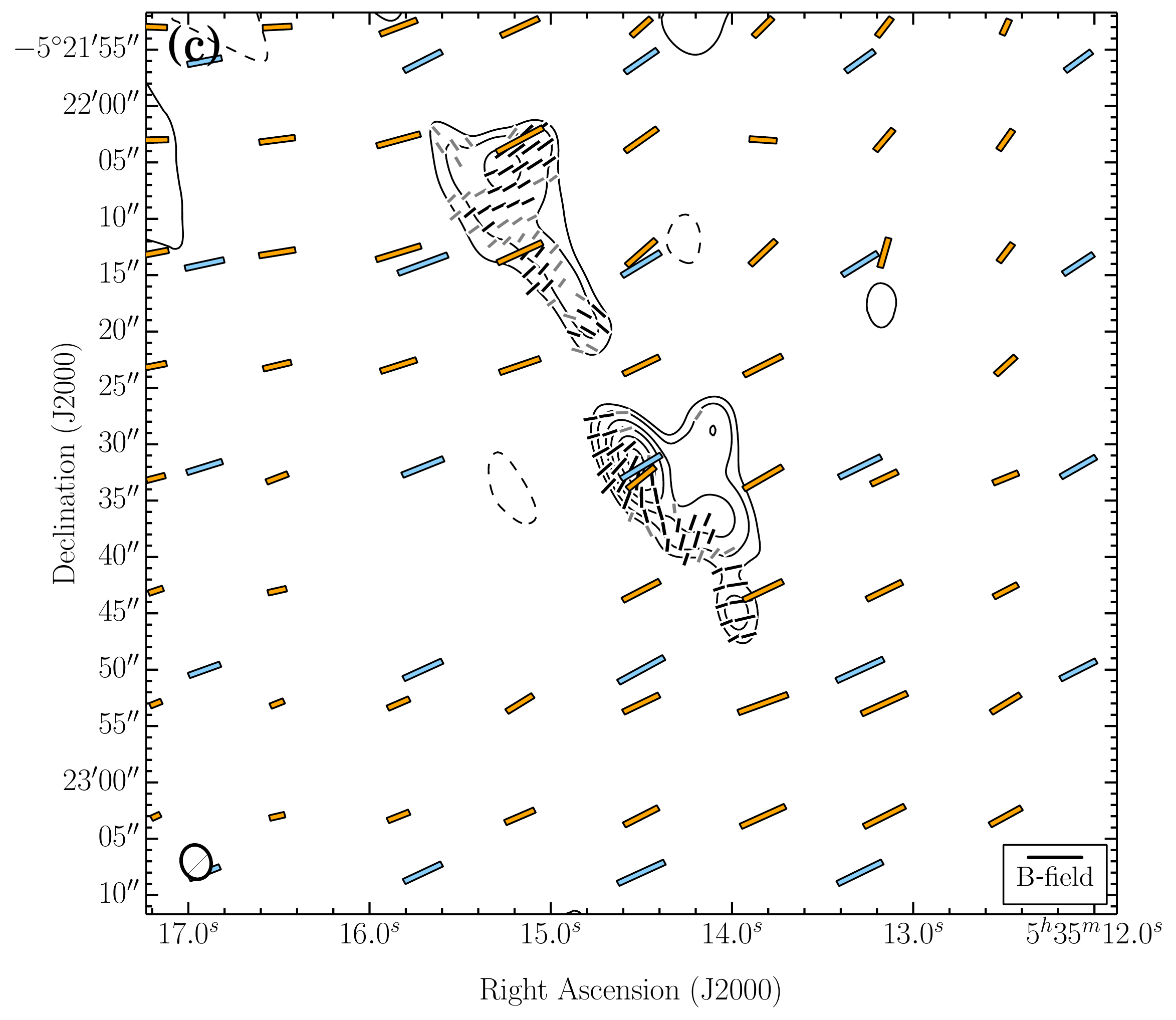}
\caption[]{ \footnotesize{
\input{OrionKL_caption.txt}
The full-width-half-max (FWHM) of the 7 mosaic pointings are plotted as gray, dotted circles.  Their
diameters ($\sim$\,44$\arcsec$) reflect the average primary beam size of the 6 and 10\,m telescopes.
There is no \textbf{(a)} plot because there were no spectral-line data to plot.
}}
\label{fig:OrionKL}
\end{center}
\end{figure*}

%%% Maps of OMC3-MMS5 and MMS6
\begin{figure*} [hbt!]
\begin{center}
\epsscale{.8}
\plottwo{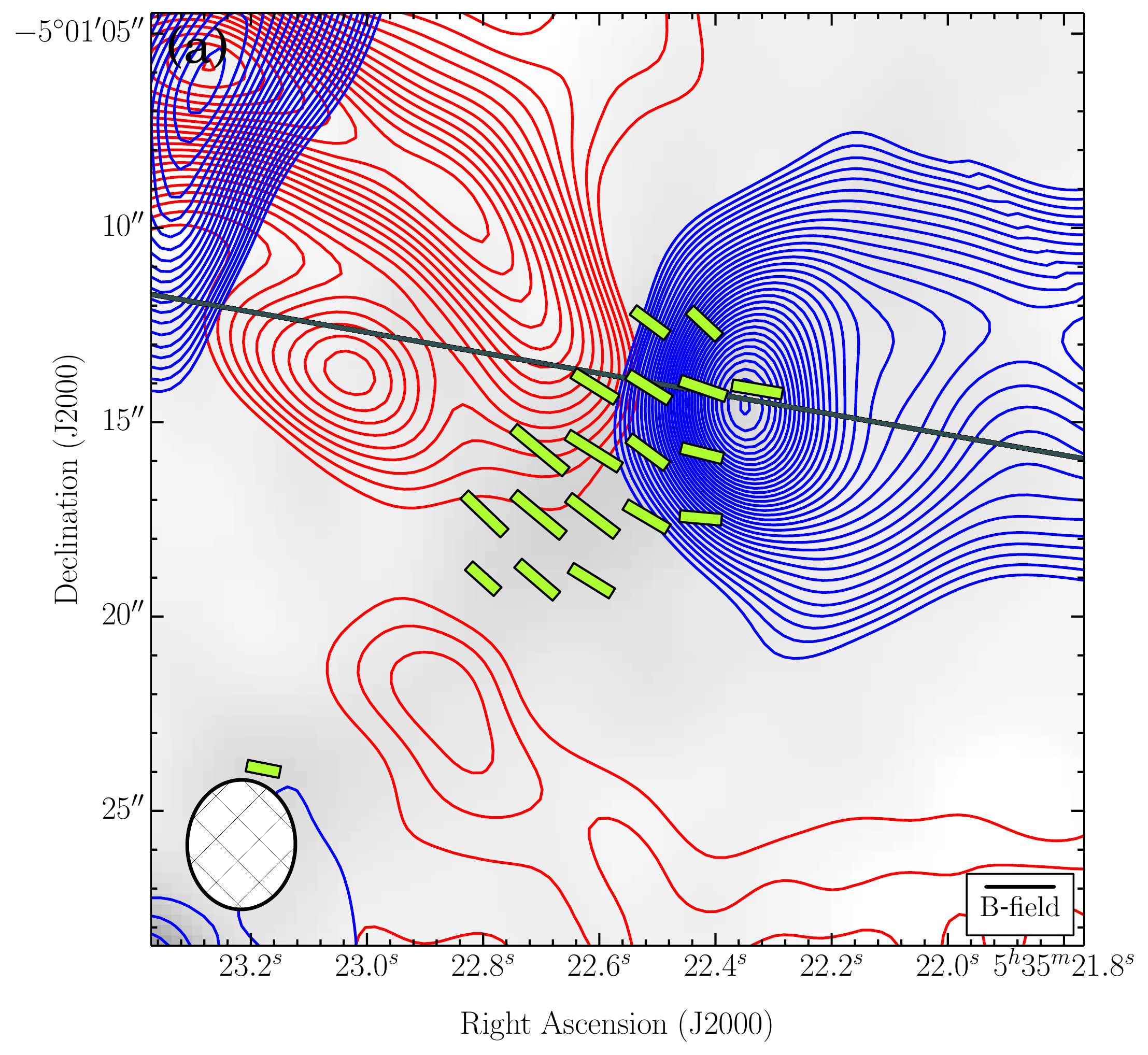}{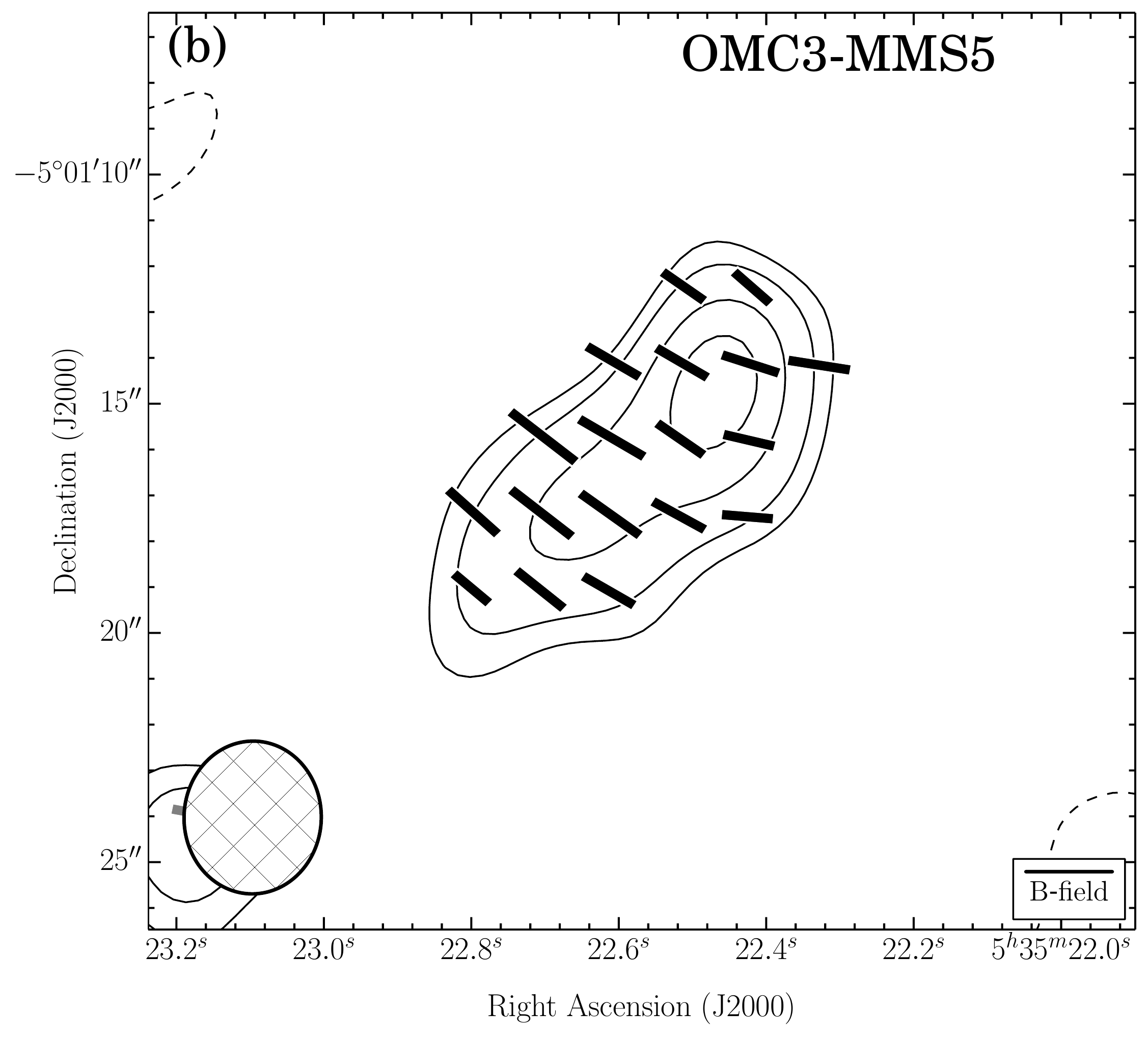}
\plottwo{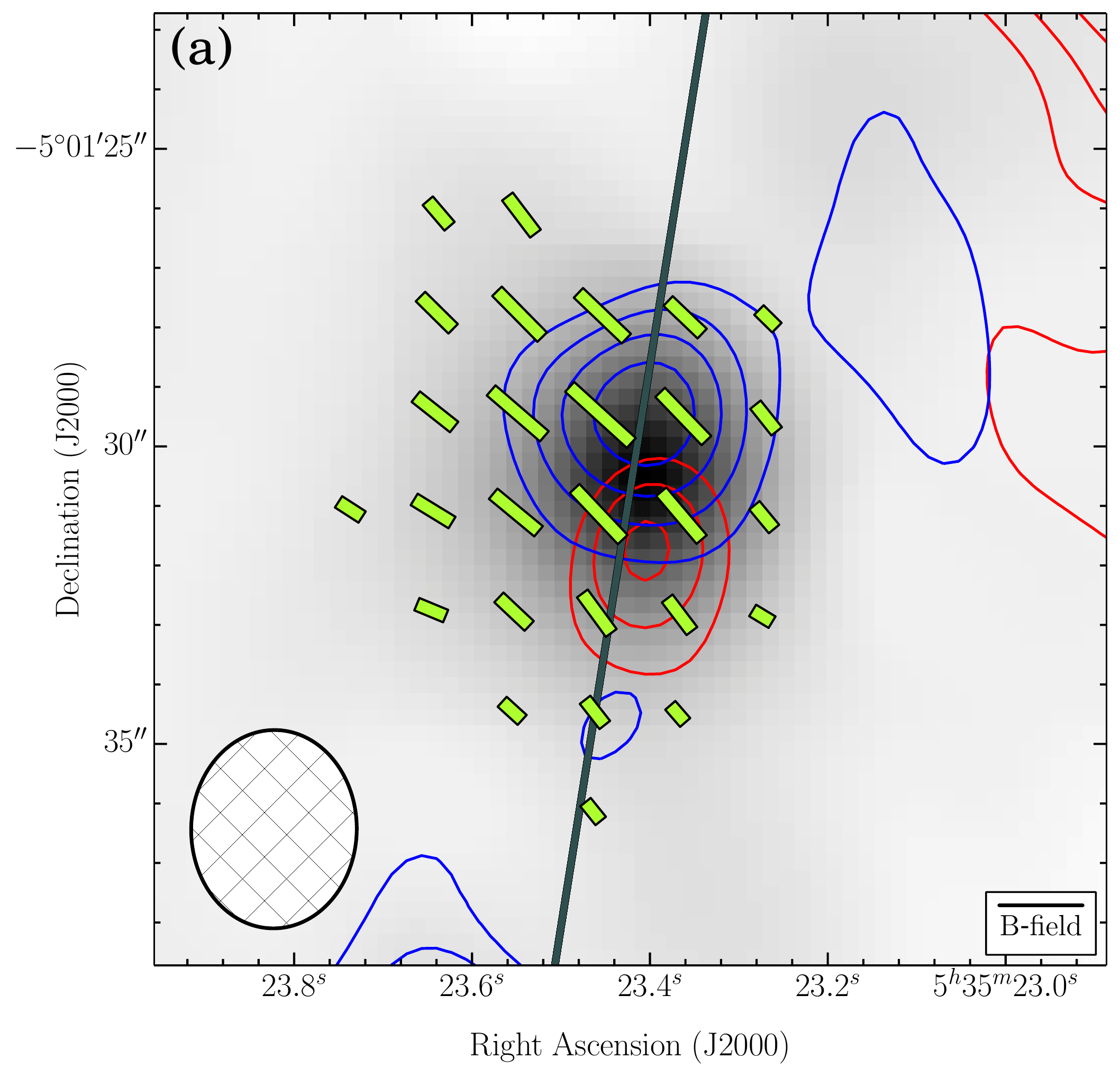}{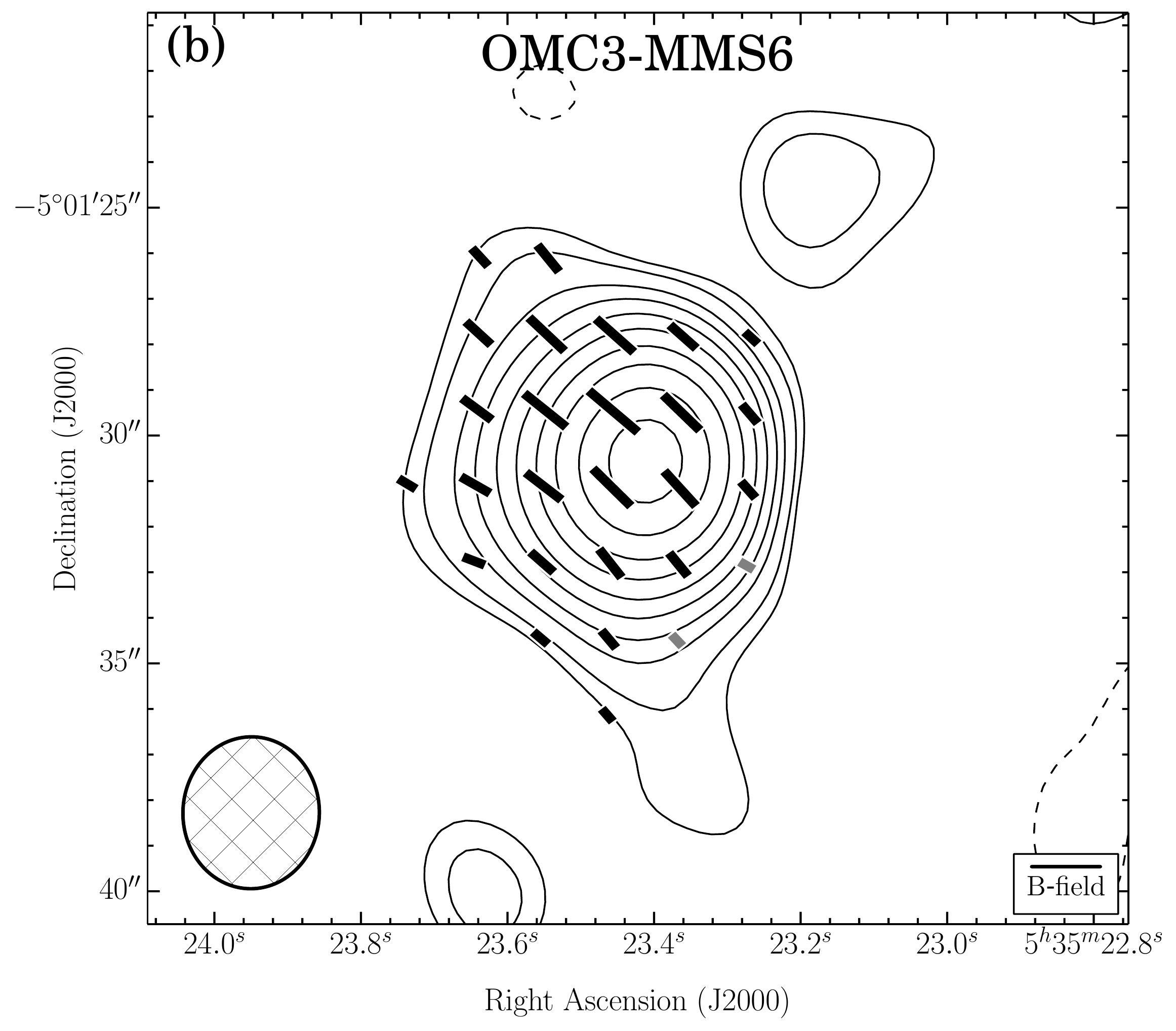}
\epsscale{0.6}
\plotone{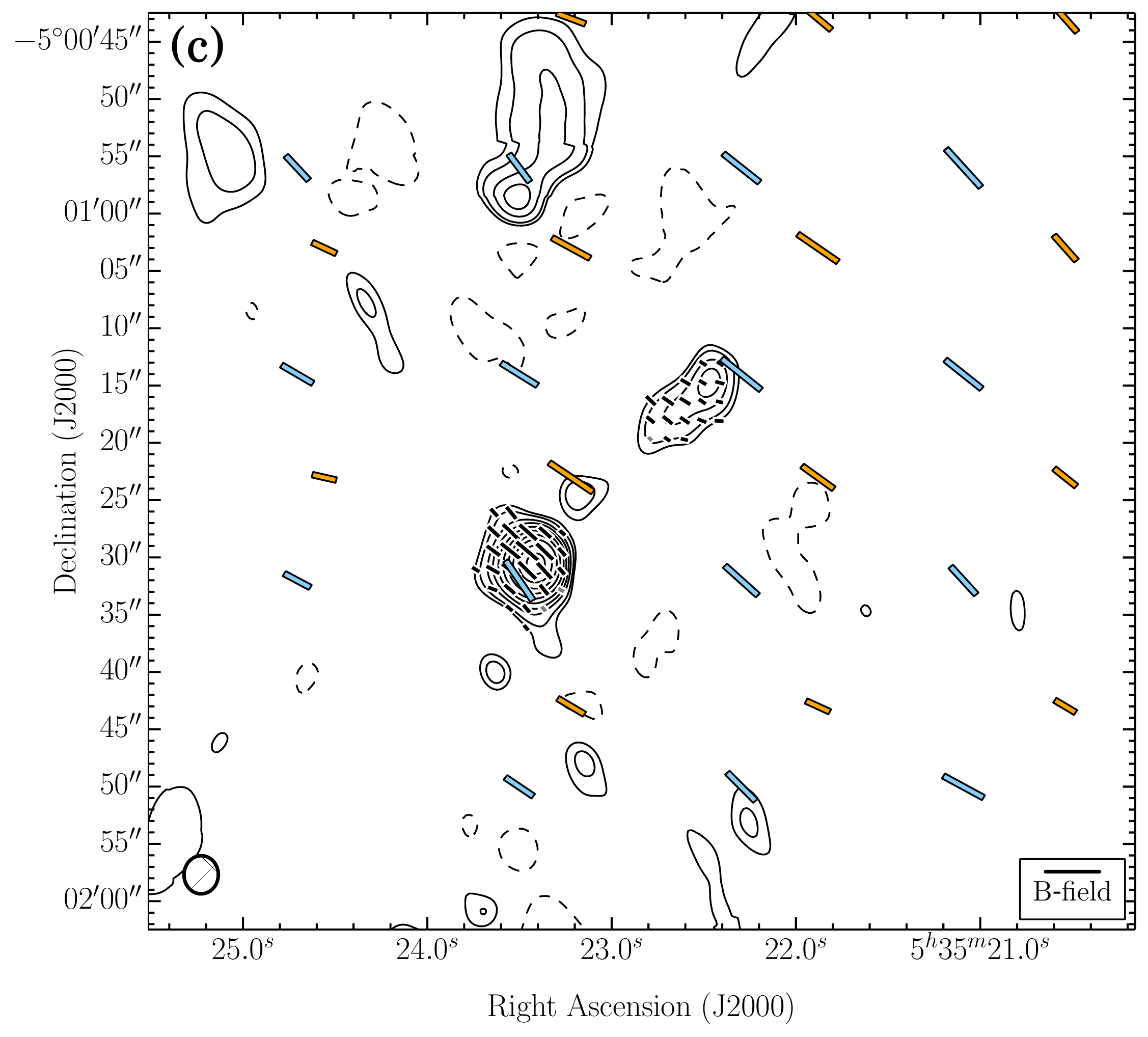}
\caption[]{ \footnotesize{
\input{MMS6_caption.txt}
}}
\label{fig:MMS6}
\end{center}
\end{figure*}

%%% Maps of OMC2-FIR3/4
\begin{figure*} [hbt!]
\begin{center}
\epsscale{1.1}
\plottwo{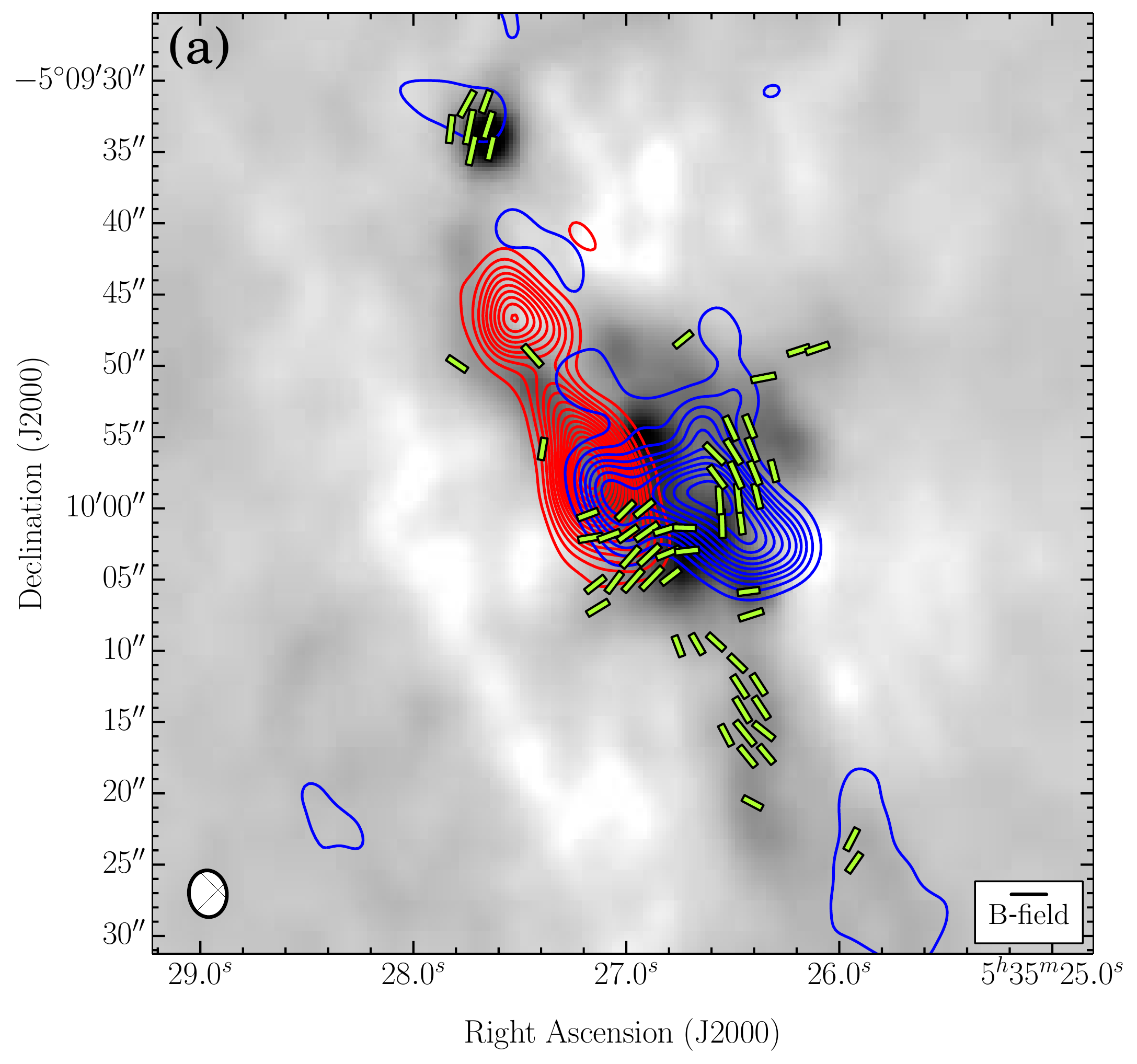}{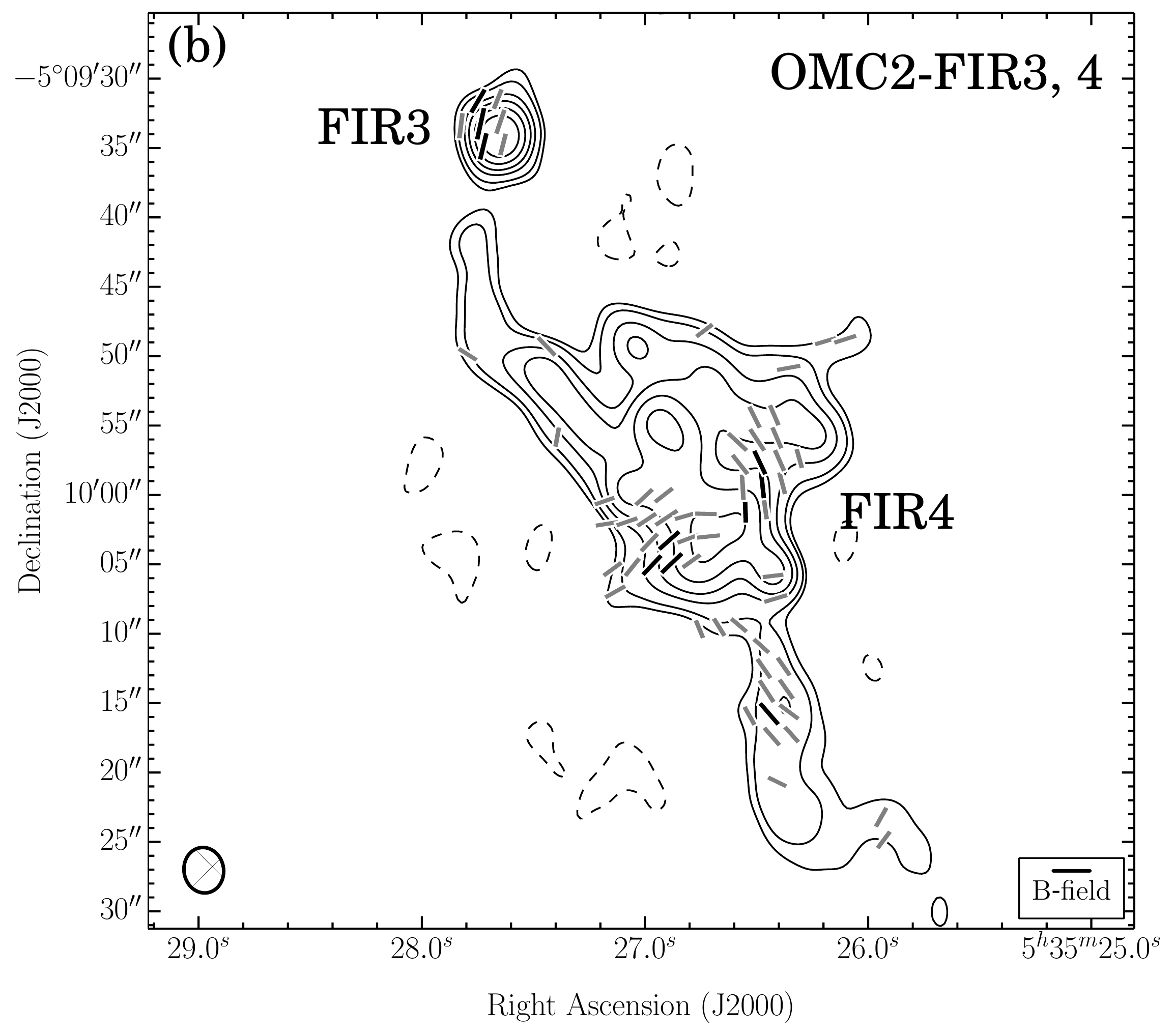}
\epsscale{0.8}
\plotone{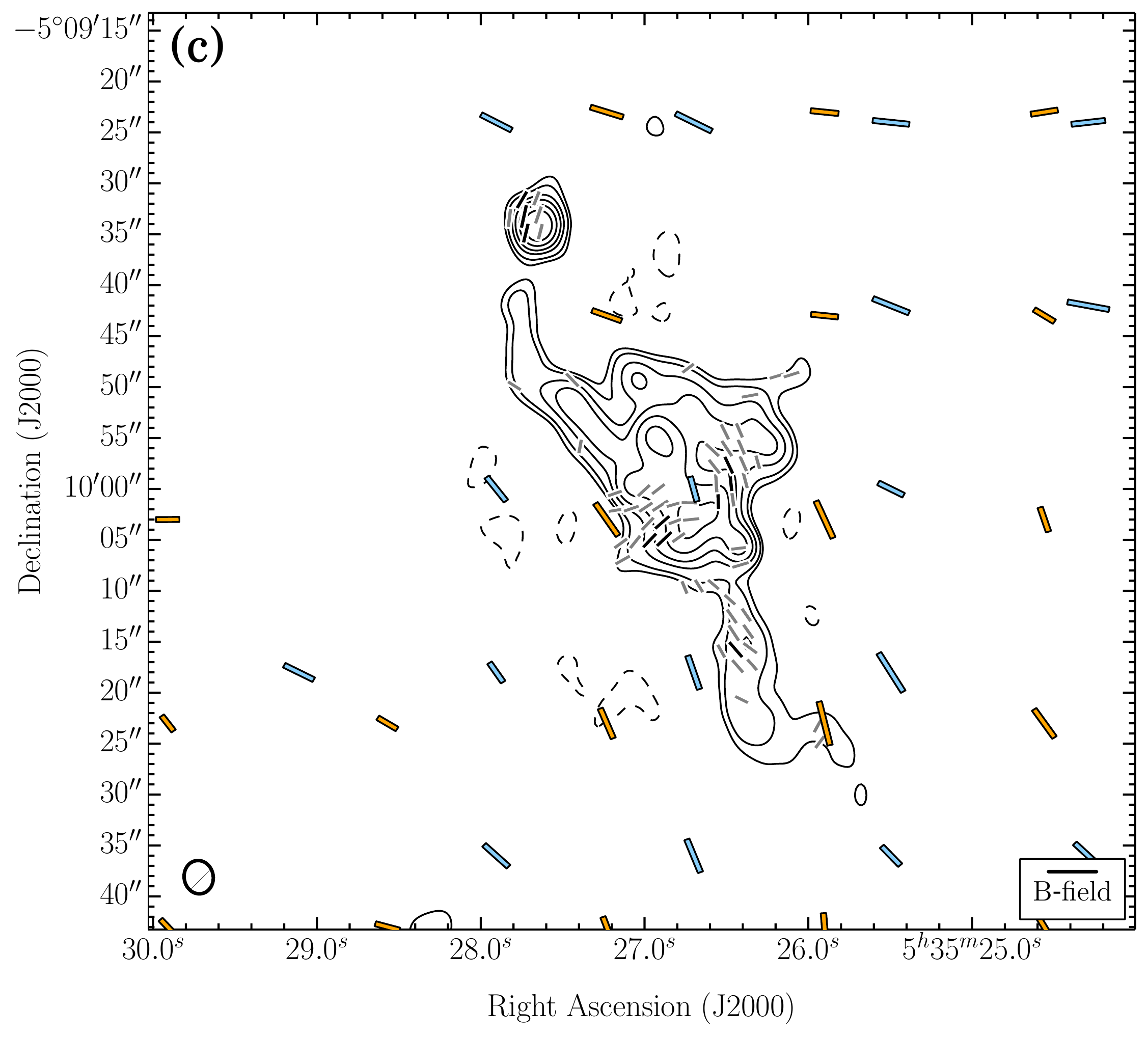}
\caption[]{ \footnotesize{
\input{FIR4_caption.txt}
}}
\label{fig:FIR4}
\end{center}
\end{figure*}

%%% Maps of CB~54
\begin{figure*} [hbt!]
\begin{center}
\epsscale{1.1}
\plottwo{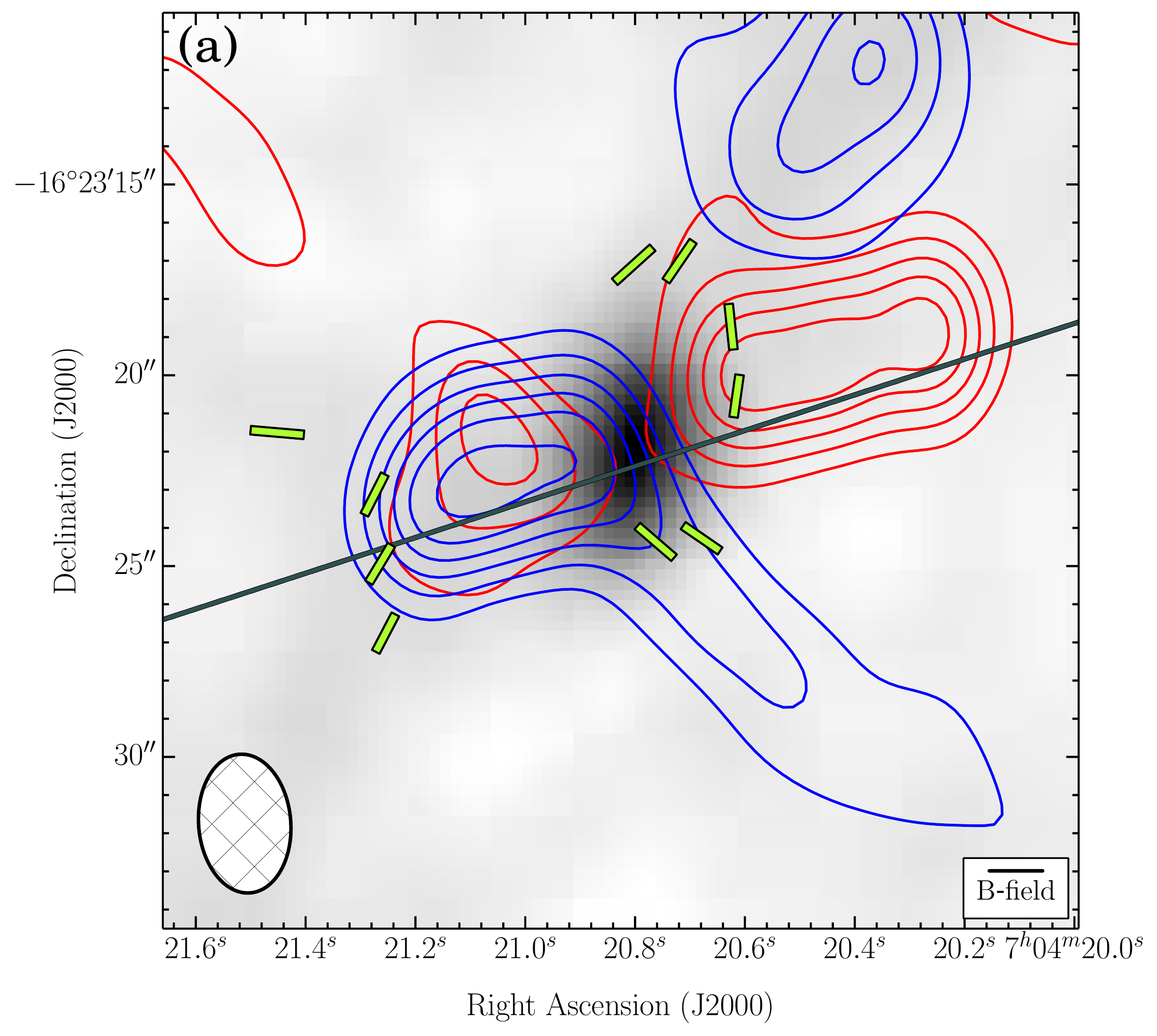}{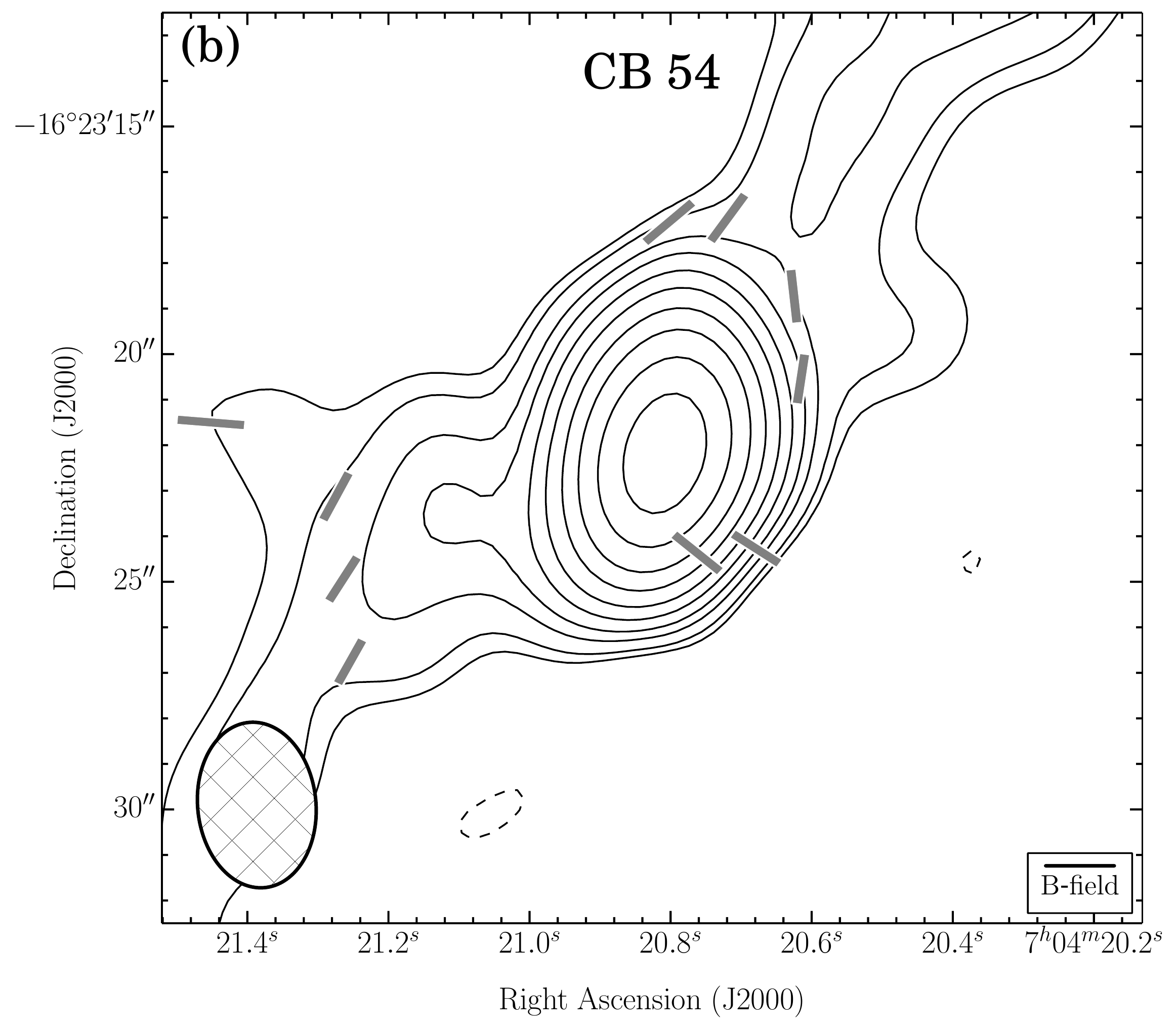}
\epsscale{0.8}
\plotone{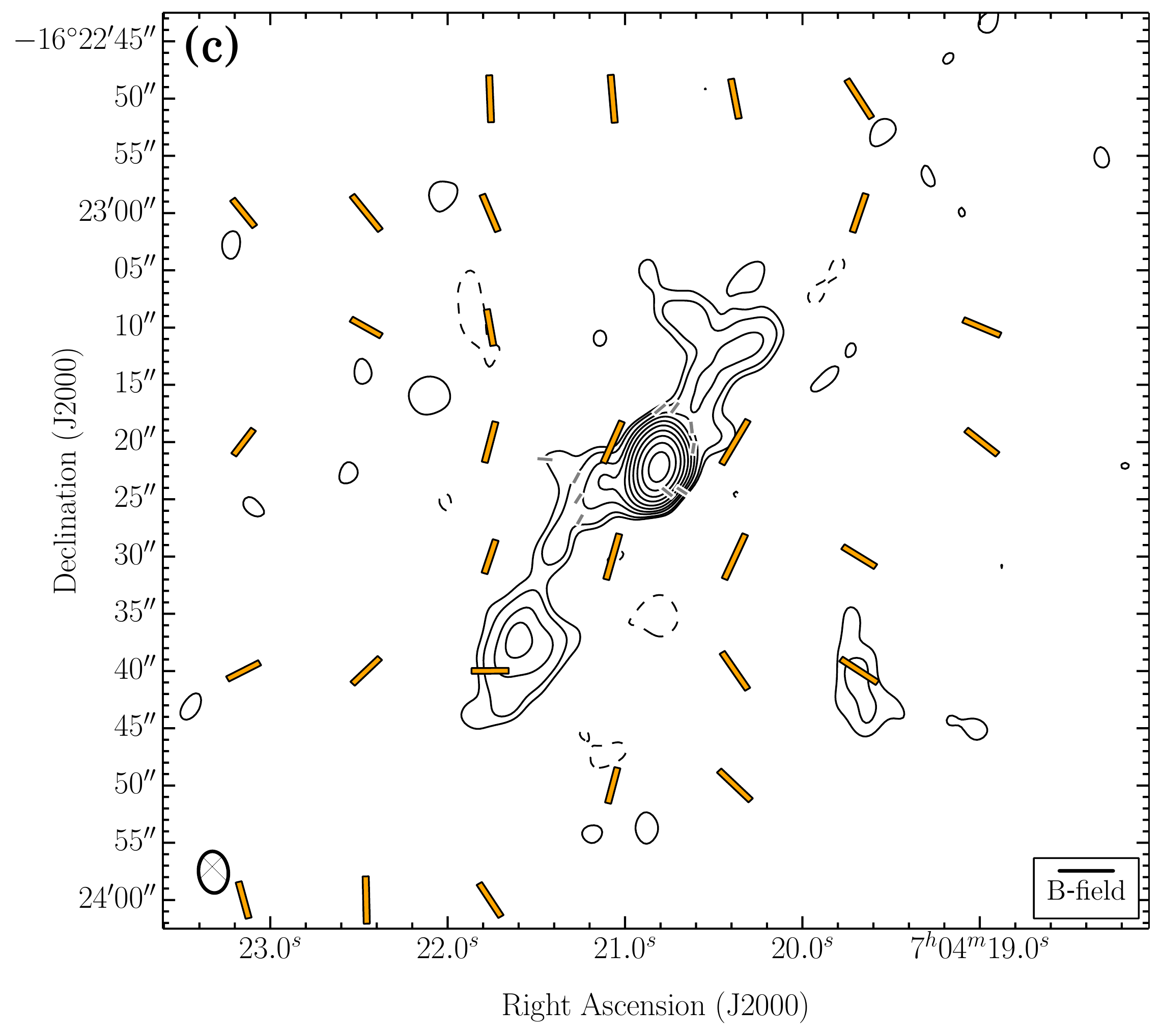}
\caption[]{ \footnotesize{
\input{CB54_caption.txt}
}}
\label{fig:CB54}
\end{center}
\end{figure*}

%%% Maps of VLA~1623
\begin{figure*} [hbt!]
\begin{center}
\epsscale{1.1}
\plottwo{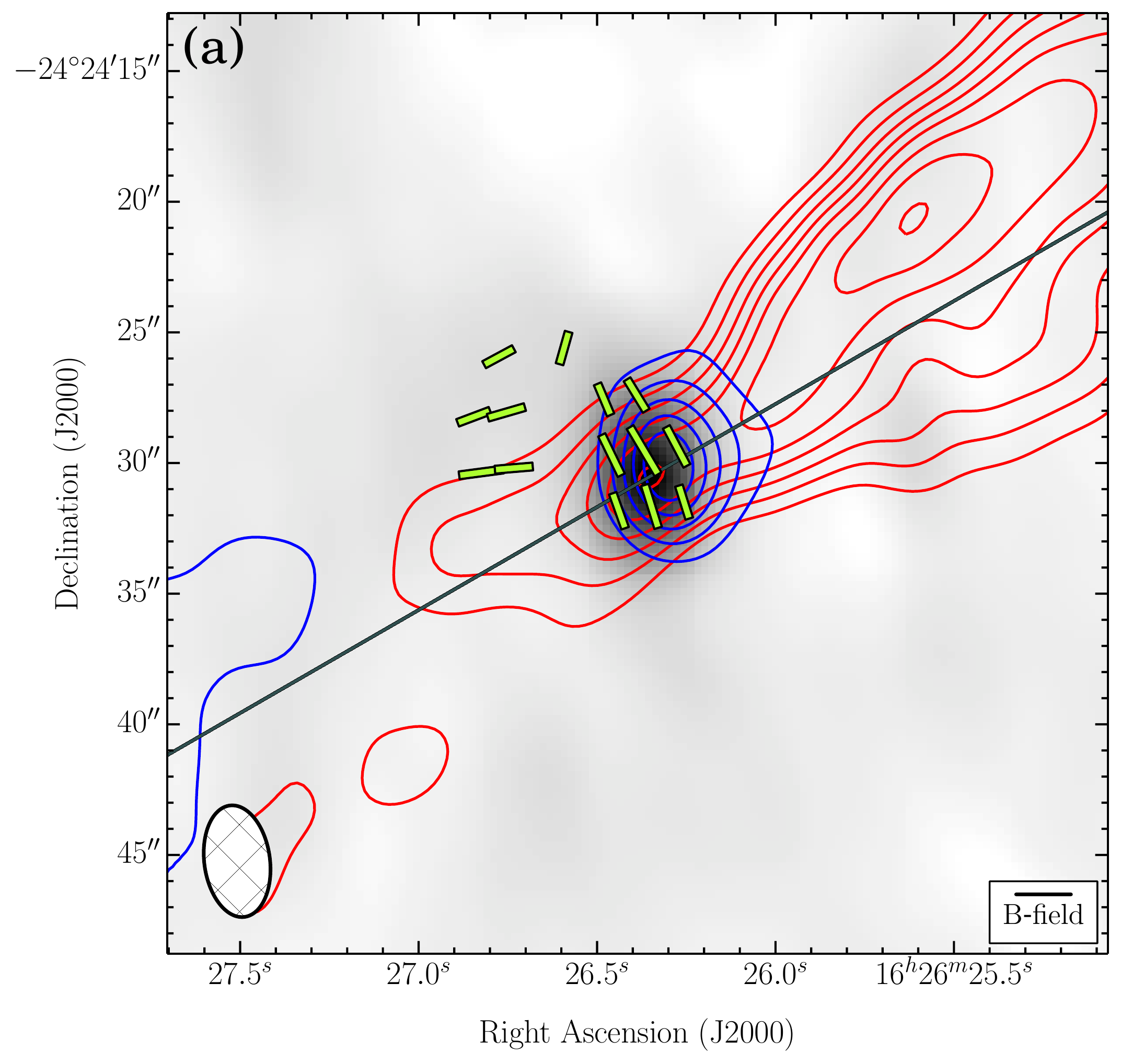}{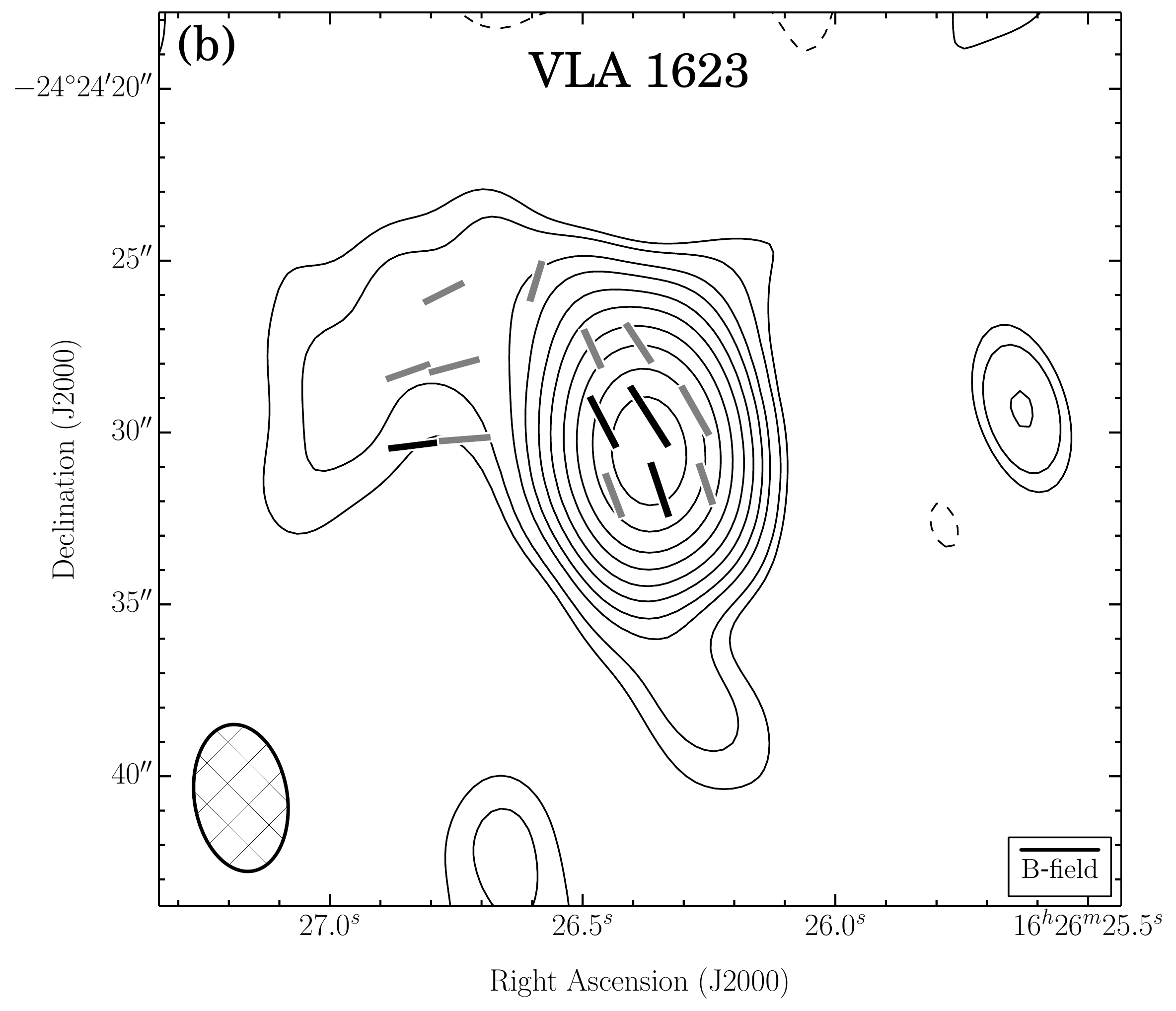}
\epsscale{0.8}
\plotone{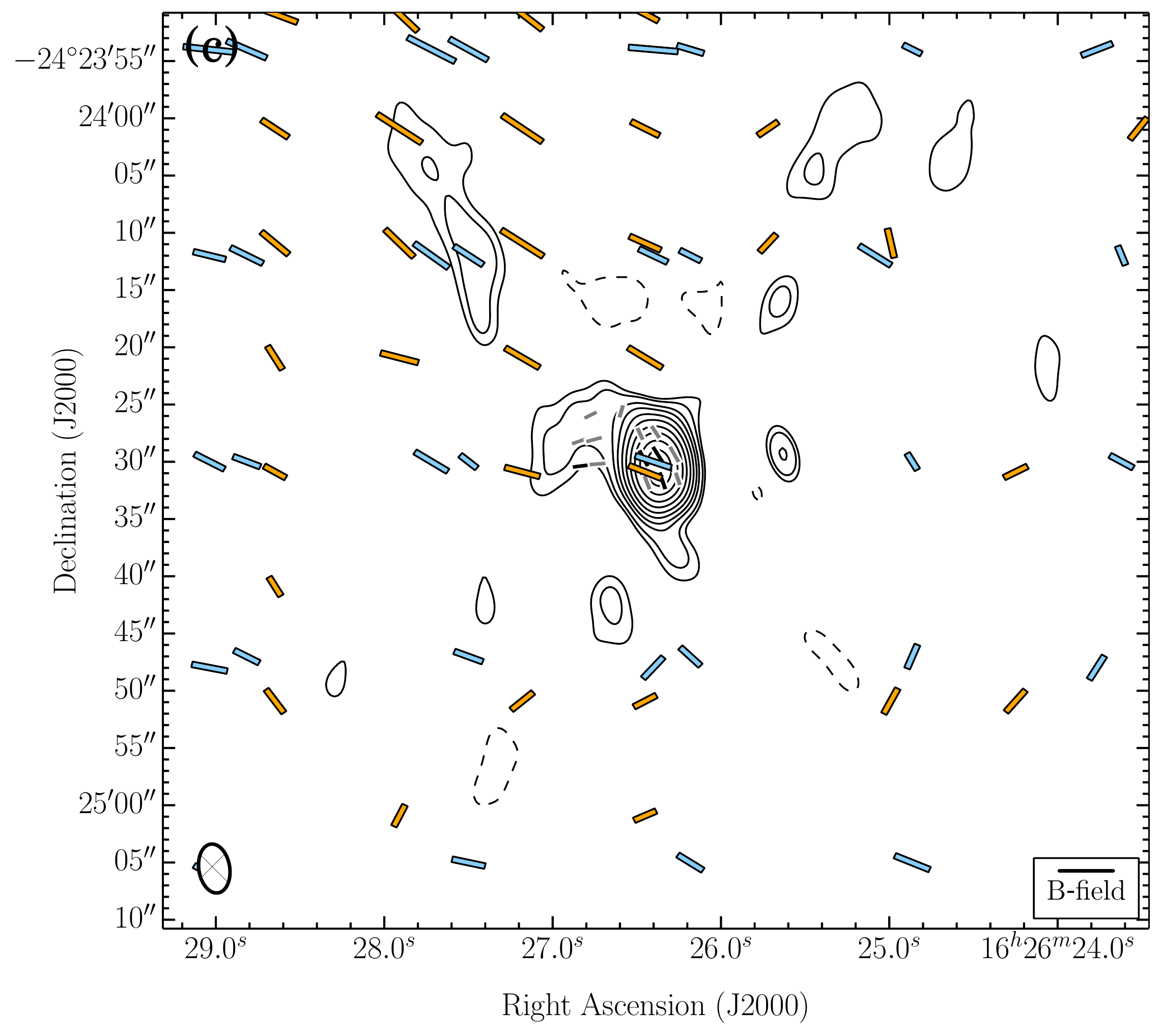}
\caption[]{ \footnotesize{
\input{VLA1623_caption.txt}
}}
\label{fig:VLA1623}
\end{center}
\end{figure*}

%%% Maps of Ser-emb~17
\begin{figure*} [hbt!]
\begin{center}
\epsscale{1.1}
\plottwo{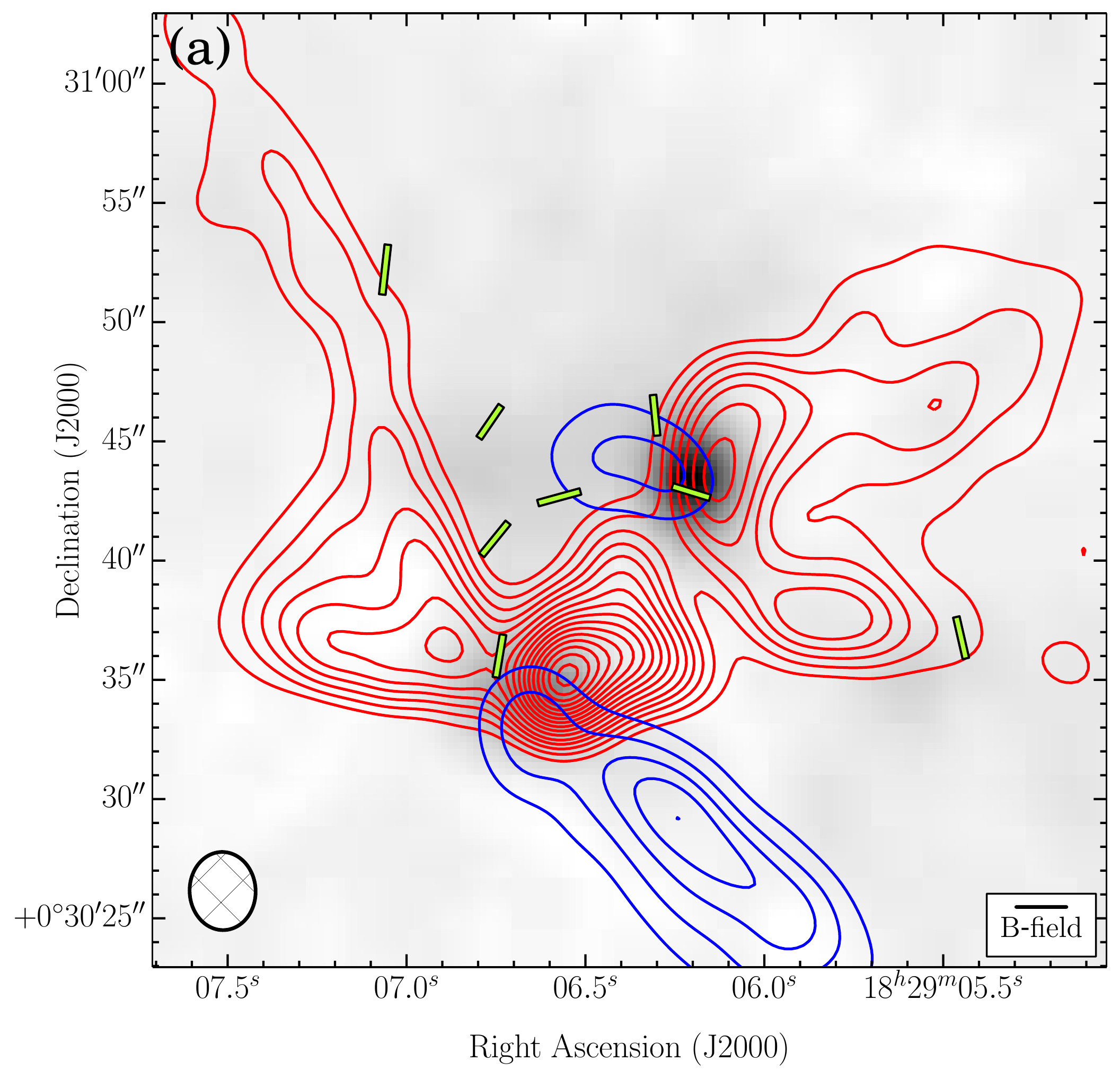}{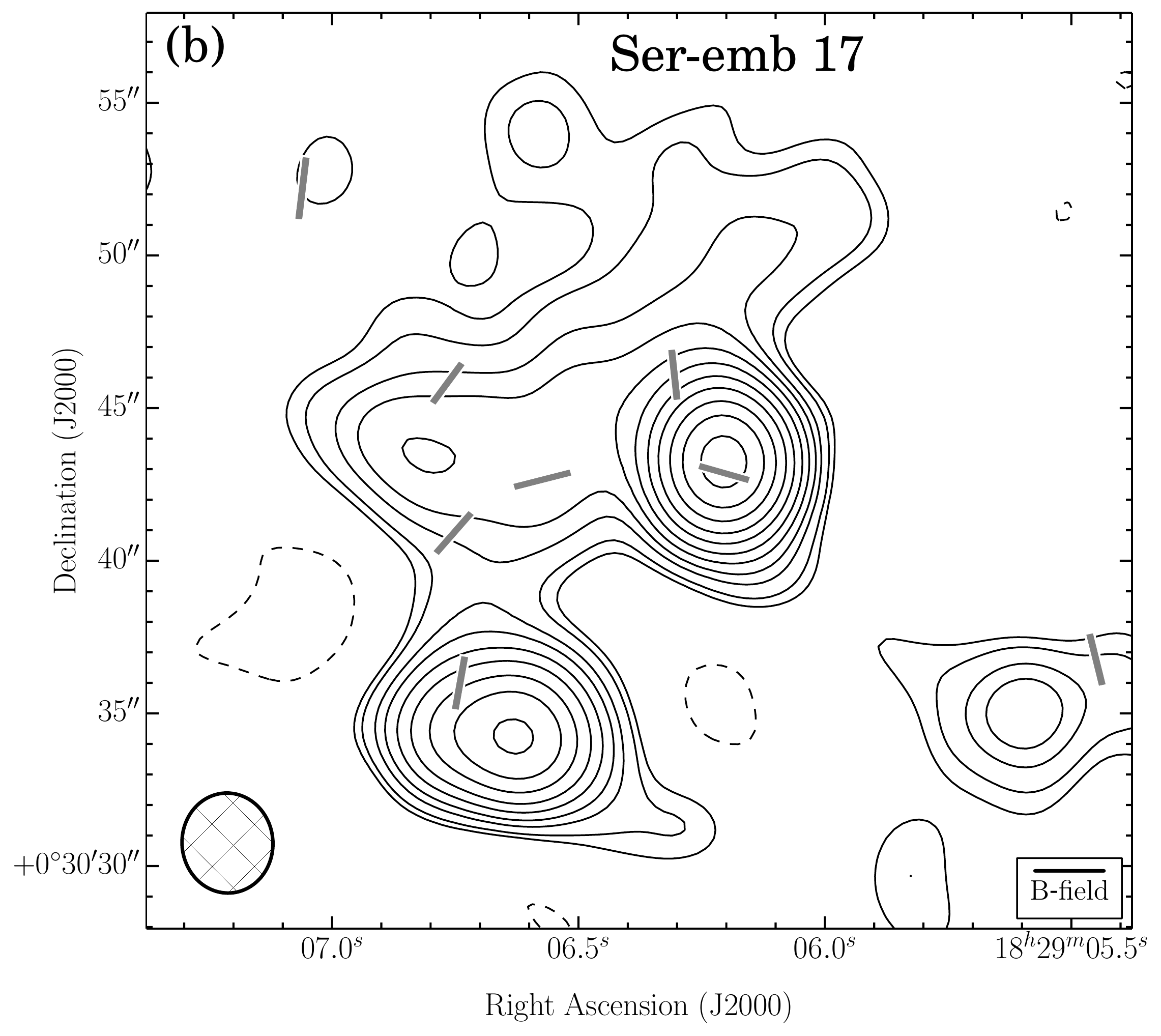}
\caption[]{ \footnotesize{
\input{SerEmb17_caption.txt}
There is no \textbf{(c)} plot because there were no SCUBA, SHARP, or Hertz data to overlay.}}
\label{fig:SerEmb17}
\end{center}
\end{figure*}

%%% Maps of Ser-emb~1
\begin{figure*} [hbt!]
\begin{center}
\epsscale{1.1}
\plottwo{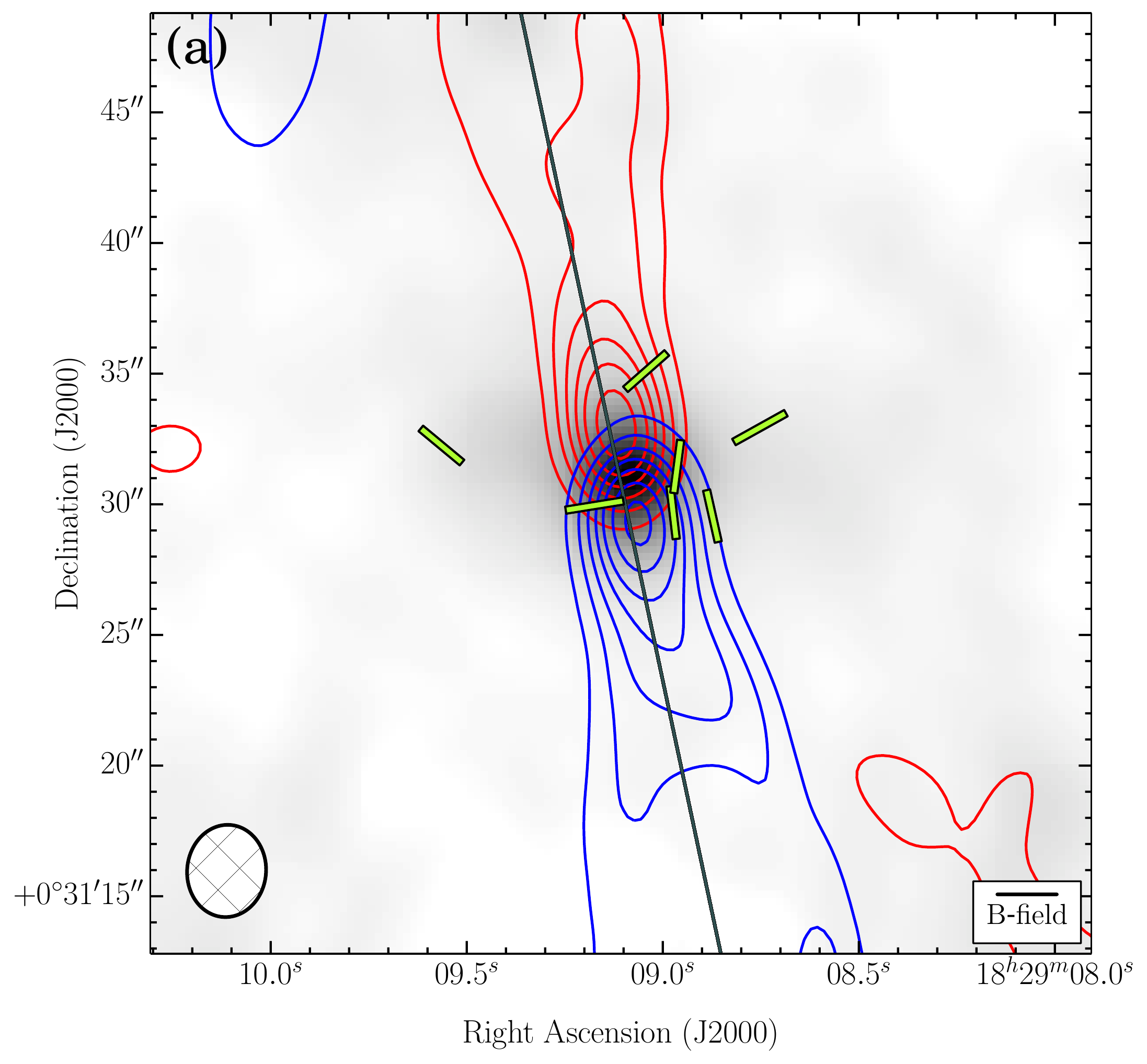}{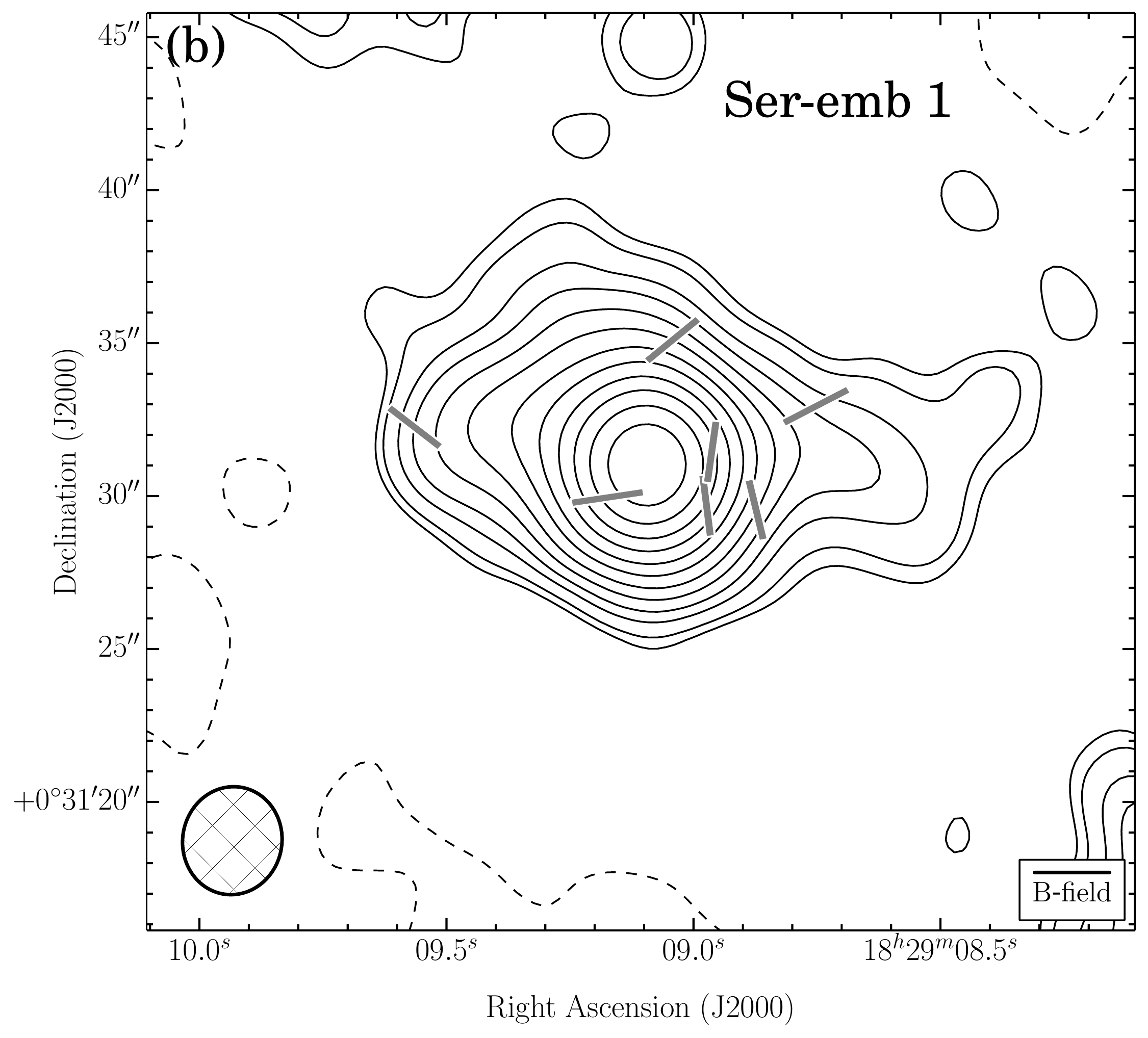}
\caption[]{ \footnotesize{
\input{SerEmb1_caption.txt}
There is no \textbf{(c)} plot because there were no SCUBA, SHARP, or Hertz data to overlay.}}
\label{fig:SerEmb1}
\end{center}
\end{figure*}

%%% Maps of Ser-emb~8 and 8(N)
\begin{figure*} [hbt!]
\begin{center}
\epsscale{1.1}
\plottwo{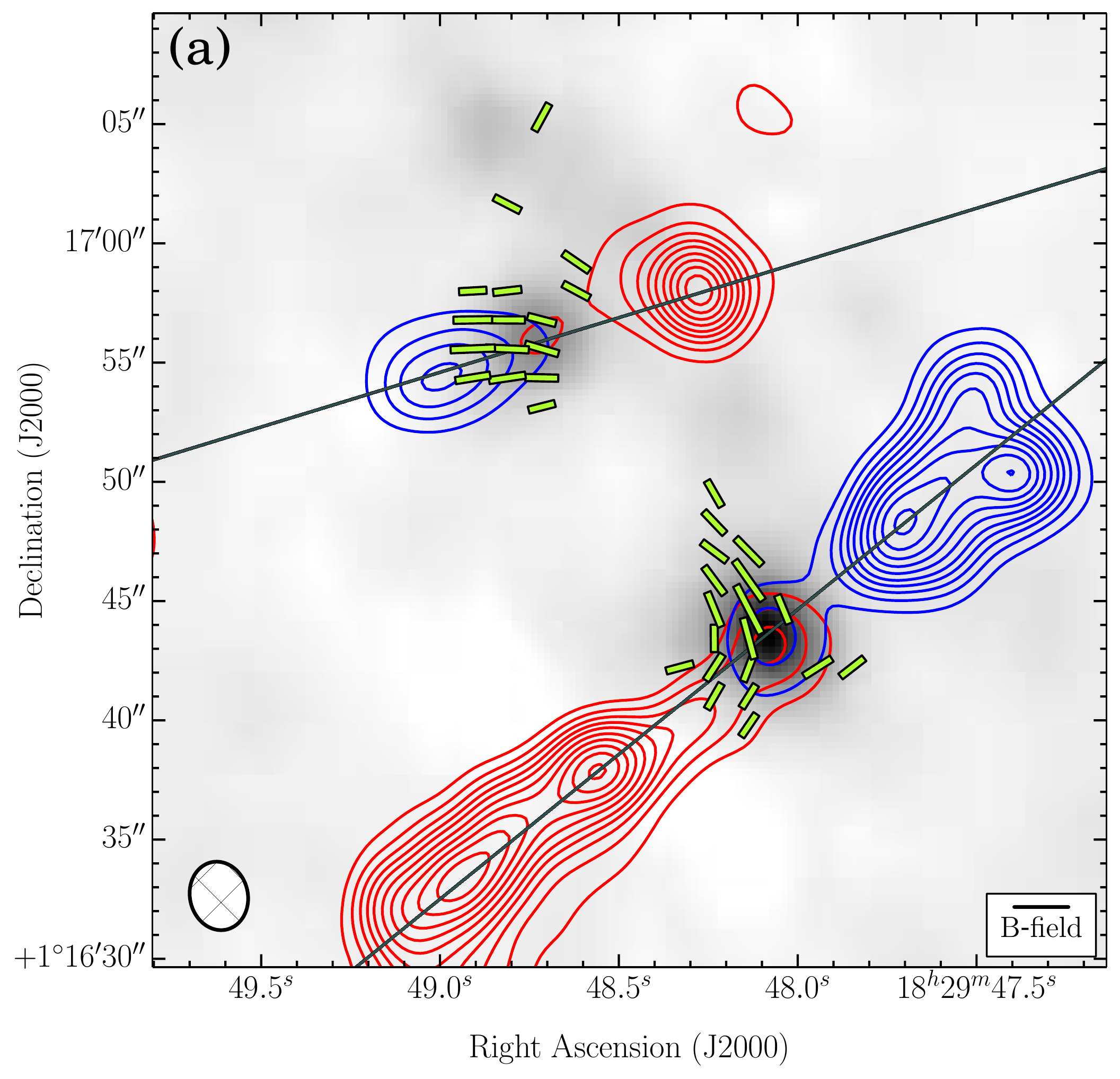}{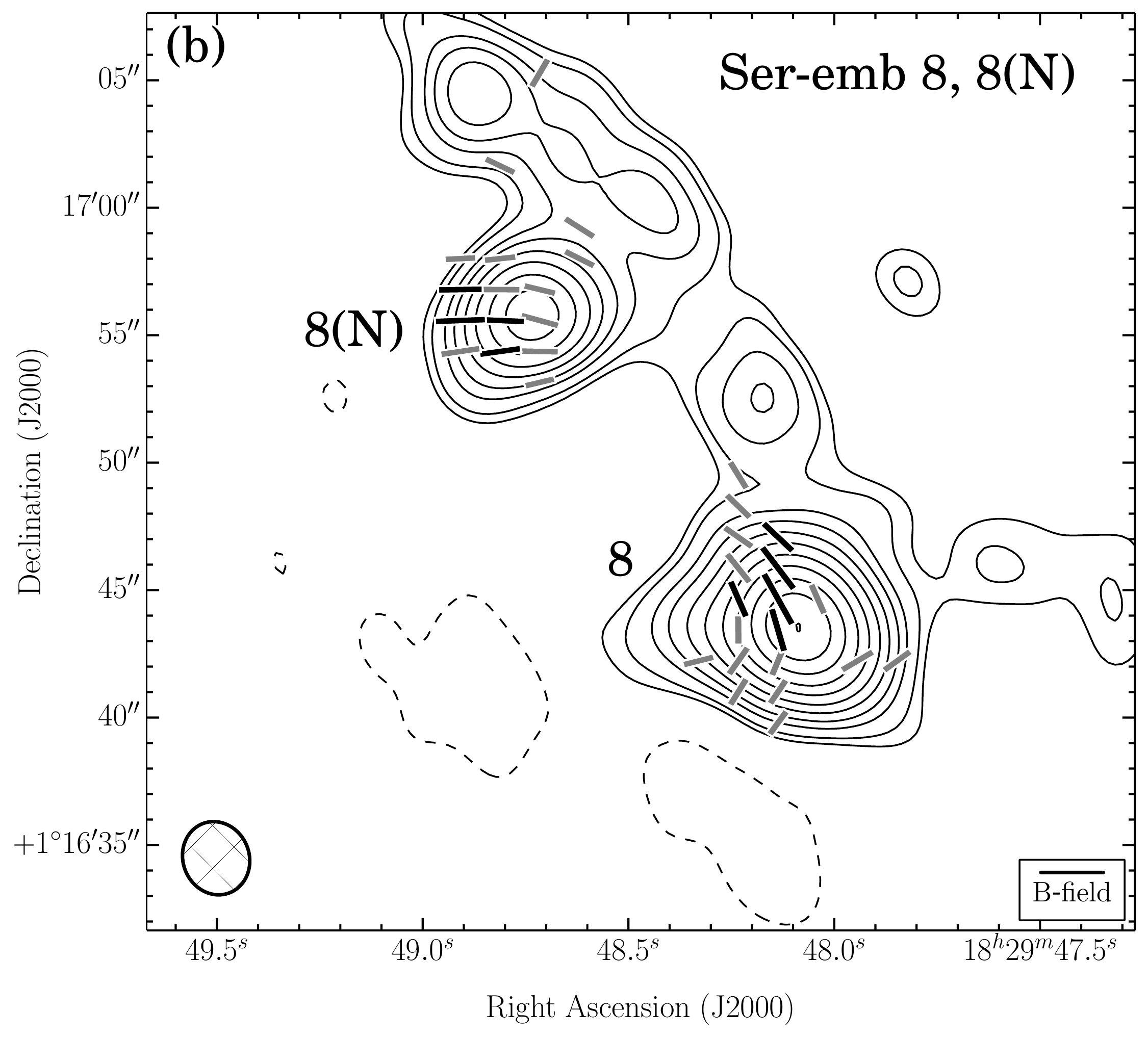}
\epsscale{0.8}
\plotone{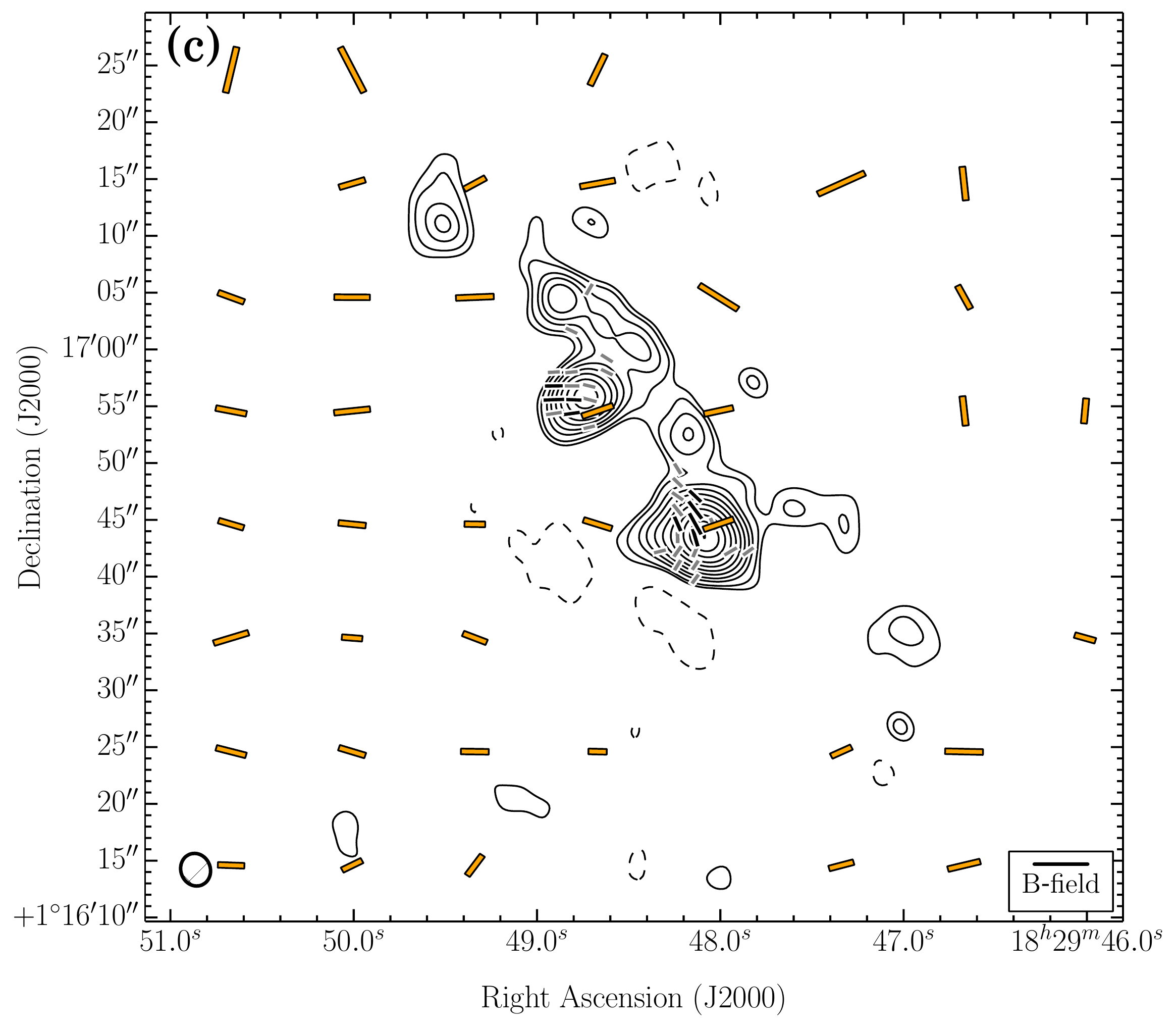}
\caption[]{ \footnotesize{
\input{SerEmb8_caption.txt}
Note: in \citet{Hull2012}, the colors of the red- and blueshifted outflow lobes were accidentally reversed.
}}
\label{fig:SerEmb8}
\end{center}
\end{figure*}

%%% Maps of Ser-emb~6
\begin{figure*} [hbt!]
\begin{center}
\epsscale{1.1}
\plottwo{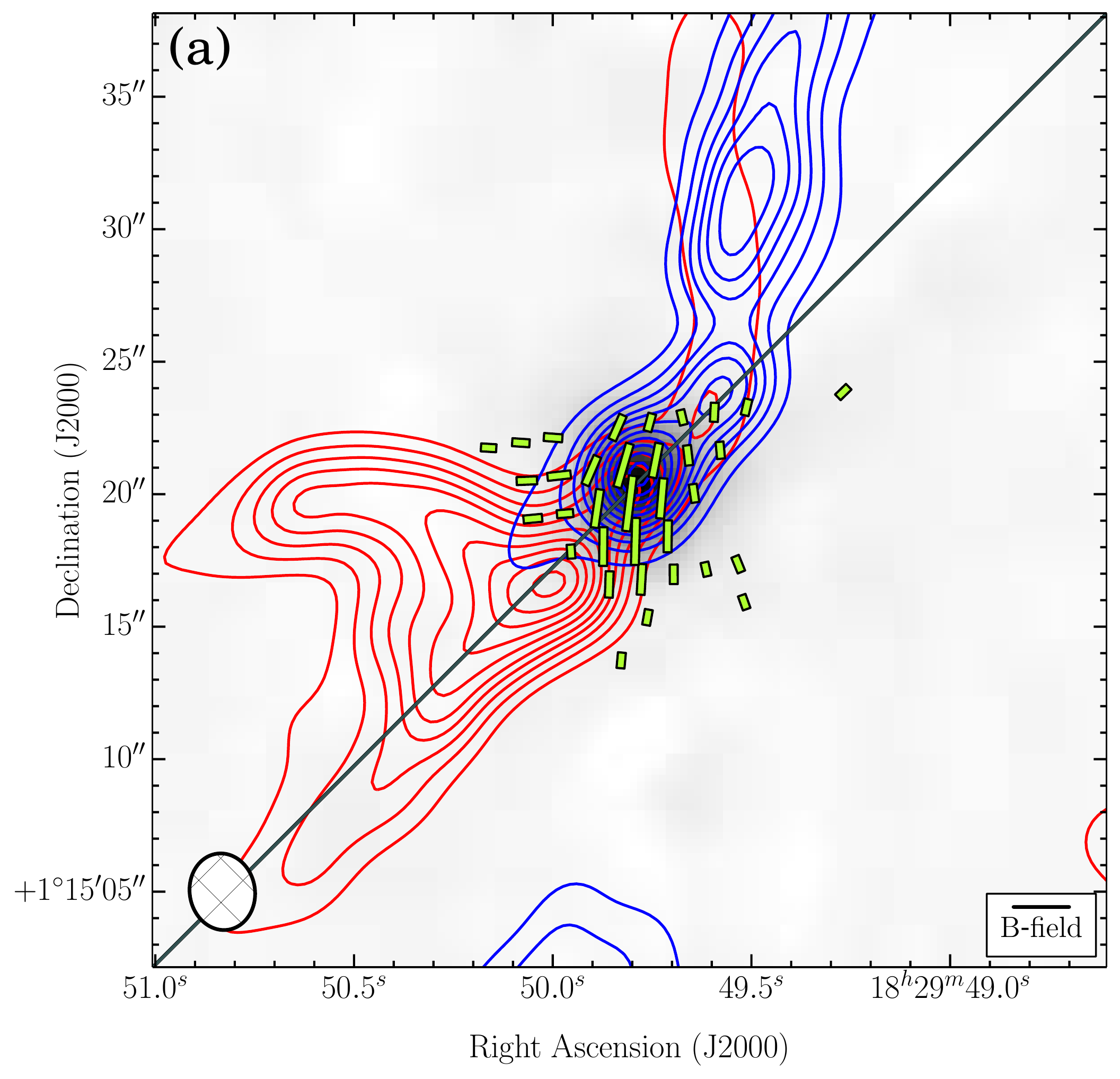}{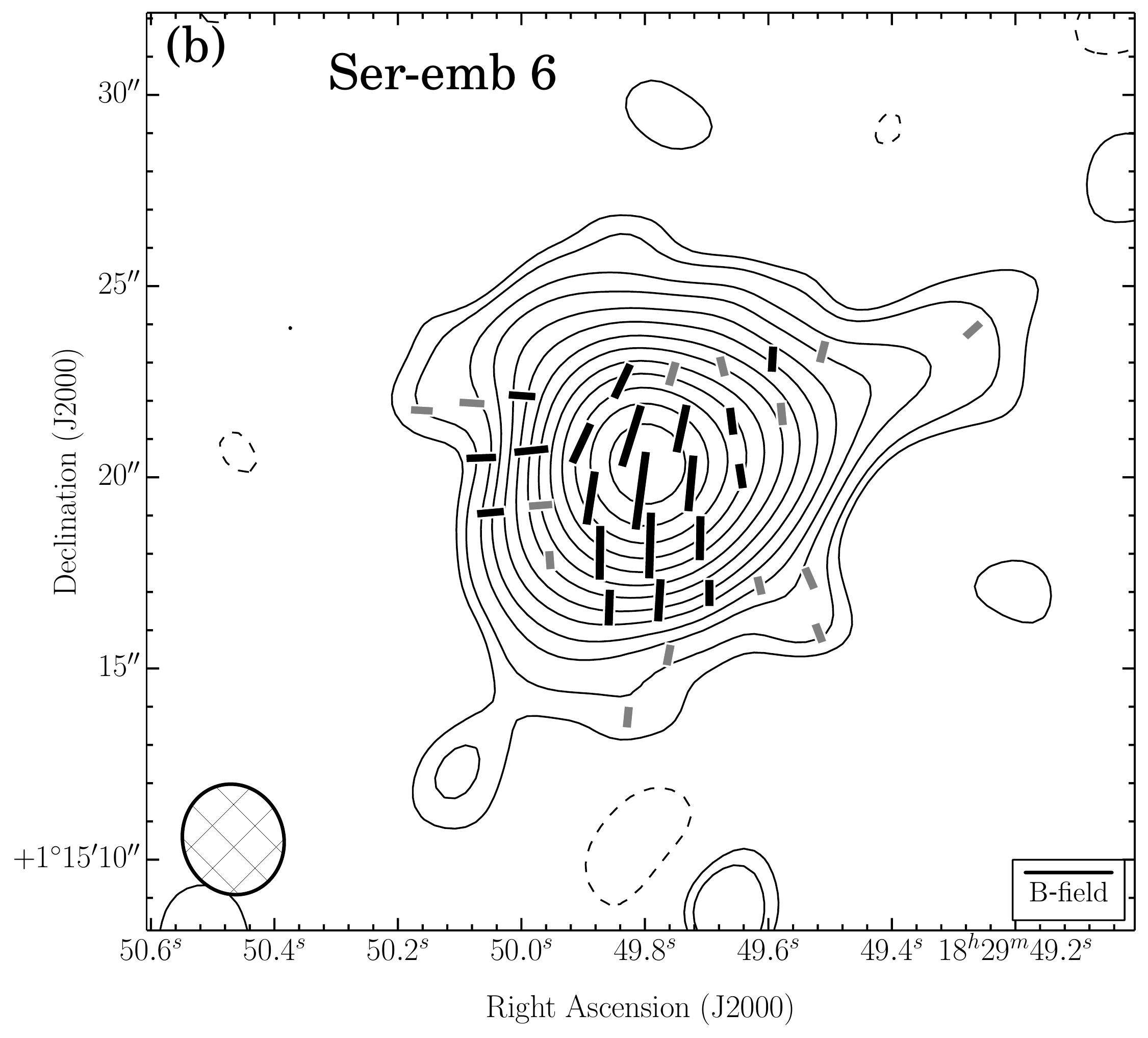}
\epsscale{0.8}
\plotone{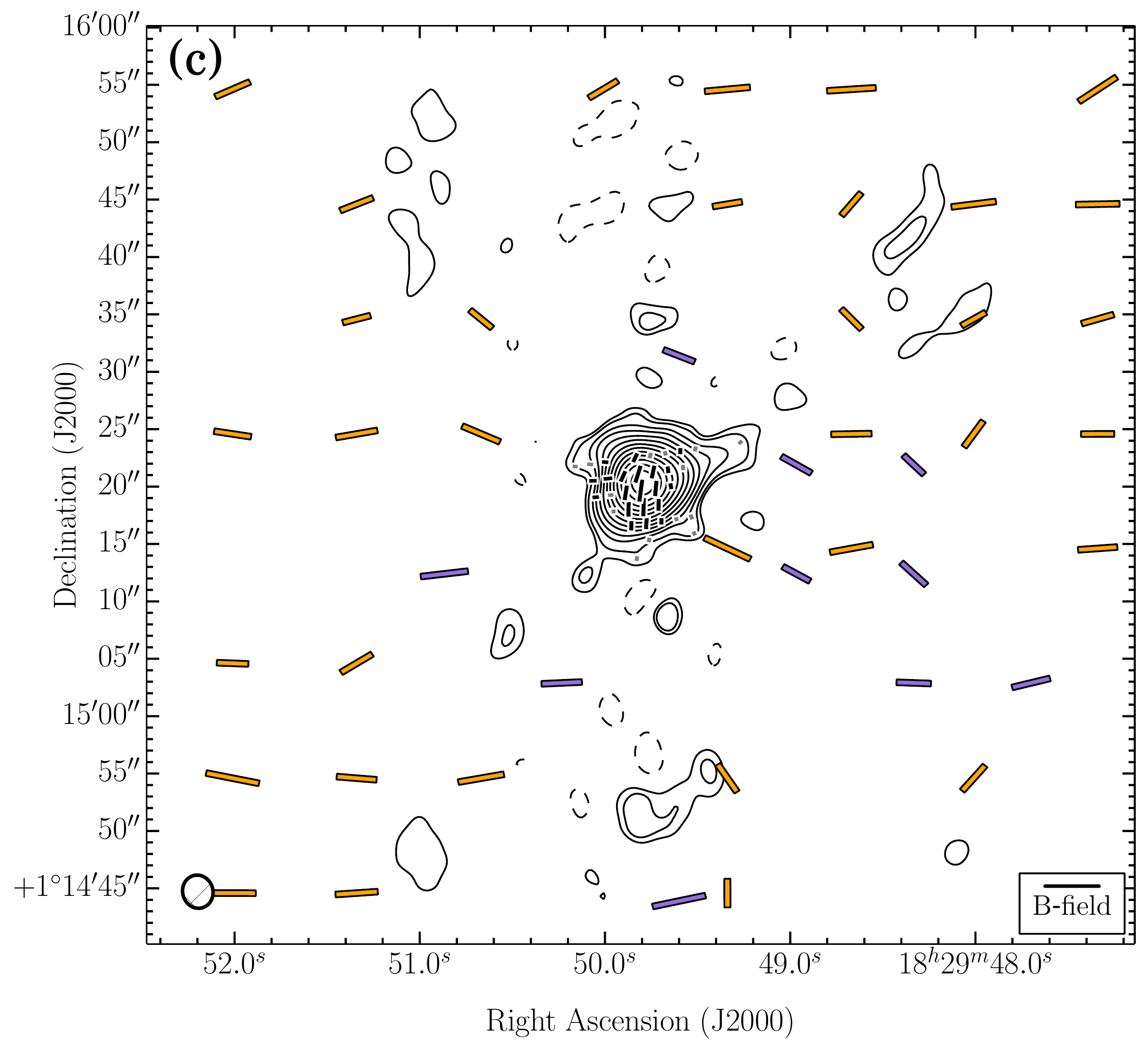}
\caption[]{ \footnotesize{
\input{SerEmb6_caption.txt}
}}
\label{fig:SerEmb6}
\end{center}
\end{figure*}

%%% Maps of HH~108~IRAS
\begin{figure*} [hbt!]
\begin{center}
\epsscale{1.1}
\plottwo{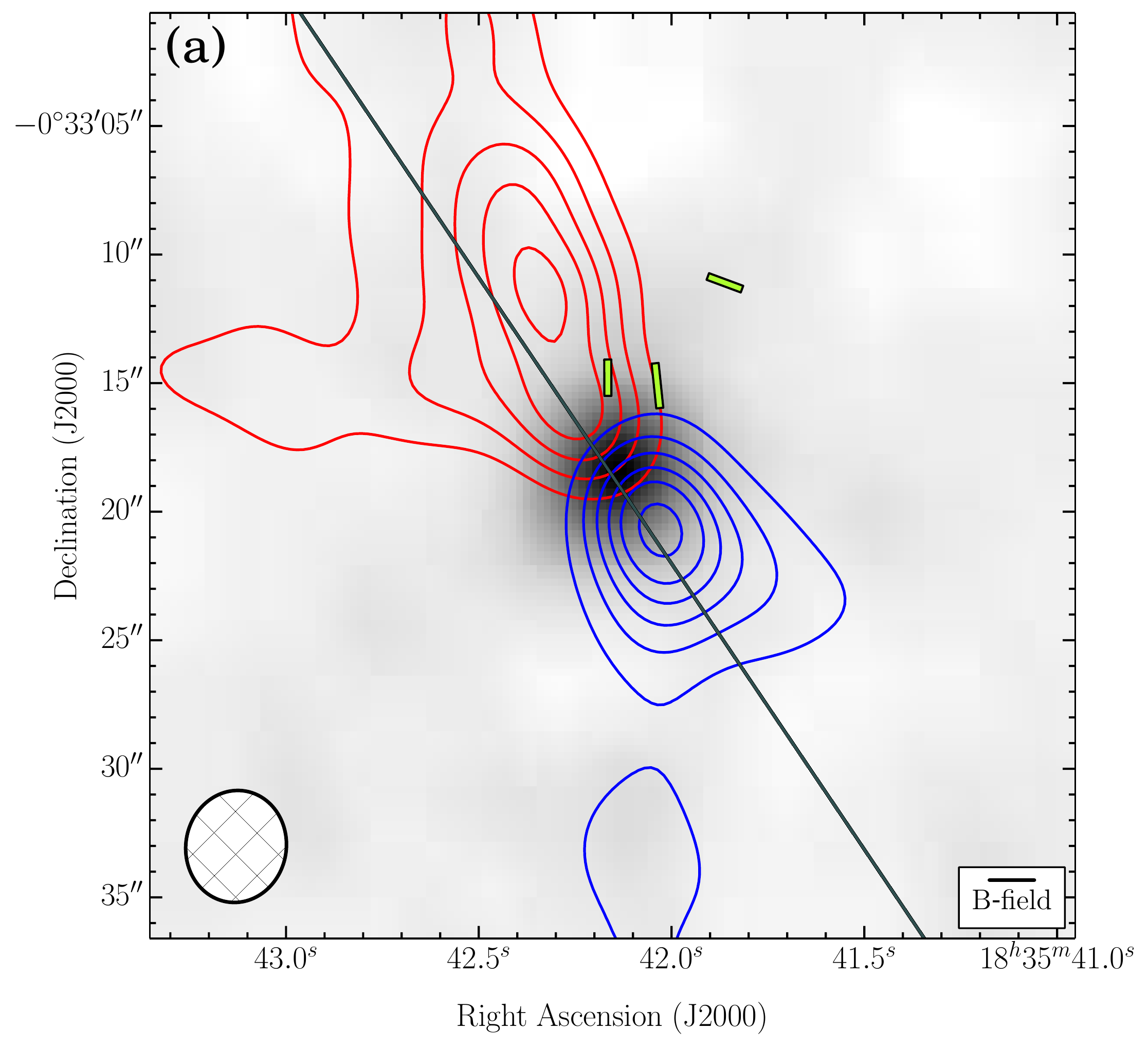}{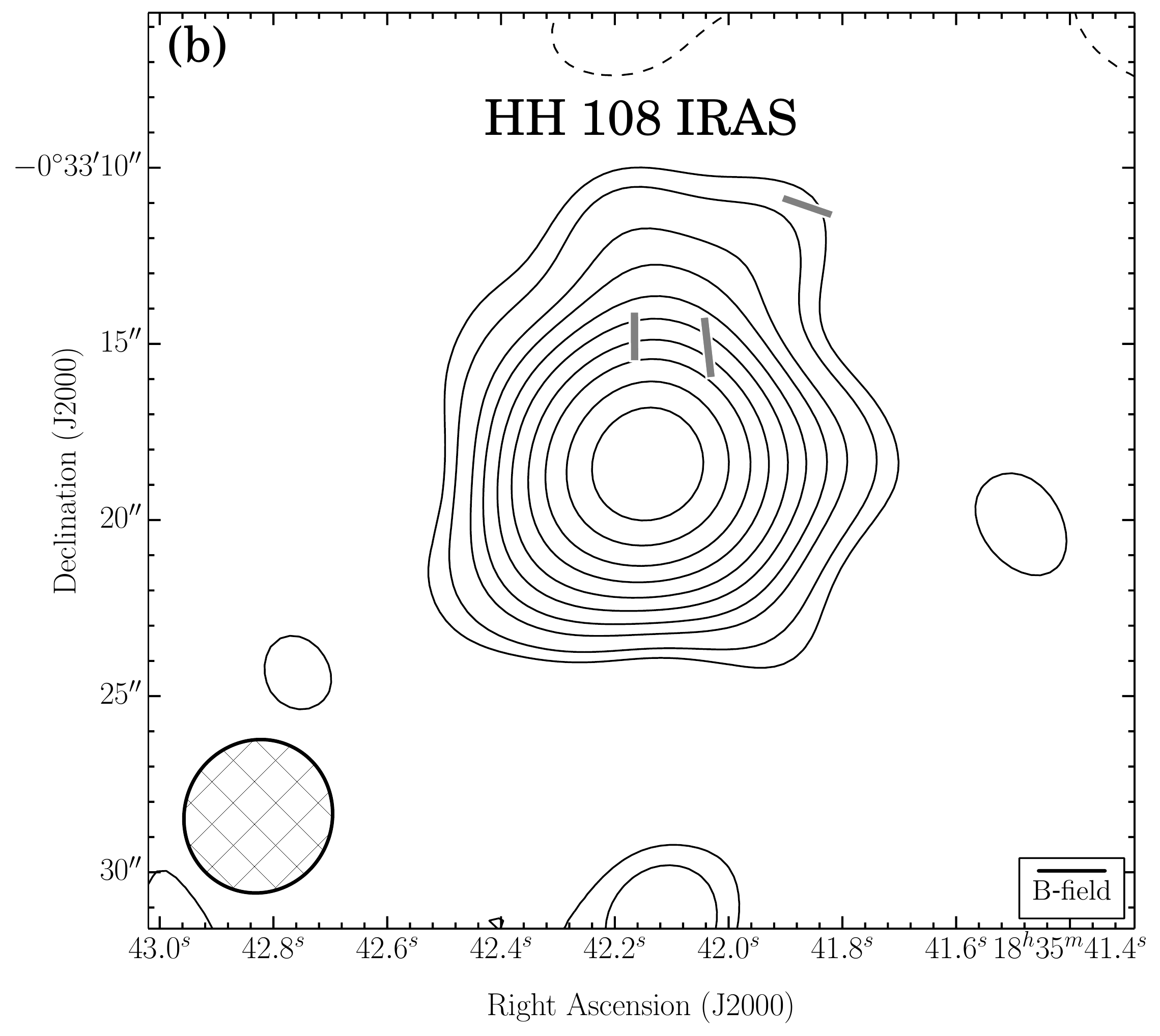}
\caption[]{ \footnotesize{
\input{HH108_caption.txt}
There is no \textbf{(c)} plot because there were no SCUBA, SHARP, or Hertz data to overlay.}}
\label{fig:HH108}
\end{center}
\end{figure*}

%%% Maps of G034.43+00.24~MM1
\begin{figure*} [hbt!]
\begin{center}
\epsscale{1.1}
\plottwo{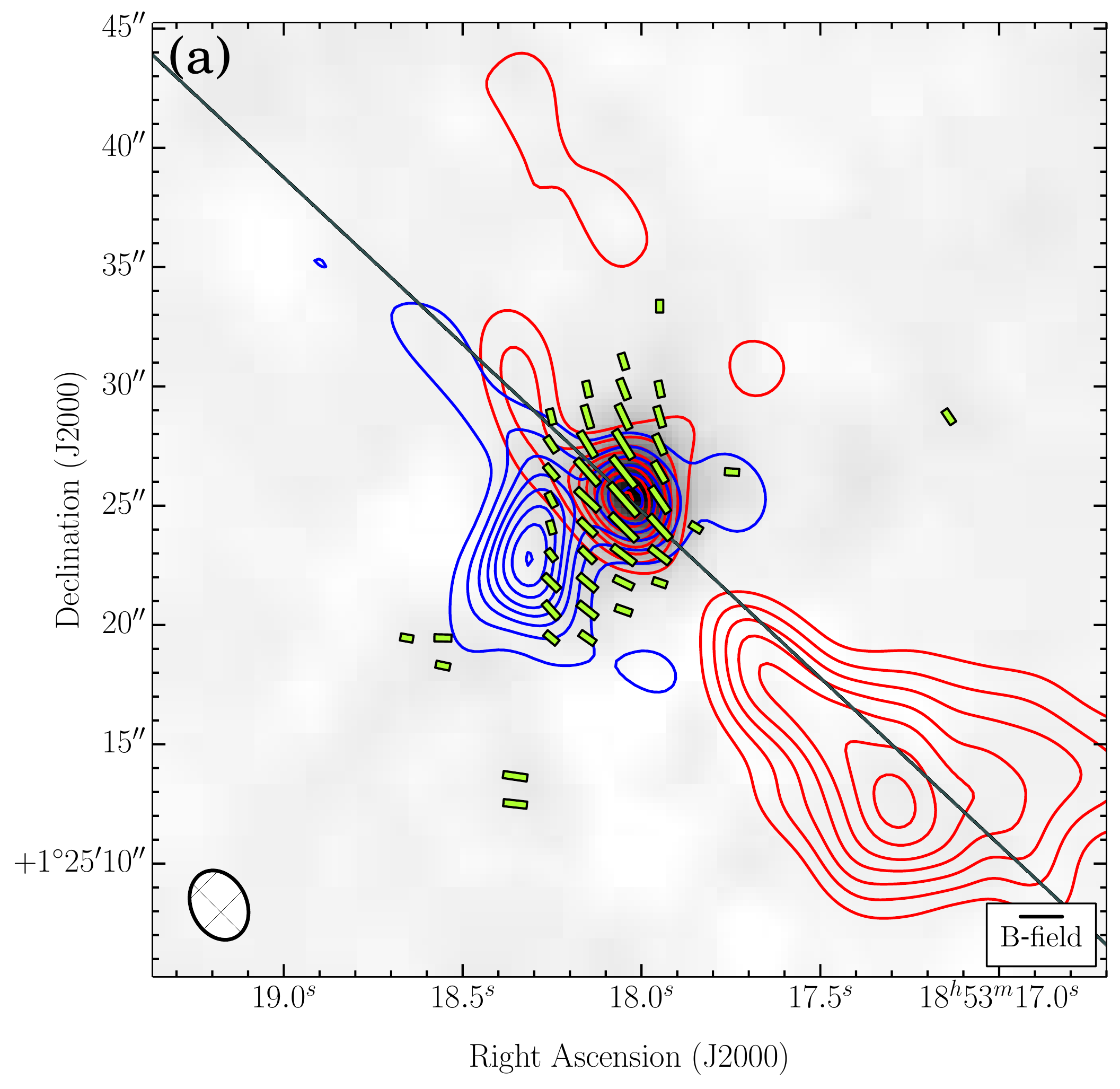}{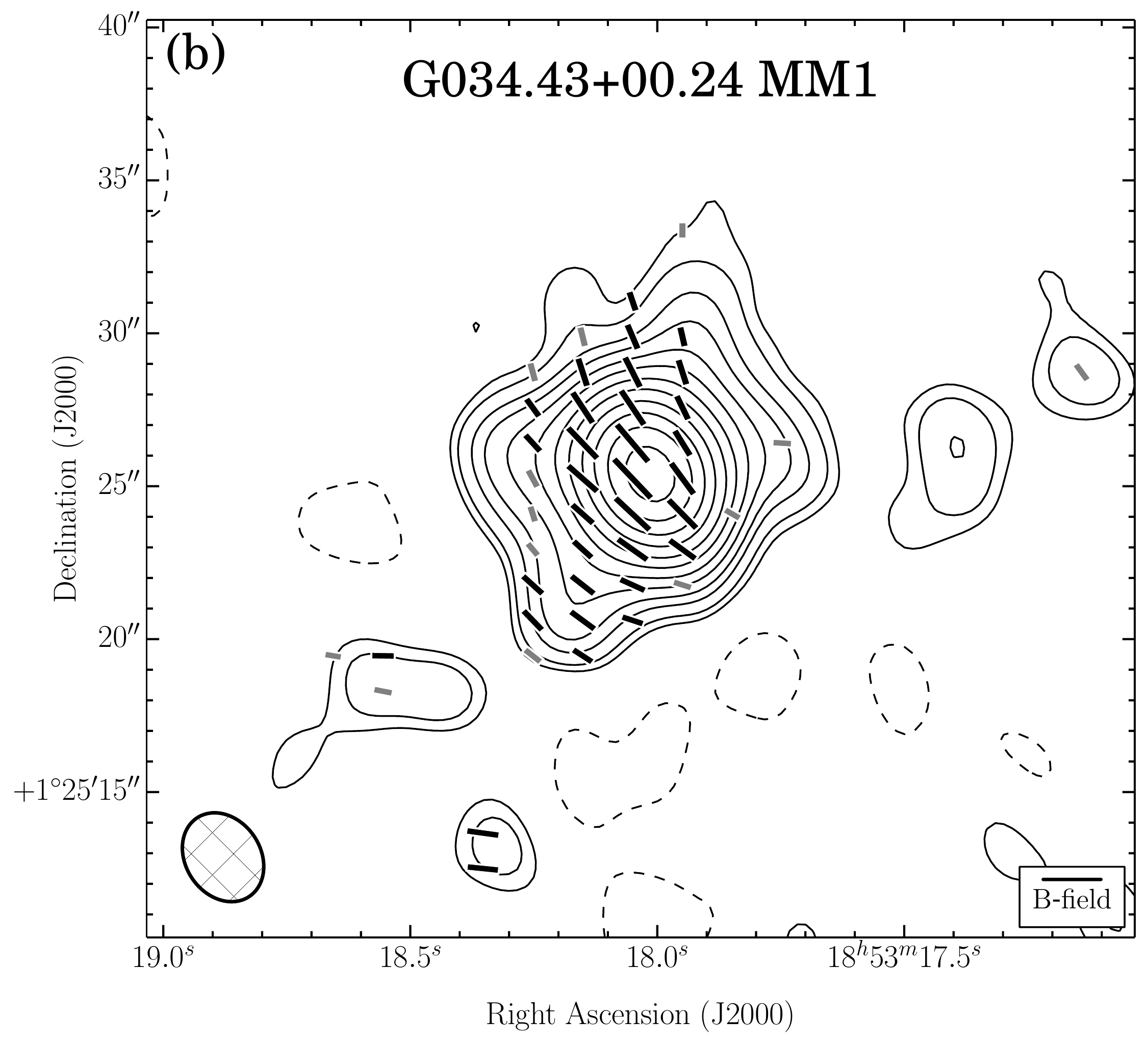}
\caption[]{ \footnotesize{
\input{G034MM1_caption.txt}
There is no \textbf{(c)} plot because there were no SCUBA, SHARP, or Hertz data to overlay.}}
\label{fig:G034MM1}
\end{center}
\end{figure*}

%%% Maps of G034.43+00.24~MM3
\begin{figure*} [hbt!]
\begin{center}
\epsscale{.9}
%\plottwo{G034MM3_pol_moments}{G034MM3_vectors}
\plotone{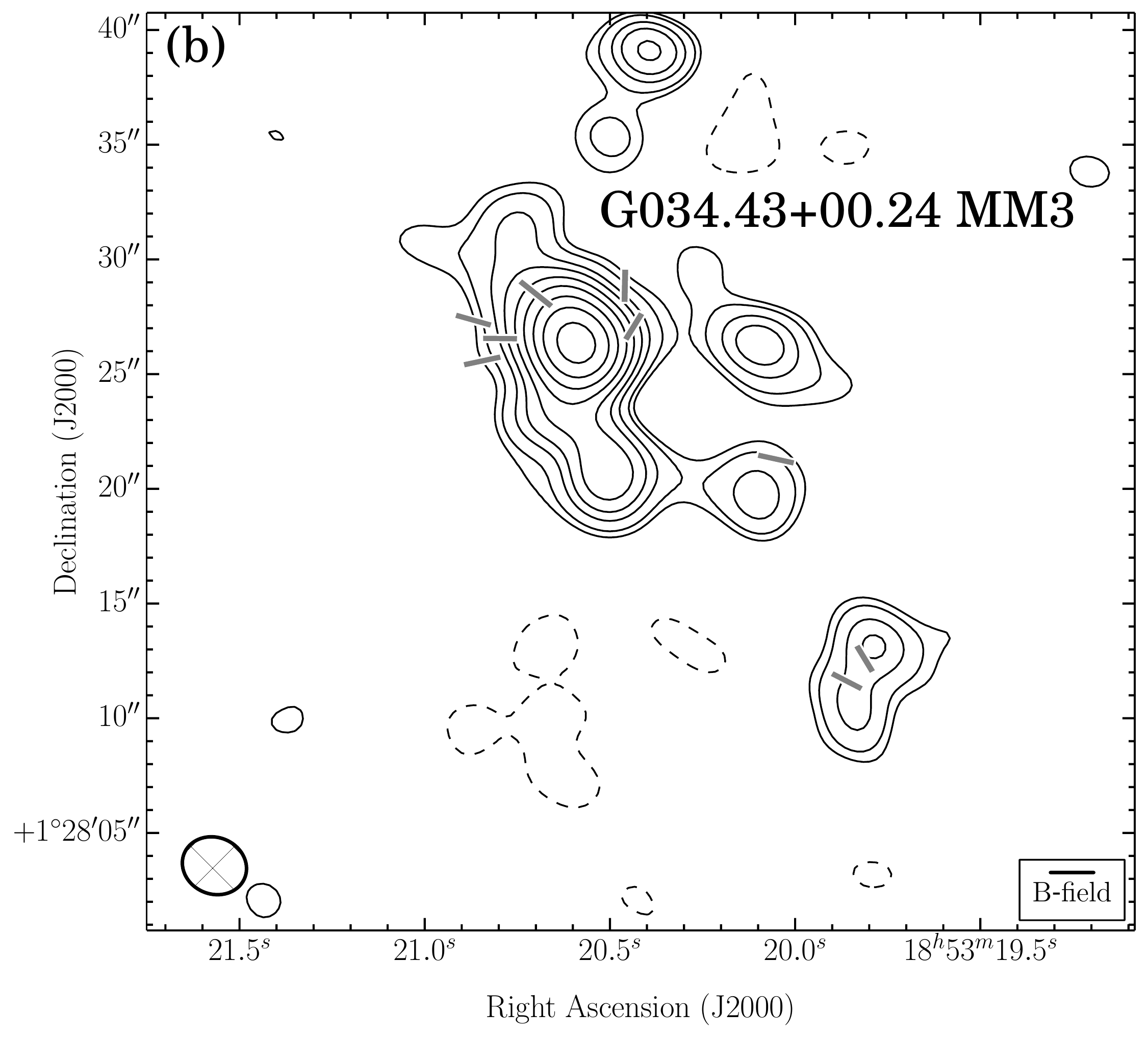}
\caption[]{ \footnotesize{
\input{G034MM3_caption.txt}
There is no \textbf{(a)} plot because there were no spectral-line data to plot.
There is no \textbf{(c)} plot because there were no SCUBA, SHARP, or Hertz data to overlay.
}}
\label{fig:G034MM3}
\end{center}
\end{figure*}

%%% Maps of B335~IRS
\begin{figure*} [hbt!]
\begin{center}
\epsscale{1.1}
\plottwo{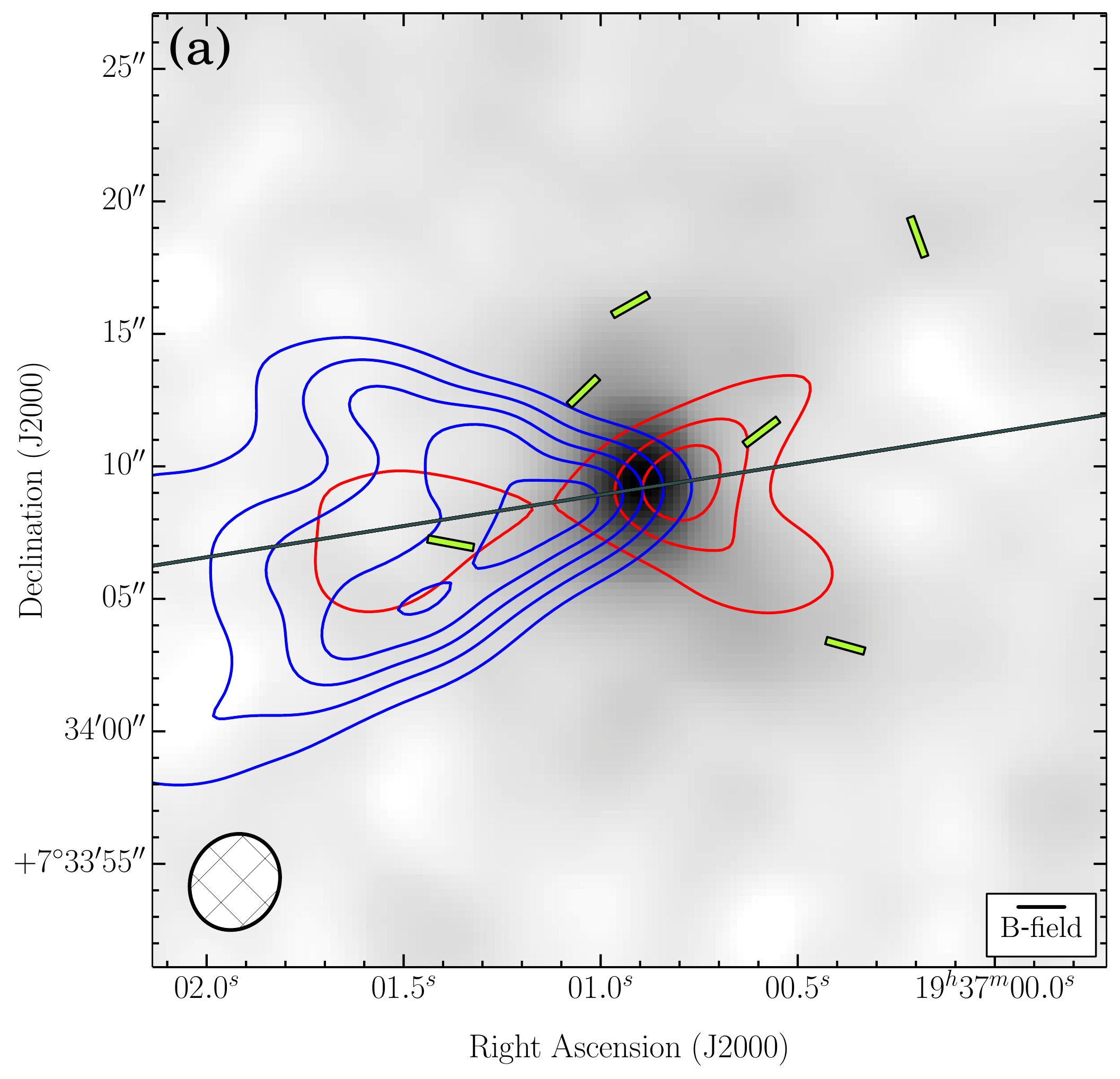}{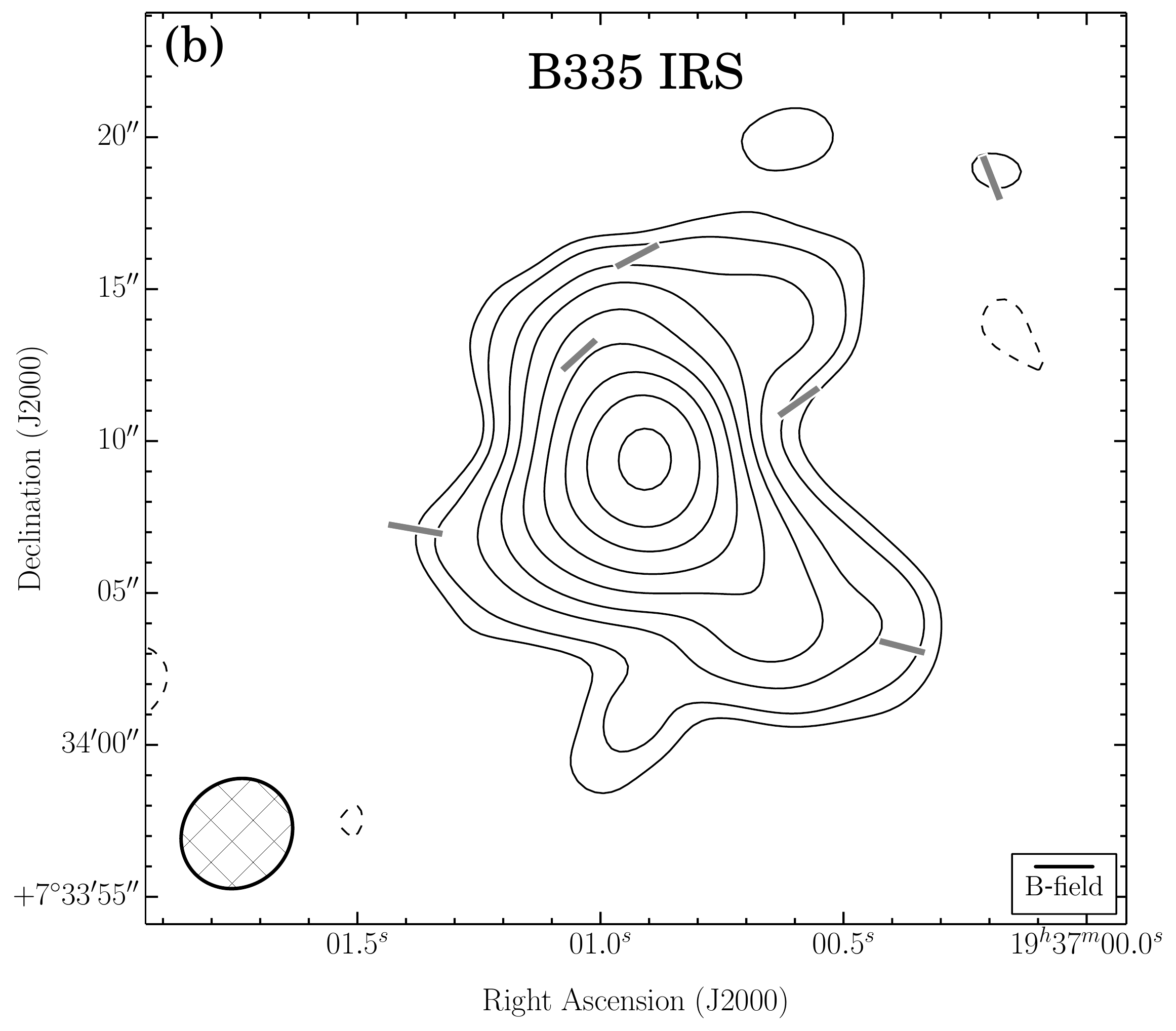}
\epsscale{0.8}
\plotone{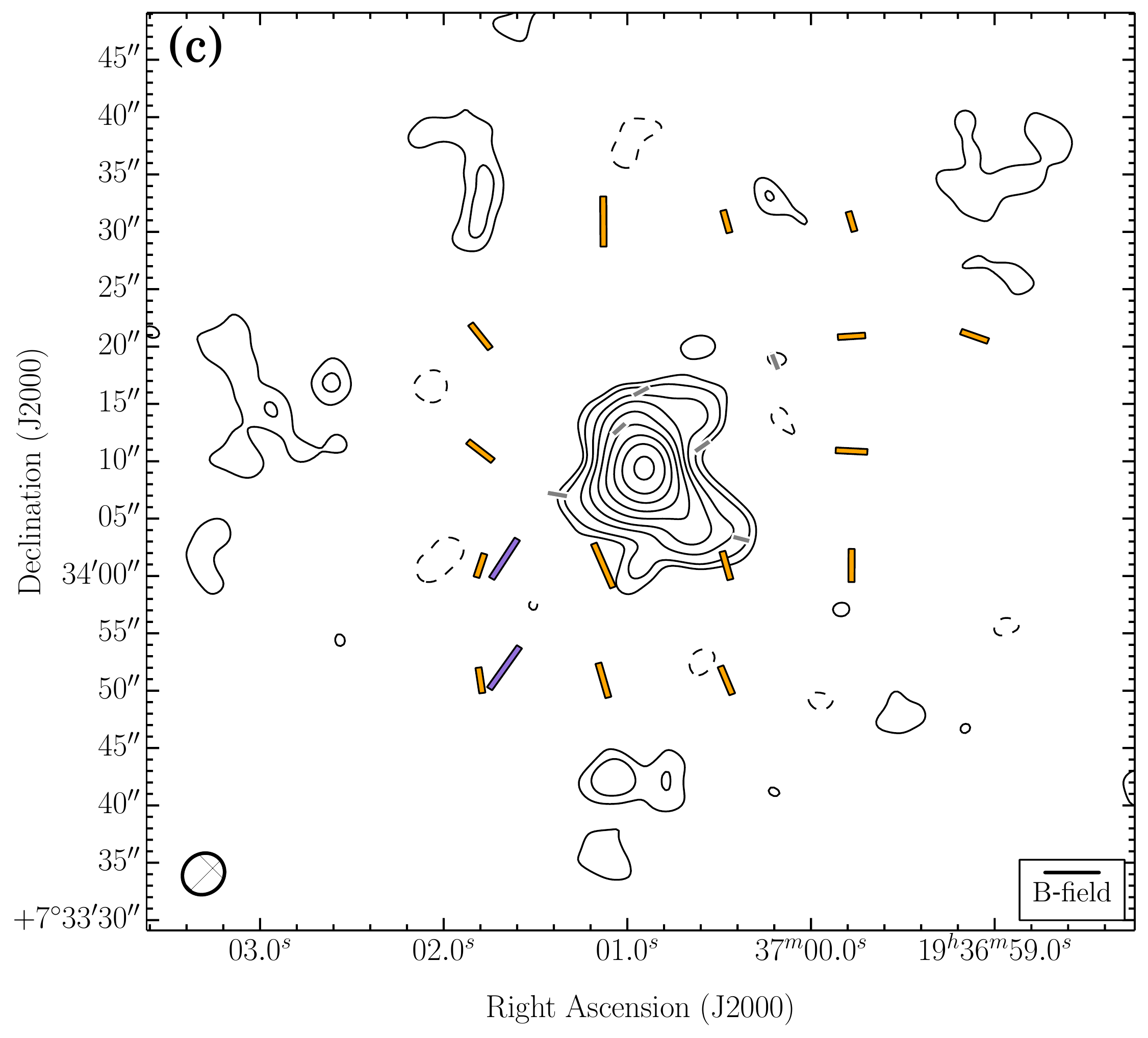}
\caption[]{ \footnotesize{
\input{B335_caption.txt}
}}
\label{fig:B335}
\end{center}
\end{figure*}

%%% Maps of DR21(OH)
\begin{figure*} [hbt!]
\begin{center}
\epsscale{1.1}
\plottwo{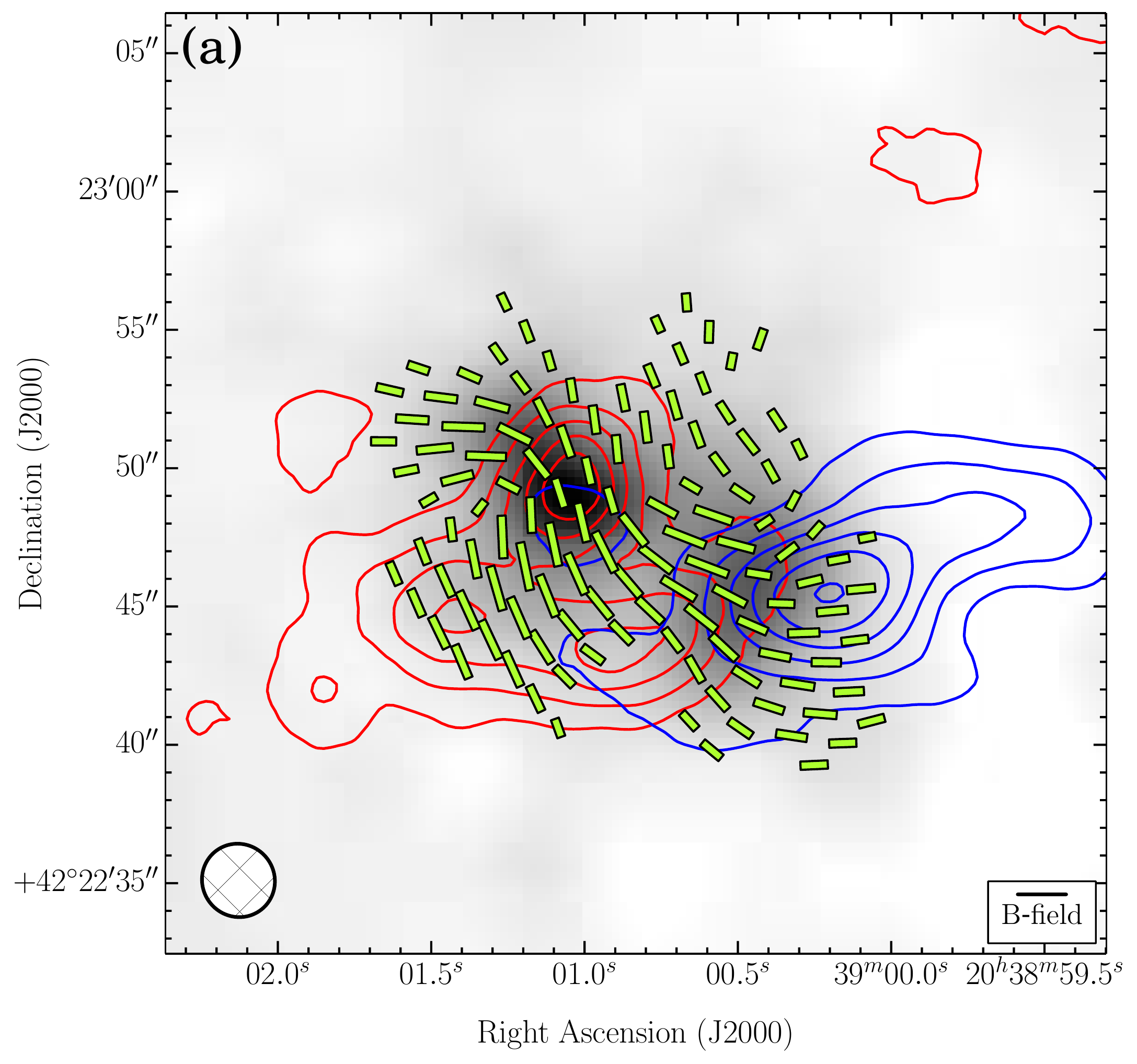}{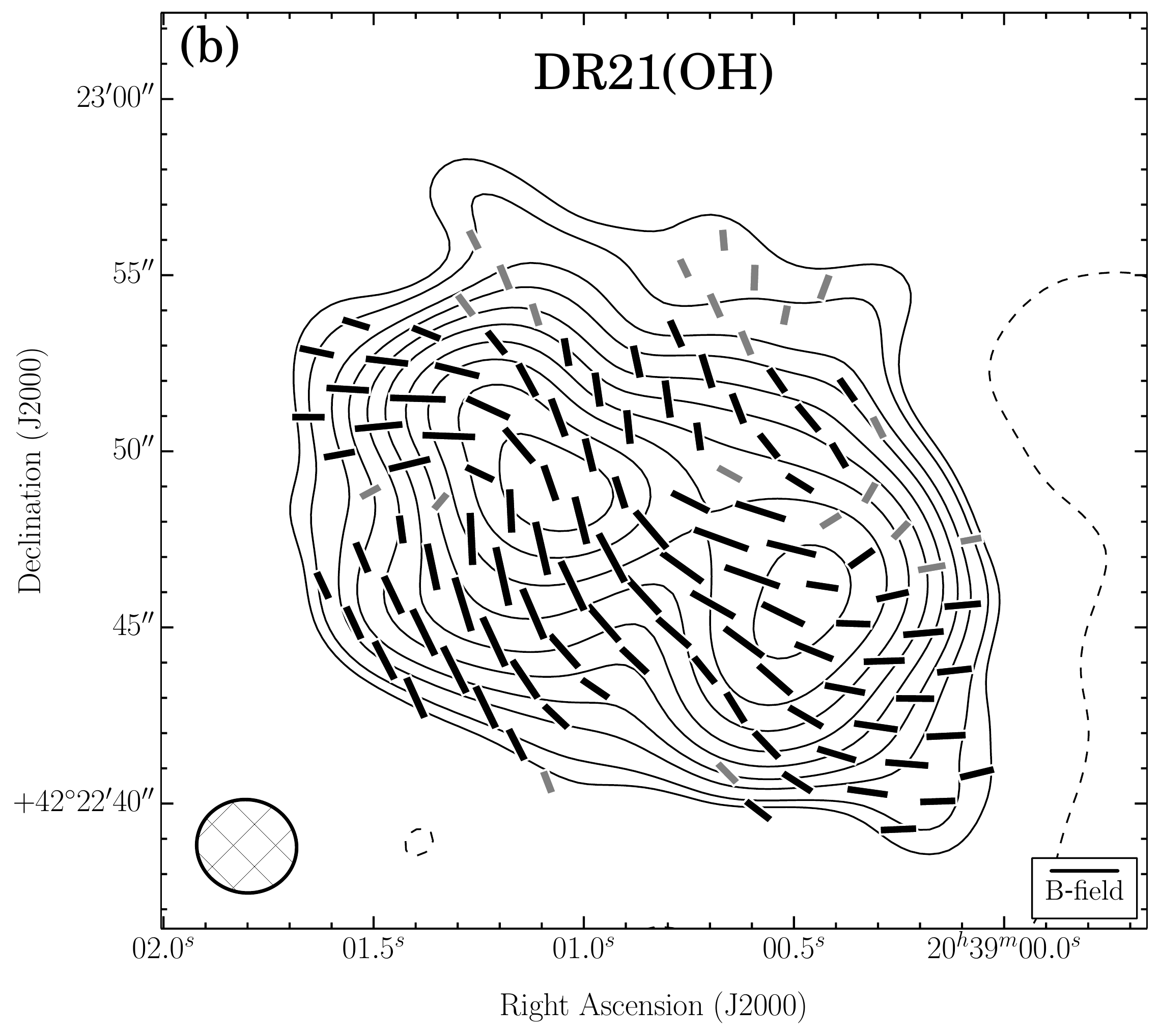}
\epsscale{0.8}
\plotone{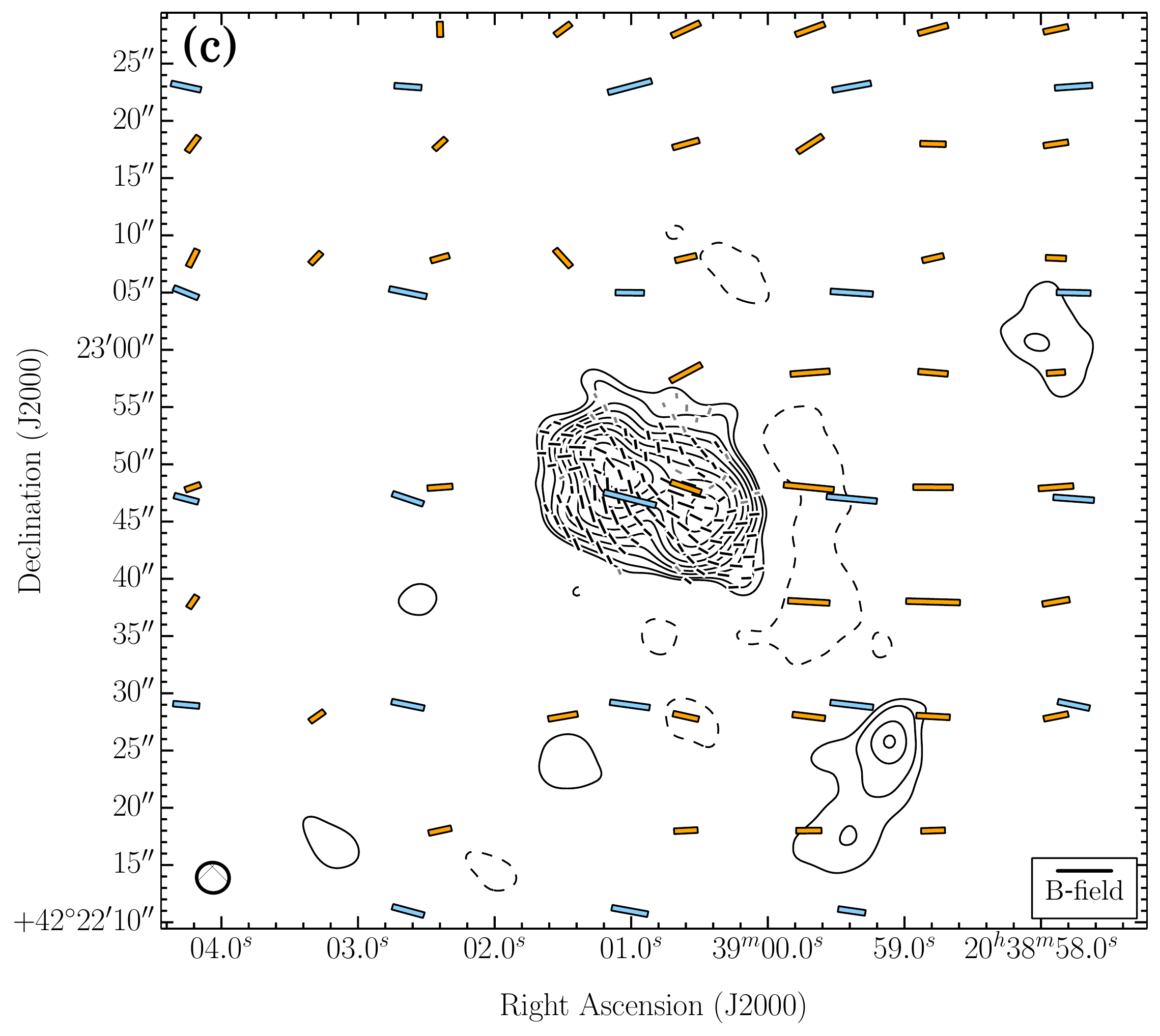}
\caption[]{ \footnotesize{
\input{DR21OH_caption.txt}
}}
\label{fig:DR21OH}
\end{center}
\end{figure*}

%%% Maps of L1157
\begin{figure*} [hbt!]
\begin{center}
\epsscale{1.1}
\plottwo{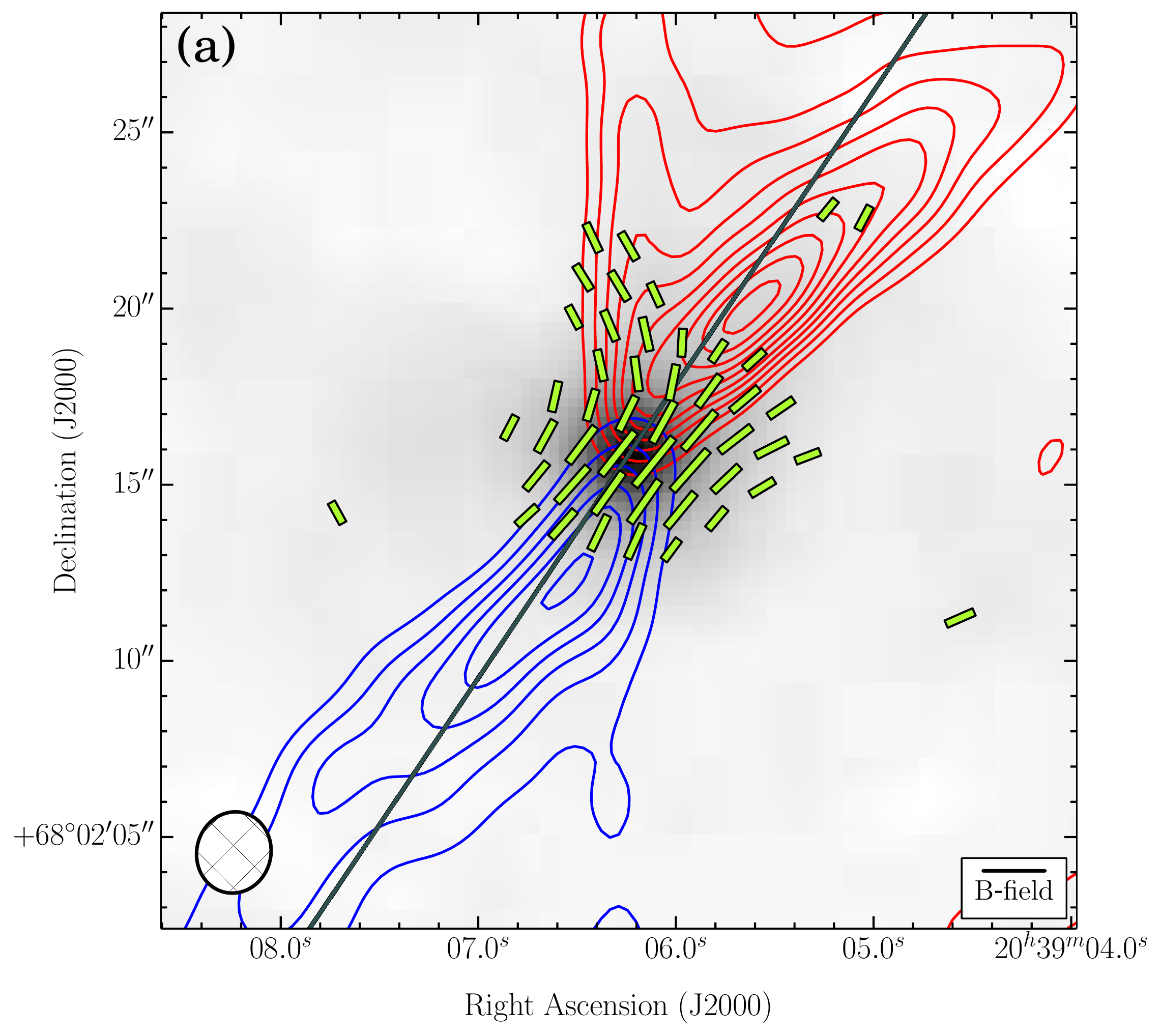}{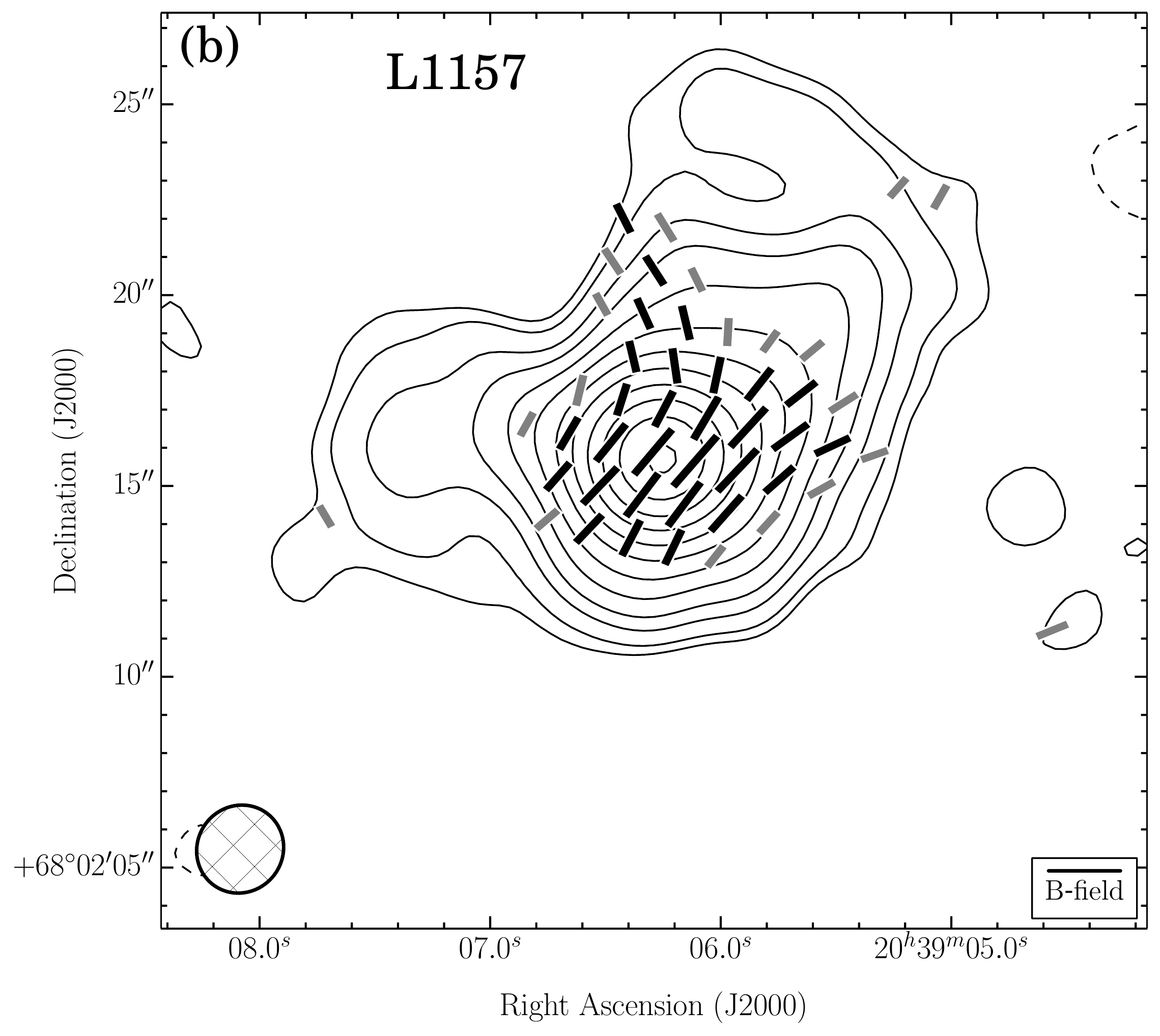}
\epsscale{0.8}
\plotone{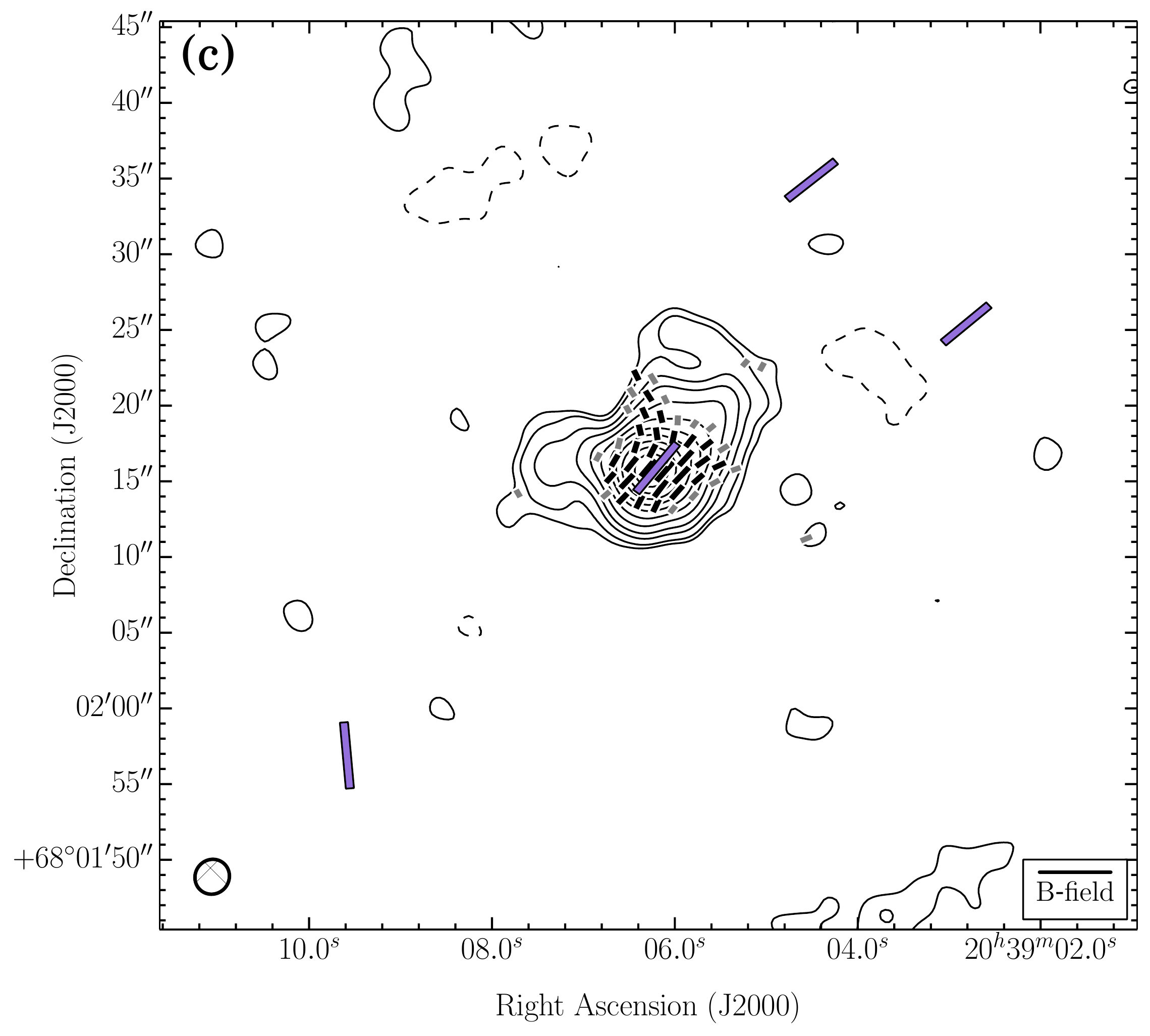}
\caption[]{ \footnotesize{
\input{L1157_caption.txt}
}}
\label{fig:L1157}
\end{center}
\end{figure*}

%%% Maps of CB~230
\begin{figure*} [hbt!]
\begin{center}
\epsscale{1.1}
\plottwo{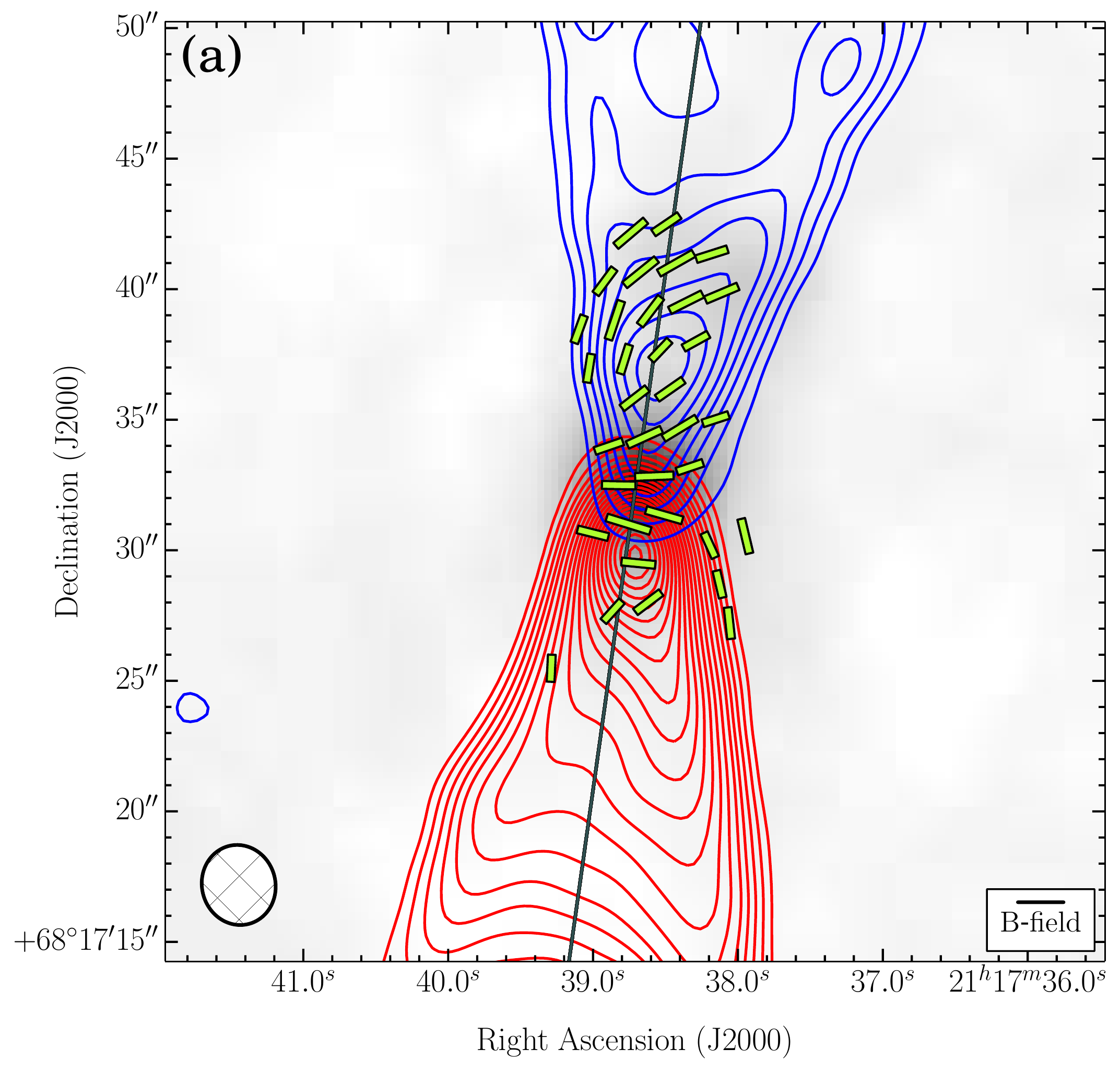}{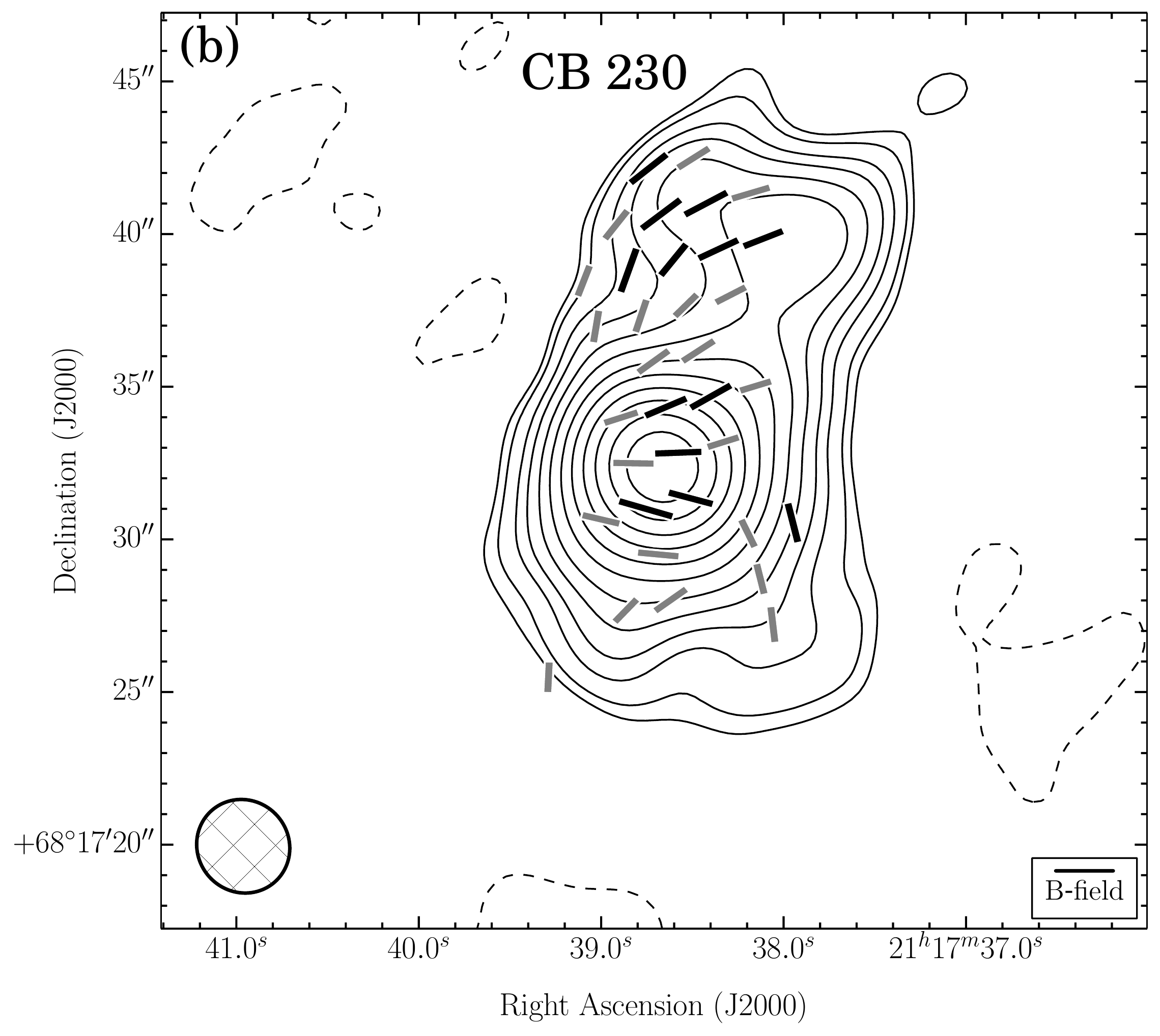}
\epsscale{0.8}
\plotone{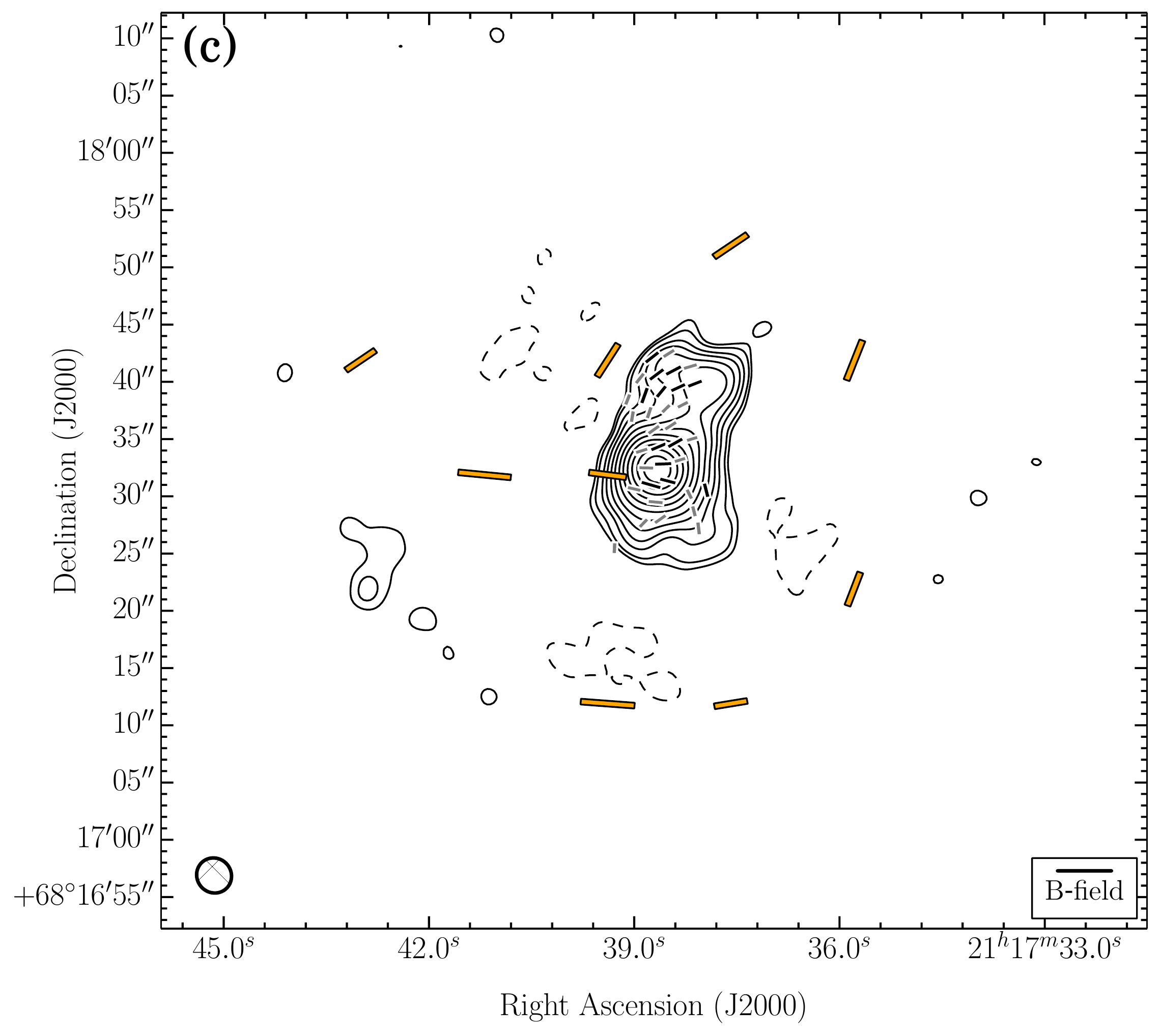}
\caption[]{ \footnotesize{
\input{CB230_caption.txt}
}}
\label{fig:CB230}
\end{center}
\end{figure*}

%%% Maps of L1165
\begin{figure*} [hbt!]
\begin{center}
\epsscale{1.1}
\plottwo{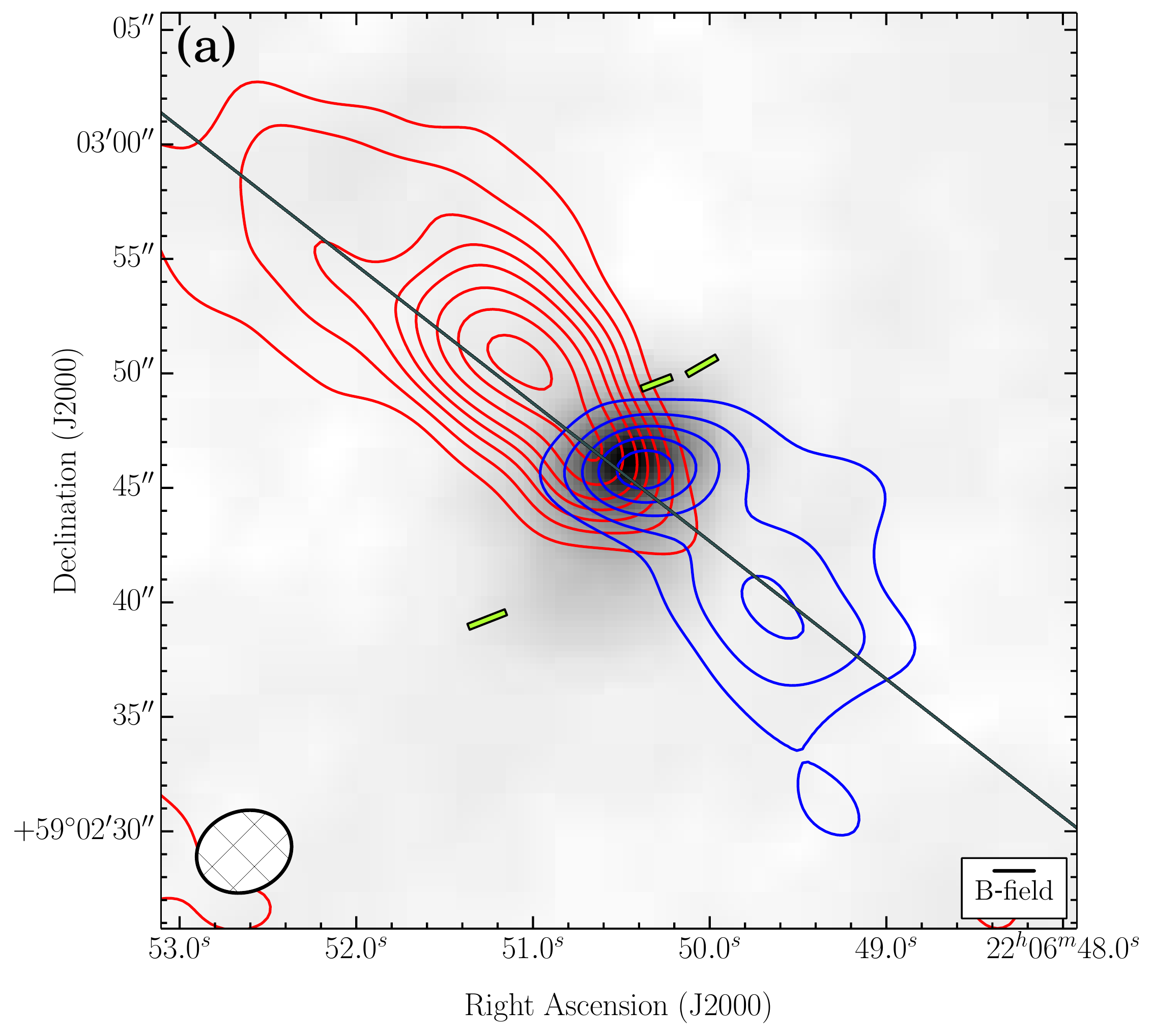}{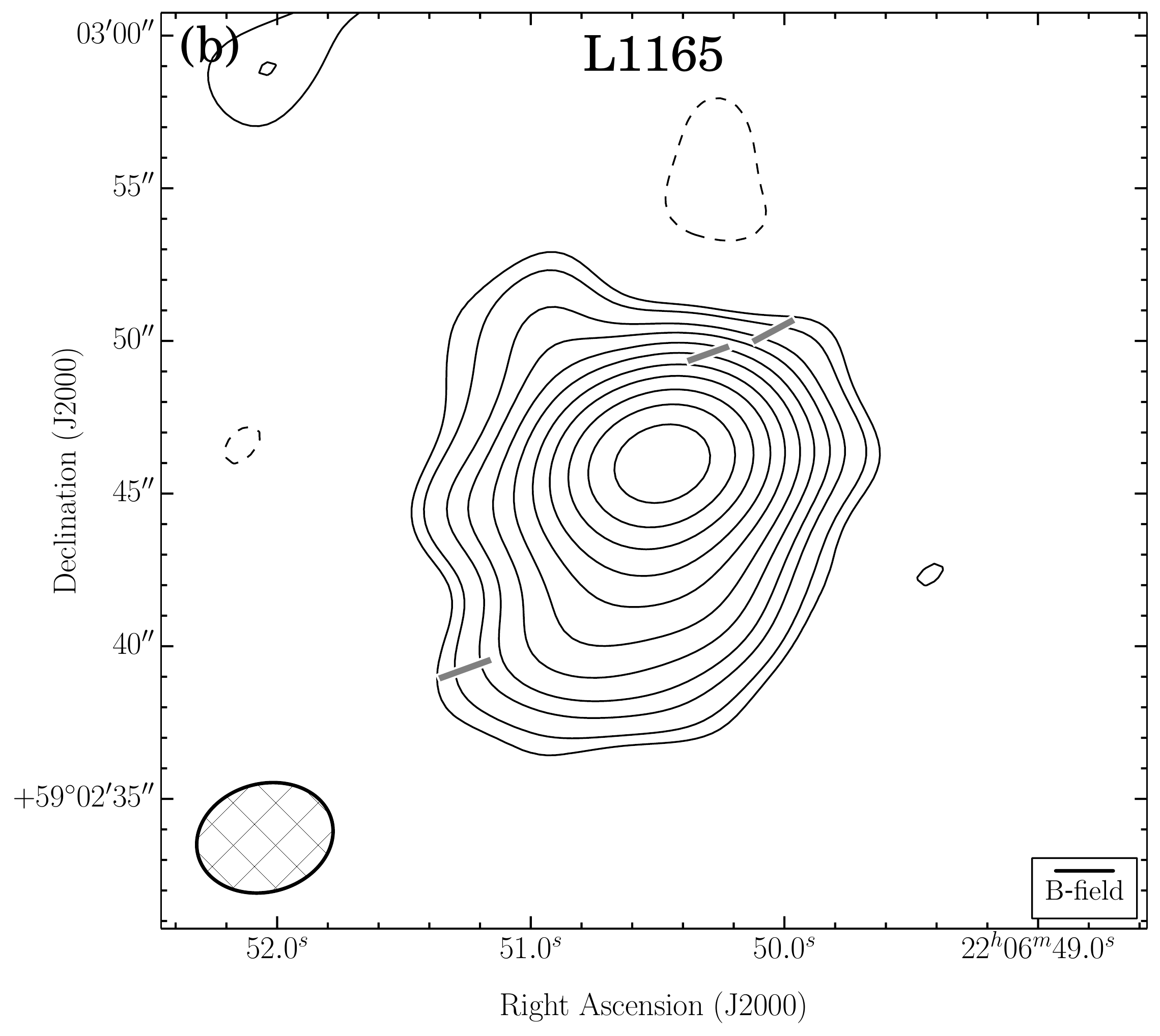}
\caption[]{ \footnotesize{
\input{L1165_caption.txt}
There is no \textbf{(c)} plot because there were no SCUBA, SHARP, or Hertz data to overlay.}}
\label{fig:L1165}
\end{center}
\end{figure*}

%%% Maps of NGC~7538~IRS~1
\begin{figure*} [hbt!]
\begin{center}
\epsscale{1.1}
\plottwo{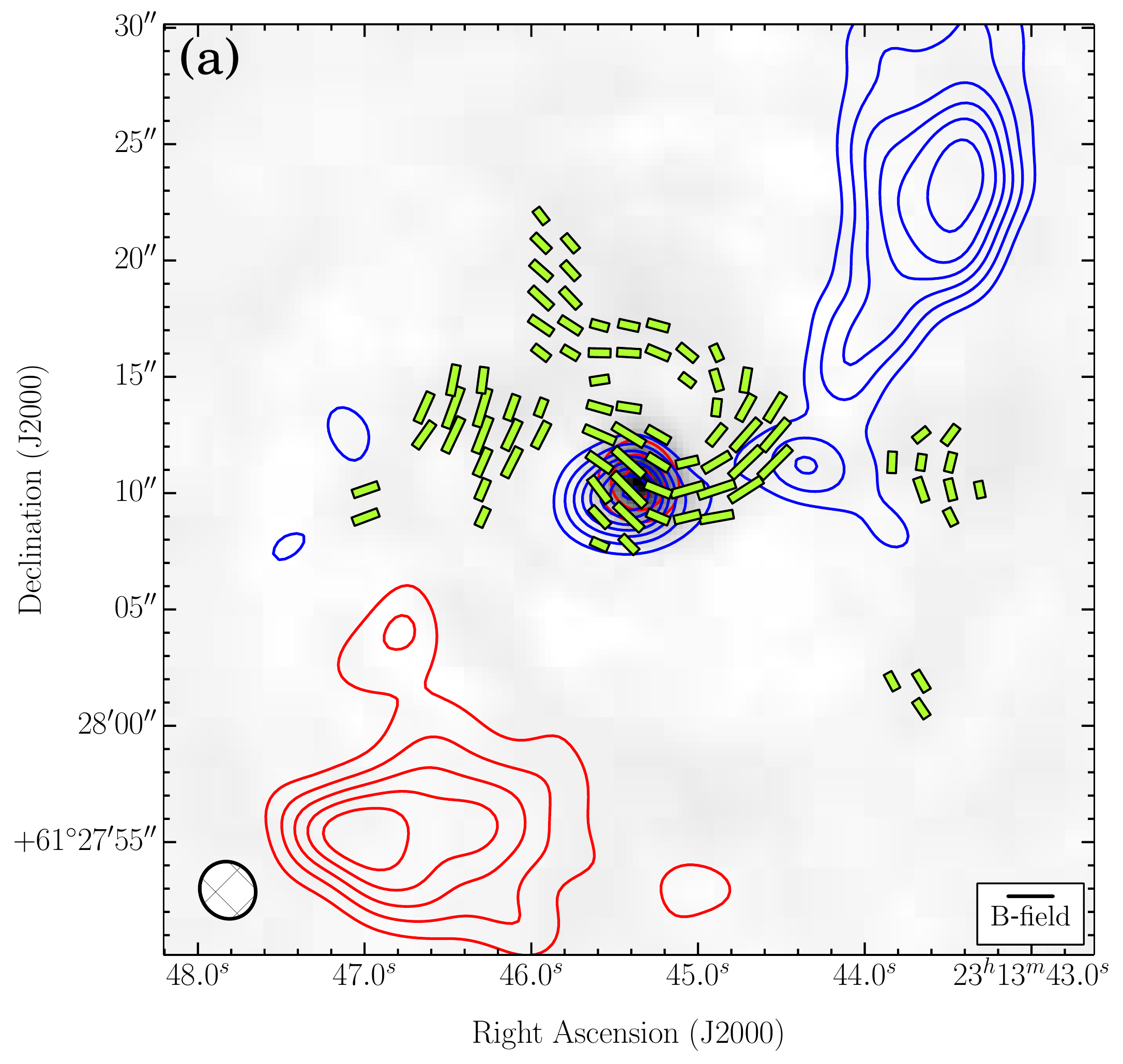}{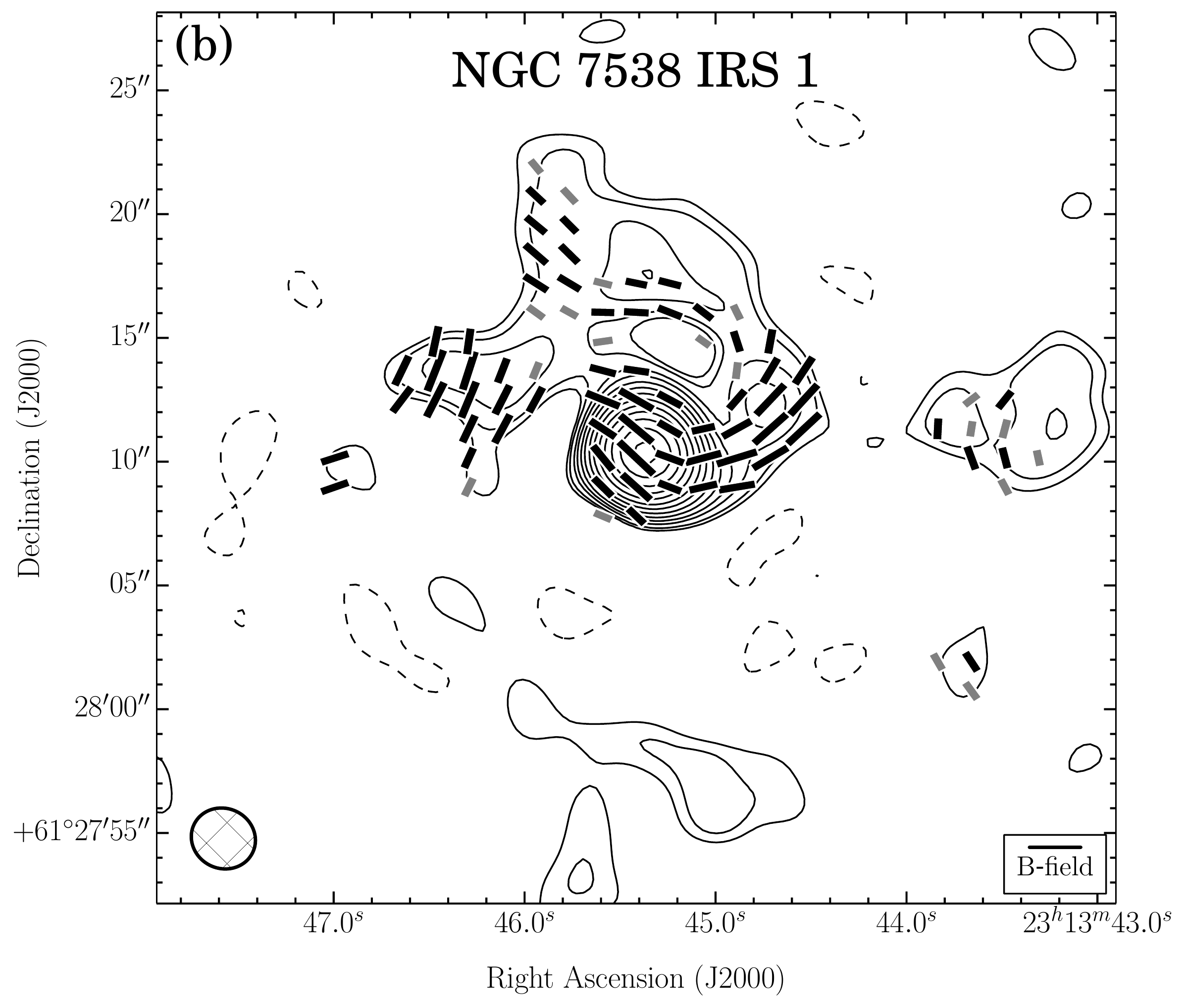}
\epsscale{0.8}
\plotone{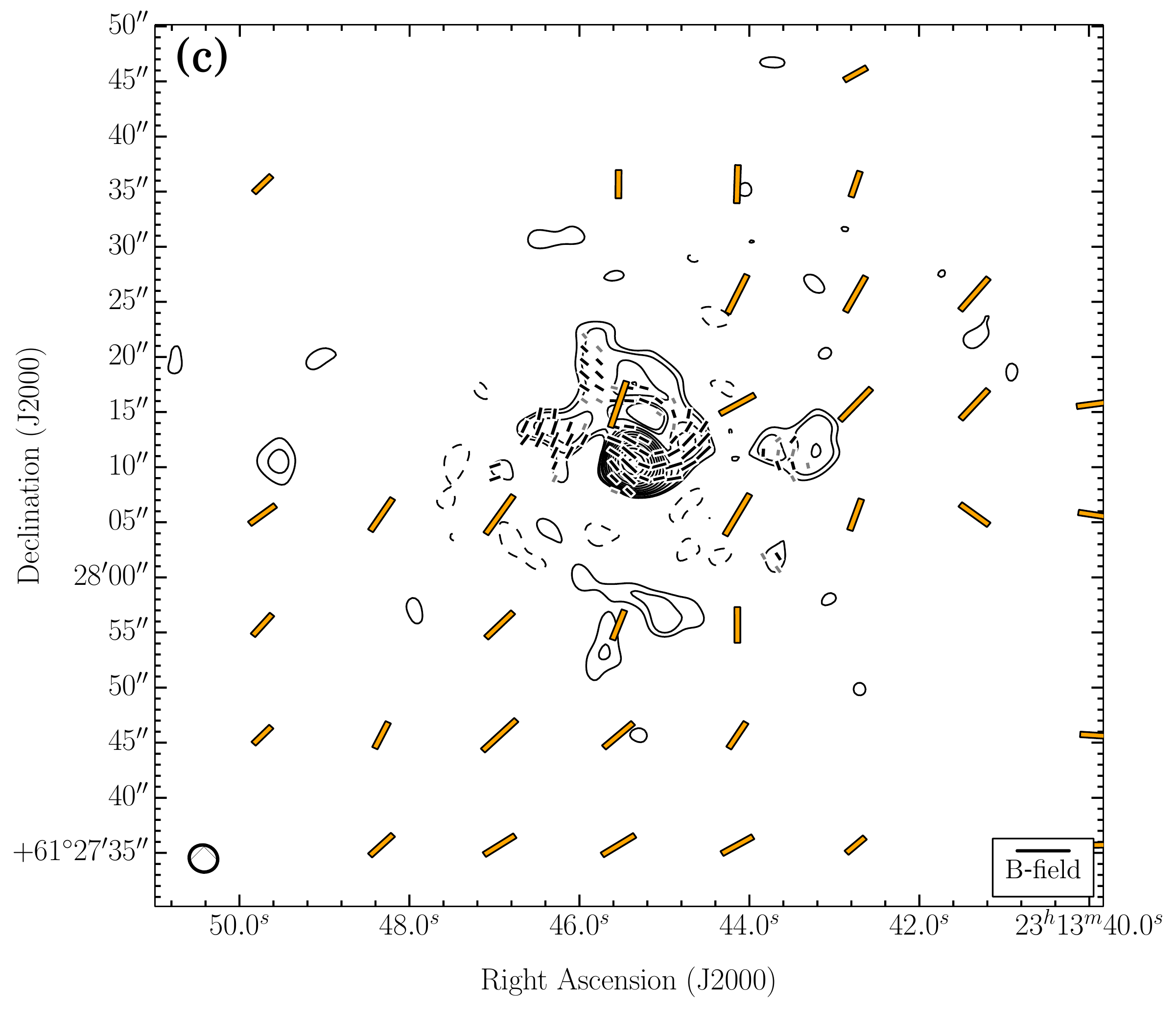}
\caption[]{ \footnotesize{
\input{NGC7538_caption.txt}
}}
\label{fig:NGC7538}
\end{center}
\end{figure*}

%%% Maps of CB~244
\begin{figure*} [hbt!]
\begin{center}
\epsscale{1.1}
\plottwo{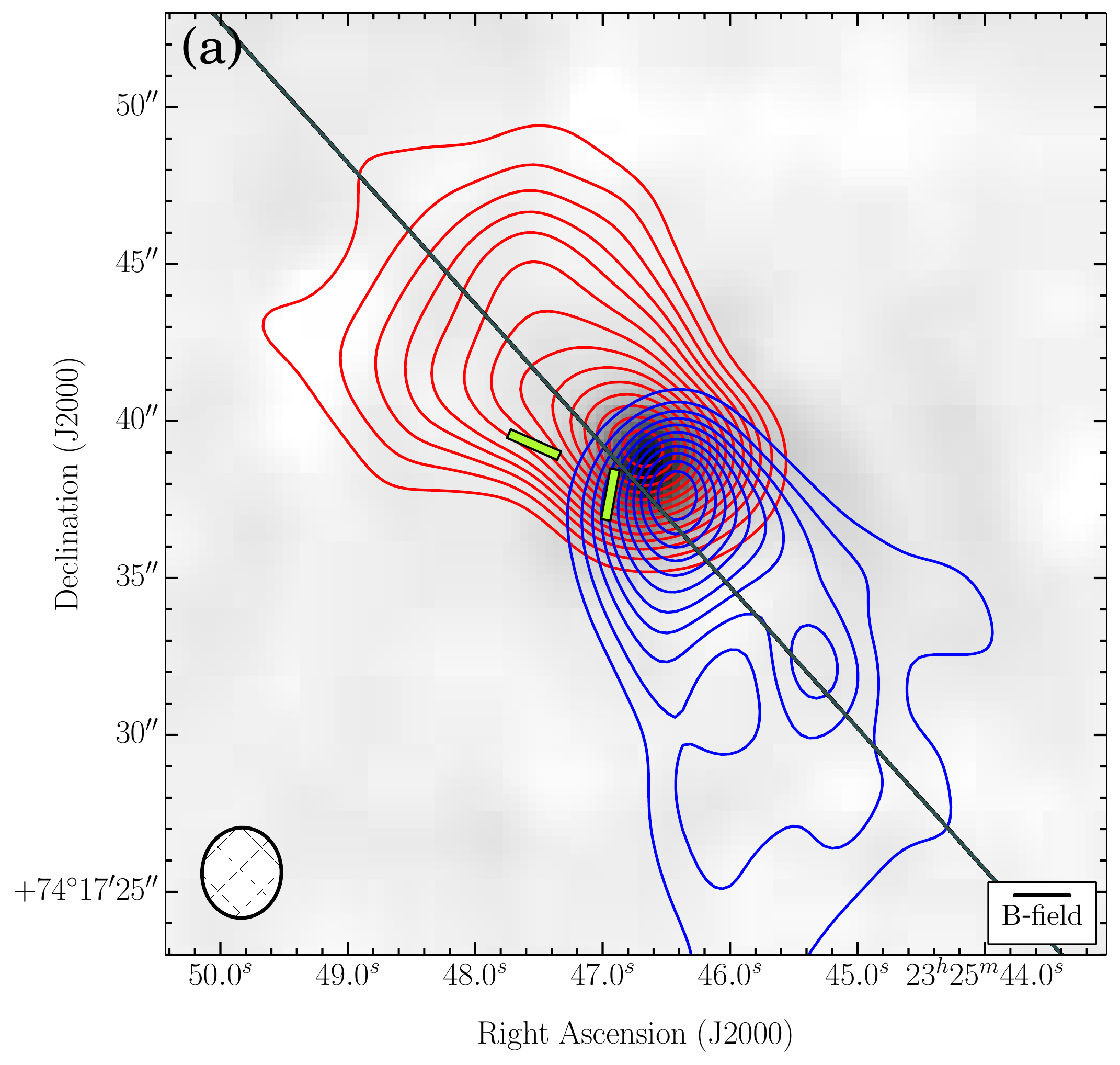}{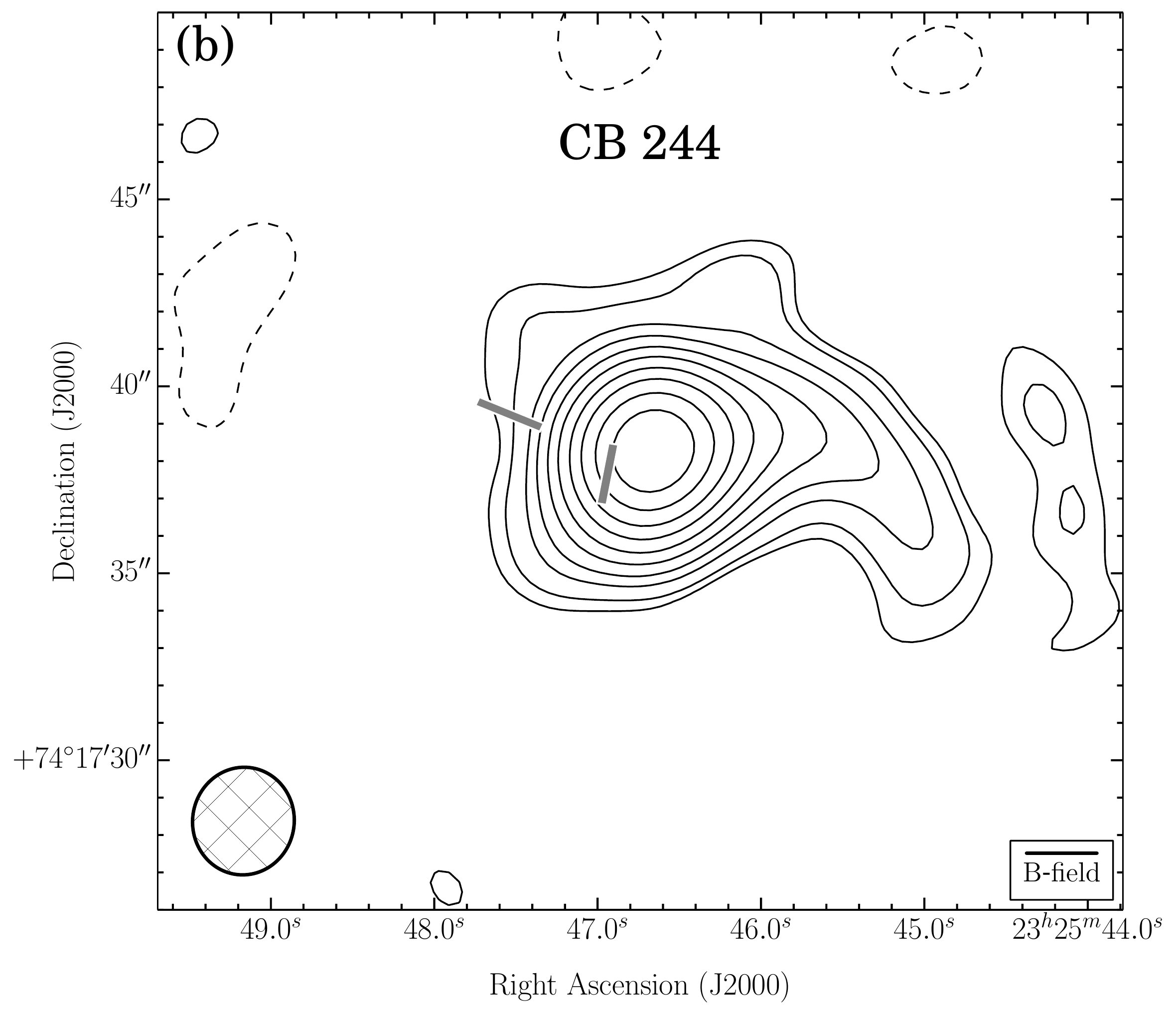}
\epsscale{0.8}
\plotone{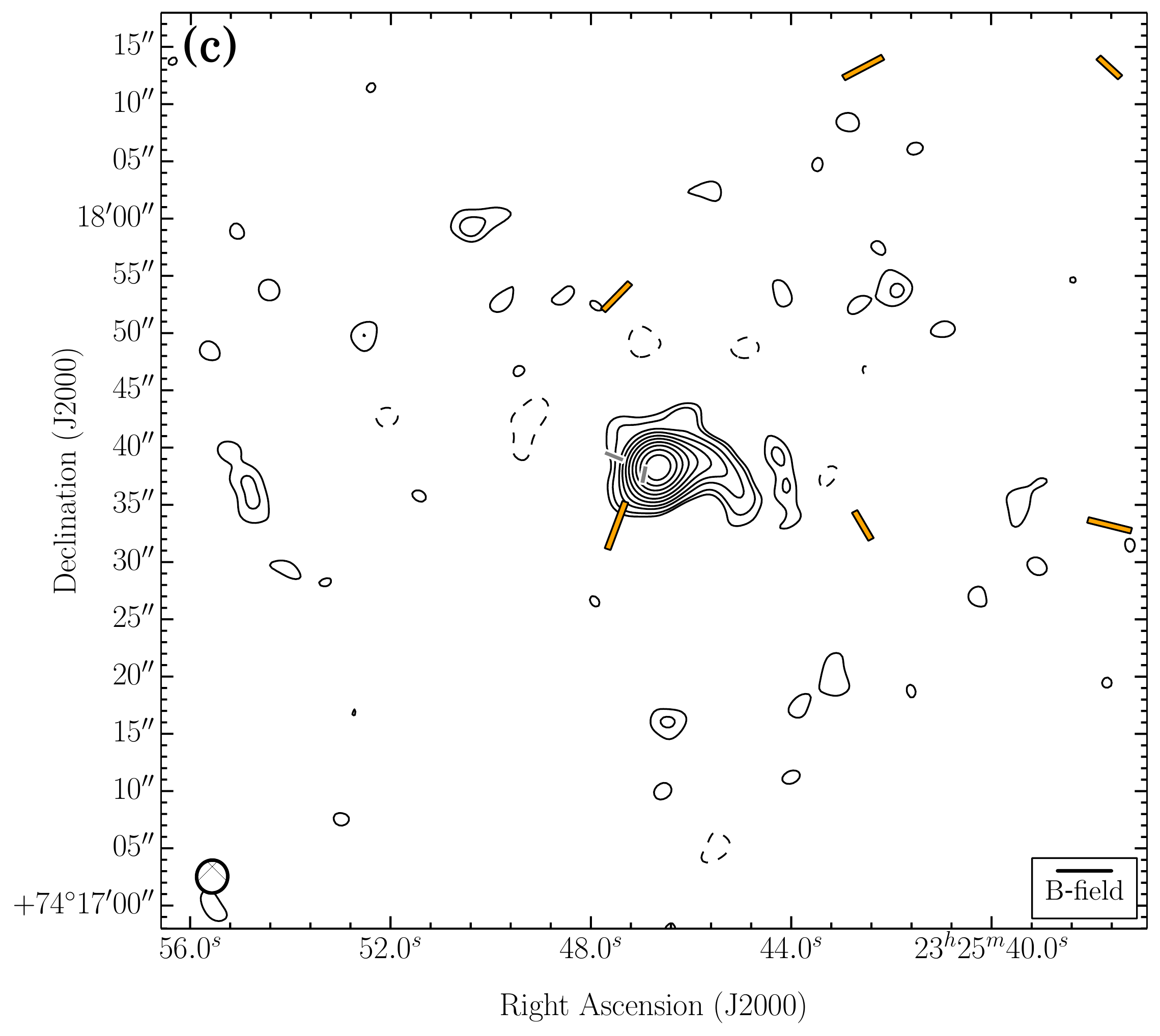}
\caption[]{ \footnotesize{
\input{CB244S_caption.txt}
}}
\label{fig:CB244S}
\end{center}
\end{figure*}

% SOURCES WITHOUT SCUBA/SHARP/Hertz DATA:
% DGTau
% SerEmb17
% SerEmb1
% HH108
% G034MM1
% L1165

% SOURCES WITHOUT SPECTRAL LINE DATA:
% OrionKL

% SOURCES WITHOUT EITHER:
% G034MM3

\clearpage

\section{\bf \large APPENDIX B: DESCRIPTION OF SOURCES}
\label{appendix:blurbs}

\subsection{W3~Main}
The W3 molecular cloud, located at a distance of 1.95\,kpc \citep{xu2006}, 
is one of the massive molecular clouds in the outer galaxy,
with an estimated total gas mass
of 3.8$\times 10^5$\,\Msol\ \citep{moore2007}.  It contains several young,
massive star-forming complexes, the most active of which is W3~Main. Early
thermal dust continuum observations identified three sources: W3~SMS1, SMS2,
and SMS3 \citep{ladd1993}. Our polarization observations are toward W3~SMS1
and are centered on the luminous infrared source IRS5 (2$\times 10^5$\,\Lsol,
\citealt{campbell1995}). 

 Discovered by \citet{wynnwilliams1972}, IRS5 is a double infrared source
\citep{howell1981}; both sources are associated with radio continuum emission
that is consistent with very young, hyper-compact \HII\ regions
\citep{vandertak2005:w3}.  Millimeter interferometer observations have resolved the
brightest dust continuum source associated with IRS5 into at least five
compact cores (MM1--MM5, \citealt{rodon2008}).  Hubble Space Telescope
observations also revealed seven near-IR sources within IRS5
\citep{megeath2005}.  Multiple outflows associated with IRS5 have also been
observed in various molecular tracers \citep{rodon2008,wang2012:w3}. It has
been proposed that IRS5 is a Trapezium cluster in the making and thus holds
valuable clues to high mass cluster formation.

Low resolution infrared and submillimeter polarization observations have revealed low
polarization, with a notable decline toward IRS~5 and a spread of values away
from the dust peak \citep{schleuning2000, Matthews2009}.  Water-maser
polarization observations have revealed an hourglass-shaped field toward IRS5
\citep{imai2003}.

The more extended structure to the west of IRS5
observed in our TADPOL image is the free-free emission associated with the
\HII\ region W3~B, better known for its infrared association IRS3
\citep{wynnwilliams1972, megeath1996}. The associated stellar source
(designated as IRS3a) is consistent with a star of spectral type O6
\citep{megeath1996}.

See Figure \ref{fig:W3Main} for maps.

\subsection{W3(OH)}

W3(OH) is another active, high-mass star formation site in the W3
molecular cloud. 
H$_2$O maser parallax measurements place the
complex at a distance of 2.04\,kpc \citep{hachisuka2006}.
W3(OH) consists of two main regions: a young,
limb-brightened ultra-compact (UC) \HII\ region with several OH masers, known as W3(OH)
\citep{dreher1981}, and a younger, massive hot core with water masers
$\sim$6\,\arcsec\ east of W3(OH) known as W3(H$_2$O) or W3(TW)
\citep{turner1984}. Both of these regions are within the TADPOL field-of-view. 
The UC \HII\ region is ionized by a massive O9 star, and has
a total luminosity of 7.1$\times10^4$\,\Lsol\ \citep{hirsch2012}. High
resolution observations have revealed dense gas in a massive
protobinary system ($\sim$22\,\Msol) towards W3(H$_2$O), without any
associated ionized emission from UC \HII\ region
\citep{wilner1999,Wyrowski1999, chen2006}. Massive, collimated
outflows and jets have been detected towards the W3(H$_2$O) system
\citep{reid1995, zapata2011}.  

SCUBA observations show significant polarization
throughout the region ($\sim$5\%) with some evidence for depolarization
towards the center \citep{Matthews2009}. Strong magnetic fields are also
implied by single-dish CN Zeeman measurements, which find a $\sim$1.1\,mG
field strength towards this region \citep{falgarone2008}.

See Figure \ref{fig:W3OH} for maps.

\subsection{L1448~IRS~2}
% \input{blurbs/L1448I.txt}
%03:25:22.4 +30:45:12.0,
%IRAS 03222+3034

L1448 IRS 2 is a
Class 0 YSO \citep{OLinger1999} located in the Perseus molecular
cloud at a distance of $\sim$\,230\, pc \citep{Hirota2011}.   Its well collimated bipolar outflow has been studied by
CO mapping \citep[e.g.,][]{WolfChase2000}, {\it Spitzer} IRAC
\citep{Tobin2007}, and molecular hydrogen mapping
\citep[e.g.,][]{Eisloffel2000}.  It is also one of the objects where
\citet{Kwon2009} found that dust grains have
grown significantly
even in the youngest protostellar stage.  The surrounding
flattened structure was studied by {\it Spitzer} observations
\citep{Tobin2010}.  Recent SHARP observations by \citet{Chapman2013}
show magnetic fields that are aligned with the bipolar outflow to within
$\sim$\,10$\degr$.

%First hydrostatic core \citep{2010ApJ...715.1344C}
%A first hydrostatic core was claimed at $50''$ east of this target
%\citep{2006AJ....131.2601O} O'Linger paper : mIR studies
%\citep{Tobin2007} Spitzer IRAC: opening angle inclination 57+6-8 degree

See Figure \ref{fig:L1448I} for maps.

\subsection{L1448N(B)}
L1448N(B) is a Class~0 YSO at the center of the L1448~IRS3 core (also called L1448N and
IRAS 03225+3034) \citep{Bachiller1986}, at a distance of $\sim$\,230\,pc \citep{Hirota2011}. 
It was first detected at 6\,cm \citep{Anglada1989}, although it is
weaker at centimeter wavelengths that its companion L1448N(A), which lies $\sim$\,7$\arcsec$ to the
northeast. L1448N(A) and L1448N(B) are suspected to be a gravitationally bound
common-envelope binary \citep{Kwon2006, Looney2000} with a separation
of $\sim$\,2000\,AU, even though they seem
to be in different evolutionary stages. L1448N(B) is the stronger source at millimeter
wavelengths \citep{Terebey1997, Looney2000}; it appears to be
younger and more embedded than its companion \citep{OLinger2006}.

CO observations of L1448N(B) show an outflow with a position angle estimated to
be 129$\arcdeg$ on large (arc-minute) scales \citep{WolfChase2000} 
and 105$\arcdeg$ on small scales \citep{Kwon2006}. The redshifted lobe is
easy to distinguish in channel maps, but the blueshifted lobe overlaps, and may
even interact with, the outflow from L1448C, $\sim$\,75$\arcsec$ to the south.
High-resolution ($0.7\arcsec\times0.5\arcsec$) maps of the 2.7\,mm continuum
emission from L1448N(B) show a protostellar envelope elongated in a direction
nearly perpendicular ($PA\sim56\arcdeg$) to the outflow \citep{Looney2000}.
Observations of the linear polarization of 1.3\,mm continuum emission made with
the BIMA interferometer at $\sim4\arcsec$ resolution \citep{Kwon2006} imply
that the magnetic field through the envelope is also approximately perpendicular
to the outflow. This orientation is consistent with lower-resolution (10\arcsec)
850\,$\mu$m polarization observations made with SCUPOL on the JCMT \citep{Matthews2009}.

See Figure \ref{fig:L1448N} for maps.

\subsection{L1448C}
L1448C is the collective name for the embedded Class~0 YSOs located
75--80$\arcsec$ southeast of L1448N, at a distance of 232\, pc \citep{Hirota2011}. L1448C has been the target of numerous
observations in the IR continuum \citep[\eg{}][]{Tobin2007}, in the (sub)millimeter
continuum \citep[\eg{}][]{Jorgensen2007}, in CO line emission \citep[\eg{}][]{Nisini2000}, 
and in SiO line emission \citep[\eg{}][]{Nisini2007} because
it is the point of origin of symmetrical, well collimated, high-velocity, and
rapidly evolving \citep{Hirano2010} outflows. The blueshifted outflow lobe
extends to the north; the westward bend in the outflow at the point where it
overlaps L1448N is strong evidence that the L1448C and L1448N outflows interact
\citep{Bachiller1995}.

The multiplicity of YSOs in L1448C was revealed by observations of millimeter continuum
emission \citep{Volgenau2004}. The strongest millimeter source, called L1448C(N)
\citep{Jorgensen2006} or L1448mm A \citep{Tobin2007}, is the likely
source of the outflows. A second source, $\sim$\,8$\arcsec$ south of L1448mm A, is
weaker in millimeter emission but prominent in maps of near- and mid-infrared emission made with
IRAC and MIPS, respectively, on Spitzer. This source is called L1448C(S)
\citep{Jorgensen2006} or L1448mm B \citep{Tobin2007}.

SCUBA maps of linearly polarized 850\,$\mu$m emission from the L1448 cloud 
only show significant polarization along the perimeter of
the clump of 850\,$\mu$m continuum emission coincident with L1448C. There is no
obvious trend in the orientation of the magnetic field lines.

See Figure \ref{fig:L1448C} for maps.

\subsection{L1455~IRS~1}
The dark cloud L1455 is located $\sim$1$\degree$ south of the active star formation region NGC~1333 at a distance of 320\,pc \citep{deZeeuw1999}.  L1455~IRS~1 (also known as L1455~FIR and IRAS 03245+3002) is the brightest far infrared source in the cloud. It is a low mass, Class I protostar, which was first detected in the far infrared with the Kuiper Airborne Observatory \citep{Davidson1984}.  High velocity CO emission was first detected by \citet{Frerking1982} and was first mapped in CO($J = 1\rightarrow0$) by \citet{Goldsmith1984}, who found extended blue- and redshifted emission over an area of more than 10$\arcmin$, indicating the presence of more than one outflow, which they thought might be powered by RNO~15  and/or  L1455~IRS~1. More recent studies \citep{Hatchell2007, Curtis2010} have identified 4 outflows in L1455, each associated with a submillimeter core. 

L1455 IRS~1 was first imaged in narrowband H$_2$S(1) emission by \citet{Davis1997}, who found three compact H$_2$ knots on the symmetry axis of IRS~1, outlining a highly collimated outflow at a position angle of 32$\degree$, while \citet{Curtis2010} determined a position angle of 42$\degree$ from their CO($J = 3\rightarrow2$) imaging. This agrees well with the TADPOL CO($J = 2\rightarrow1$) imaging, which shows a well-defined bipolar molecular outflow. Although the dust polarization is not very strong, it is still a case where the B-field appears to be perpendicular to the outflow.

See Figure \ref{fig:L1455} for maps.

\subsection{NGC~1333-IRAS~2A}
The IRAS~2 (IRAS 03258+3104) core lies approximately 11$\arcmin$ south-southwest
of the center of the NGC~1333 reflection nebula, at a distance of 320\,pc \citep{deZeeuw1999}. 
%Estimates for the distance to
%NGC~1333 range from 220~pc \citep{Cernis1990} to 350~pc \citep{Herbig1983}.}
IRAS~2 hosts at least three deeply embedded YSOs \citep{Sandell2001}.
IRAS~2A, the strongest emitter at (sub)millimeter wavelengths, is a Class 0 object near the center of the
core \citep{Lefloch1998}, at the intersection of nearly perpendicular CO
outflows \citep{Sandell1994, Engargiola1999}.
One outflow (position angle $\sim$\,104$\degree$), with a blueshifted lobe that extends 
$\sim$\,100$\arcsec$ to the west and a redshifted lobe that extends $\sim$\,85$\arcsec$ 
to the east, is highly collimated and presumably young. The other outflow
(position angle $\sim$\,25$\degree$), with blueshifted (south) and redshifted (north) lobes
that extend at least 70$\arcsec$ in either direction, is poorly-collimated and
older. The coincidence of IRAS~2A with the point of origin of the outflows 
suggests that IRAS~2A is an unresolved ($<65$\,AU) binary system 
\citep[\eg{}][]{Jorgensen2004}.

The magnetic field across the IRAS~2 core, as mapped with the SCUPOL on the JCMT
(14$\arcsec$ resolution), was described as weak with a ``random field pattern''
\citep{Curran2007a}. However, higher resolution (3$\arcsec$) data obtained by \citeauthor{Curran2007a} with
the BIMA (Berkeley Illinois Maryland Array) interferometer shows magnetic field line with a
roughly east-west orientation across most of the emitting region, which is consistent with the TADPOL observations. 
%Northeast of
%the emission peak, the field lines are oriented NW--SE; southeast of the peak,
%the lines are oriented NE--SW.

See Figure \ref{fig:IRAS2A} for maps.

\subsection{SVS~13}
SVS~13 was discovered as a near-infrared source
by \citet{Strom1976} in the NGC~1333 star forming region. Using VLBI observations of
22\,GHz H$_2$O masers, \citet{Hirota2008} found a distance of 235\,pc.
Observations at millimeter wavelengths reveal at least three continuum sources within
SVS~13. These sources, which form a straight line in the plane of the sky
from northeast to southwest,
have been named as A, B, and C, respectively \citep[][and references
therein]{Looney2000}.  Source A is a Class 0/I source coincident with the infrared/optical counterparts of SVS~13; sources B and C are Class 0 sources. 
High resolution BIMA observations revealed a weak component
of source A that is located 6\arcsec\ to the southwest of the source and is
coincident with centimeter continuum source VLA3 \citep{Rodriguez1997, Rodriguez1999}.
TADPOL observations focus on sources A and B.

Observational evidence suggests that SVS~13 is powering the well studied
chain of Herbig-Haro (HH) objects HH~7--11
\citep{Bachiller2000, Looney2000}.  However, there is some debate as to the main exciting
source of the outflow, which could be either VLA3 or SVS~13
\citep{Rodriguez1997}.  This object is known to
be one of the brightest H$_2$O maser sources among the known low-mass YSOs
\citep{Haschick1980, Claussen1996, Furuya2003}.  

%Recent $\sim$\,5$\arcsec$ ($\sim$\,1500\,AU) resolution maps of the SVS~13
%region by the CARMA Large Area Star-formation Survey (CLASSy) key project
%team in the $\rm J=1-0$ transitions of HCN, HCO+ and N$_2$H+
%\citep{Mundy2013}.

See Figure \ref{fig:SVS13} for maps.

\subsection{NGC~1333-IRAS~4A}
NGC~1333-IRAS~4A comprises two deeply embedded Class 0 YSOs at the south end of the NGC~1333
reflection nebula, located at a distance of 320\,pc \citep{deZeeuw1999}. The binarity of IRAS~4A, first detected in 0.84\,mm
CSO-JCMT baseline data \citep{Lay1995}, has been resolved interferometrically
at millimeter \citep{Looney2000}, submillimeter \citep{Jorgensen2007}, and centimeter \citep{Reipurth2002} 
wavelengths. The two components are 1.8$\arcsec$ apart (580\,AU at 320\,pc) 
and share a common envelope with an estimated mass
of 2.9\,$M_\sun$ \citep{Looney2003}. High-resolution observations of molecular
line emission from IRAS~4A have revealed both low-density \citep{Jorgensen2007} 
and high-density \citep{DiFrancesco2001} tracers with inverse P-Cygni
profiles, which have been interpreted as evidence that envelope material is
falling onto the central protostars.

The outflows emanating from IRAS~4A have been mapped in several CO transitions
\citep[\eg{}][]{Blake1995, Knee2000, Jorgensen2007, Yildiz2012}.
The outflows are well collimated but are ``bent'' in the sky plane. Close
($<0.5\arcmin$) to IRAS~4A, the outflows are oriented north-south; further from
the protostars, the outflows have a position angle or $\sim45\arcdeg$.
The redshifted lobe extends northward, and the blueshifted lobe extends
southward. The extent of the outflows on the sky ($4\arcmin$) and the large
range of line-of-sight velocities suggests that the outflow axis has an
inclination $<45\arcdeg$. 

Maps of linearly polarized dust emission from IRAS~4A have been made at
850\,$\mu$m with the SCUBA polarimeter on the JCMT \citep{Matthews2009}.
These maps imply a large-scale magnetic field that is fairly uniform in the
northeast-southwest direction across the IRAS~4 core.  \citet{Girart2006} 
also mapped IRAS~4A at high resolution with the Submillimeter Array (SMA), revealing one of the first
``hourglass'' B-field morphologies ever seen in a low-mass protostar.

See Figure \ref{fig:IRAS4A} for maps.

\subsection{NGC~1333-IRAS~4B and 4B2}
%http://simbad.u-strasbg.fr/simbad/sim-id?submit=display&bibdisplay=refsum&bibyear1=1850&bibyear2=%24currentYear&Ident=%40638419&Name=%5BKJT2007%5D+SMM+J032920%2B31131#lab_bib
% See Herczeg+2012 for distance, source/envelope mass
%\input{blurbs/IRAS4B.txt}
% This appears to be the correct source, even though I'm pretty sure the coordinates are wrong...
%http://simbad.u-strasbg.fr/simbad/sim-id?Ident=%403952765&Name=%5bKJD2006%5d%20SMM%20J032919%2b31131&submit=submit
%\input{blurbs/IRAS4B2.txt}
NGC~1333-IRAS~4B and 4B2 are Class 0 sources in Perseus at a distance of 320\,pc \citep{deZeeuw1999}, and about 30\arcsec{} 
to the southeast of  the well known Class 0 source NGC~1333-IRAS~4A.  IRAS~4B 
hosts a slow ($\sim$10 \kms{}) bipolar molecular outflow oriented north-south.
Single-dish maps from the SCUBA \citep{Matthews2009} and Hertz \citep{Dotson2010} 
polarimeters show polarization consistent with the prominent polarization detected by CARMA on the western 
edge of the core.  

Strong water lines were detected toward IRAS~4B by the Spitzer infrared spectrograph 
\citep{Watson2007} and Herschel HIFI \citep{Herczeg2012}. \citet{Watson2007} attributed 
the emission to shocked material falling from the protostellar envelope onto the dense 
surface of the circumstellar disk, which requires that the disk of IRAS~4B be oriented roughly face-on, 
thus allowing emission to escape from the cavity evacuated by the bipolar outflow.  
That assumption was called into question after VLBI measurements of the proper motions of water  
masers in the outflow of IRAS~4B that suggest that the object is in fact viewed edge-on \citep{Marvel2008}.  
The claim by \citet{Watson2007} that the water emission originates in the disk has been challenged 
by \citet{Herczeg2012}, who argue that the emission originates in shocks within the bipolar outflow cavity.

% References: 
% Chiang, Looney + 2008 (discussion of two separate objects)
% Sandell & Knee 2001 (listing of both IRAS 4BE and W) 
% di Francesco 2001 (resolved it with the PdBI)
% Choi, Panis, Evans 1999 (discovered EW 10" binary; called it 4C)
% Rodriguez, Anglada, Curiel 1999 (called 4C the source 1' NE of IRAS4A)
% Smith, Bonnell + 2000 (also call 4C the 1' NE source)
%  
% http://cdsads.u-strasbg.fr/cgi-bin/nph-bib_query?1999ApJS..122..519C&db_key=AST&nosetcookie=1
% http://cdsads.u-strasbg.fr/cgi-bin/nph-bib_query?1999ApJS..125..427R&db_key=AST&nosetcookie=1
% http://cdsads.u-strasbg.fr/cgi-bin/nph-bib_query?2008ApJ...685..285M&db_key=AST&nosetcookie=1
% http://adsabs.harvard.edu/cgi-bin/bib_query?arXiv:0803.1272
% http://cdsads.u-strasbg.fr/cgi-bin/nph-bib_query?2001ApJ...546L..49S&db_key=AST&nosetcookie=1
% http://adsabs.harvard.edu/abs/2002AJ....124.1045R
% http://adsabs.harvard.edu/abs/2000MNRAS.319..991S
% http://adsabs.harvard.edu/abs/1991ApJ...376L..17S
% http://adsabs.harvard.edu/abs/2000ApJ...529..477L

IRAS~4B2 (also called IRAS~4BE, IRAS~4B$^{\prime}$, and IRAS~4BII) is the weaker binary
companion 10$\arcsec$ to the east of IRAS~4B.  The source has been called
IRAS~4C as well \citep[e.g.,][]{Choi1999, Looney2000}, but 4C is generally used
as the name of a source $\sim$40\arcsec{} east-northeast of IRAS~4A
\citep[e.g.,][]{Rodriguez1999, Smith2000, Sandell2001}.  
In their BIMA observations, \citet{Choi1999} saw IRAS~4B2 as an unresolved
extension of continuum emission to the east of IRAS~4B.  \cite{Sandell2001}
and \citet{DiFrancesco2001} later resolved the 10$\arcsec$ IRAS~4B/IRAS~4B2 binary using
the JCMT and the PdBI, respectively.  

While prior to the TADPOL survey no spectral line emission had been detected
toward IRAS~4B2, we see a small, faint E--W outflow in CO($J = 2\rightarrow1$).

See Figure \ref{fig:IRAS4B} for maps.

\subsection{HH~211~mm}
%\input{blurbs/HH211.txt}
%03:43:56.8 
%+32:00:52.0

HH~211~mm is the Class 0 YSO \citep{Froebrich2005} launching the well-known bipolar outflow
HH 211. It is located in the IC 348 cluster at the eastern part of
the Perseus molecular cloud, at a distance of 320\,pc \citep{deZeeuw1999}.  The jet HH 211 was relatively recently
detected by near-IR H$_2$ observations \citep{McCaughrean1994}.
\citet{Gueth1999} showed that the driving object is HH 211
mm, and they distinguished between the collimated jet and the slow extended outflow
components using interferometric millimeter-continuum and CO observations.
The bipolar outflow has been studied extensively in various molecular
line transitions \citep[e.g., SiO($J = 1\rightarrow0$);][]{Chandler2001},
and recently {\it Spitzer} IRS observations showed that the bipolar outflow material is mostly molecular
\citep{Dionatos2010}.  Based on the bipolar outflow velocity and extension,
the kinematic age is estimated to be only $\sim$\,1000\,yr.
\citep[e.g,][]{Gueth1999}.  Recent submillimeter
interferometric observations have revealed that the object is a
protobinary system separated by about $0.3\arcsec$ \citep{Lee2009},
and the kinematic structure of the envelope has been studied by
CARMA N$_2$H$^+$ observations \citep{Tobin2011}.  The
SCUPOL map toward the HH 211 and IC 348 region showed polarization that is
neither aligned with nor perpendicular to the the bipolar outflow
\citep{Matthews2009}.

See Figure \ref{fig:HH211} for maps.

\subsection{DG~Tau}
DG~Tau is a Class II, 0.67\,M$_\sun$ K5-M0 T Tauri star \citep[][and references therein]{Gudel2007} 
located at a distance of roughly 140\,pc in the Taurus-Auriga star-forming association 
\citep{Kenyon1994, Torres2009}.  It is remarkable primarily for its well collimated jet, HH 158, 
and was among the first T Tauri stars known to exhibit such strong and clear accretion and 
outflow activity \citep{Mundt1983}.  The literature on its jet is correspondingly vast, as it has 
been studied extensively across the electromagnetic spectrum \citep[see, e.g.,][and 
references therein]{Schneider2013, Rodriguez2012b, Lynch2013}.  DG~Tau is properly known 
as DG~Tau~A, since it has a common proper motion companion DG~Tau~B, a Class I source 
that launches the HH 159 jet \citep[e.g.,][]{Rodriguez2012a}.

The dust disk around DG~Tau has the dubious distinction of being the most frequently observed 
by (sub)millimeter polarimeters.  It was one of the first two T Tauri disks that seemed to exhibit a 
tentative (3\,$\sigma$) detection of unresolved 850\,$\mu$m polarization using the single-dish JCMT \citep{Tamura1999}, 
apparently indicating a large-scale toroidal magnetic field threading the disk.  Follow-up observations 
at 350\,$\mu$m with the CSO did not confirm the 3\% polarization fraction, but the wavelengths were 
too widely separated to rule out a spectral dependence of the polarization fraction \citep{Krejny2009}.  
As part of the TADPOL survey the JCMT detection was followed up at 1.3\,mm using the CARMA 
polarimeter, which again resulted in a sensitive non-detection \citep{Hughes2013}.  The CARMA 
results indicate that either the JCMT detection was spurious or the polarization originates from large 
spatial scales that are filtered out by the interferometer---an envelope, perhaps---rather than from the 
circumstellar disk itself.  

See Figure \ref{fig:DGTau} for maps.

\subsection{L1551~NE}
L1551~NE is a low-mass Class I protostar first discovered with IRAS
\citep{Emerson1984} and located a few arcminutes from L1551 IRS5
and at a distance of 140 pc \citep{Kenyon1994}.
L1551~NE is a binary system with a bipolar molecular outflow and
a Keplerian circumbinary disk
% Interesting - Moriarty-Schieven et al 2000 "discovered" a binary companion
% to L1551 NE that was really a piece of a circumbinary disk. The
% actual companion was found two years later with the VLA at 3.6cm
% by Reipurth et al and this second source was coincident with
% a second source proposed much earlier by Rodriguez et al.
\citep{Moriarty1995, Rodriguez1995, Reipurth2002,
Takakuwa2012}. SCUBA 850\,$\micron$ measurements show polarization
orientations in the extended dust envelope that are mostly perpendicular to
the outflow direction, with no polarization detected at the continuum
peak \citep{Matthews2009}.  Our TADPOL observations show a clear pattern
of polarization across the continuum peak, with  B-field orientations
perpendicular to the outflow direction.

See Figure \ref{fig:L1551} for maps.

\subsection{L1527}
%\input{blurbs/L1527.txt}
%04:39:53.9 +26:03:09.7,
%IRAS 04368+2557

L1527 is a Class 0/I YSO located in the Taurus molecular
cloud at a distance of about 140\,pc \citep[e.g.,][]{Andre2000}.
Its bipolar outflow is oriented in the east-west direction and is nearly in
the plane of the sky, which makes the object an ideal target for studying
the disk and outflow structure at the earliest stage of low mass
star formation \citep[e.g.,][]{Jorgensen2007}.  Recently the
disk was revealed to have Keplerian motion, and the protostellar mass
was estimated to be $\sim$\,0.2 M$_\sun$ using CARMA $^{13}$CO observations
\citep{Tobin2012}.  In addition, detailed modeling of SMA
and CARMA continuum observations found that the disk is large (about
125\,AU in radius) and is thicker than the hydrostatic equilibrium case
\citep{Tobin2013}.  

The source's orientation is also beneficial
for magnetic field studies.  SCUPOL
detected quite irregular polarization at 850\,$\mu$m, but at the
center the B-field orientation is perpendicular to the
bipolar outflow \citep{Matthews2009}, consistent with the TADPOL observations.  In contrast, SHARP
detected polarization in the outer regions at 350\,$\mu$m that was consistent 
with B-fields that are aligned with the bipolar outflow
\citep{Davidson2011}.

%\citep{2013ApJ...768..110C}: binary systems? So far, no evidence

See Figure \ref{fig:L1527} for maps.

\subsection{CB~26}
CB~26 \citep{Clemens1988} is a Bok
globule generally accepted to be associated with the Taurus-Auriga complex a
distance of 140\,pc \citep{Launhardt2001,Henning2001}.
The embedded YSO located near the edge of the globule is
a Class I source with a luminosity of 0.5\,$L_\sun$ \citep{Stecklum2004}.
Millimeter interferometric
observations of dust continuum and molecular spectral
line emission \citep{Launhardt2001} show both an edge-on disk that sits at the center of
a near-infrared bipolar reflection nebula \citep{Stecklum2004},
as well as a bipolar outflow \citep{Launhardt2009} perpendicular
to the disk.

SCUBA 850\,$\micron$ measurements show polarization orientations 
both predominantly parallel to the disk \citep{Henning2001} 
and predominantly perpendicular to the disk \citep{Matthews2009}.
% The same goddamned data, reduced by two groups!
Our TADPOL observations detect polarization near the dust peak that 
is consistent with the latter SCUBA results.

See Figure \ref{fig:CB26} for maps.

\subsection{Orion-KL}
Orion-KL, the Kleinmann-Low Nebula in Orion, is the nearest region of high mass
star formation, 415\,pc away \citep{Menten2007}.  It lies inside Orion Molecular
Cloud 1 (OMC1), which in turn forms part of an integral shaped filamentary
cloud that is more than 7\,pc long \citep{Johnstone1999}.  Our map, a mosaic of 7
pointings, covers Orion-KL and its associated hot core, and a piece of the more
quiescent ``northern ridge'' about $25\arcsec$ to the NE.

At least two massive stars, Source~I (SrcI) and the Becklin-Neugebauer Object
(BN) are associated with Orion-KL.  Proper motion measurements show that these
two stars are recoiling from one another at 35--40~km~s$^{-1}$; they appear to
have been ejected from a multiple system just 500 years ago \citep{Gomez2008,
Goddi2011}.  This explosive event also is thought to have created a set of bow
shocks and fingers that form a poorly collimated, NW-SE high velocity outflow;
a separate, lower velocity outflow emerges from SrcI in the perpendicular
direction \citep{Plambeck2009}.

Extensive polarization maps from SCUBA, Hertz, and Stokes \citep{Dotson2000}
show that the large-scale magnetic field in OMC1 is perpendicular to the long
axis of the molecular cloud, with evidence for an hourglass-shaped pinch
centered on KL \citep{Schleuning1998}. Higher resolution 345~GHz SMA
observations show a remarkable circularly symmetric polarization pattern
centered between SrcI and BN, near the site of the putative explosive event; a
possible interpretation is that the explosion dragged the magnetic field
outward into a radial pattern \citep{Tang2010}.  However, within 500\,AU of SrcI the
magnetic field deduced from SiO v=0 
maser\footnote{These masers, in the ground
vibrational state, should not be confused with the stronger v=1 masers closer
to the star; both the intensity and the polarization of the v=1 masers are time
variable.} 
polarization observations is relatively straight.
\citep{Plambeck2003}.  The B-field orientation is highly uncertain because maser
polarization may be either parallel or perpendicular to the field, and because
of possible Faraday rotation by foreground plasma; \citeauthor{Plambeck2003} argued that
it is at PA 145$^{\circ}$, roughly perpendicular to the low velocity outflow
from SrcI.

The TADPOL map shows the radial magnetic field pattern previously detected with
the SMA.  Our map extends further north, and shows that the magnetic field
orientation in the northern ridge cloud is consistent with the large scale field,
except near the SW tip where it forms part of the radial pattern. 

See Figure \ref{fig:OrionKL} for maps.

\subsection{OMC3~MMS5 and MMS6}
MMS5 and MMS6 are condensations in Orion Molecular Cloud 3 (OMC3)---a narrow ridge or filament about 1\,pc
long and at a distance of 415\,pc \citep{Menten2007}.  MMS6 is the brightest millimeter continuum source in
OMC3, with an estimated mass of 36\,M$_\sun$ \citep{Chini1997}.  It contains a compact
core, MMS6-main, that probably is heated by an extremely young intermediate
mass Class 0 protostar \citep{Takahashi2012b}.  A bipolar outflow, with a total length
of only $4''$ (2000 AU), emerges along a N-S axis from MM6-main; the dynamical
age of the outflow is less than 100 years \citep{Takahashi2012a}.  A more
extended outflow emerges along an E-W axis from MMS5.

The large scale magnetic field orientation in OMC3 inferred from SCUBA data is
perpendicular to the long axis of the cloud.  \citet{Matthews2001} argue that
the pattern of depolarization along the central axis is best explained by a
field toroidally wrapped around a filament, rather than by a straight field
perpendicular to a sheet.  The field orientations that we measure for both the MMS5 and
MMS6 cores are closely aligned with the large-scale field.  Our results for
MMS6 agree well with previous $3.6''$ resolution observations of this source
made with BIMA \citep{Matthews2005}.

See Figure \ref{fig:MMS6} for maps.

\subsection{OMC2-FIR3 and 4}
OMC2 is an intermediate mass star-forming region located north of
the massive OMC1 complex in the so-called integral-shaped filament
of Orion~A. Located at a distance of 415\,pc \citep{Menten2007}, it
is one of the brightest regions in the Orion Nebula and is know to
harbor several protostellar objects and pre-main-sequence stars.
Earlier continuum studies at (sub)millimeter wavelengths
have established it to be in a later evolutionary stage of star formation
than the OMC3 region neighboring it to the north
\citep{Chini1997, Lis1998}. More recent studies have modeled
several of its embedded sources as infalling protostars, young stars
with disks, and binaries comprising both types of objects \citep{Adams2012}. 
More specifically, although OMC2-FIR4 is well modeled as
a Class 0 protostar of approximately $50\,L_{\sun}$ and $10^{-4}M_{\sun}\,\mathrm{yr}^{-1}$
mass infall rate \citep{Adams2012}, recent high resolution measurements
have resolved three spatially distinct sources in its core \citep{Lopez2013}. 

The polarization of the OMC2 region and the FIR3 and 4 sources was first
investigated at 350\,$\mu$m by \citet{Houde2004} with Hertz at the
CSO (see also \citealt{Dotson2010}). This region is characterized by
extremely low polarization levels as well as a strong depolarization
with increasing total intensity. For example, OMC2-FIR4 was found
to have a mean polarization of $0.35\%\pm\,0.08\%$ within region
of approximately $\pm$\,0.3$\arcmin$ from its peak (200\,Jy within a beam
of $20\arcsec$ FWHM at 350\,$\micron$), making it one of the
most weakly polarized molecular cloud complexes ever observed at that
wavelength. Although higher polarization was measured 
at 850\,$\micron$, the relatively low polarization as compared with other sources
within the integral-shaped filament was confirmed by \citet{Matthews2009} 
and \citet{Poidevin2010}. OMC2-FIR4 was also observed to
be a region of transition in the orientation of the polarization.
The mean polarization angle goes from $\sim$\,115$^{\circ}$ south of
it to $\sim$\,175$^{\circ}$ to the north at OMC2-FIR3, where the polarization
angle is relatively well aligned with the filament. This transition
region is almost coincident with a location of intense outflow activity
reported by \citet{Williams2003}.

The aforementioned low levels of polarization obtained with single-dish
measurements are in contrast with the results shown in Table \ref{table:obs} and
Figure \ref{fig:FIR4}, where mean polarization fractions of $\sim$\,6--8\% are detected
with CARMA, making FIR 3 and 4 two of the most polarized sources in our sample. This
may be reconciled by the fact that the magnetic field, as measured
at small scales with CARMA, significantly changes its orientation
within just a few arcseconds towards the center of the map (i.e.,
at OMC2-FIR4). The change of nearly $90^{\circ}$ observed could
account for the exceedingly low polarization observed with Hertz,
if the data were combined within a single Hertz beam. 
%However, the differences
%between the single-dish and CARMA observations are 
%more difficult to explain for OMC2-FIR3 (the compact source to the north
%% $\mathrm{RA\,}(\mathrm{J2000})\simeq5^{\mathrm{h}}35^{\mathrm{m}}27\fs6$,
%% $\mathrm{Dec\,}(\mathrm{J2000})\simeq-5^{\circ}09\arcmin35$ 
%in Figure \ref{fig:FIR4}), where the measured polarization orientations are
%at odds with the single-dish orientations in the vicinity.

See Figure \ref{fig:FIR4} for maps.

\subsection{CB~54}
CB~54 (LBN1042; \citealt{Lynds1965,Clemens1988})
is a Bok globule at a distance of 1.1\,kpc
\citep[and references therein]{Henning2001}.
(Sub)millimeter and mid-infrared observations
reveal two Class I YSOs coincident with source IRAS 07020-1618,
with strong dust continuum emission and one or two bipolar outflows
\citep{Yun1994,Zhou1996,Ciardi2007,deGregorio2009,Launhardt2010}.
SCUBA 850\,$\micron$ measurements show weak polarization across the continuum
source, with changing orientation and increased strength in the envelope
\citep{Henning2001,Matthews2009}. 

%Our TADPOL observations detect
%multiple outflow lobes, and an inferred B-field orientation at the dust peak 
%that is roughly consistent with that from the SCUBA map.

See Figure \ref{fig:CB54} for maps.

\subsection{VLA~1623}
VLA 1623 is the prototypical Class 0 source, discovered by \cite{Andre1993}. It is located in the southern edge of the Ophiuchus A cloud at a distance of 125\,pc \citep{Loinard2008}. It has a highly collimated outflow, as is typical of Class 0 sources, and this outflow extends over a large distance at a common position angle \citep{Andre1990}.  The polarization toward VLA 1623 was measured by \cite{Holland1996} using the single bolometer UKT14, and the B-field was found to be perpendicular to the  CO outflow.  \cite{Holland1996} note that this alignment of the field implies that the large-scale field in the cloud cannot therefore collimate the outflow.  The single dish polarization data from SCUBA \citep{Matthews2009} and Hertz \citep{Dotson2010} are consistent with this early picture, as are the CARMA data, which show that the central region has a preferred B-field orientation orthogonal to the outflow. This central region may be a flattened pseudo-disk, as discussed for other Class 0 sources by \citet{Davidson2011}.  The orientation of the field is aligned along the major axis of this pseudo-disk and orthogonal to the outflow, and even follows the emission extension to the north east. 
%The offset seen between the single-dish and CARMA data, rather than an averaged value most likely measured in the larger beams of the JCMT and CSO \textbf{(I'm not exactly sure what this sentence means)}.  
While the field orientation on small scales might alone suggest a toroidal field component, we note that in fact the CARMA field orientations are consistent with those of the large-scale emission.

See Figure \ref{fig:VLA1623} for maps.

\subsection{Ser-emb~1, 6, 8, 8(N), and 17}

Ser-emb~1, 6, 8, 8(N), and 17 are low-mass, Class 0 (Ser-emb~6, 8, 8(N), 17) and Class I (Ser-emb~1) protostars in the 
Serpens Main cluster, located at a distance of 415\,pc \citep{Dzib2010}.  
All sources are discussed in \citet{Enoch2011}, who made high-resolution
1.3\,mm maps of nine low-mass cores in Serpens using CARMA.
The results in \citet{Enoch2011} follow up on a large 1.1\,mm survey of protostars
using Bolocam on the CSO \citep{Enoch2007}.

Ser-emb~6 (also called Serp-FIR1 and Serp-SMM1) is the brightest
(sub)millimeter source in the Serpens Main cluster.  It harbors a Class 0 protostar at its
center.  Continuum observations at 6\,cm using the Karl G. Jansky Very Large Array
(VLA) resolved the continuum peak into three collinear sources, where the NW
and SE components appear to be moving away from the central source
\citep{Rodriguez1989}.  While the three sources are not obviously present in
the 1.3\,mm maps from \citet{Enoch2011}, it is possible that the
outer two are associated with the lobes of the complicated bipolar outflow,
which may be causing the disturbed B-field morphology on the E and W edges of
the TADPOL map.
The large-scale polarization properties of Ser-emb~6 are discussed at length
in \citet{Chapman2013}, who note that the polarization they measure with SHARP
is not consistent with the small-scale morphology measured by CARMA. This
could be because of projection effects, as the source and its complicated
bipolar outflow are thought to be viewed at a high inclination angle with
respect to the sky \citep{Enoch2009}.

% SIMBAD search
% http://simbad.u-strasbg.fr/simbad/sim-id?submit=display&bibdisplay=refsum&bibyear1=1850&bibyear2=%24currentYear&Ident=%402656182&Name=NAME+SERPENS+SMM+1#lab_bib

Ser-emb~8 is also called S68N by \citet{McMullin1994}, who examined the chemistry in the 
Serpens Main region. The two sources are notable for having the most distinct, well collimated 
SiO($J = 5\rightarrow4$) bipolar outflows in the entire TADPOL sample.
It is also interesting to note that despite lying only $\sim$\,10$\arcsec$ apart from one another,
the pair of cores exhibit B-fields and outflows that are both parallel (Ser-emb~8) and
perpendicular (Ser-emb~8(N)).

% References
% Ser-emb 8
% http://simbad.u-strasbg.fr/simbad/sim-id?submit=display&bibdisplay=refsum&bibyear1=1850&bibyear2=%24currentYear&Ident=%402656191&Name=%5BMAN98%5D+B2-MM13#lab_bib

% Ser-emb 8(N)

% http://simbad.u-strasbg.fr/simbad/sim-id?submit=display&bibdisplay=refsum&bibyear1=1850&bibyear2=%24currentYear&Ident=%404036319&Name=%5BWM2000%5D+S68Nc#lab_bib

See Figures \ref{fig:SerEmb17}, \ref{fig:SerEmb1}, \ref{fig:SerEmb8}, and \ref{fig:SerEmb6} for maps.

\subsection{HH~108~IRAS}
%\input{blurbs/HH108.txt}
%18:35:42.17 -00:33:18.6,
%IRAS 18331-0035
%HH 108 IRAS  
%HH 109 (= HH 108 MMS)
%L588

HH 108 IRAS (also known as IRAS 18331-0035) is a Class 0/I YSO \citep{Froebrich2005} in the Serpens
molecular cloud, at a distance of about 310\,pc \citep{deLara1991}.  The
object was identified by 1.3\,mm observations
\citep{Chini1997b} following the detection of the
Herbig-Haro object HH 108 \citep{Reipurth1992}.  A colder
millimeter object was discovered $70\arcsec$ northeast of HH 108 IRAS, and was named HH
108 MMS \citep{Chini1997b}.  An elongated structure enclosing
these two objects was also detected in submillimeter observations
\citep{Chini2001}.  Recently, \citet{Tobin2011}
reported IRAM 30\,m and CARMA data of N$_2$H$^+$ showing a velocity
gradient perpendicular to the bipolar outflow.  
\citet{Siebenmorgen2000} have detected polarization
of scattered light toward HH 108 IRAS and HH 108 MMS at 14\,$\mu$m, which
shows magnetic fields aligned with the elongated structure enclosing the
two objects.  However, no previous polarimetric observations of thermal dust emission have been
reported toward this region.

See Figure \ref{fig:HH108} for maps.

\subsection{G034.43+00.24~MM1 and MM3}
G034.43+00.24 is a massive star-forming region associated with the IRAS source
IRAS 18507$+$0121, located at a distance of 1.56\,kpc
\citep{kurayama2011}. The most prominent source in the complex is the UC \HII\
region G34.4$+$0.23, which is embedded in a massive (1000\,\Msol) dense core
\citep{miralles1994,molinari1996:hmc,bronfman1996}.  At $\sim$\,5\,$\arcsec$
resolution, 3\,mm spectral line and continuum observations revealed another
massive ($\sim$240\,\Msol) dense core north of the UC \HII\ region known
as G34.4MM. Near-infrared non-detection and the lack of a significant radio counterpart
suggested a deeply embedded, high-mass protostar \citep{shepherd2004}. Further
millimeter continuum and Spitzer mid-IR observations revealed that the region is
located in an $\sim$8\,$\arcmin$ long infrared dark cloud known as MSXDC
G034.43+00.24 \citep{rathborne2005}.  \citeauthor{rathborne2005} identified
four compact millimeter clumps labeled MM1--MM4, with MM1 corresponding to the millimeter core
G34.4MM mentioned above. TADPOL observations focused on the cores
MM1 and MM3.

The estimated masses and luminosities are 800\,\Msol\ and 32000\,\Lsol\ for
MM1, and 170\,\Msol\ and 12000\,\Lsol\, for MM3 \citep{rathborne2005}. SMA
continuum observations at much higher resolution show that MM1 is a 29\,\Msol\
unresolved core with hot-core-like line emission
\citep{rathborne2008:hot_core}.  Recent ALMA observations by \citet{Sakai2013}
show that MM3 is a hot core with a mass of $\lesssim 1.1$\,\Msol.\
%\citeauthor{rathborne2008:hot_core} suggested
%that MM1 and MM3 represent two different phases in the evolution of a
%high-mass star, with MM1 in a hot molecular core phase, and MM3 in an earlier
%protostellar phase (see also \citealt{garay2004:new_cores}).  
Water maser emission, massive CO outflows, and excess 4.5\,$\mu$m emission indicative of
shocks have all been reported toward both MM1 and MM3
\citep{wang06:masers,shepherd2007_g34,chambers09,sanhueza2010}.  
Note that despite the significantly closer distance that  
\citet{kurayama2011} found using maser parallax measurements,
all of the above works (except for \citealt{Sakai2013}) have estimated physical parameters assuming a kinematic distance of 3.7--3.9\,kpc.

The MM1 core was observed in both 3\,mm continuum and CO($J = 1\rightarrow0$) line
polarization with the BIMA array at 16\,\arcsec\ resolution
\citep{cortes2008}. Both the continuum and line polarization observations
reveal a uniform polarization pattern with an orientation perpendicular to the
major axis of the filament. TADPOL observations reveal a much more complex
polarization pattern with significantly disturbed B-fields---and even hints of hourglass morphology---in the densest parts
of the core.

See Figures \ref{fig:G034MM1} and \ref{fig:G034MM3} for maps.

\subsection{B335~IRS}
B335 IRS (Barnard 335, also called CB199 \citep{Clemens1988}) is an extremely isolated Bok
globule harboring a low-mass Class 0 protostar. 
The distance to B335 is uncertain, but a recent estimate by \citet{Stutz2008}
finds that it may be as close as 150\,pc.
\citet{Stutz2008} also derive a luminosity for B335 of 1.2\,$L_{\sun}$.
The protostellar nature of
B335 was first uncovered by \citet{keene1980,keene1983} using far-infrared
observations that were indicative of an embedded source.  \citet{Frerking1982}
found evidence for an outflow in CO ($J = 1\rightarrow0$);  
B335 also presents a convincing infall signature with
blue-asymmetric line profiles in optically thick tracers
\citep{Zhou1993}. \citet{Harvey2003} found evidence for a circumstellar disk
with a radius <\,60\,AU from modeling of interferometric visibilities.
Kinematic data from $\sim$\,1000 AU scales shown little evidence for rapid
inner envelope rotation \citep{Yen2013}.

% Mention polarimetry 

See Figure \ref{fig:B335} for maps.

\subsection{DR21(OH)}
DR21(OH) is a massive star-forming clump located in the heart of the
DR21 molecular ridge in Cygnus~X. At a distance of 1.5\,kpc
\citep{rygl2010:parallax}, Cygnus~X is one of the nearest massive star
forming complexes in our galaxy \citep{schneider2006,
  motte2007}. DR21(OH) has an estimated bolometric luminosity of
1.7$\times 10^4$\,\Lsol\ and a mass of 1800\,\Msol\
\citep{jakob2007}. Millimeter observations have revealed two massive
millimeter sources, MM1 and MM2 \citep{mangum1991, liechti1997},
each with masses of a few\,$\times$\,100\,\Msol. At
sub-arcsecond resolution, these two cores are resolved into a cluster
of $\sim$10 compact massive cores between 5--24\,\Msol\
\citep{zapata2012:dr21}. DR21(OH) is rich in methanol, water and OH maser
features \citep{plambeck1990, liechti1997, kurtz2004,
  araya2009}. Infrared observations have identified several deeply
embedded YSOs in the region \citep{kumar2007,davis2007}. High-velocity
bipolar outflows from both MM1 and MM2 have been detected in CO and
other tracers \citep{lai2003:dr21, zapata2012:dr21}. Therefore, MM1/2 are
excellent candidates for very young massive protostars. Our TADPOL
observations encompass these two massive sources. 

DR21(OH) has been part of several single-dish and
interferometric polarization studies. At low resolution, a uniform
large scale magnetic field perpendicular to the DR21 filament is
observed \citep{vallee2006,kirby2009:dr21}. At higher resolution, a
more complex polarization pattern, consistent with our TADPOL
observations, has been revealed toward DR21(OH)
\citep{lai2003:dr21, Girart2013}. The plane-of-sky component of the magnetic field is
estimated to be 0.62 and 2.1\,mG at scales of 0.34 and 0.08\,pc
respectively \citep{Girart2013}. These measurements are consistent
with previous single-dish CN Zeeman measurements, which estimate the
line-of-sight magnetic field strength to be $0.36$ and $0.71$\,mG
for MM1 and MM2, respectively \citep{crutcher1999,falgarone2008}. 

See Figure \ref{fig:DR21OH} for maps.

\subsection{L1157}
L1157-mm (IRAS 20386+6751) is a Class 0 source located in the Cepheus Flare
region. The distance to L1157 is uncertain, with estimates between 250\,pc
and 440\,pc \citep{Viotti1969, Straizys1992, Kun1998, Kun2008};
we adopt a distance of 250\,pc \citep{Looney2007}.
L1157 has a large ($\sim$\,20,000\,AU) flattened envelope structure
detected in 8\,$\mu$m absorption \citep{Looney2007} and molecular tracers
N$_2$H$^+$ and NH$_3$ \citep{Chiang2010, Tobin2011} that reveal
complex kinematics from rotation, infall, and outflow. This large scale
structure is perpendicular to the well known outflow from the central source
\citep{Gueth1996}, affording an edge-on view of the system. The dust
emission from the system has been observed at multiple wavelengths and
resolutions at CARMA \citep{Chiang2012} and was best modeled as an envelope
with a unresolved disk (<\,40\,AU). Recent high resolution
($\sim$\,0.06$\arcsec$) VLA 7.3\,mm dust continuum observations have placed a
limit on the size of the disk component of <\,20 AU \citep{Tobin2013}. The
magnetic field was shown to have a well defined hourglass shape and was
estimated to have a plane-of-sky magnitude of 1--4\,mG \citep{Stephens2013}.

See Figure \ref{fig:L1157} for maps.

\subsection{CB~230}
CB~230 \citep{Clemens1988} is an isolated globule located in the Cepheus flare
region, at a distance of 325\,pc \citep{Straizys1992}, although other more recent estimates
range from 270--515\,pc \citep{Kun2009}.
%a conservative estimate of 325 pc is usually adopted for this
%region \citep{Straizys1992}. 
The protostellar system is classified as Class
0/I \citep{Launhardt2013}.  A large envelope is detected in both N$_2$H$^+$
and ammonia and exhibits a velocity gradient suggestive of rotation
\citep{Chen2007, Tobin2011}.  The protostellar source is also identified as a
triple system \citep{Yun1996,Launhardt2013,Tobin2013b}. The
wide companion (CB~230~IRS2) is located 10\arcsec\ from CB~230~IRS1 and is not
observed at wavelengths longer than 24\,$\micron$ 
\citep{Launhardt2001a,Massi2008,Launhardt2013}; however, IRS2 does seem to drive a CO outflow
\citep{Launhardt2001b}. The close companion to IRS1 (IRS1B) was recently
discovered by \citet{Tobin2013b} and is separated from IRS1 by
0.3\arcsec\ (100\,AU).

%%%%% Mention polarization

See Figure \ref{fig:CB230} for maps.

\subsection{L1165}
The L1165 dark cloud has a L-shaped filament structure, and is located in
the Cepheus region \citep[e.g.,][]{Reipurth1997,Tobin2010}.  Two IRAS
objects are in the kink of the filamentary structure; however, only the southern
object IRAS 22051+5848 was detected at 850\,$\mu$m and classified
as a Class I YSO \citep{Visser2002}; this is the source we observe.
The distance to the object is somewhat ambiguous in the
literature: 750\,pc \citep[e.g.,][]{Reipurth1997} and 300\,pc
\citep[e.g.,][]{Dobashi1994}.  We adopt the latter 300\,pc,
which gives a reasonable luminosity for a low mass YSO and has been
used in previous (sub)millimeter studies.  The bipolar outflow has
been mapped in CO($J = 2\rightarrow1$) by the JCMT \citep[][]{Visser2002,
Parker1991}, and the near infrared nebula feature was imaged
in K-band by the University of Hawaii 2.2\,m telescope
\citep{Connelley2007}.  It has been imaged in 350\,$\mu$m continuum emission
by the SHARC-II camera at the CSO \citep{Wu2007}.
In addition, {\it Spitzer} IRAC data have revealed that the bipolar
outflow cavity and the elongated envelope structure (imaged in 8\,$\mu$m emission)
are perpendicular to each other \citep{Tobin2010}.
Recently \citet{Tobin2011} detected a velocity gradient
along the elongated envelope structure using IRAM 30\,m
observations in N$_2$H$^+$ and CARMA observations in N$_2$H$^+$ and
HCO$^+$; these observations showed that the HCO$^+$ feature can be interpreted as
rotation around a 0.5 M$_\sun$ central protostar.  There have been no previous
polarimetric observations toward this object.

See Figure \ref{fig:L1165} for maps.

\subsection{NGC~7538~IRS~1}
The hyper-compact \ion{H}{2} region NGC~7538~IRS~1 was first discovered in a 2\,$\mu$m and 20\,$\mu$m survey 
of NGC~7538 \citep{WynnWilliams1974}, which lies at a distance of 2.65\,kpc \citep{Moscadelli2009}. 
IRS~1 is an extremely young high-mass star (type $\sim$O7), which powers a N-S 
ionized thermal jet \citep{Gaume1995, Sandell2009}, drives a molecular outflow 
\citep{Scoville1986, Zhu2013}, and excites a variety of molecular masers, including 
OH, H$_2$O, H$_2$CO, CH$_3$OH, NH$_3$, and $^{15}$NH$_3$ \citep{Johnston1989}. 
It is still heavily accreting with an accretion rate of
$\sim$10$^{-3}$\,\Msolperyr, quenching the formation of an
 \ion{H}{2} region. All the observational evidence suggests that IRS\,1 should 
be surrounded by an E--W accretion disk, which has yet to be confirmed.
Imaging with the SCUBA polarimeter at 850\,$\mu$m indicates that 
the magnetic field is disturbed around IRS~1 \citep{Momose2001}. At the position
of IRS\,1 the degree of polarization is 2--3\%, with a polarization (not B-field) position angle of 90\degree; however,
immediately east the position angle is $\sim$50\degree, similar to what \citet{Flett1991}
obtained at 800\,$\mu$m. 
It is not surprising that the magnetic field around IRS\,1 is disturbed,
because the molecular cloud core is forming a young cluster \citep{Qiu2011}.
At least eight of the cluster members are seen as millimeter continuum sources and
H$_2$O masers, suggesting that they are surrounded by accretion disks and are
probably powering outflows.

See Figure \ref{fig:NGC7538} for maps.

\subsection{CB~244}
CB~244 (L1262) is an isolated globule in the Cepheus region with an estimated
distance of 200\,pc, which is the distance of the nearby cloud L1235
\citep{Snell1981, Benson1989,Stutz2010}.  CB~244 harbors both a Class 0
protostar driving an outflow \citep{Yun1994,Visser2002} and a neighboring
starless core \citep{Stutz2010}. \citet{Chen2007} observed a filamentary
distribution of N$_2$H$^+$ in the envelope surrounding the protostar; the
velocity structure of the line was suggestive of rapid
envelope rotation.

% Mention polarization

See Figure \ref{fig:CB244S} for maps.

\end{document}